\documentclass[twocolumn,aps,superscriptaddress,prL,floatfix,showpacs,longbibliography, nofootinbib]{revtex4-2}

\usepackage{graphicx}
\usepackage{dcolumn}
\usepackage{bm}

\usepackage{ifthen,amsmath,amssymb}
\usepackage{hyperref}
\usepackage{float}
\usepackage{aas_macros}
\usepackage[export]{adjustbox}
\usepackage{bookmark}
\usepackage{newtxtext,newtxmath}
\usepackage[dvipsnames,table]{xcolor}

\newcommand{\bx}{\boldsymbol x}
\newcommand{\bvel}{\boldsymbol v}
\newcommand{\bn}{\boldsymbol n}
\newcommand{\nv}{\hat{\bf n}}

\begin{document}

\preprint{APS/123-QED}

\title{Tracing cosmic gas in filaments and halos: Low-redshift insights from the kinematic Sunyaev-Zel'dovich effect}

\author{Boryana Hadzhiyska}
\email{boryanah@alumni.princeton.edu}
\affiliation{Miller Institute for Basic Research in Science, University of California, Berkeley, CA, 94720, USA}
\affiliation{Physics Division, Lawrence Berkeley National Laboratory, Berkeley, CA 94720, USA}
\affiliation{Berkeley Center for Cosmological Physics, Department of Physics, University of California, Berkeley, CA 94720, USA}

\author{Simone Ferraro}
\affiliation{Physics Division, Lawrence Berkeley National Laboratory, Berkeley, CA 94720, USA}
\affiliation{Berkeley Center for Cosmological Physics, Department of Physics, University of California, Berkeley, CA 94720, USA}

\author{Rongpu Zhou}
\affiliation{Physics Division, Lawrence Berkeley National Laboratory, Berkeley, CA 94720, USA}
\affiliation{Berkeley Center for Cosmological Physics, Department of Physics, University of California, Berkeley, CA 94720, USA}


\date{\today}

\begin{abstract}
In this work, we leverage CMB data from the Atacama Cosmology Telescope (ACT) and LSS data from the imaging survey conducted by the Dark Energy Spectroscopic Instrument (DESI) to study the distribution of gas around galaxy groups at low redshift, $z \approx 0.3$, via the kinematic Sunyaev-Zel'dovich (kSZ) effect. In particular, we perform velocity-weighted stacking on the photometric Bright Galaxy Sample (BGS) to isolate the monopole and quadrupole of the kSZ signal, orienting the stacked images along 2D filaments identified using the Hessian of the projected gravitational potential. We find a 7.2$\sigma$ detection in the monopole of the signal (i.e., the gas density profile) 
and a 4$\sigma$ detection in the quadrupole ($m = 2$), 
constituting the first measurement of the alignment between gas distribution and the cosmic web through the kSZ effect. As it is a linear probe of the local gas density, the kSZ has heightened sensitivity to the warm-hot intergalactic medium (WHIM), which is believed to house the majority of the ``missing baryons.'' Mapping out the gas density at low redshifts, as enabled by our measurements, is crucial for weak lensing surveys, for which the impact of baryons on small scales is a major impediment. We compare the anisotropic signal against two hydrodynamical simulations, TNG300-1 and Illustris, which have very different baryonic feedback prescriptions. We find that the anisotropic signal measured in the data is comparable but slightly larger and more extended compared with the simulations. This suggests that there is excess accretion and feedback taking place through the filaments, hinting at the possible presence of spin-filament alignment of the BGS objects.


\end{abstract}

\maketitle


\section{Introduction}
\label{sec:intro}

Mapping the distribution of baryons (gas) in the Universe has recently been brought to the forefront of a number of cosmological and astrophysical studies. The closer we get to completing this mapping, the closer we will be to unraveling galaxy formation and evolution and disentangling it from cosmological effects on small scales \citep{2005ApJ...634..964O,2007ApJ...668....1N,2007ARA&A..45..117M,2010ApJ...725...91B}. Garnering a full census of baryons via direct detection remains elusive, because a substantial fraction of the gas in the late Universe is in a diffuse and warm state that resides in the low-density regions of the Universe, known as the warm-hot intergalactic medium (WHIM). On the other hand, the fractional contribution of baryons to the total energy budget of the Universe can be inferred from early Universe probes such as the cosmic microwave background (CMB), which yields a larger baryon fraction and thus motivates the search for the `missing baryons' via direct detection \citep{1992MNRAS.258P..14P,1999ApJ...514....1C,2004IAUS..220..227F,2012ApJ...759...23S,2017ARA&A..55..343B,2017JCAP...11..040B}. Moreover, the baryon-dark matter connection poses a significant challenge for cosmological analyses such as weak lensing studies, for which the effect of baryons remains a major unknown \citep{2022PhRvD.105b3514A,2023A&A...678A.109A}. 

Constraining the state of the WHIM allows us to disentangle astrophysical effects from manifestations of exotic dark energy and dark matter models, since the gas is a tracer of low-redshift cosmic web structure. However, astrophysical models offer a large range of predictions for the WHIM evolution depending on the implemented feedback model \citep{2001ApJ...552..473D}.  Because the WHIM is sensitive to baryonic feedback, it can be used to constrain the physics of active galactic nuclei (AGN) and supernova explosions, both of which are active areas of research and key to unraveling black hole and stellar evolution \citep{2013ARA&A..51..511K,2014ARA&A..52..529Y,2015ARA&A..53...51S}. AGN feedback is also of huge importance to cosmology, as the redistribution of baryons due to astrophysical processes is considered one of the biggest impediments to fully exploiting the power of current and future weak gravitational lensing surveys \citep{2015MNRAS.450.1212H}.  One of the interesting open questions is the interplay between gas feedback and the cosmic web, which lets us understand whether the distribution of gas around galaxy groups and clusters is largely isotropic or instead strongly aligned with filamentary structures in the Universe. Answering this question would have profound consequences on our modeling of the formation and evolution of galaxies.

Fortunately, the WHIM induces a weak signal on the CMB map through the thermal and kinematic Sunyaev-Zel’dovich (SZ) effects (see \citep{1972CoASP...4..173S,1999PhR...310...97B,2019SSRv..215...17M} for a review on the SZ effect). The thermal SZ (tSZ) effect, parameterized by the Compton-y parameter, is sensitive to gas pressure (i.e., the product between free-electron gas density and temperature) along the line-of-sight. As such the signal is dominated by galaxy clusters and is orders-of-magnitude higher than the signal from intergalactic filaments. On the other hand, the kinematic SZ (kSZ) effect depends on the gas density and bulk velocity. While it is smaller in magnitude than the tSZ effect, the kSZ effect has emerged as a promising probe if the WHIM, as it allows us to trace the spatial distribution of baryons, even in the outskirts of galaxy groups and clusters due to its linear dependence on the gas density \citep{2016PhRvL.117e1301H,2016PhRvD..94l3526F,2021PhRvD.104d3518K,2021PhRvD.103f3513S,2024arXiv240707152H}. To map out the WHIM via the SZ effects, one thus needs to adopt signal-enhancing techniques such as stacking, which lead to a decrease in the noise by a factor of square-root of the number of stacked objects \citep{2010ApJ...716.1118P,2011ApJ...736...39H,2011ApJ...732...44S,2013ApJ...767...38S,2013A&A...550A.131P,2021PhRvD.103f3514A}.

The late Universe is characterized by an intricate `cosmic web' consisting of an interconnected network of voids, filaments and knots, which evolved from the near-uniform field of Gaussian random density fluctuations in the early Universe \citep{1996Natur.380..603B}. The large-scale pattern of the cosmic web can be predicted from features in the primordial density field, namely, the locations of the peaks and troughs as well as the moments of the density field and the tidal forces. Meanwhile, the small-scale details are the result of complex local gravitational interactions in the late-time Universe that led to the virialization of structure and the formation of galaxy groups and clusters. Mapping the connection between the gas and the cosmic web can be a highly potent tool in our hands for understanding structure evolution \citep{2022ApJ...933..134L,2023MNRAS.523.1346L,2022MNRAS.510.3335H,2024arXiv241014404I}. 

On small scales, the orientation of the gas and galaxies embedded in the cosmic web is determined by highly complex dynamical processes such as mergers, tidal disruption events, friction and splashback, which make it very challenging to probe the nature of dark energy and dark matter independently of the unknown astrophysics. Studying how gas in the galaxy group or cluster is oriented with respect to the filaments surrounding it could provide insight into these questions and give us a glimpse of the processes by which the cosmic web feeds cold gas into galaxies, the direction in which the gas is ejected from the galaxy core, and the distance which it reaches before `raining back' onto the galaxy. In the quest of constructing a complete cosmological paradigm, baryonic feedback processes play a huge role. The complex small-scale processes that move baryons out of dark matter halos and change the 3D gas density and temperature distribution are not well understood, and must be constrained before we can glean cosmological information from this observational regime. At the same time, understanding feedback processes is essential to assembling a full picture of how black holes and massive stars operate.

In this work, we provide a first look at the gas distribution along filaments in the low-redshift sample of bright galaxies (BGS) produced by the DESI imaging survey. While previous works, such as Ref.~\citep{2022ApJ...933..134L}, have focused on the gas pressure anisotropy, here we study the gas density through the kSZ effect. As argued above, such localized measurements of the gas density anisotropy are motivated by various science cases and thus of broad interest to both astrophysicists and cosmologists. While a significant fraction of the anisotropic kSZ signal stems from group pairs, i.e. we are detecting the gas density inside adjacent halos lined up along filaments, we demonstrate that compared with other probes, more of the kSZ signal is sourced by gas in the WHIM, since we are most sensitive to the halo outskirt regime (1 to 2 Mpc$/h$) and contributions from nearby groups and clusters on average cancels thanks to the velocity dependence of the kSZ. Importantly, we address the question of feedback modeling by comparing the observational signal with several simulations with very different feedback prescriptions and put constraints on plausible models. Echoing the results of previous works \citep{2016PhRvD..93h2002S,2021PhRvD.103f3513S,2024arXiv240707152H} that have studied the stacked kSZ effect, we find evidence for large baryonic feedback, which affects most significantly lower-mass halos ($10^{12.5} M_\odot$) and less so higher-mass halos ($10^{14} M_\odot$).

This paper is organized as follows. In Section~\ref{sec:data} data, we present the CMB and galaxy survey data used in this work as well as the two hydrodynamical simulations. We then describe the techniques used for performing an oriented stacking of the kSZ effect as well as for defining the filament orientation. In Section~\ref{sec:res}, we detail our main findings regarding the anisotropy and mass evolution of the kSZ signal. Finally, we summarize our results and discuss their implications in Section~\ref{sec:conc}.

\section{Data}
\label{sec:data}

In this section, we summarize the observational and simulation data sets used in this study.

\subsection{Dark Energy Spectroscopic Instrument}
\label{sec:desi}

The Dark Energy Spectroscopic Instrument is a robotic, fiber-fed, highly multiplexed spectroscopic telescope that operates on the Mayall 4-meter telescope at Kitt Peak National Observatory \citep{2022AJ....164..207D}. DESI can obtain simultaneous spectra of almost 5000 objects over a $\sim$$3^\circ$ field \citep{2016arXiv161100037D,2023AJ....165....9S,2023arXiv230606310M} and is currently conducting a five-year dark energy survey of about a third of the sky \citep{2013arXiv1308.0847L}. The end product will comprise spectra for approximately 40 million galaxies and quasars \citep{2016arXiv161100036D}.


Here, we use the publicly available photometric sample of the Bright Galaxy Sample (BGS) 
\cite{2023AJ....165..253H,2023AJ....165...58Z,Zhou:2023gji} from the DESI Legacy Imaging Survey, which was used to select DESI targets from three telescopes: Blanco for Dark Energy Camera Legacy Survey (DECaLS), Mayall for the Mayall $z$-band Legacy Survey (MzLS), and Bok for the Beijing–Arizona Sky Survey (BASS). We also employ the photometric redshifts presented in \citet{Zhou:2023gji} for Data Release 9 (DR9), which have been calibrated via DESI spectroscopic redshifts and thus have only a small fraction of contaminants ($\lesssim$0.1\%) and a mean error of $\sigma_z/(1+z) \lesssim 0.02$. Furthermore, we impose a cut on the stellar masses (estimated via the photometric bands \citep{2023AJ....165...58Z}) of $M_\ast > 10^{10.5} \ M_\odot$, so as to boost the signal-to-noise ratio (SNR) of the kSZ measurement, which is roughly proportional to the host halo mass.



\subsection{Atacama Cosmology Telescope}
\label{sec:act}
This paper uses the public, single-frequency coadded CMB maps \cite{2020JCAP...12..046N}
from the Data Release 5 (DR5) of the Atacama Cosmology Telescope (ACT), corresponding to data collected over a decade 2008-2018\footnote{The CMB maps used here are publicly available at \url{https://lambda.gsfc.nasa.gov/product/act/actpol_prod_table.html}}. ACT used a 6m diameter telescope, located in the Atacama Desert of Chile and measured the CMB from 2007 to 2022. We will primarily use the coadded f090 map, with a central frequency of 98 GHz and covering about 18,000 sq deg, and the f150 map, centered at 150 GHz. For this work, we use only the night-time portion of the data, and where point sources have been subtracted.  
The ACT maps are produced in the plate-carr\'{e}e (CAR) projection scheme, and are naturally convolved with the ACT beam\footnote{Approximately a Gaussian with Full Width at half maximum (FWHM) $\approx$ 2.1 arcmin for the f090 map and FWHM of 1.4 arcmin for the f150 map.}, so that any interpretation or comparison with simulations will need to take this into account. 

We apply a mask to the ACT maps that removes all galaxies within 10 arcmin of a point source or a cluster \citep{2024ApJ...962..112Q}. In addition, we remove galaxies around which the measured filtered temperature decrements are more than a $5\sigma$ outlier in any of the bins. That way, we minimize the impact of especially bright objects and avoid introducing bias to the stacked profiles.

\subsection{Oriented stacking} 
\label{sec:stack}

To obtain the oriented stacked profiles of BGS objects through the kSZ effect, we need two ingredients: an estimate of the line-of-sight velocity, and a measurement of the CMB temperature decrement at the location of each galaxy.

We create cutouts of the CMB temperature map, $\mathcal{T}_i(\hat n)$, around each galaxy $i$, and similarly to Ref.~\citep{2022ApJ...933..134L}, compute the cosine series coefficients as a function of radius, $C_m^i(r)$, from the center of the galaxy, as follows:
\begin{equation}
\label{eq:C_m}
C_m^i(r) = \frac{1}{p \pi} \int_0^{2\pi} d \theta {T}_i(\hat n) \cos(m \theta) ,
\end{equation}
where the normalization is $p = 1$ if $m = 1$ and $p = 2$, otherwise 
We compute the monopole ($m = 0$, the isotropic signal) and the quadrupole ($m = 2$). Since the mean redshift of the BGS is $\bar z \approx 0.3$, 1 arcmin corresponds to a comoving transverse scale of 0.25 Mpc$/h$. 
Similarly to previous works \citep{2016PhRvD..93h2002S,2021PhRvD.103f3513S,2024arXiv240707152H}, for the monopole, we adopt the compensated aperture photometry (CAP) filter to report the radial dependence of the signal for some radial bin, $r_j$:
\begin{equation}
    W_{j} =
\left\{
\begin{aligned}
1& &  &\text{for} \, r < r_j \,, \\
-1& &  &\text{for} \, r_j \leq r \leq \sqrt{2} r_j \,, \\
0& & &\text{otherwise}. \\
\end{aligned}
\right.
\label{eq:W_CAP}
\end{equation}
In other words, at $r_j$ we compute the integrated signal within a disk and subtract the integrated signal in an annulus of the same area around it.
As demonstrated previously \citep{2023MNRAS.526..369H}, the CAP filter features the favorable property of suppressing the spurious contributions along the line-of-sight from random uncorrelated structure as well as mitigating the primary CMB contamination. To make the anisotropic measurement, we use a simple mean filter in which instead of subtracting the `ring' from the `disk' signal for each aperture, we simply taking the average of the signal in the `ring.' The reason we do not need to adopt the CAP filter in this case is that the weighting by $\cos(2\theta)$ naturally filters out isotropic contributions such as the large-scale primary CMB modes and uncorrelated contaminants. In addition, the `mean ring' signal is easier to interpret, as it corresponds to simple azimuthal averages. For the CAP measurement, we use 9 radial bins, ranging between 1 and 8 arcmin, whereas for the $m = 2$ measurement, we use 7 radial bins, ranging between 2 and 14 arcmin, as there we are interested in the gas distribution surrounding galaxy groups in the DESI BGS sample.

We estimate the reconstructed velocity field by solving the linearized continuity equation in redshift space:
\begin{equation}
    \nabla \cdot \bvel + \frac{f}{b} \nabla \cdot [(\bvel \cdot \hat \bn) \hat \bn] = -a H f \frac{\delta_g}{b}
\end{equation}
where $\delta_g$ is the observed galaxy number overdensity, $H(z)$ is the redshift-dependent Hubble parameter, $f$ is the logarithmic growth rate, defined as $f \equiv d \ln(D)/d \ln(a)$ with $D(a)$ the growth factor and $a$ the scale factor. Here, we assume that the galaxy overdensity $\delta_g$ is related to the matter overdensity in redshift space, $\delta$, by a linear bias factor, $b$, such that $\delta_g = b \delta$.

To estimate the individual galaxy velocities, we adopt the standard reconstruction method used in many BAO analyses, which yields the first-order galaxy displacement field, $\psi(\bx)$. In particular, we apply the \texttt{MultiGrid} implementation to solve the linearized system of partial differential equations via the package `pyrecon'\footnote{\url{https://github.com/cosmodesi/pyrecon}} \cite{2015MNRAS.450.3822W}. We then calculate the velocity via:
\begin{equation}
  \bvel^{\rm rec}(\bx) = f(z) a(z) H(z) \psi(\bx) .
\end{equation}
In principle, this gives us access to the 3D reconstructed velocity field, but the component that matters for kSZ analysis is only the line-of-sight one, which we denote by $v^{\rm rec} \equiv v_\parallel^{\rm rec}$.  

Once we have attained the reconstructed line-of-sight velocity and the cosine series coefficients for each galaxy, we apply the velocity-weighted, uniform-mean estimator from \cite{2021PhRvD.103f3513S}:
\begin{equation}
    \hat{\tau}_{\rm kSZ}(\theta_d) = -
    \frac{1}{r}
    \frac{v_{\rm rms}^{\rm rec}}{c}
    \frac{\sum_i \mathcal{T}_i(\theta_d) (v_{{\rm rec}, i}/c)}{\sum_i (v_{{\rm rec}, i}/c)^2} T_{\rm CMB},
    \label{eq:kSZ_est}
\end{equation}
where the sum is over all galaxies, $T_{\rm CMB} = 2.7255 \ {\rm K}$ is the mean CMB temperature, $v_{\rm rms}^{\rm rec}$ is the rms of the radial component of the reconstructed velocities, $v_{{\rm rec}, i}$, $c$ is the speed of light, and $r \equiv \langle v^{\rm rec} v^{\rm true} \rangle/ (v_{\rm rms}^{\rm rec} v_{\rm rms}^{\rm true})$ is the cross-correlation coefficient between the reconstructed and true velocity, with the averaging being done over all galaxies. When evaluating the CAP signal, $\mathcal{T}_i$ is the CAP-filtered (see Eq.~\ref{eq:W_CAP}) CMB temperature cutout at the location of the $i^{\rm th}$ galaxy, and in the case of the anisotropic measurement, it is the product between the CMB temperature map and the respective cosine phase (see Eq.~\ref{eq:C_m}; $m = 2$ for the default case). Typically \citep{2021PhRvD.103f3513S,2024arXiv240707152H}, this coefficient is evaluated from mock simulations, and the value Ref.~\citep{2024PhRvD.109j3534H} found for a photometric survey with DESI-like specifications is $r = 0.3 \pm 0.03$. If this normalization is not applied, the measured profiles and their covariance will be underestimated by $1/r$ and $1/r^2$, respectively, relative to the true optical depth. 

Similarly to Ref.~\citep{2024arXiv240707152H}, we 
ensure that the number of galaxies in each velocity bin is symmetric around the mean by randomly downsampling, which avoids unwanted biases from massive clusters and guarantees that in the absence of a kSZ signal, our estimator yields zero mean signal. Since we sum over an equal number of galaxies with a positive and a negative sign per absolute value bin in the velocity, additive contaminants such as the cosmic infrared background (CIB) and the tSZ effect cancel, as has been demonstrated in Refs.~\citep{2016PhRvD..93h2002S,2021PhRvD.103f3513S,2024arXiv240707152H}. We also demonstrate this through the null test shown in the bottom of Fig.~\ref{fig:m2_null}.


We estimate the covariance matrix for the stacked profiles by calculating the covariance between the radial bins of each of the cosine coefficients and then dividing by $\sqrt{N_g}$ to get the average covariance for the full sample. We note that this calculation underestimates the true covariance by about 5-10\% because it assumes that the CMB temperature cutouts are all independent of each other (when in reality we know that some are overlapping) \citep{2021PhRvD.103f3513S,2024arXiv240113033C}. 

\subsection{Simulations}
\label{sec:sims}

To put our findings into context, we compare the gas density anisotropy we observe in the data against two hydrodynamical simulations: IllustrisTNG \citep{2019ComAC...6....2N} and Illustris \citep{2015A&C....13...12N}, which have very different AGN and supernova feedback prescriptions. To make the measurements in the simulations, we construct 2D maps of kSZ, then convolve them with a 1.4 arcmin Gaussian (corresponding to the f150 map beam) to roughly match the ACT beam effects, 
and finally perform the stacking by reorienting each image along the cosmic web filament as described in Section~\ref{sec:stack}.



Each simulation outputs a number of useful quantities: gas mass ($m$), electron abundance ($x$), and internal energy ($\epsilon$) for each gas particle $i$. Assuming a primordial hydrogen mass fraction of $X_H=0.76$, we compute the volume-weighted electron number density, $n_{\rm e}$ as
\begin{equation}
V_i n_{{\rm e},i} = x_i m_i \frac{X_H}{m_p}
\end{equation}
where $m_p$ the proton mass. We then compute the 2D maps of the kSZ and the optical depth by binning the gas particles into a (10000, 10000) grid, so that the optical depth in cell $j$ is given by:
\begin{equation}
\tau_j = \sigma_T A_j^{-1} \sum_{i \in A_j} V_i n_{{\rm e},i} ,
\end{equation}
and the momentum of the electron density is:
\begin{equation}
b_j = \sigma_T A_j^{-1} \sum_{i \in A_j} V_i n_{{\rm e},i} v_i/c,
\end{equation}
where $A_j$ is the area of each grid cell (of size $\sim$0.01 Mpc), $\sigma_T$ is the Thomson cross section, and $c$ is the speed of light. Analogously, we can compute the Y-Compton map relevant to the tSZ effect and the Thomson optical depth $\tau$ map relevant to e.g., patchy screening.

\subsection{Filament classification}
\label{sec:filament}

To perform an oriented stacking of the signal and reveal the alignment between gas density and the cosmic web, crucially we need to define and identify filaments in the cosmic web. To do so, we adopt the formalism of Ref.~\citep{2016MNRAS.460..256A}. Specifically, we will solve the Poisson equation, $\nabla^2\phi=\delta_g$, on the sphere to obtain the rescaled Newtonian potential, $\phi\equiv\bar{\phi}/(4\pi G\bar{\rho})$, where $\delta_g$ is the galaxy overdensity field and $\bar{\rho}$ the mean density in the Universe. Typically, the galaxy density field is smoothed with some characteristic smoothing scale to mitigate shot noise, filter out non-linear effects, or focus on a specific scale of interest. We will then define the tidal tensor field as $t_{ij}=\partial_i\partial_j\phi$, so that $\delta_g={\rm Tr}(\hat{t})$, and diagonalize the tidal tensor in order to get a sense of the directions of expansion and contraction, which will be used when performing the oriented stacking of the kSZ effect. 

We provide a step-by-step summary of the procedure below.
\begin{itemize}
\item The observable in a photometric survey is the projected galaxy overdensity field $\delta_g(\nv)$, which encodes the fluctuations in the angular number density of galaxies with respect to the mean and in practice can be computed by counting the number of galaxies in each pixel $N_p$ and dividing by the normalized average number of randoms per pixel $\bar{N}$:
\begin{equation}
  \delta_{g,p} = \frac{N_p}{\bar{N}} - 1.
\end{equation}
To mitigate shot-noise effects and non-linearities, we apply a Gaussian smoothing kernel with standard deviation of $\theta_G = 0.03 \ {\rm deg}$ (1.8 arcmin). Here, we adopt the HEALPix pixelization scheme \cite{2005ApJ...622..759G} with a resolution parameter $N_{\rm side}=2048$, corresponding to an approximate pixel size of $1.7 \ {\rm arcmin}$. At the median redshift of our data ($\bar{z}\sim0.3$), this corresponds to a comoving scale of $\sim$0.42 ${\rm Mpc}/h$.

\begin{figure}
    \centering
    \includegraphics[width=0.48\textwidth]{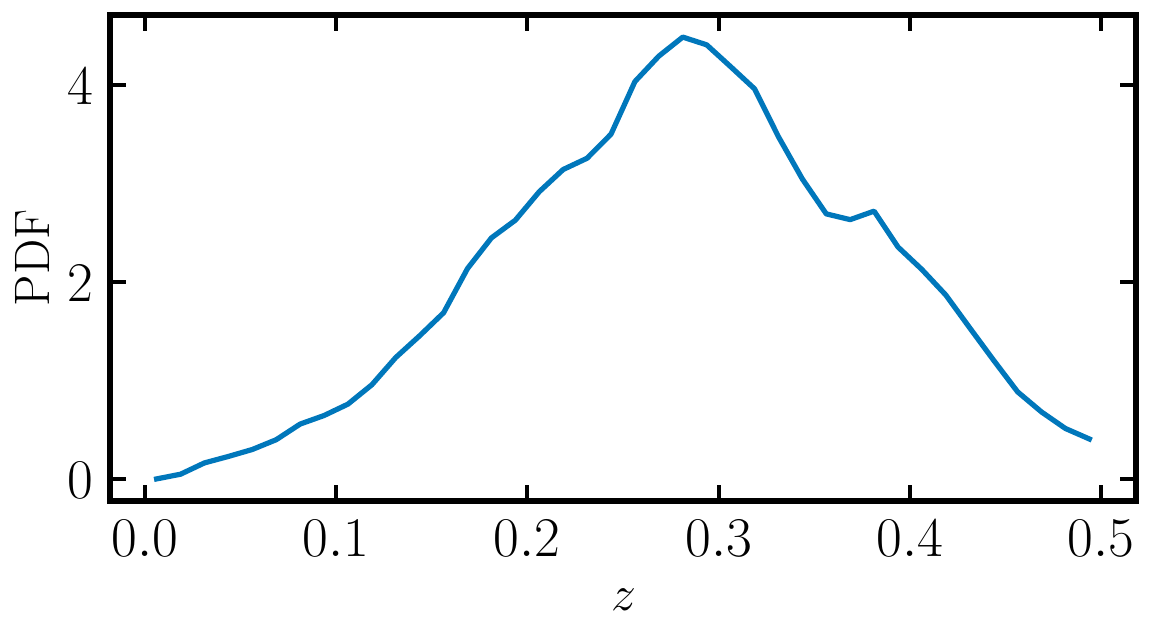}
    \includegraphics[width=0.48\textwidth]{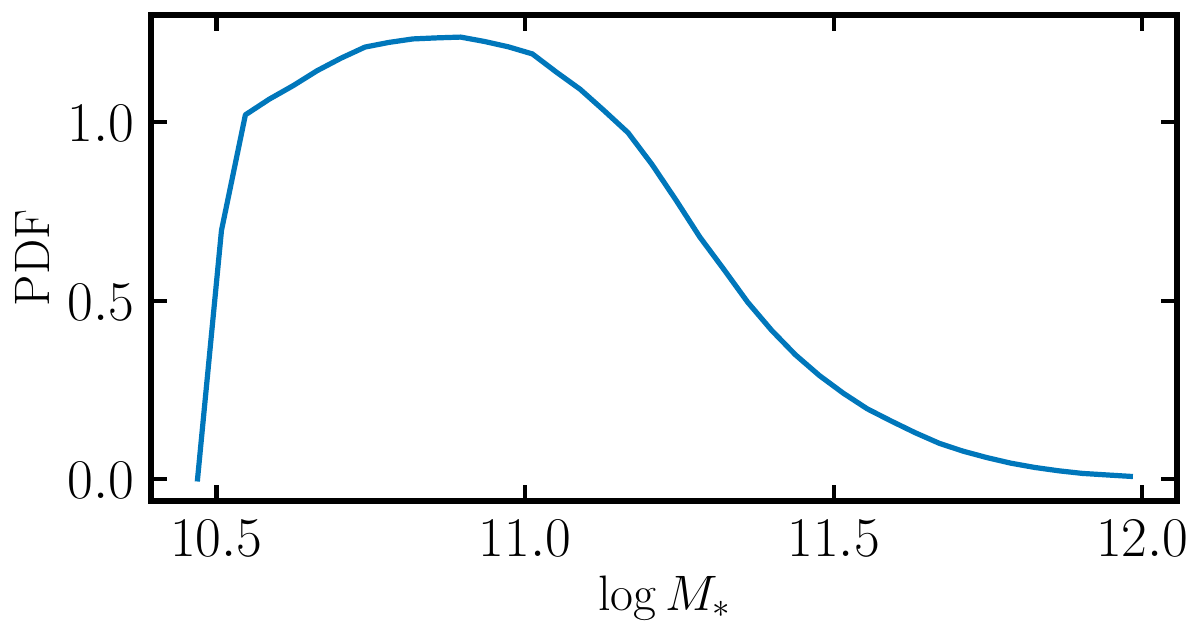}
    \caption{{\it Top:} Photometric redshift distribution of the BGS objects used in this study, with mean and median redshift of 0.29. 
    In total, there are 7.3 million BGS objects in the overlap between the DESI footprint and ACT, which survive the various cuts imposed to make the kSZ measurement (see Section~\ref{sec:stack}). 
{\it Bottom:} Stellar mass distribution of the BGS objects used in this study. Masses are inferred from the photometric band measurements via a machine learning algorithm trained on the CMASS galaxy sample \citep{2023AJ....165...58Z}.
    We impose a hard threshold of $\log M_\ast = 10.5$ in order to isolate galaxies residing in massive halos for which the SNR of the kSZ signal is higher. The mass is expressed in units of $M_\odot$.} 
    \label{fig:z_logm_bgs}
\end{figure}

\begin{figure*}
    \centering
    \includegraphics[width=0.59\textwidth]{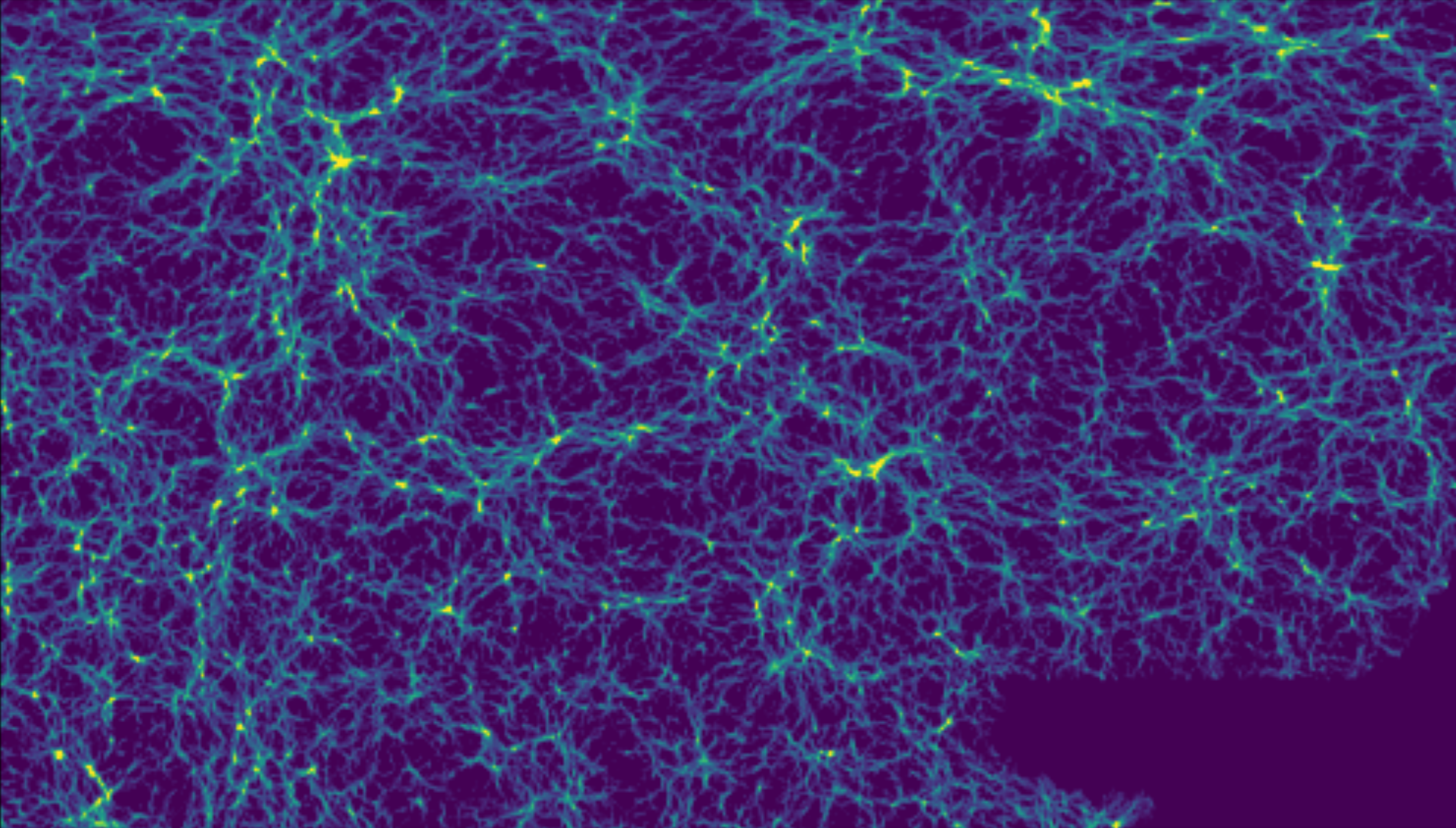}
\ \ \ \
    \includegraphics[width=0.342\textwidth]{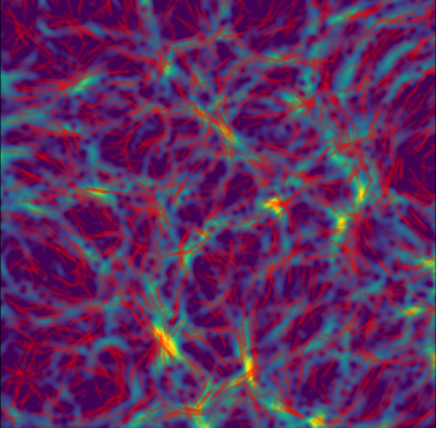}
    \caption{{\it Left:} Map displaying the largest eigenvalue in each pixel of the BGS tidal tensor field. The galaxy density field is smoothed with a Gaussian kernel with 0.03 deg standard deviation, and the size of the cutout region is 20 deg by 40 deg.
{\it Right:} Zoom-in of the largest eigenvalue map of size 10 deg by 10 deg
with the smallest eigenvector superimposed on the map at the location of each pixel. As expected, the smallest eigenvector, which physically corresponds to the direction of maximal expansion, is well-aligned with the filaments visible on the eigenvalue map.} 
    \label{fig:cw_bgs}
\end{figure*}

\item We can then solve for the 2D potential $\phi$ via the Poisson equation. Namely, we invert the covariant Laplacian on the sphere in harmonic space, since the harmonic coefficients of the two quantities are proportional to each other:
\begin{equation}
  \phi_{\ell m} = -\frac{\delta_{\ell m}}{\ell(\ell+1)}.
\end{equation}

\item To obtain the 2D tidal tensor $t_{ab}$, we need to apply the covariant Hessian to the 2D potential, i.e., $t_{ab}\equiv H_{ab}\,\phi$, where the covariant Hessian operator is given by
\begin{equation}
\label{eq:cov_hess}
\hat{H}\equiv\left(
\begin{array}{ccc}
\partial_\theta^2 & \partial_\theta(\partial_\varphi/\sin\theta) \\
\partial_\theta(\partial_\varphi/\sin\theta) &
\partial_\varphi^2/\sin^2\theta+\cot\theta\partial_\theta
\end{array}
\right).
\end{equation}

\item We then diagonalize the estimated tidal tensor in each pixel. The eigenvectors hold information about the direction of maximal stretching and maximal contraction, and the eigenvalues hold information about the amount of stretching/contraction. We use the smallest eigenvector to identify the orientation of the filaments, $e_0(\nv)$ (see Fig.~\ref{fig:cw_bgs}). When performing the kSZ stacking, we reorient the CMB temperature cutout so that it is aligned with $e_0(\nv)$ in the pixel corresponding to the galaxy center.
\end{itemize}

\subsection{Signal and noise}
Throughout this paper, we quote two estimates of detection significance for the isotropic and anisotropic signals, $\chi^2_{\rm null}$ and signal-to-noise ratio (SNR). Below we detail our procedure for combining the two frequency channels of the ACT temperature measurements as well as our definition of the SNR. 

We obtain the joint covariance and joint signal as follows. For a given aperture $r$ and a given galaxy, $i$, we compute the filtered profile, 
\begin{equation}
\mathbf{\tau}_{\rm joint}^i(r) = \frac{\mathbf{1}^T C^{-1} \mathbf{\tau}_{\nu}^i(r)}{\mathbf{1}^T C^{-1} \mathbf{1}}
\end{equation}
where $\nu$ corresponds to either of the two frequency channels (f090 or f150) and $C$ is the combined covariance of measured signals with the two frequency channels. We note that to make this evaluation more stable numerically, we ignore the off diagonal terms of $C$. The joint signal and covariance are finally obtained by finding the mean and covariance of the quantity, $\mathbf{\tau}_{\rm joint}^i(r)$, over the galaxy sample.

We obtain the signal-to-noise ratio as follows:
\begin{equation}
    {\rm SNR} = \sqrt{\chi^2_{\rm null} - \chi^2_{\rm best-fit}},
    \label{eq:SNR_null}
\end{equation}
where the $\chi^2$ statistic is defined as
\begin{equation}
    \chi^2_{\rm model} = (d - m)^\top {\rm C}^{-1} \hspace{0.1 cm} (d - m).
    \label{eq:chi_squared}
\end{equation}
In the case of $\chi^2_{\rm null}$, $m = 0$, whereas for the best-fit model, we adopt a rescaling of the simulation curves extracted from the original Illustris simulation, which have previously been shown to provide a good match to the gas density profiles of red galaxies on large scales \citep{2024arXiv240707152H}.

\section{Results}
\label{sec:res}

In this section, we describe the BGS sample and our oriented stacked measurement of the kSZ. Finally, we compare our findings with the output from hydrodynamical simulations.

\subsection{The Bright Galaxy Sample}
\label{sec:bgs}

In this section, we present several relevant properties of the BGS sample, including their mass and redshift distributions, as well as the filamentary structures they form.

In Fig.~\ref{fig:z_logm_bgs}, we show the distribution of photometric redshifts of the DESI BGS. The total number of galaxies in the overlap between the DESI footprint and ACT, which are left after the various cuts imposed to make the kSZ measurement (see Section~\ref{sec:stack}), is 7.3 million. The mean and median of this distribution is 0.29. 
The photometric redshift error has been estimated to be about $\sigma_z/(1+z) \approx 0.02$ \citep{Zhou:2023gji}. Relative to a spectroscopic sample, this uncertainty leads to a lower SNR of about a factor of 2 \citep{2024PhRvD.109j3534H,2024PhRvD.109j3533R}. We note that the cut in stellar mass that we impose in this study, $M_\ast > 10^{10.5} \ M_\odot$, selects bright objects with slightly higher redshifts than the complete BGS. We see that at redshifts higher than $z \gtrsim 0.4$, the redshift distribution sharply drops off. This is also the lower redshift threshold of the sample of Luminous Red Galaxies (LRGs). As such, our measurements of the gas density around lower-redshift galaxies are highly complementary to previous measurements of LRGs such as Refs.~\citep{2021PhRvD.103f3513S,2024arXiv240707152H} and give us an important leverage arm for understanding the redshift evolution of AGN feedback and the effect of baryons on weak lensing analysis. 

Fig.~\ref{fig:z_logm_bgs} displays the distribution of estimated galaxy stellar masses of the DESI BGS. Masses are inferred from the photometric band measurements via a machine learning algorithm based on CMASS galaxies \citep{2023AJ....165...58Z}. 
Since the optical depth is roughly proportional to the halo mass, $\tau \propto M_{\rm halo}$, we can maximize the SNR by selecting galaxies residing in more massive halos. Although we cannot estimate the halo mass directly, we know from hydrodynamical simulations roughly what the relationship between stellar mass and halo mass is, also known as the stellar-to-halo-mass ratio \citep{2018ARA&A..56..435W}. Thus, by selecting massive galaxies, we are also predominantly selecting massive halos, thus increasing the SNR of the measurement. We impose a hard threshold of $M_\ast > 10^{10.5} \ M_\odot$ on the stellar mass, which results in selecting halos with masses roughly above $\gtrsim$$10^{12} \ M_\odot$ and a mean mass of several times $10^{13} \ M_\odot$ \citep{2024MNRAS.530..947Y}.

Fig.~\ref{fig:cw_bgs} shows the filaments of the cosmic web formed by the DESI BGS. In particular, it displays the largest eigenvalue in each pixel of the tidal tensor field computed from the smoothed BGS density field with a Gaussian kernel of width 0.03 deg. Physically, the largest eigenvalue tells us about the amount of stretching and contraction that this pixel is undergoing. The larger its value, the more the contraction. We also show a zoom-in region of this map, superimposing the value of the smallest eigenvector in each pixel. As expected, the smallest eigenvector, which physically corresponds to the direction of maximal expansion, is well-aligned with the filaments visible on the eigenvalue map.


\subsection{Isotropic gas density profile}
\label{sec:iso}

\begin{figure}
    \centering
    \includegraphics[width=0.5\textwidth]{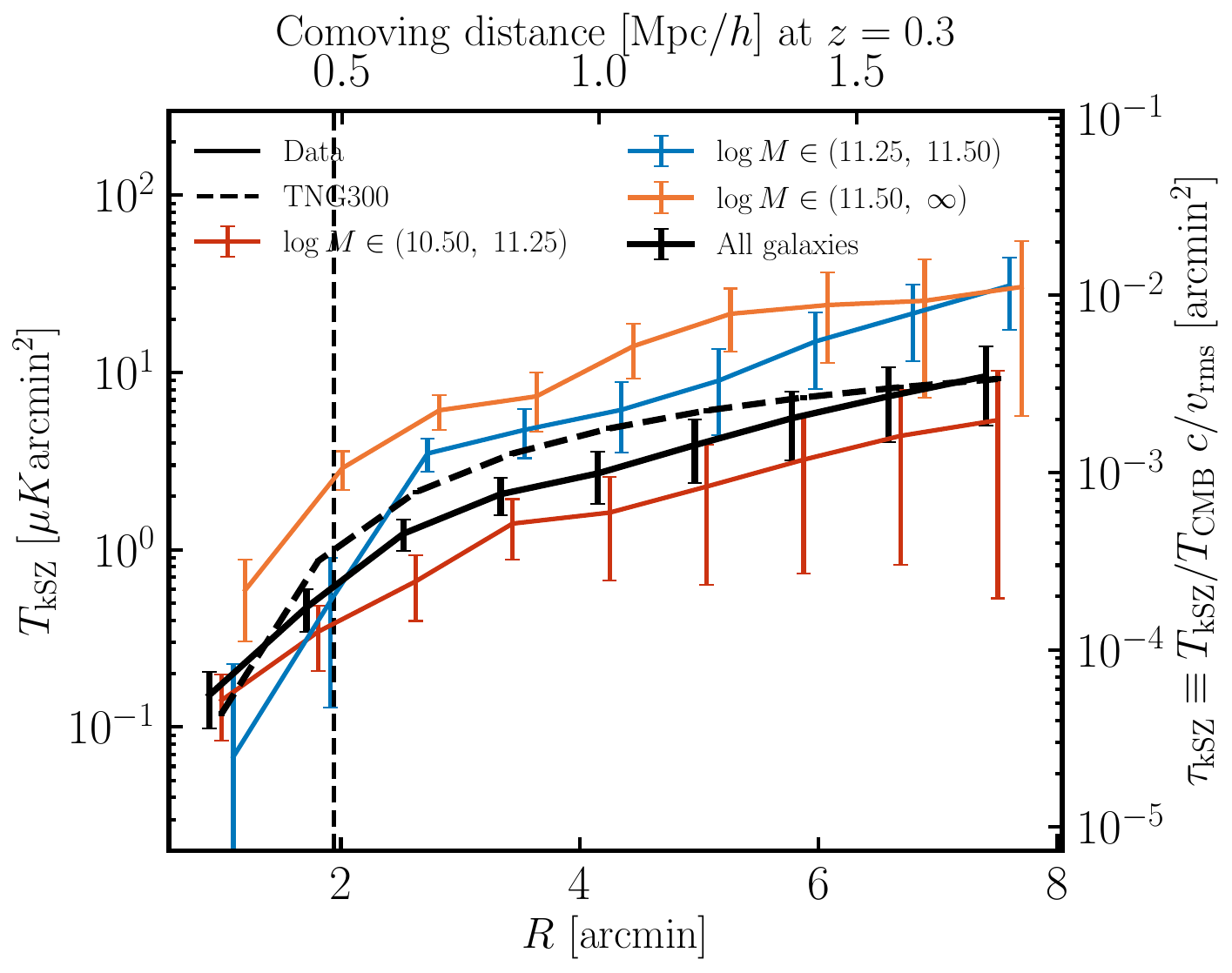} \\
    \includegraphics[width=0.3\textwidth]{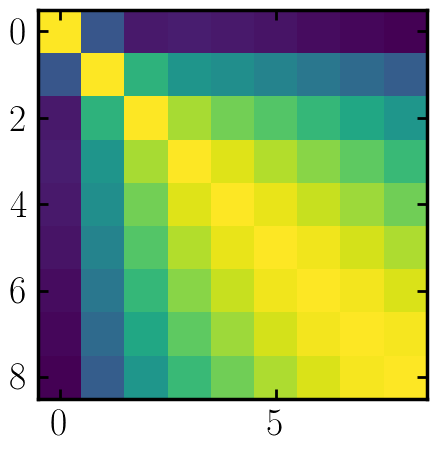}
    \caption{{\it Top:} Stacked kSZ signal of the photometric BGS as a function of the radius of the CAP filter,  which approximately corresponds to the cumulative gas density profile as a function of the distance from the galaxy group center associated with the BGS objects. The kSZ signal is obtained by stacking the 
    ACT DR5 maps 
    and is detected at 7.2$\sigma$ relative to the null, corresponding to a $\chi^2_{\rm null} \approx 65.0$ with 9 dof. 
 The vertical dashed line indicates the mean virial radius of the host halos. 
 Consistently with previous works \citep{2021PhRvD.103f3513S,2021PhRvD.103f3514A,2024arXiv240707152H}, we see that the gas profile is well extended beyond the virial radius,
suggesting that the feedback activity in BGS-hosting halos is strong enough to push much of the gas out of the galaxy group center. In our comparison with the TNG300-1 simulation (dashed line), we find an overall good match between the measured shapes, though we see a discrepancy at intermediate apertures. Qualitatively, the samples split into stellar mass bins behave as expected: halos hosting massive galaxies have larger optical depths, on average. We leave a more detailed analysis and comparison with simulations for the future. The top $x$ axis and the right $y$ axis show the comoving distance and the optical depth, respectively. {\it Bottom:} Full covariance of the data shown as the correlation matrix. Noticeably, despite the CAP filter suppressing the primary CMB contributions, the points at large aperture, where the kSZ signal and the primary CMB are hard to disentangle from each other, are significantly correlated.
}
    \label{fig:ksz_cap}
\end{figure}

\begin{figure}
    \centering
    \includegraphics[width=0.5\textwidth]{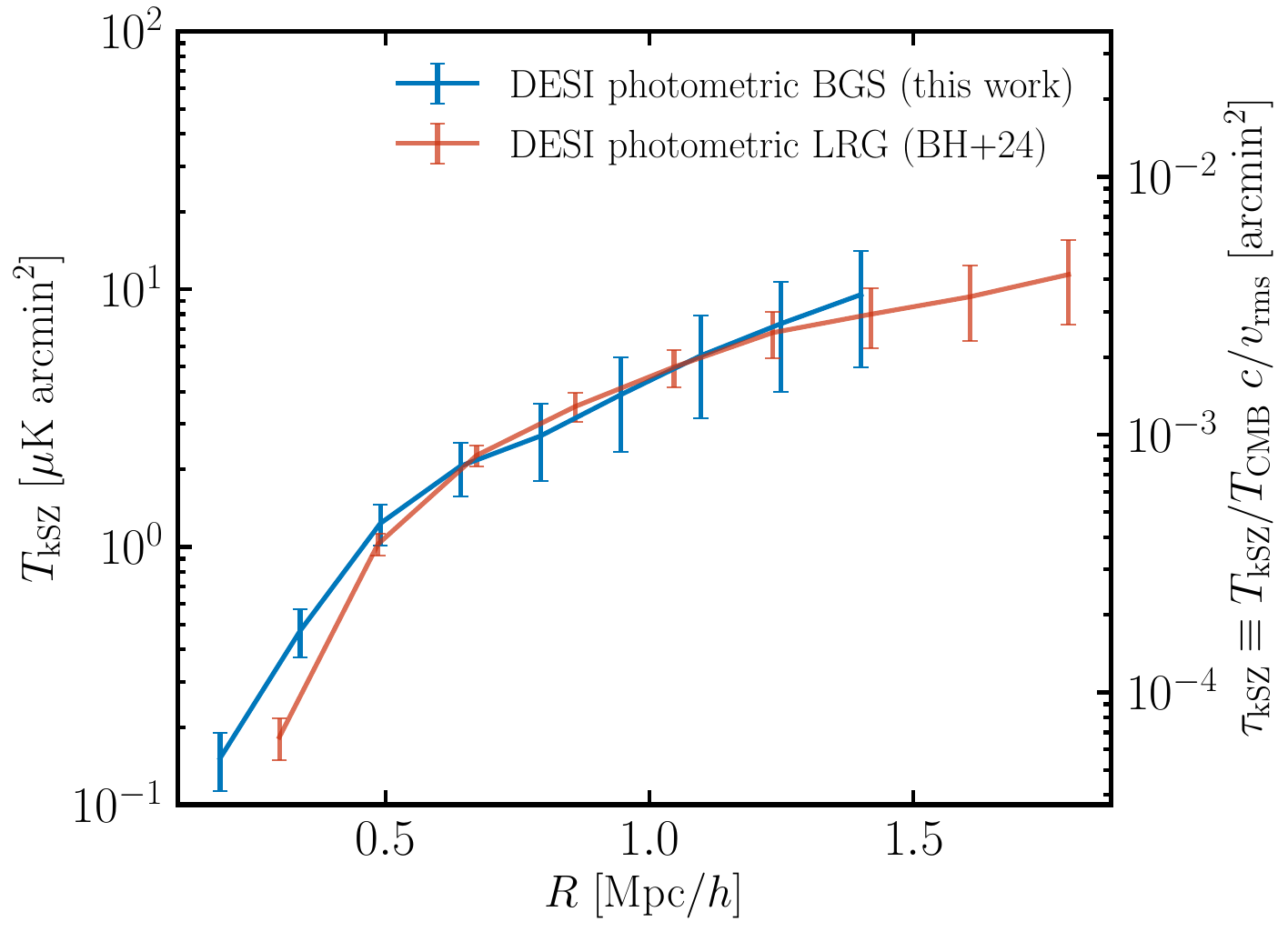}
    \caption{Comparison between the CAP measured around photometric DESI LRGs with ACT DR6 and photometric DESI BGS with ACT DR5 (joint `f090' and `f150' channels). We show the kSZ signal as a function of proper distance from the center of the galaxy groups. While the SNR of the LRGs is higher (due to the larger number of objects and the lower primary CMB noise at a smaller aperture, as well as the deeper and more uniform ACT maps), we see that the two curves are quite comparable in their shape and amplitude. The reason is that the host halo masses are roughly the same in both cases (a few times $10^{13} \ M_\odot$). The noise properties, however, are different because the two samples are at different redshifts and therefore subtend a different angle on the sky.}
    \label{fig:ksz_cap_lrg}
\end{figure}

In this section, we show our measurements of the isotropic kSZ signal of the photometric BGS. 

We start by displaying the usual quantity relevant to stacked kSZ analysis in Fig.~\ref{fig:ksz_cap}, $T_{\rm kSZ}$, which is the CAP-filtered gas density profile inferred from the kSZ signal, as described in Section~\ref{sec:stack} and Eq.~\ref{eq:kSZ_est}, i.e. the CAP-filtered monopole ($m = 0$) from Eq.~\ref{eq:C_m}. It can be thought of as the cumulative gas density profile, and on the $x$ axis we show the distance to the galaxy group center. To guide the eye, we indicate roughly the virial radius typical of the BGS halos ($\sim$$0.5 \ {\rm Mpc}/h$), and the corresponding optical depth. We see that the profiles are extended well beyond the virial radius, indicating that baryonic feedback pushes a large fraction of the gas outside the virial radius. In order to recover the full cosmic gas fraction, one needs to encompass the gas in a sphere several times the virial radius. The detection significance of this measurement is 7.2$\sigma$, with a $\chi^2_{\rm null} \approx 65.0$ and 9 degrees-of-freedom (dof). 

We also show a comparison between the density curve of the total sample and the TNG300 simulation (normalized to match the data at the largest aperture). The slight gap at intermediate apertures suggests the possibility of stronger feedback in the BGS sample compared with the simulated sample, though we defer further analysis for the future, where we would be able to make a higher-significance measurement of these profiles with improved CMB maps and spectroscopic BGS objects. In addition, we break the sample into 3 stellar mass bins: [10.5, 11.25), [11.25, 11.5), and [11.5, $\infty$) and while the detection significance is modest in each of the mass bins, we find qualitatively reassuring results: namely, galaxy groups housing more massive galaxies tend to also have higher halo masses and are thus more optically thick. 

We also compare the profile with the measurement of the kSZ signal around photometric DESI LRGs from Ref.~\citep{2024arXiv240707152H} in Fig.~\ref{fig:ksz_cap_lrg}. We note that the LRG measurement uses the ACT DR6 maps, which have better sensitivity and uniformity compared with the ACT DR5 maps used to make the BGS measurement here. We show that as a function of proper distance from the center of the galaxy groups. We opt for this choice, as the LRG and BGS objects have different angular sizes due to the large difference in their mean redshifts ($z = 0.3$ for BGS and $z = 0.7$ for LRGs). While the SNR of the LRGs is higher (due to the larger number of objects and the lower primary CMB noise at a smaller aperture), we see that the two curves are quite comparable in their shape and amplitude. The similarity in the shapes suggests that the feedback processes expelling gas from the BGS objects are similar to those of the LRGs. The reason is that the host halo masses are roughly the same in both cases due to two effects: 1) the number density of BGS is higher, implying that at fixed redshift, we are selecting lower mass halo hosts; 2) halos at $z = 0.3$ (BGS) are on average more massive than the halos at $z = 0.7$ (LRGs). More detailed analysis involving lensing is needed to more accurately infer the mean halo mass. 


Besides the averaged profile, we also show the covariance matrix at the bottom of Fig.~\ref{fig:ksz_cap}. As expected, we see that at large apertures the noise is highly correlated, as in this regime the measurement is dominated by the primary CMB. In the smallest radial bin, the noise is coming predominantly from the ACT beam. The least amount of correlation between the radial bins is found in the intermediate regime, roughly between 2 and 4 arcmin, where the SNR is also highest (see top panel).

\subsection{Anisotropic gas density profile}
\label{sec:aniso}

\begin{figure}
    \centering
    \includegraphics[width=0.4\textwidth]{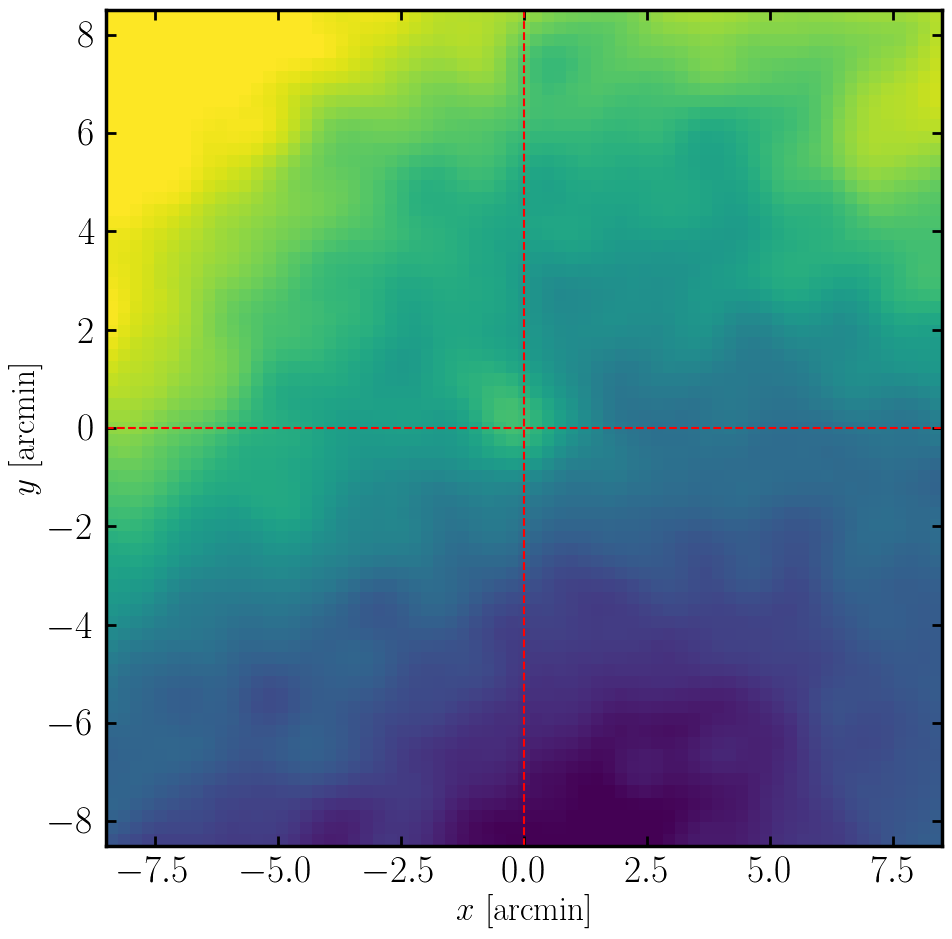}
    \includegraphics[width=0.4\textwidth]{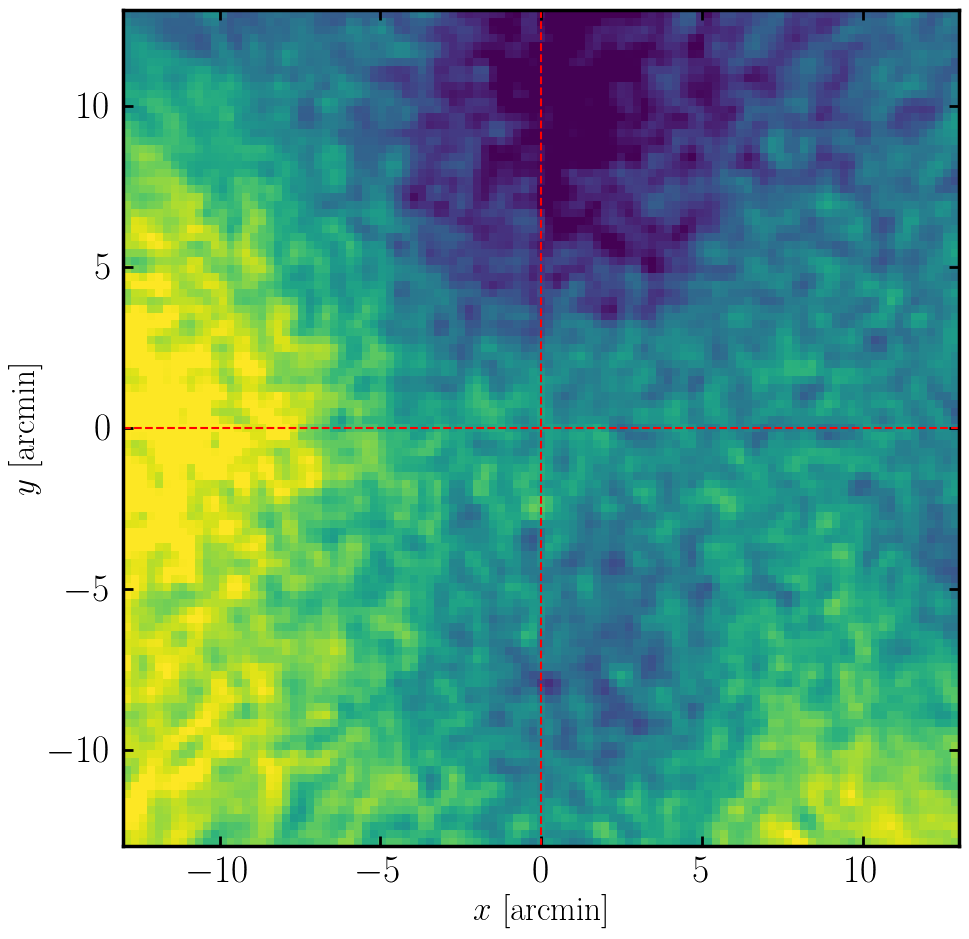}
    \caption{Stacked images of the photometric DESI BGS with random orientations ({\it top}), and oriented along the filaments ({\it bottom:}). {\it Top:} We see an excess in the middle of the image due to the average optical depth of the stacked galaxy groups. The large-scale noise is predominantly sourced by the primary CMB fluctuations. {\it Bottom:} 
We see an excess signal along the horizontal, corresponding to the direction of the filament, and a deficiency in the perpendicular direction. The stacked group in the center of the image is difficult to see by eye at this scale. We thus expect to detect a positive $m = 2$ signal, which is indeed the case, as seen in Fig.~\ref{fig:m2}. The granularity of the image is due to the fact that no smoothing has been performed apart from a bilinear interpolation to a map of twice the resolution.}
    \label{fig:ksz_ani_stack}
\end{figure}

\begin{figure}
    \centering
    \includegraphics[width=0.5\textwidth]{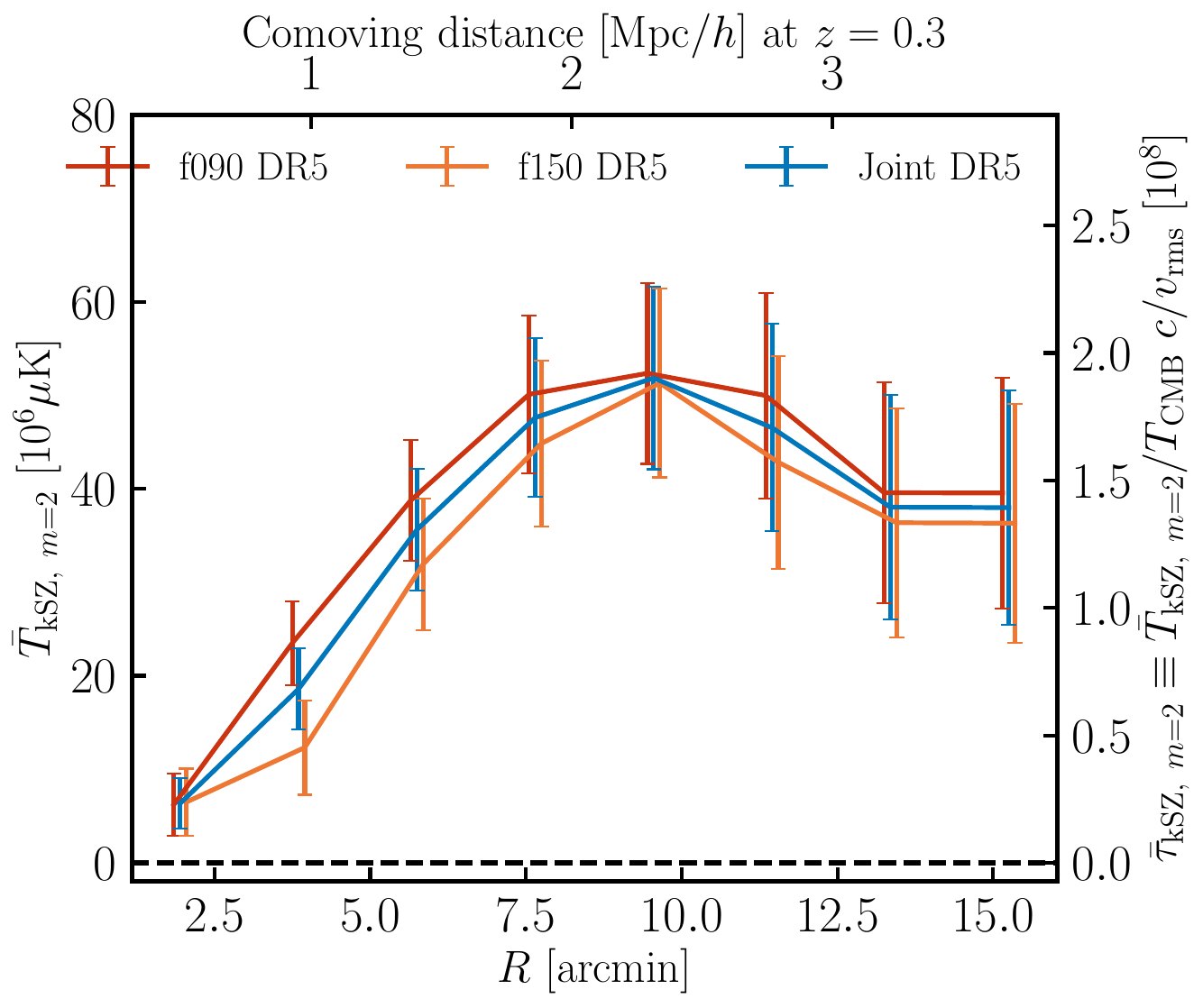} \\
    \includegraphics[width=0.3\textwidth]{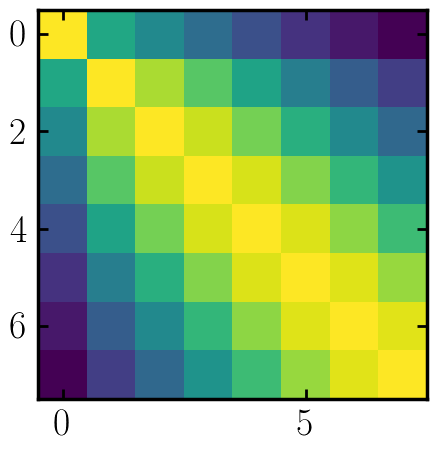}
    \caption{{\it Top:} Anisotropic kSZ signal of the photometric BGS as a function of radius from the center of the galaxy group. To make this measurement, we perform an oriented stacking by rotating each galaxy image such that it is aligned with the filament direction at the central pixel (i.e., the RA and DEC of the object). We detect the signal at 4.0$\sigma$  ($\chi^2_{\rm null} = 37.3$, 7 dof). 
 {\it Bottom:}  Covariance of the measurement shown as a correlation matrix. At large aperture, the points are highly correlated. Similarly to the CAP filter, the long-wavelength contribution gets canceled out thanks to the cosine-weighted averaging in Eq.~\ref{eq:C_m}.}
    \label{fig:m2}
\end{figure}

\begin{figure}
    \centering
    \includegraphics[width=0.5\textwidth]{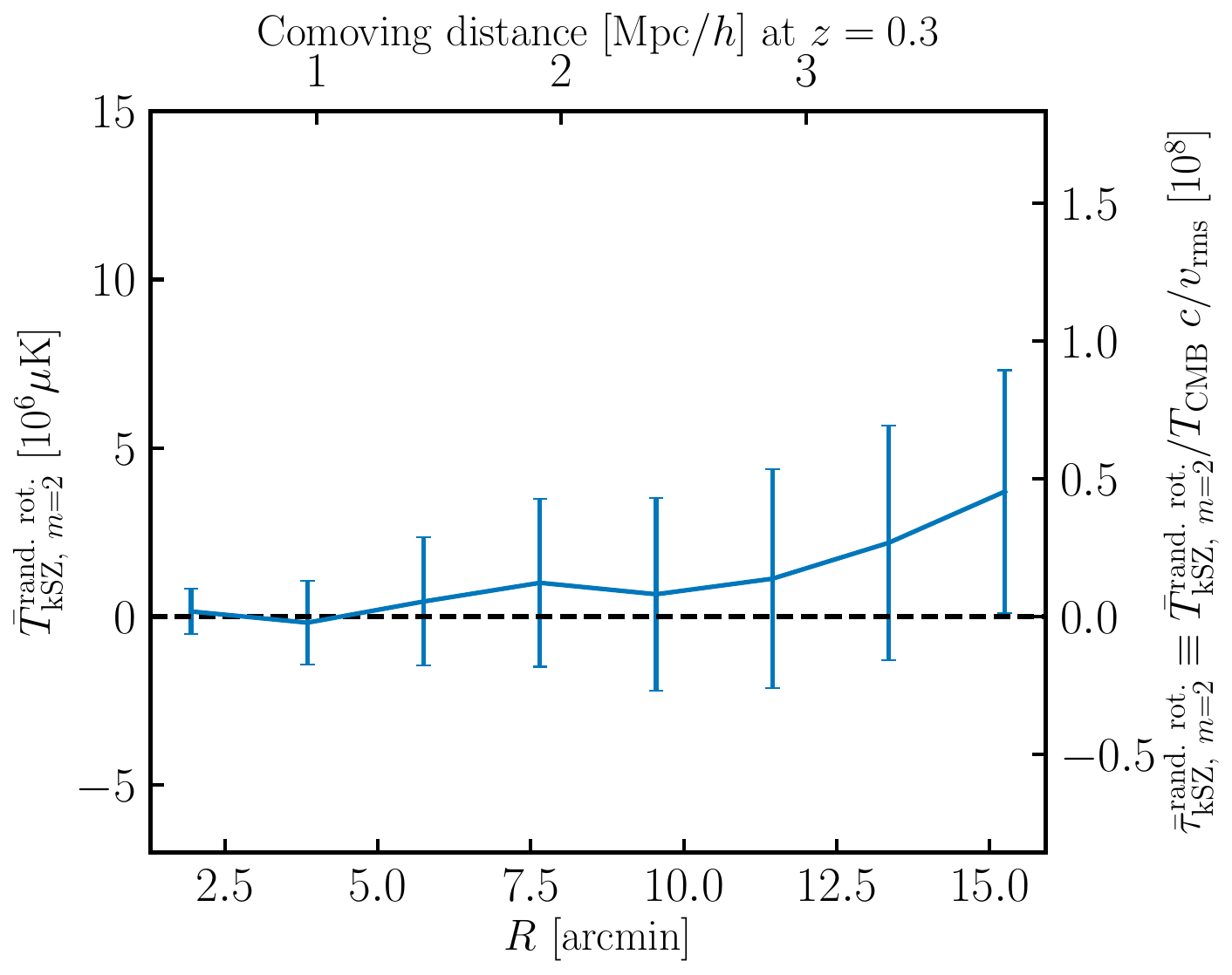}\\
    \includegraphics[width=0.5\textwidth]{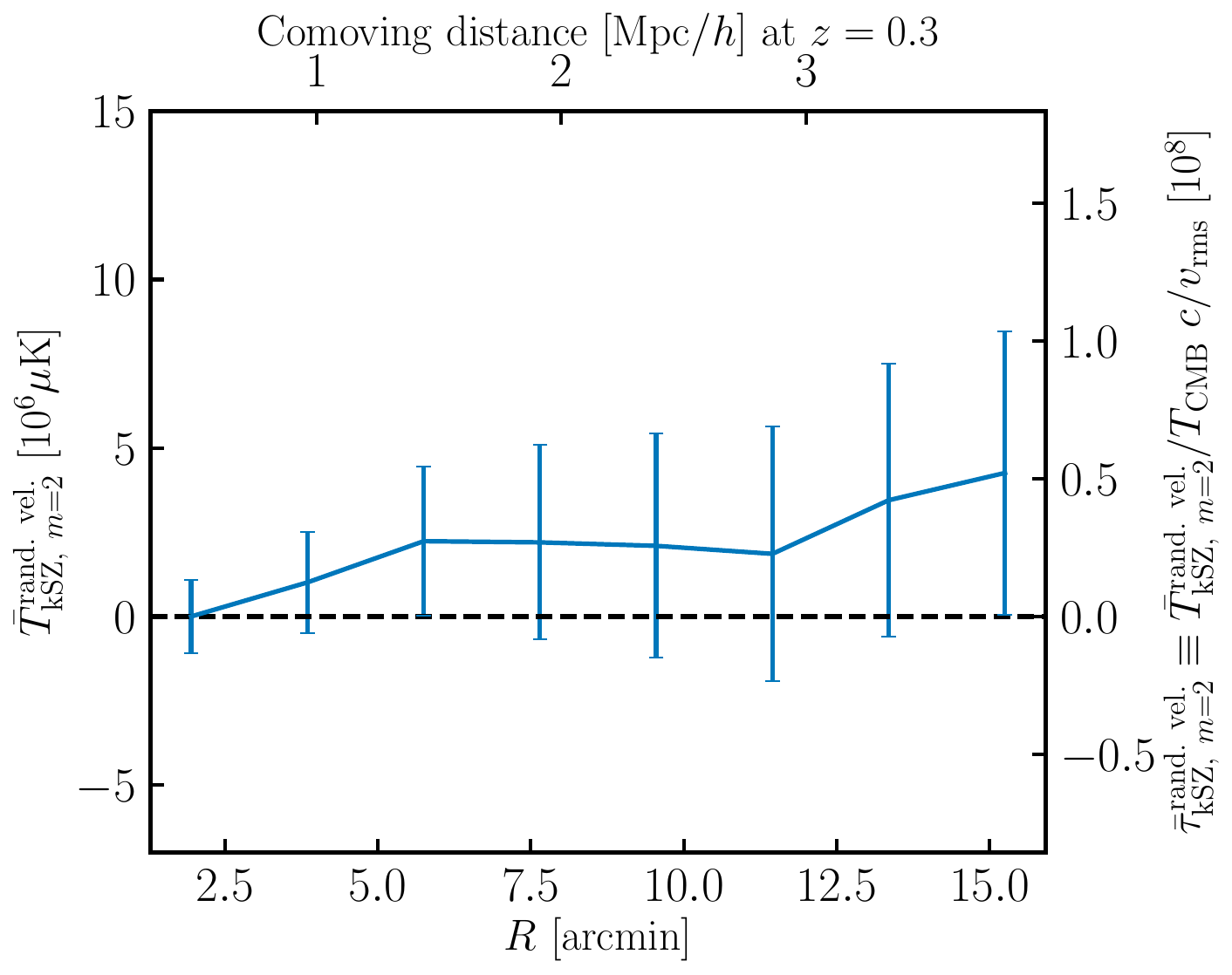}
    \caption{Tests of the robustness of the detected anisotropy in the stacked kSZ signal. Test 1 ({\it top}): We rotate the individual galaxy images by a random angle before stacking instead of aligning them along the identified filaments (Fig.~\ref{fig:m2}). We see that the measurement is consistent with zero, as expected, with a $\chi_{\rm null}^2$ value of 5.5 and PTE of 0.7. 
Test 2 ({\it bottom}): We multiply the individual cutouts of the galaxies by a random velocity before aligning them along the filaments to obtain the oriented stacks. Reassuringly, this test is also consistent with zero at $\chi^2_{\rm null}$ of 6.0 and PTE of 0.64, which strongly disfavors foregrounds as a viable explanation for the detected signal. 
}
    \label{fig:m2_null}
\end{figure}

We now move on to the study of the anisotropic gas profiles through the oriented stacks method described in Section~\ref{sec:stack}. We excise an image of each galaxy, multiply it by the reconstructed velocity, and align it with the direction of the 2D filament (see Fig.~\ref{fig:cw_bgs}). We display the oriented stacked image of the photometric BGS in the bottom of Fig.~\ref{fig:ksz_ani_stack}. We can clearly see excess signal along the horizontal, corresponding to the direction of the filament. We thus expect to detect a positive $m = 2$ signal, as indeed confirmed by Fig.~\ref{fig:m2}. 

We study the anisotropic kSZ signal in Fig.~\ref{fig:m2} and detect alignment between the gas density and cosmic web at 4.0$\sigma$ ($\chi^2_{\rm null} = 37.3$, 7 dof), 
suggesting that a large fraction of the gas gets fed into the halo via the large-scale filaments, as filaments act as primary sites for the transportation of material, in agreement with our understanding of structure formation. To obtain the filamentary structures, we smooth the BGS density field with a Gaussian smoothing kernel of 0.03 deg (1.8 arcmin). A finer smoothing would be limited by the beam size, whereas a coarser smoothing tends to wash out the signal, as it defers it to a larger characteristic radius, which is dominated by primary CMB fluctuations where kSZ stacking is too noisy. This is the first measurement of the anisotropic distribution of the gas density using the kSZ effect, and the use of low-redshift intermediate-mass galaxies as targets can in principle reveal a lot about the flow properties of gas and the alignment of relativistic jets from AGN with the cosmic web. In addition, it allows us to probe the baryon content of filamentary structures in the WHIM, which are typically inaccessible due to their lower number density and lower temperature. 

Besides the averaged profile, we also show the covariance matrix on the bottom of Fig.~\ref{fig:m2}. As in the case of the CAP-filter isotropic measurement of Fig.~\ref{fig:ksz_cap}, we also show the covariance matrix, which has a similar structure: strong covariance between the radial bins at large apertures due to the primary CMB signal, making the stacked kSZ measurement of gas density less sensitive on large scales. However, we note that the off-diagonal terms appear to be smaller than in the CAP case. On small scales, the noise comes predominantly from the ACT beam. 

We also perform several null tests to test the credibility of our result, shown in Fig.~\ref{fig:m2_null}. For the first test, we rotate the CMB cutouts at the location of the DESI galaxies by a random angle before stacking them together (and weighting each of them by the estimated reconstructed velocity at its 3D position). We find that the `randomized angle' measurement is consistent with zero, with a $\chi^2_{\rm null}$ of 5.5 and PTE of 0.70. This lends further credence to our main findings, shown in Fig.~\ref{fig:m2}. For the second test, we multiply each of the cutouts by a randomly selected velocity, to test whether the anisotropic signal persists, which would indicate that instead of it being due to the kSZ, it is dominated by foregrounds. Fortunately, this test also yields a measurement consistent with zero at $\chi^2_{\rm null}$ of 6.0 and PTE of 0.64, reassuring us that the anisotropic signal of Fig.~\ref{fig:m2} is most consistent with being the result of the kSZ effect. 

\subsection{Comparison with simulations}
\label{sec:comparison}

\begin{figure*}
    \centering
    \includegraphics[width=0.3\textwidth]{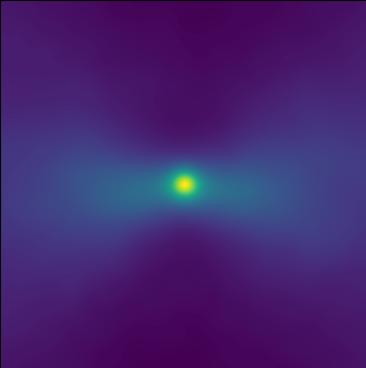}
    \includegraphics[width=0.3\textwidth]{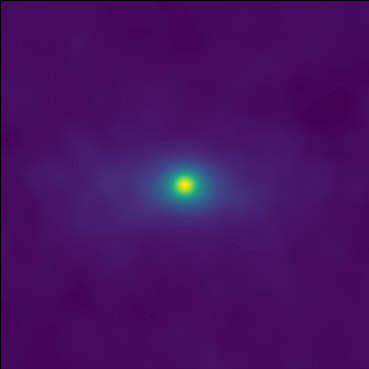}
    \includegraphics[width=0.3\textwidth]{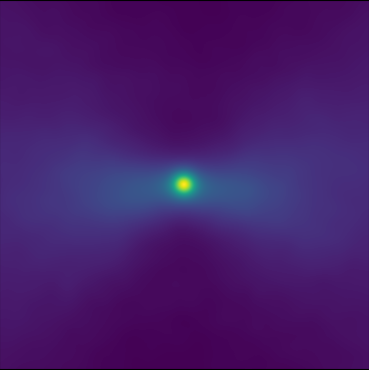}
    \caption{Stacked images (25'$\times$25') of the gas density, kSZ and dark matter maps around BGS-like objects in TNG300-1, oriented along the filaments identified through the procedure described in Section~\ref{sec:filament}. The map is generated at $z = 0.3$, smoothed with $R = 0.45 \ {\rm Mpc}/h$, which roughly corresponds to the mean redshift of the data and the smoothing scale we apply to it, i.e., 0.03 deg.}
    \label{fig:stack_sim}
\end{figure*}

\begin{figure}
    \centering
    \includegraphics[width=0.5\textwidth]{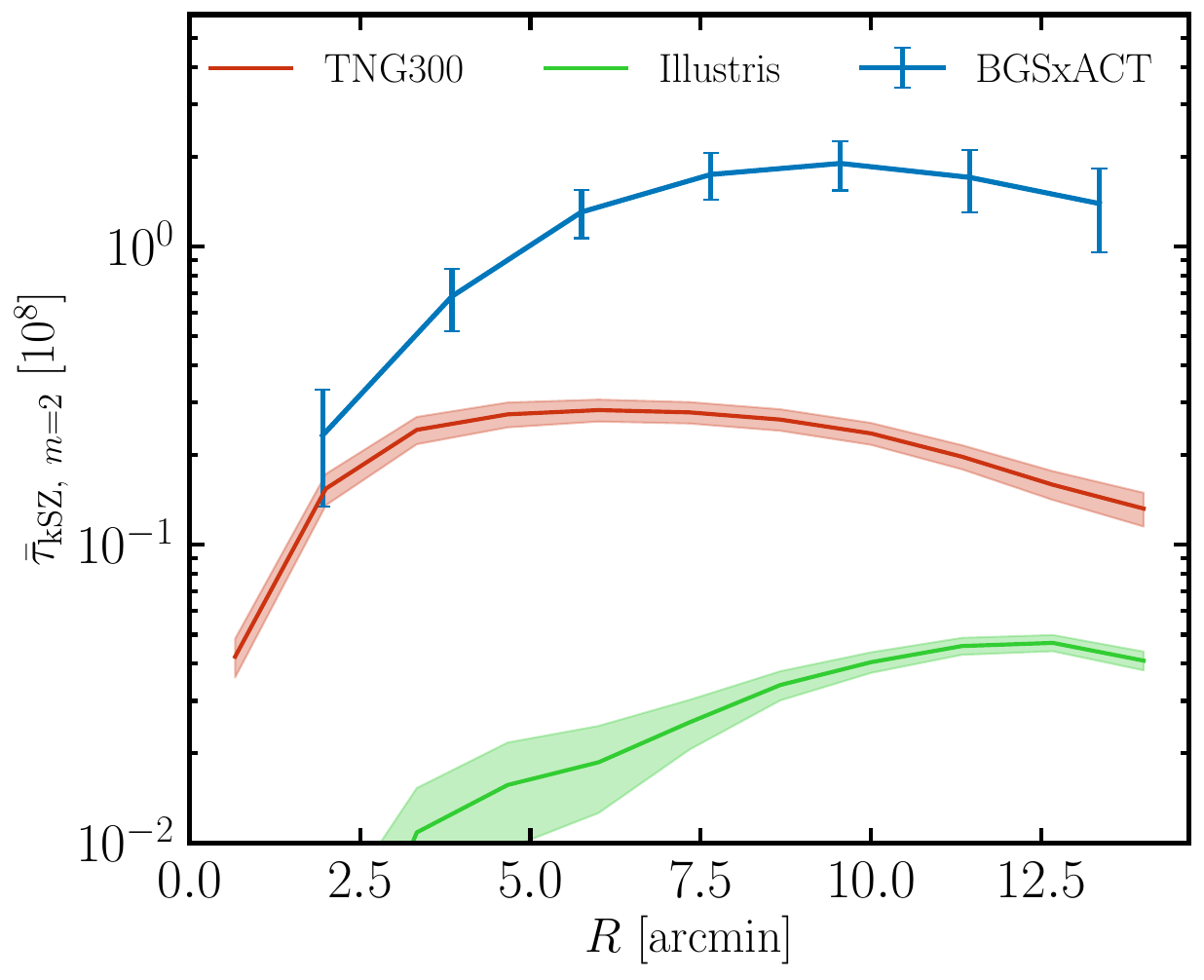}
    \caption{Comparison between the anisotropy of gas density measured via the BGS of the DESI imaging data and that measured in hydrodynamical simulations TNG300-1 and Illustris at $z = 0.3$. We see that while the overall amplitude and shape are similar in simulations and observations, there are some differences: the measured anisotropic signal appears to peak at larger apertures compared with TNG300-1, and its amplitude is significantly larger than Illustris, which is known to have a particularly isotropic gas distribution. This suggests strongly that the gas distribution around BGS is much more aligned with the underlying filamentary structure compared with the predictions from simulations, as a result of both inflow and outflow.} 
    \label{fig:m2_comparison}
\end{figure}

In this section, we are interested in comparing the amount of anisotropy observed in the Universe via the photometric BGS sample with predictions from hydrodynamical simulations. After creating the optical depth, kSZ, and dark matter maps as outlined in Section~\ref{sec:sims}, we perform oriented stacking in the same manner as done on the real data: namely, we create map cutouts around all BGS-like galaxies in the 2D simulated maps and rotate each of the cutouts so as to align them along the filaments identified through the procedure described in Section~\ref{sec:filament}, before stacking. In the case of the kSZ map, we multiply the cutout around a given BGS object by its host halo velocity along the line of sight. The result of the stacking is shown in Fig.~\ref{fig:stack_sim} for all three types of maps. We select BGS-like objects through a cut in the stellar mass of subhalos such that the number density of the selected objects nears that of the BGS subsample used in the analysis, i.e. $n_{\rm gal} = 1 \times 10^{-3} \ ({\rm Mpc}/h)^3$. We check that our results are qualitatively insensitive to this choice, with the shape and amplitude of the signal being largely independent of the mass threshold.

There are several noteworthy observations one can make. First, the dark matter anisotropy appears more tightly aligned with the filaments than the gas anisotropy, which can be explained by the fact that a larger fraction of the gas is gravitationally unbound or weakly bound, making its distribution more diffuse. Second, while the kSZ traces (second image) the same underlying quantity as the first image, the optical depth, it is evident that the anisotropic signal is much less pronounced in the kSZ stacked image. 
This finding might at first be somewhat puzzling since as previously shown in \citep{2023MNRAS.526..369H}, when measuring the isotropic profiles using CAP, we obtain identical profiles from both the optical depth and the kSZ fields up to $\sim$10 arcmin, which would naively imply that the anisotropic signal should also be identical in optical depth and kSZ. The reason it is not the case is that the kSZ signal contributed from nearby (in 2D) structures cancel on average, as their velocities are uncorrelated with that of the host halo bulk velocity.  The CAP filter, among other beneficial properties, removes these spurious contributions along the line-of-sight, leaving the kSZ signal unchanged and reducing the amplitude of the optical depth signal, such that the two are now sensitive to the targeted halo and, therefore, match each other very well.

Another way to see this is to imagine the extreme scenario of having perfectly isotropic gas distribution around each galaxy host. Then when performing the stacking, we would still get an excess signal along the filaments in the case of the optical depth (and thus non-zero $m = 2$ signal), since the contribution from nearby halos is always positive, but for the kSZ it would cancel (yielding zero $m = 2$ signal) unless these nearby structures move with a coherent velocity to the host halo, which we know to only be the case for close-by objects in 3D \footnote{We test this by performing a tomographic identification of filaments (in slabs of thickness 20 {\rm Mpc}/h along the $z$ axis) finding that the optical depth and kSZ signal become almost identical. Small differences appear beyond $\sim$10 arcmin where the velocity starts to decorrelate.} 
We note that this still preserves the $m = 0$ signal, which we expect to be identical for optical depth and kSZ. The only subtlety there is that in the case of the optical depth, we get contributions from random uncorrelated structures along the line-of-sight, but this is exactly what the CAP filter takes care of. 

In Fig.~\ref{fig:m2_comparison}, we compare the measured anisotropic signal, $m = 2$, in data and in the two simulations: TNG300-1 and Illustris. In the case of the simulations, we show the kSZ signal at $z = 0.3$, $\tau_{\rm kSZ}$. The DESI curve appears to increase with distance from the group center until $\sim$8 arcmin, after which it starts to decrease. We attribute this to the fact that the velocity starts to decorrelate slowly on these scales, reducing the correlation between the reconstructed galaxy velocity and the true velocity of nearby structures. In addition, this behavior is sensitive to the choice of smoothing scale in the process of identifying filaments. As a reminder, we aim to resolve the smallest possible scales as our focus is on charting out the gas in the WHIM region. 
Similar is the case with the simulations, where close to the group center we see little evidence of anisotropy, but as we moved further away from the halo virial radius the filamentary distribution of the gas becomes more evident. Further away from the center, the gas density decreases, the velocities decorrelate, and the effect becomes harder to measure.

Thus, the scale at which the transition between increase and decrease happens is tied to physical processes, such as how much of the gas is fed through the filaments, how much of the gas is expelled into the filaments, and how fast the velocity decorrelates, as well as the choice of smoothing scale, which roughly dictates the characteristic scale at which the $m = 2$ signal peaks. 

We see that the overall shape and amplitude of the signal in simulations and data are similar, but there are some notable differences. While the amplitude of the TNG300-1 curve is comparable, its shape differs in that it falls off faster with aperture. If the conjecture that feedback in the Universe is stronger than predicted from the TNG model \cite{2024arXiv240707152H}, then the difference in shape and amplitude might provide another glimpse at the same effect. Namely, the gas outflow contributes to the anisotropic signal both by increasing its amplitude (especially if there is an alignment between AGN jets and filaments; see below) and by increasing its reach. This is consistent with what we find for the anisotropic signal in Illustris. Illustris has much stronger feedback, which leads to very extended gas bubbles forming around the galaxies. This would explain a) the shape of the signal and more specifically, the fact that it falls off very slowly (beyond the scales displayed in the figure), b) the lower amplitude of the signal, coming from the fact that the baryonic feedback model implemented in Illustris is isotropic (i.e, it is not tied with another physical property such as the angular momentum of the galaxy) and due to its strength, it tends to erase the knowledge of the directionality of gas flow. Additionally, the fact that Illustris lacks massive halos also leads to a naturally smaller amplitude, and we have tested that both of these effects contribute equally.

Speculatively, this can also be partly contributed by an alignment between the direction in which gas is expelled and the direction of the cosmic web filaments. In this mass range, the type of feedback that dominates comes from AGN, which is most active in star-forming blue (spiral) galaxies. Intriguingly, recent intrinsic alignment studies seem to indicate that unlike their red counterparts, which tend to be aligned with the cosmic web along their most elongated dimension, blue galaxies seem to be preferentially aligned such that their disks are perpendicular to the filament direction \citep{2007ApJ...671.1248L,2013MNRAS.428.1827T,2015MNRAS.448.3391C,2018ApJ...866..138W}, and their angular momenta point along the filament. Since the angular momentum of the galaxy is expected to be correlated with the angular momentum of the supermassive black hole in its center, we might naturally expect azimuthally symmetric jets throwing out gas preferentially along the filaments. This would inevitably boost the $m = 2$ signal and might explain the strong anisotropic signal we find in the DESI and ACT data. In other words, we find evidence for strong baryonic feedback in the filament direction due to the alignment between galaxy spin and cosmic filaments.

One of the main caveats of this comparison against simulations is the effect of the cosmic web projection and sample selection, which are both somewhat sensitive to the size of the simulation and would therefore need to be done more carefully in the future especially if we would like to quantitatively understand (at the percent level) the amount of preferential alignment of the gas with the filaments. More detailed checks will be performed in the future using large-volume full-sky light cone simulations such as those of \textsc{AbacusSummit} \cite{2021MNRAS.508.4017M} with realistic galaxies and gas properties painted on them. Nonetheless, in the current paper, we argue that both the data and the simulations are biased low as a result the misidentification of the filament direction (since we identify filaments in 2D vs. 3D), and that bias is similar between the two because the kSZ effect is very localized to the region around the object on which we are stacking: any contribution to the signal coming from beyond a cylinder of depth of $\lesssim$$40 \ {\rm Mpc}/h$ would just cancel out in both simulations and data, and as such, as long as the simulation is larger than that, the bias in data and simulations should be similar. To test this, we perform a simple test of splitting the simulations into thin slabs of $20 \ {\rm Mpc}/h$ when identifying the cosmic web direction for the oriented stack, finding a small effect on the obtained curves.

\section{Discussion and conclusions}
\label{sec:conc}

In this work, we use publicly available data from the galaxy survey DESI and the CMB experiment ACT to measure the anisotropic distribution of diffuse gas in filaments, observed through the kSZ effect. These measurements are essential for providing long-needed tests for the astrophysical feedback models implemented in state-of-the-art hydrodynamical simulations as well as for finding the baryons `missing' from the cosmic census. In this work, using the kSZ effect, we make the first measurement of the gas density profiles of galaxies at low redshifts ($z \approx 0.3$) and the first measurement of the alignment between gas density and cosmic filaments. The benefit of performing the ``oriented stacking'' through the kSZ effect lies in the fact that the kSZ effect is sensitive to the main halo and the WHIM around it, canceling out contributions from nearby groups and clusters, and additionally, it is linearly proportional to the gas density, making it ideal for studying directly the gas distribution out to and beyond the virial radius. 

Here, we adopt the BGS of the DESI imaging survey with a stellar mass threshold of $\log M > 10.5$, reconstruct its velocity field via the provided low-noise photometric redshifts ($\sigma_z/(1+z) \approx 0.02$), and identify 2D cosmic filaments employing the projected tidal field method of Ref.~\citep{2016MNRAS.460..256A}. We then make a cutout around each of the BGS objects on the CMB map, rotate it so the cutout is aligned with the filament direction at the location of the galaxy, multiply it by the estimated velocity, and perform a stacking to obtain the average map, shown in Fig.~\ref{fig:ksz_ani_stack}. This procedure is summarized in Section~\ref{sec:stack}.

First, we display the radially averaged isotropic kSZ signal in Fig.~\ref{fig:ksz_cap}, which gives us a way of reconstructing the gas density profile of the BGS host halos at 7.2$\sigma$ ($\chi^2_{\rm null} = 65.0$, 9 dof). Similarly to previous works 
 \citep{2021PhRvD.103f3513S,2021PhRvD.103f3514A,2024arXiv240707152H}, we see that the gas density is extended far beyond the virial radius and appears more extended than the dark matter. This implies that baryons might play a larger than expected role in cosmological conundrums like the `low $S_8$' tension (e.g., see \citep{2022JHEAp..34...49A} and references therein). It is, thus, crucial to take into account the effect of baryons on the total matter distribution in order to accurately model cosmological probes that depend on it, such as the weak lensing signal. In Fig.~\ref{fig:ksz_cap_lrg}, we compare the photometric BGS signal obtained in this work with the photometric LRG signal obtained in Ref.~\citep{2024arXiv240707152H} and find that the two are very comparable in their shape and amplitude. This suggests that the mean halo mass and strength of the feedback is very similar, which is reassuring and shows consistency of the measured amount of feedback across redshift. Follow-up studies involving lensing can help pinpoint more accurately the host halo mass and recover the mass as well as gas properties of the sample.

Next, we move on to the anisotropic kSZ signal we detect around the BGS objects, which we evaluate through the second order of the cosine series expansion ($m = 2$ of Eq.~\ref{eq:C_m}). We find a 4.0$\sigma$ ($\chi^2_{\rm null} = 37.3$, 7 dof) 
anisotropy in the gas density signal around BGS objects, as demonstrated in Fig.~\ref{fig:m2}, indicating that the gas density is higher along filamentary regions than perpendicular to them. We offer two astrophysical explanations for this phenomenon: 1) gas, along with dark matter, is fed into a galaxy group or cluster preferentially through the filaments due to gravitational tidal forces; 2) blue galaxies may be expelling more gas in the direction of the filament due to the intrinsic alignment of the disks of actively star-forming galaxies perpendicular to the tidal field. Thus, echoing the conjecture that the angular momentum of the galaxy is correlated with the spin of the actively accreting supermassive black hole in the center (i.e., AGN) and the cosmic filament orientation \citep{2007ApJ...671.1248L,2013MNRAS.428.1827T,2015MNRAS.448.3391C,2018ApJ...866..138W}, we expect to find energetic jets expelling gas more readily along the filaments than perpendicular to them. 

We perform a number of null tests and cross-checks in order to verify our findings, two of which are shown in Fig.~\ref{fig:m2_null}. In the first test, we randomly rotate the galaxy cutouts before stacking them to find that the detected signal goes away, and in the second, we multiply the galaxy cutouts by a random velocity component along the line-of-sight before stacking them, which yields an answer consistent with null (PTEs of 0.70 and 0.64, respectively). 
 This reassures us that our detected anisotropicity is unlikely to be the result of foreground contamination or non-kSZ related sources.

Finally, to put our findings into context, we compare the oriented stacked profiles from the data with two hydrodynamical simulations: TNG300-1 and Illustris. Through Fig.~\ref{fig:stack_sim}, we visually corroborate the claim that unlike other stacked probes, such as patchy screening and tSZ, the kSZ oriented stacking method isolates the signal in the halo and the surrounding WHIM, receiving less contribution from nearby halos (i.e., close in 2D but far in 3D), which offers a glimpse of the gas distribution in the most diffuse and underdense regions. This is also seen quantitatively in Fig.~\ref{fig:m2_comparison}. Interestingly, we find evidence for more pronounced anisotropy of the kSZ signal in the real Universe compared with simulations, implying a possible excess of gas inflow or outflow along the filament direction on larger scales than expected by the two hydrodynamical simulations considered in this work.

Studying gas properties through versatile probes like the oriented stacking of the kSZ effect is a uniquely powerful tool for mapping out the 3D distribution of gas in the Universe, bringing us one step closer to fully unraveling the thermodynamics of galaxy groups and clusters as well as the complex processes of galaxy formation and evolution.


\acknowledgements

We thank Martine Lokken, Joanne Cohn, Adam Hincks, Bernardita Ried Guachalla, Emmanuel Schaan, Edward Wollack and  for useful discussions. B.H. thanks the Miller Institute for financially supporting her postdoctoral research. S.F. and R.Z. are supported by Lawrence Berkeley National Laboratory and the Director, Office of Science, Office of High Energy Physics of the U.S. Department of Energy under Contract No.\ DE-AC02-05CH11231.

\bibliography{apssamp}

\providecommand{\noopsort}[1]{}\providecommand{\singleletter}[1]{#1}%
\begin{thebibliography}{61}%
\makeatletter
\providecommand \@ifxundefined [1]{%
 \@ifx{#1\undefined}
}%
\providecommand \@ifnum [1]{%
 \ifnum #1\expandafter \@firstoftwo
 \else \expandafter \@secondoftwo
 \fi
}%
\providecommand \@ifx [1]{%
 \ifx #1\expandafter \@firstoftwo
 \else \expandafter \@secondoftwo
 \fi
}%
\providecommand \natexlab [1]{#1}%
\providecommand \enquote  [1]{``#1''}%
\providecommand \bibnamefont  [1]{#1}%
\providecommand \bibfnamefont [1]{#1}%
\providecommand \citenamefont [1]{#1}%
\providecommand \href@noop [0]{\@secondoftwo}%
\providecommand \href [0]{\begingroup \@sanitize@url \@href}%
\providecommand \@href[1]{\@@startlink{#1}\@@href}%
\providecommand \@@href[1]{\endgroup#1\@@endlink}%
\providecommand \@sanitize@url [0]{\catcode `\\12\catcode `\$12\catcode
  `\&12\catcode `\#12\catcode `\^12\catcode `\_12\catcode `\%12\relax}%
\providecommand \@@startlink[1]{}%
\providecommand \@@endlink[0]{}%
\providecommand \url  [0]{\begingroup\@sanitize@url \@url }%
\providecommand \@url [1]{\endgroup\@href {#1}{\urlprefix }}%
\providecommand \urlprefix  [0]{URL }%
\providecommand \Eprint [0]{\href }%
\providecommand \doibase [0]{https://doi.org/}%
\providecommand \selectlanguage [0]{\@gobble}%
\providecommand \bibinfo  [0]{\@secondoftwo}%
\providecommand \bibfield  [0]{\@secondoftwo}%
\providecommand \translation [1]{[#1]}%
\providecommand \BibitemOpen [0]{}%
\providecommand \bibitemStop [0]{}%
\providecommand \bibitemNoStop [0]{.\EOS\space}%
\providecommand \EOS [0]{\spacefactor3000\relax}%
\providecommand \BibitemShut  [1]{\csname bibitem#1\endcsname}%
\let\auto@bib@innerbib\@empty
\bibitem [{\citenamefont {{Ostriker}}\ \emph {et~al.}(2005)\citenamefont
  {{Ostriker}}, \citenamefont {{Bode}},\ and\ \citenamefont
  {{Babul}}}]{2005ApJ...634..964O}%
  \BibitemOpen
  \bibfield  {author} {\bibinfo {author} {\bibfnamefont {J.~P.}\ \bibnamefont
  {{Ostriker}}}, \bibinfo {author} {\bibfnamefont {P.}~\bibnamefont {{Bode}}},\
  and\ \bibinfo {author} {\bibfnamefont {A.}~\bibnamefont {{Babul}}},\
  }\bibfield  {title} {\bibinfo {title} {{A Simple and Accurate Model for
  Intracluster Gas}},\ }\href {https://doi.org/10.1086/497122} {\bibfield
  {journal} {\bibinfo  {journal} {\apj}\ }\textbf {\bibinfo {volume} {634}},\
  \bibinfo {pages} {964} (\bibinfo {year} {2005})},\ \Eprint
  {https://arxiv.org/abs/astro-ph/0504334} {arXiv:astro-ph/0504334 [astro-ph]}
  \BibitemShut {NoStop}%
\bibitem [{\citenamefont {{Nagai}}\ \emph {et~al.}(2007)\citenamefont
  {{Nagai}}, \citenamefont {{Kravtsov}},\ and\ \citenamefont
  {{Vikhlinin}}}]{2007ApJ...668....1N}%
  \BibitemOpen
  \bibfield  {author} {\bibinfo {author} {\bibfnamefont {D.}~\bibnamefont
  {{Nagai}}}, \bibinfo {author} {\bibfnamefont {A.~V.}\ \bibnamefont
  {{Kravtsov}}},\ and\ \bibinfo {author} {\bibfnamefont {A.}~\bibnamefont
  {{Vikhlinin}}},\ }\bibfield  {title} {\bibinfo {title} {{Effects of Galaxy
  Formation on Thermodynamics of the Intracluster Medium}},\ }\href
  {https://doi.org/10.1086/521328} {\bibfield  {journal} {\bibinfo  {journal}
  {\apj}\ }\textbf {\bibinfo {volume} {668}},\ \bibinfo {pages} {1} (\bibinfo
  {year} {2007})},\ \Eprint {https://arxiv.org/abs/astro-ph/0703661}
  {arXiv:astro-ph/0703661 [astro-ph]} \BibitemShut {NoStop}%
\bibitem [{\citenamefont {{McNamara}}\ and\ \citenamefont
  {{Nulsen}}(2007)}]{2007ARA&A..45..117M}%
  \BibitemOpen
  \bibfield  {author} {\bibinfo {author} {\bibfnamefont {B.~R.}\ \bibnamefont
  {{McNamara}}}\ and\ \bibinfo {author} {\bibfnamefont {P.~E.~J.}\ \bibnamefont
  {{Nulsen}}},\ }\bibfield  {title} {\bibinfo {title} {{Heating Hot Atmospheres
  with Active Galactic Nuclei}},\ }\href
  {https://doi.org/10.1146/annurev.astro.45.051806.110625} {\bibfield
  {journal} {\bibinfo  {journal} {\araa}\ }\textbf {\bibinfo {volume} {45}},\
  \bibinfo {pages} {117} (\bibinfo {year} {2007})},\ \Eprint
  {https://arxiv.org/abs/0709.2152} {arXiv:0709.2152 [astro-ph]} \BibitemShut
  {NoStop}%
\bibitem [{\citenamefont {{Battaglia}}\ \emph {et~al.}(2010)\citenamefont
  {{Battaglia}}, \citenamefont {{Bond}}, \citenamefont {{Pfrommer}},
  \citenamefont {{Sievers}},\ and\ \citenamefont
  {{Sijacki}}}]{2010ApJ...725...91B}%
  \BibitemOpen
  \bibfield  {author} {\bibinfo {author} {\bibfnamefont {N.}~\bibnamefont
  {{Battaglia}}}, \bibinfo {author} {\bibfnamefont {J.~R.}\ \bibnamefont
  {{Bond}}}, \bibinfo {author} {\bibfnamefont {C.}~\bibnamefont {{Pfrommer}}},
  \bibinfo {author} {\bibfnamefont {J.~L.}\ \bibnamefont {{Sievers}}},\ and\
  \bibinfo {author} {\bibfnamefont {D.}~\bibnamefont {{Sijacki}}},\ }\bibfield
  {title} {\bibinfo {title} {{Simulations of the Sunyaev-Zel'dovich Power
  Spectrum with Active Galactic Nucleus Feedback}},\ }\href
  {https://doi.org/10.1088/0004-637X/725/1/91} {\bibfield  {journal} {\bibinfo
  {journal} {\apj}\ }\textbf {\bibinfo {volume} {725}},\ \bibinfo {pages} {91}
  (\bibinfo {year} {2010})},\ \Eprint {https://arxiv.org/abs/1003.4256}
  {arXiv:1003.4256 [astro-ph.CO]} \BibitemShut {NoStop}%
\bibitem [{\citenamefont {{Persic}}\ and\ \citenamefont
  {{Salucci}}(1992)}]{1992MNRAS.258P..14P}%
  \BibitemOpen
  \bibfield  {author} {\bibinfo {author} {\bibfnamefont {M.}~\bibnamefont
  {{Persic}}}\ and\ \bibinfo {author} {\bibfnamefont {P.}~\bibnamefont
  {{Salucci}}},\ }\bibfield  {title} {\bibinfo {title} {{The baryon content of
  the universe}},\ }\href {https://doi.org/10.1093/mnras/258.1.14P} {\bibfield
  {journal} {\bibinfo  {journal} {\mnras}\ }\textbf {\bibinfo {volume} {258}},\
  \bibinfo {pages} {14P} (\bibinfo {year} {1992})},\ \Eprint
  {https://arxiv.org/abs/astro-ph/0502178} {arXiv:astro-ph/0502178 [astro-ph]}
  \BibitemShut {NoStop}%
\bibitem [{\citenamefont {{Cen}}\ and\ \citenamefont
  {{Ostriker}}(1999)}]{1999ApJ...514....1C}%
  \BibitemOpen
  \bibfield  {author} {\bibinfo {author} {\bibfnamefont {R.}~\bibnamefont
  {{Cen}}}\ and\ \bibinfo {author} {\bibfnamefont {J.~P.}\ \bibnamefont
  {{Ostriker}}},\ }\bibfield  {title} {\bibinfo {title} {{Where Are the
  Baryons?}},\ }\href {https://doi.org/10.1086/306949} {\bibfield  {journal}
  {\bibinfo  {journal} {\apj}\ }\textbf {\bibinfo {volume} {514}},\ \bibinfo
  {pages} {1} (\bibinfo {year} {1999})},\ \Eprint
  {https://arxiv.org/abs/astro-ph/9806281} {arXiv:astro-ph/9806281 [astro-ph]}
  \BibitemShut {NoStop}%
\bibitem [{\citenamefont {{Fukugita}}(2004)}]{2004IAUS..220..227F}%
  \BibitemOpen
  \bibfield  {author} {\bibinfo {author} {\bibfnamefont {M.}~\bibnamefont
  {{Fukugita}}},\ }\bibfield  {title} {\bibinfo {title} {{Cosmic Matter
  Distribution: Cosmic Baryon Budget Revisited}},\ }in\ \href
  {https://doi.org/10.48550/arXiv.astro-ph/0312517} {\emph {\bibinfo
  {booktitle} {Dark Matter in Galaxies}}},\ \bibinfo {series} {IAU Symposium},
  Vol.\ \bibinfo {volume} {220},\ \bibinfo {editor} {edited by\ \bibinfo
  {editor} {\bibfnamefont {S.}~\bibnamefont {{Ryder}}}, \bibinfo {editor}
  {\bibfnamefont {D.}~\bibnamefont {{Pisano}}}, \bibinfo {editor}
  {\bibfnamefont {M.}~\bibnamefont {{Walker}}},\ and\ \bibinfo {editor}
  {\bibfnamefont {K.}~\bibnamefont {{Freeman}}}}\ (\bibinfo {year} {2004})\ p.\
  \bibinfo {pages} {227},\ \Eprint {https://arxiv.org/abs/astro-ph/0312517}
  {arXiv:astro-ph/0312517 [astro-ph]} \BibitemShut {NoStop}%
\bibitem [{\citenamefont {{Shull}}\ \emph {et~al.}(2012)\citenamefont
  {{Shull}}, \citenamefont {{Smith}},\ and\ \citenamefont
  {{Danforth}}}]{2012ApJ...759...23S}%
  \BibitemOpen
  \bibfield  {author} {\bibinfo {author} {\bibfnamefont {J.~M.}\ \bibnamefont
  {{Shull}}}, \bibinfo {author} {\bibfnamefont {B.~D.}\ \bibnamefont
  {{Smith}}},\ and\ \bibinfo {author} {\bibfnamefont {C.~W.}\ \bibnamefont
  {{Danforth}}},\ }\bibfield  {title} {\bibinfo {title} {{The Baryon Census in
  a Multiphase Intergalactic Medium: 30\% of the Baryons May Still be
  Missing}},\ }\href {https://doi.org/10.1088/0004-637X/759/1/23} {\bibfield
  {journal} {\bibinfo  {journal} {\apj}\ }\textbf {\bibinfo {volume} {759}},\
  \bibinfo {eid} {23} (\bibinfo {year} {2012})},\ \Eprint
  {https://arxiv.org/abs/1112.2706} {arXiv:1112.2706 [astro-ph.CO]}
  \BibitemShut {NoStop}%
\bibitem [{\citenamefont {{Bullock}}\ and\ \citenamefont
  {{Boylan-Kolchin}}(2017)}]{2017ARA&A..55..343B}%
  \BibitemOpen
  \bibfield  {author} {\bibinfo {author} {\bibfnamefont {J.~S.}\ \bibnamefont
  {{Bullock}}}\ and\ \bibinfo {author} {\bibfnamefont {M.}~\bibnamefont
  {{Boylan-Kolchin}}},\ }\bibfield  {title} {\bibinfo {title} {{Small-Scale
  Challenges to the {\ensuremath{\Lambda}}CDM Paradigm}},\ }\href
  {https://doi.org/10.1146/annurev-astro-091916-055313} {\bibfield  {journal}
  {\bibinfo  {journal} {\araa}\ }\textbf {\bibinfo {volume} {55}},\ \bibinfo
  {pages} {343} (\bibinfo {year} {2017})},\ \Eprint
  {https://arxiv.org/abs/1707.04256} {arXiv:1707.04256 [astro-ph.CO]}
  \BibitemShut {NoStop}%
\bibitem [{\citenamefont {{Battaglia}}\ \emph {et~al.}(2017)\citenamefont
  {{Battaglia}}, \citenamefont {{Ferraro}}, \citenamefont {{Schaan}},\ and\
  \citenamefont {{Spergel}}}]{2017JCAP...11..040B}%
  \BibitemOpen
  \bibfield  {author} {\bibinfo {author} {\bibfnamefont {N.}~\bibnamefont
  {{Battaglia}}}, \bibinfo {author} {\bibfnamefont {S.}~\bibnamefont
  {{Ferraro}}}, \bibinfo {author} {\bibfnamefont {E.}~\bibnamefont
  {{Schaan}}},\ and\ \bibinfo {author} {\bibfnamefont {D.~N.}\ \bibnamefont
  {{Spergel}}},\ }\bibfield  {title} {\bibinfo {title} {{Future constraints on
  halo thermodynamics from combined Sunyaev-Zel'dovich measurements}},\ }\href
  {https://doi.org/10.1088/1475-7516/2017/11/040} {\bibfield  {journal}
  {\bibinfo  {journal} {\jcap}\ }\textbf {\bibinfo {volume} {2017}},\ \bibinfo
  {eid} {040} (\bibinfo {year} {2017})},\ \Eprint
  {https://arxiv.org/abs/1705.05881} {arXiv:1705.05881 [astro-ph.CO]}
  \BibitemShut {NoStop}%
\bibitem [{\citenamefont {{Amon}}\ \emph {et~al.}(2022)\citenamefont {{Amon}},
  \citenamefont {{Gruen}}, \citenamefont {{Troxel}}, \citenamefont
  {{MacCrann}}, \citenamefont {{Dodelson}}, \citenamefont {{Choi}},
  \citenamefont {{Doux}}, \citenamefont {{Secco}}, \citenamefont {{Samuroff}},
  \citenamefont {{Krause}}, \citenamefont {{Cordero}}, \citenamefont {{Myles}},
  \citenamefont {{DeRose}}, \citenamefont {{Wechsler}}, \citenamefont
  {{Gatti}}, \citenamefont {{Navarro-Alsina}}, \citenamefont {{Bernstein}},
  \citenamefont {{Jain}}, \citenamefont {{Blazek}}, \citenamefont {{Alarcon}},
  \citenamefont {{Fert{\'e}}}, \citenamefont {{Lemos}}, \citenamefont
  {{Raveri}}, \citenamefont {{Campos}}, \citenamefont {{Prat}}, \citenamefont
  {{S{\'a}nchez}}, \citenamefont {{Jarvis}}, \citenamefont {{Alves}},
  \citenamefont {{Andrade-Oliveira}}, \citenamefont {{Baxter}}, \citenamefont
  {{Bechtol}}, \citenamefont {{Becker}}, \citenamefont {{Bridle}},
  \citenamefont {{Camacho}}, \citenamefont {{Carnero Rosell}}, \citenamefont
  {{Carrasco Kind}}, \citenamefont {{Cawthon}}, \citenamefont {{Chang}},
  \citenamefont {{Chen}}, \citenamefont {{Chintalapati}}, \citenamefont
  {{Crocce}}, \citenamefont {{Davis}}, \citenamefont {{Diehl}}, \citenamefont
  {{Drlica-Wagner}}, \citenamefont {{Eckert}}, \citenamefont {{Eifler}},
  \citenamefont {{Elvin-Poole}}, \citenamefont {{Everett}}, \citenamefont
  {{Fang}}, \citenamefont {{Fosalba}}, \citenamefont {{Friedrich}},
  \citenamefont {{Gaztanaga}}, \citenamefont {{Giannini}}, \citenamefont
  {{Gruendl}}, \citenamefont {{Harrison}}, \citenamefont {{Hartley}},
  \citenamefont {{Herner}}, \citenamefont {{Huang}}, \citenamefont {{Huff}},
  \citenamefont {{Huterer}}, \citenamefont {{Kuropatkin}}, \citenamefont
  {{Leget}}, \citenamefont {{Liddle}}, \citenamefont {{McCullough}},
  \citenamefont {{Muir}}, \citenamefont {{Pandey}}, \citenamefont {{Park}},
  \citenamefont {{Porredon}}, \citenamefont {{Refregier}}, \citenamefont
  {{Rollins}}, \citenamefont {{Roodman}}, \citenamefont {{Rosenfeld}},
  \citenamefont {{Ross}}, \citenamefont {{Rykoff}}, \citenamefont {{Sanchez}},
  \citenamefont {{Sevilla-Noarbe}}, \citenamefont {{Sheldon}}, \citenamefont
  {{Shin}}, \citenamefont {{Troja}}, \citenamefont {{Tutusaus}}, \citenamefont
  {{Tutusaus}}, \citenamefont {{Varga}}, \citenamefont {{Weaverdyck}},
  \citenamefont {{Yanny}}, \citenamefont {{Yin}}, \citenamefont {{Zhang}},
  \citenamefont {{Zuntz}}, \citenamefont {{Aguena}}, \citenamefont {{Allam}},
  \citenamefont {{Annis}}, \citenamefont {{Bacon}}, \citenamefont {{Bertin}},
  \citenamefont {{Bhargava}}, \citenamefont {{Brooks}}, \citenamefont
  {{Buckley-Geer}}, \citenamefont {{Burke}}, \citenamefont {{Carretero}},
  \citenamefont {{Costanzi}}, \citenamefont {{da Costa}}, \citenamefont
  {{Pereira}}, \citenamefont {{De Vicente}}, \citenamefont {{Desai}},
  \citenamefont {{Dietrich}}, \citenamefont {{Doel}}, \citenamefont
  {{Ferrero}}, \citenamefont {{Flaugher}}, \citenamefont {{Frieman}},
  \citenamefont {{Garc{\'\i}a-Bellido}}, \citenamefont {{Gaztanaga}},
  \citenamefont {{Gerdes}}, \citenamefont {{Giannantonio}}, \citenamefont
  {{Gschwend}}, \citenamefont {{Gutierrez}}, \citenamefont {{Hinton}},
  \citenamefont {{Hollowood}}, \citenamefont {{Honscheid}}, \citenamefont
  {{Hoyle}}, \citenamefont {{James}}, \citenamefont {{Kron}}, \citenamefont
  {{Kuehn}}, \citenamefont {{Lahav}}, \citenamefont {{Lima}}, \citenamefont
  {{Lin}}, \citenamefont {{Maia}}, \citenamefont {{Marshall}}, \citenamefont
  {{Martini}}, \citenamefont {{Melchior}}, \citenamefont {{Menanteau}},
  \citenamefont {{Miquel}}, \citenamefont {{Mohr}}, \citenamefont {{Morgan}},
  \citenamefont {{Ogando}}, \citenamefont {{Palmese}}, \citenamefont
  {{Paz-Chinch{\'o}n}}, \citenamefont {{Petravick}}, \citenamefont {{Pieres}},
  \citenamefont {{Romer}}, \citenamefont {{Sanchez}}, \citenamefont
  {{Scarpine}}, \citenamefont {{Schubnell}}, \citenamefont {{Serrano}},
  \citenamefont {{Smith}}, \citenamefont {{Soares-Santos}}, \citenamefont
  {{Tarle}}, \citenamefont {{Thomas}}, \citenamefont {{To}}, \citenamefont
  {{Weller}},\ and\ \citenamefont {{DES Collaboration}}}]{2022PhRvD.105b3514A}%
  \BibitemOpen
  \bibfield  {author} {\bibinfo {author} {\bibfnamefont {A.}~\bibnamefont
  {{Amon}}}, \bibinfo {author} {\bibfnamefont {D.}~\bibnamefont {{Gruen}}},
  \bibinfo {author} {\bibfnamefont {M.~A.}\ \bibnamefont {{Troxel}}}, \bibinfo
  {author} {\bibfnamefont {N.}~\bibnamefont {{MacCrann}}}, \bibinfo {author}
  {\bibfnamefont {S.}~\bibnamefont {{Dodelson}}}, \bibinfo {author}
  {\bibfnamefont {A.}~\bibnamefont {{Choi}}}, \bibinfo {author} {\bibfnamefont
  {C.}~\bibnamefont {{Doux}}}, \bibinfo {author} {\bibfnamefont {L.~F.}\
  \bibnamefont {{Secco}}}, \bibinfo {author} {\bibfnamefont {S.}~\bibnamefont
  {{Samuroff}}}, \bibinfo {author} {\bibfnamefont {E.}~\bibnamefont
  {{Krause}}}, \bibinfo {author} {\bibfnamefont {J.}~\bibnamefont {{Cordero}}},
  \bibinfo {author} {\bibfnamefont {J.}~\bibnamefont {{Myles}}}, \bibinfo
  {author} {\bibfnamefont {J.}~\bibnamefont {{DeRose}}}, \bibinfo {author}
  {\bibfnamefont {R.~H.}\ \bibnamefont {{Wechsler}}}, \bibinfo {author}
  {\bibfnamefont {M.}~\bibnamefont {{Gatti}}}, \bibinfo {author} {\bibfnamefont
  {A.}~\bibnamefont {{Navarro-Alsina}}}, \bibinfo {author} {\bibfnamefont
  {G.~M.}\ \bibnamefont {{Bernstein}}}, \bibinfo {author} {\bibfnamefont
  {B.}~\bibnamefont {{Jain}}}, \bibinfo {author} {\bibfnamefont
  {J.}~\bibnamefont {{Blazek}}}, \bibinfo {author} {\bibfnamefont
  {A.}~\bibnamefont {{Alarcon}}}, \bibinfo {author} {\bibfnamefont
  {A.}~\bibnamefont {{Fert{\'e}}}}, \bibinfo {author} {\bibfnamefont
  {P.}~\bibnamefont {{Lemos}}}, \bibinfo {author} {\bibfnamefont
  {M.}~\bibnamefont {{Raveri}}}, \bibinfo {author} {\bibfnamefont
  {A.}~\bibnamefont {{Campos}}}, \bibinfo {author} {\bibfnamefont
  {J.}~\bibnamefont {{Prat}}}, \bibinfo {author} {\bibfnamefont
  {C.}~\bibnamefont {{S{\'a}nchez}}}, \bibinfo {author} {\bibfnamefont
  {M.}~\bibnamefont {{Jarvis}}}, \bibinfo {author} {\bibfnamefont
  {O.}~\bibnamefont {{Alves}}}, \bibinfo {author} {\bibfnamefont
  {F.}~\bibnamefont {{Andrade-Oliveira}}}, \bibinfo {author} {\bibfnamefont
  {E.}~\bibnamefont {{Baxter}}}, \bibinfo {author} {\bibfnamefont
  {K.}~\bibnamefont {{Bechtol}}}, \bibinfo {author} {\bibfnamefont {M.~R.}\
  \bibnamefont {{Becker}}}, \bibinfo {author} {\bibfnamefont {S.~L.}\
  \bibnamefont {{Bridle}}}, \bibinfo {author} {\bibfnamefont {H.}~\bibnamefont
  {{Camacho}}}, \bibinfo {author} {\bibfnamefont {A.}~\bibnamefont {{Carnero
  Rosell}}}, \bibinfo {author} {\bibfnamefont {M.}~\bibnamefont {{Carrasco
  Kind}}}, \bibinfo {author} {\bibfnamefont {R.}~\bibnamefont {{Cawthon}}},
  \bibinfo {author} {\bibfnamefont {C.}~\bibnamefont {{Chang}}}, \bibinfo
  {author} {\bibfnamefont {R.}~\bibnamefont {{Chen}}}, \bibinfo {author}
  {\bibfnamefont {P.}~\bibnamefont {{Chintalapati}}}, \bibinfo {author}
  {\bibfnamefont {M.}~\bibnamefont {{Crocce}}}, \bibinfo {author}
  {\bibfnamefont {C.}~\bibnamefont {{Davis}}}, \bibinfo {author} {\bibfnamefont
  {H.~T.}\ \bibnamefont {{Diehl}}}, \bibinfo {author} {\bibfnamefont
  {A.}~\bibnamefont {{Drlica-Wagner}}}, \bibinfo {author} {\bibfnamefont
  {K.}~\bibnamefont {{Eckert}}}, \bibinfo {author} {\bibfnamefont {T.~F.}\
  \bibnamefont {{Eifler}}}, \bibinfo {author} {\bibfnamefont {J.}~\bibnamefont
  {{Elvin-Poole}}}, \bibinfo {author} {\bibfnamefont {S.}~\bibnamefont
  {{Everett}}}, \bibinfo {author} {\bibfnamefont {X.}~\bibnamefont {{Fang}}},
  \bibinfo {author} {\bibfnamefont {P.}~\bibnamefont {{Fosalba}}}, \bibinfo
  {author} {\bibfnamefont {O.}~\bibnamefont {{Friedrich}}}, \bibinfo {author}
  {\bibfnamefont {E.}~\bibnamefont {{Gaztanaga}}}, \bibinfo {author}
  {\bibfnamefont {G.}~\bibnamefont {{Giannini}}}, \bibinfo {author}
  {\bibfnamefont {R.~A.}\ \bibnamefont {{Gruendl}}}, \bibinfo {author}
  {\bibfnamefont {I.}~\bibnamefont {{Harrison}}}, \bibinfo {author}
  {\bibfnamefont {W.~G.}\ \bibnamefont {{Hartley}}}, \bibinfo {author}
  {\bibfnamefont {K.}~\bibnamefont {{Herner}}}, \bibinfo {author}
  {\bibfnamefont {H.}~\bibnamefont {{Huang}}}, \bibinfo {author} {\bibfnamefont
  {E.~M.}\ \bibnamefont {{Huff}}}, \bibinfo {author} {\bibfnamefont
  {D.}~\bibnamefont {{Huterer}}}, \bibinfo {author} {\bibfnamefont
  {N.}~\bibnamefont {{Kuropatkin}}}, \bibinfo {author} {\bibfnamefont
  {P.}~\bibnamefont {{Leget}}}, \bibinfo {author} {\bibfnamefont {A.~R.}\
  \bibnamefont {{Liddle}}}, \bibinfo {author} {\bibfnamefont {J.}~\bibnamefont
  {{McCullough}}}, \bibinfo {author} {\bibfnamefont {J.}~\bibnamefont
  {{Muir}}}, \bibinfo {author} {\bibfnamefont {S.}~\bibnamefont {{Pandey}}},
  \bibinfo {author} {\bibfnamefont {Y.}~\bibnamefont {{Park}}}, \bibinfo
  {author} {\bibfnamefont {A.}~\bibnamefont {{Porredon}}}, \bibinfo {author}
  {\bibfnamefont {A.}~\bibnamefont {{Refregier}}}, \bibinfo {author}
  {\bibfnamefont {R.~P.}\ \bibnamefont {{Rollins}}}, \bibinfo {author}
  {\bibfnamefont {A.}~\bibnamefont {{Roodman}}}, \bibinfo {author}
  {\bibfnamefont {R.}~\bibnamefont {{Rosenfeld}}}, \bibinfo {author}
  {\bibfnamefont {A.~J.}\ \bibnamefont {{Ross}}}, \bibinfo {author}
  {\bibfnamefont {E.~S.}\ \bibnamefont {{Rykoff}}}, \bibinfo {author}
  {\bibfnamefont {J.}~\bibnamefont {{Sanchez}}}, \bibinfo {author}
  {\bibfnamefont {I.}~\bibnamefont {{Sevilla-Noarbe}}}, \bibinfo {author}
  {\bibfnamefont {E.}~\bibnamefont {{Sheldon}}}, \bibinfo {author}
  {\bibfnamefont {T.}~\bibnamefont {{Shin}}}, \bibinfo {author} {\bibfnamefont
  {A.}~\bibnamefont {{Troja}}}, \bibinfo {author} {\bibfnamefont
  {I.}~\bibnamefont {{Tutusaus}}}, \bibinfo {author} {\bibfnamefont
  {I.}~\bibnamefont {{Tutusaus}}}, \bibinfo {author} {\bibfnamefont {T.~N.}\
  \bibnamefont {{Varga}}}, \bibinfo {author} {\bibfnamefont {N.}~\bibnamefont
  {{Weaverdyck}}}, \bibinfo {author} {\bibfnamefont {B.}~\bibnamefont
  {{Yanny}}}, \bibinfo {author} {\bibfnamefont {B.}~\bibnamefont {{Yin}}},
  \bibinfo {author} {\bibfnamefont {Y.}~\bibnamefont {{Zhang}}}, \bibinfo
  {author} {\bibfnamefont {J.}~\bibnamefont {{Zuntz}}}, \bibinfo {author}
  {\bibfnamefont {M.}~\bibnamefont {{Aguena}}}, \bibinfo {author}
  {\bibfnamefont {S.}~\bibnamefont {{Allam}}}, \bibinfo {author} {\bibfnamefont
  {J.}~\bibnamefont {{Annis}}}, \bibinfo {author} {\bibfnamefont
  {D.}~\bibnamefont {{Bacon}}}, \bibinfo {author} {\bibfnamefont
  {E.}~\bibnamefont {{Bertin}}}, \bibinfo {author} {\bibfnamefont
  {S.}~\bibnamefont {{Bhargava}}}, \bibinfo {author} {\bibfnamefont
  {D.}~\bibnamefont {{Brooks}}}, \bibinfo {author} {\bibfnamefont
  {E.}~\bibnamefont {{Buckley-Geer}}}, \bibinfo {author} {\bibfnamefont
  {D.~L.}\ \bibnamefont {{Burke}}}, \bibinfo {author} {\bibfnamefont
  {J.}~\bibnamefont {{Carretero}}}, \bibinfo {author} {\bibfnamefont
  {M.}~\bibnamefont {{Costanzi}}}, \bibinfo {author} {\bibfnamefont {L.~N.}\
  \bibnamefont {{da Costa}}}, \bibinfo {author} {\bibfnamefont {M.~E.~S.}\
  \bibnamefont {{Pereira}}}, \bibinfo {author} {\bibfnamefont {J.}~\bibnamefont
  {{De Vicente}}}, \bibinfo {author} {\bibfnamefont {S.}~\bibnamefont
  {{Desai}}}, \bibinfo {author} {\bibfnamefont {J.~P.}\ \bibnamefont
  {{Dietrich}}}, \bibinfo {author} {\bibfnamefont {P.}~\bibnamefont {{Doel}}},
  \bibinfo {author} {\bibfnamefont {I.}~\bibnamefont {{Ferrero}}}, \bibinfo
  {author} {\bibfnamefont {B.}~\bibnamefont {{Flaugher}}}, \bibinfo {author}
  {\bibfnamefont {J.}~\bibnamefont {{Frieman}}}, \bibinfo {author}
  {\bibfnamefont {J.}~\bibnamefont {{Garc{\'\i}a-Bellido}}}, \bibinfo {author}
  {\bibfnamefont {E.}~\bibnamefont {{Gaztanaga}}}, \bibinfo {author}
  {\bibfnamefont {D.~W.}\ \bibnamefont {{Gerdes}}}, \bibinfo {author}
  {\bibfnamefont {T.}~\bibnamefont {{Giannantonio}}}, \bibinfo {author}
  {\bibfnamefont {J.}~\bibnamefont {{Gschwend}}}, \bibinfo {author}
  {\bibfnamefont {G.}~\bibnamefont {{Gutierrez}}}, \bibinfo {author}
  {\bibfnamefont {S.~R.}\ \bibnamefont {{Hinton}}}, \bibinfo {author}
  {\bibfnamefont {D.~L.}\ \bibnamefont {{Hollowood}}}, \bibinfo {author}
  {\bibfnamefont {K.}~\bibnamefont {{Honscheid}}}, \bibinfo {author}
  {\bibfnamefont {B.}~\bibnamefont {{Hoyle}}}, \bibinfo {author} {\bibfnamefont
  {D.~J.}\ \bibnamefont {{James}}}, \bibinfo {author} {\bibfnamefont
  {R.}~\bibnamefont {{Kron}}}, \bibinfo {author} {\bibfnamefont
  {K.}~\bibnamefont {{Kuehn}}}, \bibinfo {author} {\bibfnamefont
  {O.}~\bibnamefont {{Lahav}}}, \bibinfo {author} {\bibfnamefont
  {M.}~\bibnamefont {{Lima}}}, \bibinfo {author} {\bibfnamefont
  {H.}~\bibnamefont {{Lin}}}, \bibinfo {author} {\bibfnamefont {M.~A.~G.}\
  \bibnamefont {{Maia}}}, \bibinfo {author} {\bibfnamefont {J.~L.}\
  \bibnamefont {{Marshall}}}, \bibinfo {author} {\bibfnamefont
  {P.}~\bibnamefont {{Martini}}}, \bibinfo {author} {\bibfnamefont
  {P.}~\bibnamefont {{Melchior}}}, \bibinfo {author} {\bibfnamefont
  {F.}~\bibnamefont {{Menanteau}}}, \bibinfo {author} {\bibfnamefont
  {R.}~\bibnamefont {{Miquel}}}, \bibinfo {author} {\bibfnamefont {J.~J.}\
  \bibnamefont {{Mohr}}}, \bibinfo {author} {\bibfnamefont {R.}~\bibnamefont
  {{Morgan}}}, \bibinfo {author} {\bibfnamefont {R.~L.~C.}\ \bibnamefont
  {{Ogando}}}, \bibinfo {author} {\bibfnamefont {A.}~\bibnamefont {{Palmese}}},
  \bibinfo {author} {\bibfnamefont {F.}~\bibnamefont {{Paz-Chinch{\'o}n}}},
  \bibinfo {author} {\bibfnamefont {D.}~\bibnamefont {{Petravick}}}, \bibinfo
  {author} {\bibfnamefont {A.}~\bibnamefont {{Pieres}}}, \bibinfo {author}
  {\bibfnamefont {A.~K.}\ \bibnamefont {{Romer}}}, \bibinfo {author}
  {\bibfnamefont {E.}~\bibnamefont {{Sanchez}}}, \bibinfo {author}
  {\bibfnamefont {V.}~\bibnamefont {{Scarpine}}}, \bibinfo {author}
  {\bibfnamefont {M.}~\bibnamefont {{Schubnell}}}, \bibinfo {author}
  {\bibfnamefont {S.}~\bibnamefont {{Serrano}}}, \bibinfo {author}
  {\bibfnamefont {M.}~\bibnamefont {{Smith}}}, \bibinfo {author} {\bibfnamefont
  {M.}~\bibnamefont {{Soares-Santos}}}, \bibinfo {author} {\bibfnamefont
  {G.}~\bibnamefont {{Tarle}}}, \bibinfo {author} {\bibfnamefont
  {D.}~\bibnamefont {{Thomas}}}, \bibinfo {author} {\bibfnamefont
  {C.}~\bibnamefont {{To}}}, \bibinfo {author} {\bibfnamefont {J.}~\bibnamefont
  {{Weller}}},\ and\ \bibinfo {author} {\bibnamefont {{DES Collaboration}}},\
  }\bibfield  {title} {\bibinfo {title} {{Dark Energy Survey Year 3 results:
  Cosmology from cosmic shear and robustness to data calibration}},\ }\href
  {https://doi.org/10.1103/PhysRevD.105.023514} {\bibfield  {journal} {\bibinfo
   {journal} {\prd}\ }\textbf {\bibinfo {volume} {105}},\ \bibinfo {eid}
  {023514} (\bibinfo {year} {2022})},\ \Eprint
  {https://arxiv.org/abs/2105.13543} {arXiv:2105.13543 [astro-ph.CO]}
  \BibitemShut {NoStop}%
\bibitem [{\citenamefont {{Aric{\`o}}}\ \emph {et~al.}(2023)\citenamefont
  {{Aric{\`o}}}, \citenamefont {{Angulo}}, \citenamefont {{Zennaro}},
  \citenamefont {{Contreras}}, \citenamefont {{Chen}},\ and\ \citenamefont
  {{Hern{\'a}ndez-Monteagudo}}}]{2023A&A...678A.109A}%
  \BibitemOpen
  \bibfield  {author} {\bibinfo {author} {\bibfnamefont {G.}~\bibnamefont
  {{Aric{\`o}}}}, \bibinfo {author} {\bibfnamefont {R.~E.}\ \bibnamefont
  {{Angulo}}}, \bibinfo {author} {\bibfnamefont {M.}~\bibnamefont {{Zennaro}}},
  \bibinfo {author} {\bibfnamefont {S.}~\bibnamefont {{Contreras}}}, \bibinfo
  {author} {\bibfnamefont {A.}~\bibnamefont {{Chen}}},\ and\ \bibinfo {author}
  {\bibfnamefont {C.}~\bibnamefont {{Hern{\'a}ndez-Monteagudo}}},\ }\bibfield
  {title} {\bibinfo {title} {{DES Y3 cosmic shear down to small scales:
  Constraints on cosmology and baryons}},\ }\href
  {https://doi.org/10.1051/0004-6361/202346539} {\bibfield  {journal} {\bibinfo
   {journal} {\aap}\ }\textbf {\bibinfo {volume} {678}},\ \bibinfo {eid} {A109}
  (\bibinfo {year} {2023})},\ \Eprint {https://arxiv.org/abs/2303.05537}
  {arXiv:2303.05537 [astro-ph.CO]} \BibitemShut {NoStop}%
\bibitem [{\citenamefont {{Dav{\'e}}}\ \emph {et~al.}(2001)\citenamefont
  {{Dav{\'e}}}, \citenamefont {{Cen}}, \citenamefont {{Ostriker}},
  \citenamefont {{Bryan}}, \citenamefont {{Hernquist}}, \citenamefont {{Katz}},
  \citenamefont {{Weinberg}}, \citenamefont {{Norman}},\ and\ \citenamefont
  {{O'Shea}}}]{2001ApJ...552..473D}%
  \BibitemOpen
  \bibfield  {author} {\bibinfo {author} {\bibfnamefont {R.}~\bibnamefont
  {{Dav{\'e}}}}, \bibinfo {author} {\bibfnamefont {R.}~\bibnamefont {{Cen}}},
  \bibinfo {author} {\bibfnamefont {J.~P.}\ \bibnamefont {{Ostriker}}},
  \bibinfo {author} {\bibfnamefont {G.~L.}\ \bibnamefont {{Bryan}}}, \bibinfo
  {author} {\bibfnamefont {L.}~\bibnamefont {{Hernquist}}}, \bibinfo {author}
  {\bibfnamefont {N.}~\bibnamefont {{Katz}}}, \bibinfo {author} {\bibfnamefont
  {D.~H.}\ \bibnamefont {{Weinberg}}}, \bibinfo {author} {\bibfnamefont
  {M.~L.}\ \bibnamefont {{Norman}}},\ and\ \bibinfo {author} {\bibfnamefont
  {B.}~\bibnamefont {{O'Shea}}},\ }\bibfield  {title} {\bibinfo {title}
  {{Baryons in the Warm-Hot Intergalactic Medium}},\ }\href
  {https://doi.org/10.1086/320548} {\bibfield  {journal} {\bibinfo  {journal}
  {\apj}\ }\textbf {\bibinfo {volume} {552}},\ \bibinfo {pages} {473} (\bibinfo
  {year} {2001})},\ \Eprint {https://arxiv.org/abs/astro-ph/0007217}
  {arXiv:astro-ph/0007217 [astro-ph]} \BibitemShut {NoStop}%
\bibitem [{\citenamefont {{Kormendy}}\ and\ \citenamefont
  {{Ho}}(2013)}]{2013ARA&A..51..511K}%
  \BibitemOpen
  \bibfield  {author} {\bibinfo {author} {\bibfnamefont {J.}~\bibnamefont
  {{Kormendy}}}\ and\ \bibinfo {author} {\bibfnamefont {L.~C.}\ \bibnamefont
  {{Ho}}},\ }\bibfield  {title} {\bibinfo {title} {{Coevolution (Or Not) of
  Supermassive Black Holes and Host Galaxies}},\ }\href
  {https://doi.org/10.1146/annurev-astro-082708-101811} {\bibfield  {journal}
  {\bibinfo  {journal} {\araa}\ }\textbf {\bibinfo {volume} {51}},\ \bibinfo
  {pages} {511} (\bibinfo {year} {2013})},\ \Eprint
  {https://arxiv.org/abs/1304.7762} {arXiv:1304.7762 [astro-ph.CO]}
  \BibitemShut {NoStop}%
\bibitem [{\citenamefont {{Yuan}}\ and\ \citenamefont
  {{Narayan}}(2014)}]{2014ARA&A..52..529Y}%
  \BibitemOpen
  \bibfield  {author} {\bibinfo {author} {\bibfnamefont {F.}~\bibnamefont
  {{Yuan}}}\ and\ \bibinfo {author} {\bibfnamefont {R.}~\bibnamefont
  {{Narayan}}},\ }\bibfield  {title} {\bibinfo {title} {{Hot Accretion Flows
  Around Black Holes}},\ }\href
  {https://doi.org/10.1146/annurev-astro-082812-141003} {\bibfield  {journal}
  {\bibinfo  {journal} {\araa}\ }\textbf {\bibinfo {volume} {52}},\ \bibinfo
  {pages} {529} (\bibinfo {year} {2014})},\ \Eprint
  {https://arxiv.org/abs/1401.0586} {arXiv:1401.0586 [astro-ph.HE]}
  \BibitemShut {NoStop}%
\bibitem [{\citenamefont {{Somerville}}\ and\ \citenamefont
  {{Dav{\'e}}}(2015)}]{2015ARA&A..53...51S}%
  \BibitemOpen
  \bibfield  {author} {\bibinfo {author} {\bibfnamefont {R.~S.}\ \bibnamefont
  {{Somerville}}}\ and\ \bibinfo {author} {\bibfnamefont {R.}~\bibnamefont
  {{Dav{\'e}}}},\ }\bibfield  {title} {\bibinfo {title} {{Physical Models of
  Galaxy Formation in a Cosmological Framework}},\ }\href
  {https://doi.org/10.1146/annurev-astro-082812-140951} {\bibfield  {journal}
  {\bibinfo  {journal} {\araa}\ }\textbf {\bibinfo {volume} {53}},\ \bibinfo
  {pages} {51} (\bibinfo {year} {2015})},\ \Eprint
  {https://arxiv.org/abs/1412.2712} {arXiv:1412.2712 [astro-ph.GA]}
  \BibitemShut {NoStop}%
\bibitem [{\citenamefont {{Harnois-D{\'e}raps}}\ \emph
  {et~al.}(2015)\citenamefont {{Harnois-D{\'e}raps}}, \citenamefont {{van
  Waerbeke}}, \citenamefont {{Viola}},\ and\ \citenamefont
  {{Heymans}}}]{2015MNRAS.450.1212H}%
  \BibitemOpen
  \bibfield  {author} {\bibinfo {author} {\bibfnamefont {J.}~\bibnamefont
  {{Harnois-D{\'e}raps}}}, \bibinfo {author} {\bibfnamefont {L.}~\bibnamefont
  {{van Waerbeke}}}, \bibinfo {author} {\bibfnamefont {M.}~\bibnamefont
  {{Viola}}},\ and\ \bibinfo {author} {\bibfnamefont {C.}~\bibnamefont
  {{Heymans}}},\ }\bibfield  {title} {\bibinfo {title} {{Baryons, neutrinos,
  feedback and weak gravitational lensing}},\ }\href
  {https://doi.org/10.1093/mnras/stv646} {\bibfield  {journal} {\bibinfo
  {journal} {\mnras}\ }\textbf {\bibinfo {volume} {450}},\ \bibinfo {pages}
  {1212} (\bibinfo {year} {2015})},\ \Eprint {https://arxiv.org/abs/1407.4301}
  {arXiv:1407.4301 [astro-ph.CO]} \BibitemShut {NoStop}%
\bibitem [{\citenamefont {{Sunyaev}}\ and\ \citenamefont
  {{Zeldovich}}(1972)}]{1972CoASP...4..173S}%
  \BibitemOpen
  \bibfield  {author} {\bibinfo {author} {\bibfnamefont {R.~A.}\ \bibnamefont
  {{Sunyaev}}}\ and\ \bibinfo {author} {\bibfnamefont {Y.~B.}\ \bibnamefont
  {{Zeldovich}}},\ }\bibfield  {title} {\bibinfo {title} {{The Observations of
  Relic Radiation as a Test of the Nature of X-Ray Radiation from the Clusters
  of Galaxies}},\ }\href@noop {} {\bibfield  {journal} {\bibinfo  {journal}
  {Comments on Astrophysics and Space Physics}\ }\textbf {\bibinfo {volume}
  {4}},\ \bibinfo {pages} {173} (\bibinfo {year} {1972})}\BibitemShut {NoStop}%
\bibitem [{\citenamefont {{Birkinshaw}}(1999)}]{1999PhR...310...97B}%
  \BibitemOpen
  \bibfield  {author} {\bibinfo {author} {\bibfnamefont {M.}~\bibnamefont
  {{Birkinshaw}}},\ }\bibfield  {title} {\bibinfo {title} {{The
  Sunyaev-Zel'dovich effect}},\ }\href
  {https://doi.org/10.1016/S0370-1573(98)00080-5} {\bibfield  {journal}
  {\bibinfo  {journal} {\physrep}\ }\textbf {\bibinfo {volume} {310}},\
  \bibinfo {pages} {97} (\bibinfo {year} {1999})},\ \Eprint
  {https://arxiv.org/abs/astro-ph/9808050} {arXiv:astro-ph/9808050 [astro-ph]}
  \BibitemShut {NoStop}%
\bibitem [{\citenamefont {{Mroczkowski}}\ \emph {et~al.}(2019)\citenamefont
  {{Mroczkowski}}, \citenamefont {{Nagai}}, \citenamefont {{Basu}},
  \citenamefont {{Chluba}}, \citenamefont {{Sayers}}, \citenamefont {{Adam}},
  \citenamefont {{Churazov}}, \citenamefont {{Crites}}, \citenamefont {{Di
  Mascolo}}, \citenamefont {{Eckert}}, \citenamefont {{Macias-Perez}},
  \citenamefont {{Mayet}}, \citenamefont {{Perotto}}, \citenamefont
  {{Pointecouteau}}, \citenamefont {{Romero}}, \citenamefont {{Ruppin}},
  \citenamefont {{Scannapieco}},\ and\ \citenamefont
  {{ZuHone}}}]{2019SSRv..215...17M}%
  \BibitemOpen
  \bibfield  {author} {\bibinfo {author} {\bibfnamefont {T.}~\bibnamefont
  {{Mroczkowski}}}, \bibinfo {author} {\bibfnamefont {D.}~\bibnamefont
  {{Nagai}}}, \bibinfo {author} {\bibfnamefont {K.}~\bibnamefont {{Basu}}},
  \bibinfo {author} {\bibfnamefont {J.}~\bibnamefont {{Chluba}}}, \bibinfo
  {author} {\bibfnamefont {J.}~\bibnamefont {{Sayers}}}, \bibinfo {author}
  {\bibfnamefont {R.}~\bibnamefont {{Adam}}}, \bibinfo {author} {\bibfnamefont
  {E.}~\bibnamefont {{Churazov}}}, \bibinfo {author} {\bibfnamefont
  {A.}~\bibnamefont {{Crites}}}, \bibinfo {author} {\bibfnamefont
  {L.}~\bibnamefont {{Di Mascolo}}}, \bibinfo {author} {\bibfnamefont
  {D.}~\bibnamefont {{Eckert}}}, \bibinfo {author} {\bibfnamefont
  {J.}~\bibnamefont {{Macias-Perez}}}, \bibinfo {author} {\bibfnamefont
  {F.}~\bibnamefont {{Mayet}}}, \bibinfo {author} {\bibfnamefont
  {L.}~\bibnamefont {{Perotto}}}, \bibinfo {author} {\bibfnamefont
  {E.}~\bibnamefont {{Pointecouteau}}}, \bibinfo {author} {\bibfnamefont
  {C.}~\bibnamefont {{Romero}}}, \bibinfo {author} {\bibfnamefont
  {F.}~\bibnamefont {{Ruppin}}}, \bibinfo {author} {\bibfnamefont
  {E.}~\bibnamefont {{Scannapieco}}},\ and\ \bibinfo {author} {\bibfnamefont
  {J.}~\bibnamefont {{ZuHone}}},\ }\bibfield  {title} {\bibinfo {title}
  {{Astrophysics with the Spatially and Spectrally Resolved Sunyaev-Zeldovich
  Effects. A Millimetre/Submillimetre Probe of the Warm and Hot Universe}},\
  }\href {https://doi.org/10.1007/s11214-019-0581-2} {\bibfield  {journal}
  {\bibinfo  {journal} {\ssr}\ }\textbf {\bibinfo {volume} {215}},\ \bibinfo
  {eid} {17} (\bibinfo {year} {2019})},\ \Eprint
  {https://arxiv.org/abs/1811.02310} {arXiv:1811.02310 [astro-ph.CO]}
  \BibitemShut {NoStop}%
\bibitem [{\citenamefont {{Hill}}\ \emph {et~al.}(2016)\citenamefont {{Hill}},
  \citenamefont {{Ferraro}}, \citenamefont {{Battaglia}}, \citenamefont
  {{Liu}},\ and\ \citenamefont {{Spergel}}}]{2016PhRvL.117e1301H}%
  \BibitemOpen
  \bibfield  {author} {\bibinfo {author} {\bibfnamefont {J.~C.}\ \bibnamefont
  {{Hill}}}, \bibinfo {author} {\bibfnamefont {S.}~\bibnamefont {{Ferraro}}},
  \bibinfo {author} {\bibfnamefont {N.}~\bibnamefont {{Battaglia}}}, \bibinfo
  {author} {\bibfnamefont {J.}~\bibnamefont {{Liu}}},\ and\ \bibinfo {author}
  {\bibfnamefont {D.~N.}\ \bibnamefont {{Spergel}}},\ }\bibfield  {title}
  {\bibinfo {title} {{Kinematic Sunyaev-Zel'dovich Effect with Projected
  Fields: A Novel Probe of the Baryon Distribution with Planck, WMAP, and WISE
  Data}},\ }\href {https://doi.org/10.1103/PhysRevLett.117.051301} {\bibfield
  {journal} {\bibinfo  {journal} {\prl}\ }\textbf {\bibinfo {volume} {117}},\
  \bibinfo {eid} {051301} (\bibinfo {year} {2016})},\ \Eprint
  {https://arxiv.org/abs/1603.01608} {arXiv:1603.01608 [astro-ph.CO]}
  \BibitemShut {NoStop}%
\bibitem [{\citenamefont {{Ferraro}}\ \emph {et~al.}(2016)\citenamefont
  {{Ferraro}}, \citenamefont {{Hill}}, \citenamefont {{Battaglia}},
  \citenamefont {{Liu}},\ and\ \citenamefont
  {{Spergel}}}]{2016PhRvD..94l3526F}%
  \BibitemOpen
  \bibfield  {author} {\bibinfo {author} {\bibfnamefont {S.}~\bibnamefont
  {{Ferraro}}}, \bibinfo {author} {\bibfnamefont {J.~C.}\ \bibnamefont
  {{Hill}}}, \bibinfo {author} {\bibfnamefont {N.}~\bibnamefont {{Battaglia}}},
  \bibinfo {author} {\bibfnamefont {J.}~\bibnamefont {{Liu}}},\ and\ \bibinfo
  {author} {\bibfnamefont {D.~N.}\ \bibnamefont {{Spergel}}},\ }\bibfield
  {title} {\bibinfo {title} {{Kinematic Sunyaev-Zel'dovich effect with
  projected fields. II. Prospects, challenges, and comparison with
  simulations}},\ }\href {https://doi.org/10.1103/PhysRevD.94.123526}
  {\bibfield  {journal} {\bibinfo  {journal} {\prd}\ }\textbf {\bibinfo
  {volume} {94}},\ \bibinfo {eid} {123526} (\bibinfo {year} {2016})},\ \Eprint
  {https://arxiv.org/abs/1605.02722} {arXiv:1605.02722 [astro-ph.CO]}
  \BibitemShut {NoStop}%
\bibitem [{\citenamefont {{Kusiak}}\ \emph {et~al.}(2021)\citenamefont
  {{Kusiak}}, \citenamefont {{Bolliet}}, \citenamefont {{Ferraro}},
  \citenamefont {{Hill}},\ and\ \citenamefont
  {{Krolewski}}}]{2021PhRvD.104d3518K}%
  \BibitemOpen
  \bibfield  {author} {\bibinfo {author} {\bibfnamefont {A.}~\bibnamefont
  {{Kusiak}}}, \bibinfo {author} {\bibfnamefont {B.}~\bibnamefont {{Bolliet}}},
  \bibinfo {author} {\bibfnamefont {S.}~\bibnamefont {{Ferraro}}}, \bibinfo
  {author} {\bibfnamefont {J.~C.}\ \bibnamefont {{Hill}}},\ and\ \bibinfo
  {author} {\bibfnamefont {A.}~\bibnamefont {{Krolewski}}},\ }\bibfield
  {title} {\bibinfo {title} {{Constraining the baryon abundance with the
  kinematic Sunyaev-Zel'dovich effect: Projected-field detection using P l a n
  c k , W M A P , and u n W I S E}},\ }\href
  {https://doi.org/10.1103/PhysRevD.104.043518} {\bibfield  {journal} {\bibinfo
   {journal} {\prd}\ }\textbf {\bibinfo {volume} {104}},\ \bibinfo {eid}
  {043518} (\bibinfo {year} {2021})},\ \Eprint
  {https://arxiv.org/abs/2102.01068} {arXiv:2102.01068 [astro-ph.CO]}
  \BibitemShut {NoStop}%
\bibitem [{\citenamefont {{Schaan}}\ \emph {et~al.}(2021)\citenamefont
  {{Schaan}}, \citenamefont {{Ferraro}}, \citenamefont {{Amodeo}},
  \citenamefont {{Battaglia}}, \citenamefont {{Aiola}}, \citenamefont
  {{Austermann}}, \citenamefont {{Beall}}, \citenamefont {{Bean}},
  \citenamefont {{Becker}}, \citenamefont {{Bond}}, \citenamefont
  {{Calabrese}}, \citenamefont {{Calafut}}, \citenamefont {{Choi}},
  \citenamefont {{Denison}}, \citenamefont {{Devlin}}, \citenamefont {{Duff}},
  \citenamefont {{Duivenvoorden}}, \citenamefont {{Dunkley}}, \citenamefont
  {{D{\"u}nner}}, \citenamefont {{Gallardo}}, \citenamefont {{Guan}},
  \citenamefont {{Han}}, \citenamefont {{Hill}}, \citenamefont {{Hilton}},
  \citenamefont {{Hilton}}, \citenamefont {{Hlo{\v{z}}ek}}, \citenamefont
  {{Hubmayr}}, \citenamefont {{Huffenberger}}, \citenamefont {{Hughes}},
  \citenamefont {{Koopman}}, \citenamefont {{MacInnis}}, \citenamefont
  {{McMahon}}, \citenamefont {{Madhavacheril}}, \citenamefont {{Moodley}},
  \citenamefont {{Mroczkowski}}, \citenamefont {{Naess}}, \citenamefont
  {{Nati}}, \citenamefont {{Newburgh}}, \citenamefont {{Niemack}},
  \citenamefont {{Page}}, \citenamefont {{Partridge}}, \citenamefont
  {{Salatino}}, \citenamefont {{Sehgal}}, \citenamefont {{Schillaci}},
  \citenamefont {{Sif{\'o}n}}, \citenamefont {{Smith}}, \citenamefont
  {{Spergel}}, \citenamefont {{Staggs}}, \citenamefont {{Storer}},
  \citenamefont {{Trac}}, \citenamefont {{Ullom}}, \citenamefont {{Van Lanen}},
  \citenamefont {{Vale}}, \citenamefont {{van Engelen}}, \citenamefont
  {{Maga{\~n}a}}, \citenamefont {{Vavagiakis}}, \citenamefont {{Wollack}},
  \citenamefont {{Xu}},\ and\ \citenamefont {{Atacama Cosmology Telescope
  Collaboration}}}]{2021PhRvD.103f3513S}%
  \BibitemOpen
  \bibfield  {author} {\bibinfo {author} {\bibfnamefont {E.}~\bibnamefont
  {{Schaan}}}, \bibinfo {author} {\bibfnamefont {S.}~\bibnamefont {{Ferraro}}},
  \bibinfo {author} {\bibfnamefont {S.}~\bibnamefont {{Amodeo}}}, \bibinfo
  {author} {\bibfnamefont {N.}~\bibnamefont {{Battaglia}}}, \bibinfo {author}
  {\bibfnamefont {S.}~\bibnamefont {{Aiola}}}, \bibinfo {author} {\bibfnamefont
  {J.~E.}\ \bibnamefont {{Austermann}}}, \bibinfo {author} {\bibfnamefont
  {J.~A.}\ \bibnamefont {{Beall}}}, \bibinfo {author} {\bibfnamefont
  {R.}~\bibnamefont {{Bean}}}, \bibinfo {author} {\bibfnamefont {D.~T.}\
  \bibnamefont {{Becker}}}, \bibinfo {author} {\bibfnamefont {R.~J.}\
  \bibnamefont {{Bond}}}, \bibinfo {author} {\bibfnamefont {E.}~\bibnamefont
  {{Calabrese}}}, \bibinfo {author} {\bibfnamefont {V.}~\bibnamefont
  {{Calafut}}}, \bibinfo {author} {\bibfnamefont {S.~K.}\ \bibnamefont
  {{Choi}}}, \bibinfo {author} {\bibfnamefont {E.~V.}\ \bibnamefont
  {{Denison}}}, \bibinfo {author} {\bibfnamefont {M.~J.}\ \bibnamefont
  {{Devlin}}}, \bibinfo {author} {\bibfnamefont {S.~M.}\ \bibnamefont
  {{Duff}}}, \bibinfo {author} {\bibfnamefont {A.~J.}\ \bibnamefont
  {{Duivenvoorden}}}, \bibinfo {author} {\bibfnamefont {J.}~\bibnamefont
  {{Dunkley}}}, \bibinfo {author} {\bibfnamefont {R.}~\bibnamefont
  {{D{\"u}nner}}}, \bibinfo {author} {\bibfnamefont {P.~A.}\ \bibnamefont
  {{Gallardo}}}, \bibinfo {author} {\bibfnamefont {Y.}~\bibnamefont {{Guan}}},
  \bibinfo {author} {\bibfnamefont {D.}~\bibnamefont {{Han}}}, \bibinfo
  {author} {\bibfnamefont {J.~C.}\ \bibnamefont {{Hill}}}, \bibinfo {author}
  {\bibfnamefont {G.~C.}\ \bibnamefont {{Hilton}}}, \bibinfo {author}
  {\bibfnamefont {M.}~\bibnamefont {{Hilton}}}, \bibinfo {author}
  {\bibfnamefont {R.}~\bibnamefont {{Hlo{\v{z}}ek}}}, \bibinfo {author}
  {\bibfnamefont {J.}~\bibnamefont {{Hubmayr}}}, \bibinfo {author}
  {\bibfnamefont {K.~M.}\ \bibnamefont {{Huffenberger}}}, \bibinfo {author}
  {\bibfnamefont {J.~P.}\ \bibnamefont {{Hughes}}}, \bibinfo {author}
  {\bibfnamefont {B.~J.}\ \bibnamefont {{Koopman}}}, \bibinfo {author}
  {\bibfnamefont {A.}~\bibnamefont {{MacInnis}}}, \bibinfo {author}
  {\bibfnamefont {J.}~\bibnamefont {{McMahon}}}, \bibinfo {author}
  {\bibfnamefont {M.~S.}\ \bibnamefont {{Madhavacheril}}}, \bibinfo {author}
  {\bibfnamefont {K.}~\bibnamefont {{Moodley}}}, \bibinfo {author}
  {\bibfnamefont {T.}~\bibnamefont {{Mroczkowski}}}, \bibinfo {author}
  {\bibfnamefont {S.}~\bibnamefont {{Naess}}}, \bibinfo {author} {\bibfnamefont
  {F.}~\bibnamefont {{Nati}}}, \bibinfo {author} {\bibfnamefont {L.~B.}\
  \bibnamefont {{Newburgh}}}, \bibinfo {author} {\bibfnamefont {M.~D.}\
  \bibnamefont {{Niemack}}}, \bibinfo {author} {\bibfnamefont {L.~A.}\
  \bibnamefont {{Page}}}, \bibinfo {author} {\bibfnamefont {B.}~\bibnamefont
  {{Partridge}}}, \bibinfo {author} {\bibfnamefont {M.}~\bibnamefont
  {{Salatino}}}, \bibinfo {author} {\bibfnamefont {N.}~\bibnamefont
  {{Sehgal}}}, \bibinfo {author} {\bibfnamefont {A.}~\bibnamefont
  {{Schillaci}}}, \bibinfo {author} {\bibfnamefont {C.}~\bibnamefont
  {{Sif{\'o}n}}}, \bibinfo {author} {\bibfnamefont {K.~M.}\ \bibnamefont
  {{Smith}}}, \bibinfo {author} {\bibfnamefont {D.~N.}\ \bibnamefont
  {{Spergel}}}, \bibinfo {author} {\bibfnamefont {S.}~\bibnamefont {{Staggs}}},
  \bibinfo {author} {\bibfnamefont {E.~R.}\ \bibnamefont {{Storer}}}, \bibinfo
  {author} {\bibfnamefont {H.}~\bibnamefont {{Trac}}}, \bibinfo {author}
  {\bibfnamefont {J.~N.}\ \bibnamefont {{Ullom}}}, \bibinfo {author}
  {\bibfnamefont {J.}~\bibnamefont {{Van Lanen}}}, \bibinfo {author}
  {\bibfnamefont {L.~R.}\ \bibnamefont {{Vale}}}, \bibinfo {author}
  {\bibfnamefont {A.}~\bibnamefont {{van Engelen}}}, \bibinfo {author}
  {\bibfnamefont {M.~V.}\ \bibnamefont {{Maga{\~n}a}}}, \bibinfo {author}
  {\bibfnamefont {E.~M.}\ \bibnamefont {{Vavagiakis}}}, \bibinfo {author}
  {\bibfnamefont {E.~J.}\ \bibnamefont {{Wollack}}}, \bibinfo {author}
  {\bibfnamefont {Z.}~\bibnamefont {{Xu}}},\ and\ \bibinfo {author}
  {\bibnamefont {{Atacama Cosmology Telescope Collaboration}}},\ }\bibfield
  {title} {\bibinfo {title} {{Atacama Cosmology Telescope: Combined kinematic
  and thermal Sunyaev-Zel'dovich measurements from BOSS CMASS and LOWZ
  halos}},\ }\href {https://doi.org/10.1103/PhysRevD.103.063513} {\bibfield
  {journal} {\bibinfo  {journal} {\prd}\ }\textbf {\bibinfo {volume} {103}},\
  \bibinfo {eid} {063513} (\bibinfo {year} {2021})},\ \Eprint
  {https://arxiv.org/abs/2009.05557} {arXiv:2009.05557 [astro-ph.CO]}
  \BibitemShut {NoStop}%
\bibitem [{\citenamefont {{Hadzhiyska}}\ \emph
  {et~al.}(2024{\natexlab{a}})\citenamefont {{Hadzhiyska}}, \citenamefont
  {{Ferraro}}, \citenamefont {{Ried Guachalla}}, \citenamefont {{Schaan}},
  \citenamefont {{Aguilar}}, \citenamefont {{Battaglia}}, \citenamefont
  {{Bond}}, \citenamefont {{Brooks}}, \citenamefont {{Calabrese}},
  \citenamefont {{Choi}}, \citenamefont {{Claybaugh}}, \citenamefont
  {{Coulton}}, \citenamefont {{Dawson}}, \citenamefont {{Devlin}},
  \citenamefont {{Dey}}, \citenamefont {{Doel}}, \citenamefont
  {{Duivenvoorden}}, \citenamefont {{Dunkley}}, \citenamefont {{Farren}},
  \citenamefont {{Font-Ribera}}, \citenamefont {{Forero-Romero}}, \citenamefont
  {{Gallardo}}, \citenamefont {{Gazta{\~n}aga}}, \citenamefont {{Gontcho
  Gontcho}}, \citenamefont {{Gralla}}, \citenamefont {{Le Guillou}},
  \citenamefont {{Gutierrez}}, \citenamefont {{Guy}}, \citenamefont {{Hill}},
  \citenamefont {{Hlo{\v{z}}ek}}, \citenamefont {{Honscheid}}, \citenamefont
  {{Juneau}}, \citenamefont {{Kisner}}, \citenamefont {{Kremin}}, \citenamefont
  {{Landriau}}, \citenamefont {{Liu}}, \citenamefont {{Louis}}, \citenamefont
  {{MacCrann}}, \citenamefont {{de Macorra}}, \citenamefont {{Madhavacheril}},
  \citenamefont {{Manera}}, \citenamefont {{Meisner}}, \citenamefont
  {{Miquel}}, \citenamefont {{Moodley}}, \citenamefont {{Moustakas}},
  \citenamefont {{Mroczkowski}}, \citenamefont {{Naess}}, \citenamefont
  {{Newman}}, \citenamefont {{Niemack}}, \citenamefont {{Niz}}, \citenamefont
  {{Page}}, \citenamefont {{Palanque-Delabrouille}}, \citenamefont
  {{Partridge}}, \citenamefont {{Percival}}, \citenamefont {{Prada}},
  \citenamefont {{Qu}}, \citenamefont {{Rossi}}, \citenamefont {{Sanchez}},
  \citenamefont {{Schlegel}}, \citenamefont {{Schubnell}}, \citenamefont
  {{Sehgal}}, \citenamefont {{Seo}}, \citenamefont {{Sif{\'o}n}}, \citenamefont
  {{Spergel}}, \citenamefont {{Sprayberry}}, \citenamefont {{Staggs}},
  \citenamefont {{Tarl{\'e}}}, \citenamefont {{Vargas}}, \citenamefont
  {{Vavagiakis}}, \citenamefont {{Weaver}}, \citenamefont {{Wollack}},
  \citenamefont {{Zhou}},\ and\ \citenamefont {{Zou}}}]{2024arXiv240707152H}%
  \BibitemOpen
  \bibfield  {author} {\bibinfo {author} {\bibfnamefont {B.}~\bibnamefont
  {{Hadzhiyska}}}, \bibinfo {author} {\bibfnamefont {S.}~\bibnamefont
  {{Ferraro}}}, \bibinfo {author} {\bibfnamefont {B.}~\bibnamefont {{Ried
  Guachalla}}}, \bibinfo {author} {\bibfnamefont {E.}~\bibnamefont {{Schaan}}},
  \bibinfo {author} {\bibfnamefont {J.}~\bibnamefont {{Aguilar}}}, \bibinfo
  {author} {\bibfnamefont {N.}~\bibnamefont {{Battaglia}}}, \bibinfo {author}
  {\bibfnamefont {J.~R.}\ \bibnamefont {{Bond}}}, \bibinfo {author}
  {\bibfnamefont {D.}~\bibnamefont {{Brooks}}}, \bibinfo {author}
  {\bibfnamefont {E.}~\bibnamefont {{Calabrese}}}, \bibinfo {author}
  {\bibfnamefont {S.~K.}\ \bibnamefont {{Choi}}}, \bibinfo {author}
  {\bibfnamefont {T.}~\bibnamefont {{Claybaugh}}}, \bibinfo {author}
  {\bibfnamefont {W.~R.}\ \bibnamefont {{Coulton}}}, \bibinfo {author}
  {\bibfnamefont {K.}~\bibnamefont {{Dawson}}}, \bibinfo {author}
  {\bibfnamefont {M.}~\bibnamefont {{Devlin}}}, \bibinfo {author}
  {\bibfnamefont {B.}~\bibnamefont {{Dey}}}, \bibinfo {author} {\bibfnamefont
  {P.}~\bibnamefont {{Doel}}}, \bibinfo {author} {\bibfnamefont {A.~J.}\
  \bibnamefont {{Duivenvoorden}}}, \bibinfo {author} {\bibfnamefont
  {J.}~\bibnamefont {{Dunkley}}}, \bibinfo {author} {\bibfnamefont {G.~S.}\
  \bibnamefont {{Farren}}}, \bibinfo {author} {\bibfnamefont {A.}~\bibnamefont
  {{Font-Ribera}}}, \bibinfo {author} {\bibfnamefont {J.~E.}\ \bibnamefont
  {{Forero-Romero}}}, \bibinfo {author} {\bibfnamefont {P.~A.}\ \bibnamefont
  {{Gallardo}}}, \bibinfo {author} {\bibfnamefont {E.}~\bibnamefont
  {{Gazta{\~n}aga}}}, \bibinfo {author} {\bibfnamefont {S.}~\bibnamefont
  {{Gontcho Gontcho}}}, \bibinfo {author} {\bibfnamefont {M.}~\bibnamefont
  {{Gralla}}}, \bibinfo {author} {\bibfnamefont {L.}~\bibnamefont {{Le
  Guillou}}}, \bibinfo {author} {\bibfnamefont {G.}~\bibnamefont
  {{Gutierrez}}}, \bibinfo {author} {\bibfnamefont {J.}~\bibnamefont {{Guy}}},
  \bibinfo {author} {\bibfnamefont {J.~C.}\ \bibnamefont {{Hill}}}, \bibinfo
  {author} {\bibfnamefont {R.}~\bibnamefont {{Hlo{\v{z}}ek}}}, \bibinfo
  {author} {\bibfnamefont {K.}~\bibnamefont {{Honscheid}}}, \bibinfo {author}
  {\bibfnamefont {S.}~\bibnamefont {{Juneau}}}, \bibinfo {author}
  {\bibfnamefont {T.}~\bibnamefont {{Kisner}}}, \bibinfo {author}
  {\bibfnamefont {A.}~\bibnamefont {{Kremin}}}, \bibinfo {author}
  {\bibfnamefont {M.}~\bibnamefont {{Landriau}}}, \bibinfo {author}
  {\bibfnamefont {R.~H.}\ \bibnamefont {{Liu}}}, \bibinfo {author}
  {\bibfnamefont {T.}~\bibnamefont {{Louis}}}, \bibinfo {author} {\bibfnamefont
  {N.}~\bibnamefont {{MacCrann}}}, \bibinfo {author} {\bibfnamefont
  {A.}~\bibnamefont {{de Macorra}}}, \bibinfo {author} {\bibfnamefont
  {M.}~\bibnamefont {{Madhavacheril}}}, \bibinfo {author} {\bibfnamefont
  {M.}~\bibnamefont {{Manera}}}, \bibinfo {author} {\bibfnamefont
  {A.}~\bibnamefont {{Meisner}}}, \bibinfo {author} {\bibfnamefont
  {R.}~\bibnamefont {{Miquel}}}, \bibinfo {author} {\bibfnamefont
  {K.}~\bibnamefont {{Moodley}}}, \bibinfo {author} {\bibfnamefont
  {J.}~\bibnamefont {{Moustakas}}}, \bibinfo {author} {\bibfnamefont
  {T.}~\bibnamefont {{Mroczkowski}}}, \bibinfo {author} {\bibfnamefont
  {S.}~\bibnamefont {{Naess}}}, \bibinfo {author} {\bibfnamefont
  {J.}~\bibnamefont {{Newman}}}, \bibinfo {author} {\bibfnamefont {M.~D.}\
  \bibnamefont {{Niemack}}}, \bibinfo {author} {\bibfnamefont {G.}~\bibnamefont
  {{Niz}}}, \bibinfo {author} {\bibfnamefont {L.}~\bibnamefont {{Page}}},
  \bibinfo {author} {\bibfnamefont {N.}~\bibnamefont
  {{Palanque-Delabrouille}}}, \bibinfo {author} {\bibfnamefont
  {B.}~\bibnamefont {{Partridge}}}, \bibinfo {author} {\bibfnamefont {W.~J.}\
  \bibnamefont {{Percival}}}, \bibinfo {author} {\bibfnamefont
  {F.}~\bibnamefont {{Prada}}}, \bibinfo {author} {\bibfnamefont {F.~J.}\
  \bibnamefont {{Qu}}}, \bibinfo {author} {\bibfnamefont {G.}~\bibnamefont
  {{Rossi}}}, \bibinfo {author} {\bibfnamefont {E.}~\bibnamefont {{Sanchez}}},
  \bibinfo {author} {\bibfnamefont {D.}~\bibnamefont {{Schlegel}}}, \bibinfo
  {author} {\bibfnamefont {M.}~\bibnamefont {{Schubnell}}}, \bibinfo {author}
  {\bibfnamefont {N.}~\bibnamefont {{Sehgal}}}, \bibinfo {author}
  {\bibfnamefont {H.}~\bibnamefont {{Seo}}}, \bibinfo {author} {\bibfnamefont
  {C.}~\bibnamefont {{Sif{\'o}n}}}, \bibinfo {author} {\bibfnamefont
  {D.}~\bibnamefont {{Spergel}}}, \bibinfo {author} {\bibfnamefont
  {D.}~\bibnamefont {{Sprayberry}}}, \bibinfo {author} {\bibfnamefont
  {S.}~\bibnamefont {{Staggs}}}, \bibinfo {author} {\bibfnamefont
  {G.}~\bibnamefont {{Tarl{\'e}}}}, \bibinfo {author} {\bibfnamefont
  {C.}~\bibnamefont {{Vargas}}}, \bibinfo {author} {\bibfnamefont {E.~M.}\
  \bibnamefont {{Vavagiakis}}}, \bibinfo {author} {\bibfnamefont {B.~A.}\
  \bibnamefont {{Weaver}}}, \bibinfo {author} {\bibfnamefont {E.~J.}\
  \bibnamefont {{Wollack}}}, \bibinfo {author} {\bibfnamefont {R.}~\bibnamefont
  {{Zhou}}},\ and\ \bibinfo {author} {\bibfnamefont {H.}~\bibnamefont
  {{Zou}}},\ }\bibfield  {title} {\bibinfo {title} {{Evidence for large
  baryonic feedback at low and intermediate redshifts from kinematic
  Sunyaev-Zel'dovich observations with ACT and DESI photometric galaxies}},\
  }\href {https://doi.org/10.48550/arXiv.2407.07152} {\bibfield  {journal}
  {\bibinfo  {journal} {arXiv e-prints}\ ,\ \bibinfo {eid} {arXiv:2407.07152}}
  (\bibinfo {year} {2024}{\natexlab{a}})},\ \Eprint
  {https://arxiv.org/abs/2407.07152} {arXiv:2407.07152 [astro-ph.CO]}
  \BibitemShut {NoStop}%
\bibitem [{\citenamefont {{Plagge}}\ \emph {et~al.}(2010)\citenamefont
  {{Plagge}}, \citenamefont {{Benson}}, \citenamefont {{Ade}}, \citenamefont
  {{Aird}}, \citenamefont {{Bleem}}, \citenamefont {{Carlstrom}}, \citenamefont
  {{Chang}}, \citenamefont {{Cho}}, \citenamefont {{Crawford}}, \citenamefont
  {{Crites}}, \citenamefont {{de Haan}}, \citenamefont {{Dobbs}}, \citenamefont
  {{George}}, \citenamefont {{Hall}}, \citenamefont {{Halverson}},
  \citenamefont {{Holder}}, \citenamefont {{Holzapfel}}, \citenamefont
  {{Hrubes}}, \citenamefont {{Joy}}, \citenamefont {{Keisler}}, \citenamefont
  {{Knox}}, \citenamefont {{Lee}}, \citenamefont {{Leitch}}, \citenamefont
  {{Lueker}}, \citenamefont {{Marrone}}, \citenamefont {{McMahon}},
  \citenamefont {{Mehl}}, \citenamefont {{Meyer}}, \citenamefont {{Mohr}},
  \citenamefont {{Montroy}}, \citenamefont {{Padin}}, \citenamefont {{Pryke}},
  \citenamefont {{Reichardt}}, \citenamefont {{Ruhl}}, \citenamefont
  {{Schaffer}}, \citenamefont {{Shaw}}, \citenamefont {{Shirokoff}},
  \citenamefont {{Spieler}}, \citenamefont {{Stalder}}, \citenamefont
  {{Staniszewski}}, \citenamefont {{Stark}}, \citenamefont {{Vanderlinde}},
  \citenamefont {{Vieira}}, \citenamefont {{Williamson}},\ and\ \citenamefont
  {{Zahn}}}]{2010ApJ...716.1118P}%
  \BibitemOpen
  \bibfield  {author} {\bibinfo {author} {\bibfnamefont {T.}~\bibnamefont
  {{Plagge}}}, \bibinfo {author} {\bibfnamefont {B.~A.}\ \bibnamefont
  {{Benson}}}, \bibinfo {author} {\bibfnamefont {P.~A.~R.}\ \bibnamefont
  {{Ade}}}, \bibinfo {author} {\bibfnamefont {K.~A.}\ \bibnamefont {{Aird}}},
  \bibinfo {author} {\bibfnamefont {L.~E.}\ \bibnamefont {{Bleem}}}, \bibinfo
  {author} {\bibfnamefont {J.~E.}\ \bibnamefont {{Carlstrom}}}, \bibinfo
  {author} {\bibfnamefont {C.~L.}\ \bibnamefont {{Chang}}}, \bibinfo {author}
  {\bibfnamefont {H.~M.}\ \bibnamefont {{Cho}}}, \bibinfo {author}
  {\bibfnamefont {T.~M.}\ \bibnamefont {{Crawford}}}, \bibinfo {author}
  {\bibfnamefont {A.~T.}\ \bibnamefont {{Crites}}}, \bibinfo {author}
  {\bibfnamefont {T.}~\bibnamefont {{de Haan}}}, \bibinfo {author}
  {\bibfnamefont {M.~A.}\ \bibnamefont {{Dobbs}}}, \bibinfo {author}
  {\bibfnamefont {E.~M.}\ \bibnamefont {{George}}}, \bibinfo {author}
  {\bibfnamefont {N.~R.}\ \bibnamefont {{Hall}}}, \bibinfo {author}
  {\bibfnamefont {N.~W.}\ \bibnamefont {{Halverson}}}, \bibinfo {author}
  {\bibfnamefont {G.~P.}\ \bibnamefont {{Holder}}}, \bibinfo {author}
  {\bibfnamefont {W.~L.}\ \bibnamefont {{Holzapfel}}}, \bibinfo {author}
  {\bibfnamefont {J.~D.}\ \bibnamefont {{Hrubes}}}, \bibinfo {author}
  {\bibfnamefont {M.}~\bibnamefont {{Joy}}}, \bibinfo {author} {\bibfnamefont
  {R.}~\bibnamefont {{Keisler}}}, \bibinfo {author} {\bibfnamefont
  {L.}~\bibnamefont {{Knox}}}, \bibinfo {author} {\bibfnamefont {A.~T.}\
  \bibnamefont {{Lee}}}, \bibinfo {author} {\bibfnamefont {E.~M.}\ \bibnamefont
  {{Leitch}}}, \bibinfo {author} {\bibfnamefont {M.}~\bibnamefont {{Lueker}}},
  \bibinfo {author} {\bibfnamefont {D.}~\bibnamefont {{Marrone}}}, \bibinfo
  {author} {\bibfnamefont {J.~J.}\ \bibnamefont {{McMahon}}}, \bibinfo {author}
  {\bibfnamefont {J.}~\bibnamefont {{Mehl}}}, \bibinfo {author} {\bibfnamefont
  {S.~S.}\ \bibnamefont {{Meyer}}}, \bibinfo {author} {\bibfnamefont {J.~J.}\
  \bibnamefont {{Mohr}}}, \bibinfo {author} {\bibfnamefont {T.~E.}\
  \bibnamefont {{Montroy}}}, \bibinfo {author} {\bibfnamefont {S.}~\bibnamefont
  {{Padin}}}, \bibinfo {author} {\bibfnamefont {C.}~\bibnamefont {{Pryke}}},
  \bibinfo {author} {\bibfnamefont {C.~L.}\ \bibnamefont {{Reichardt}}},
  \bibinfo {author} {\bibfnamefont {J.~E.}\ \bibnamefont {{Ruhl}}}, \bibinfo
  {author} {\bibfnamefont {K.~K.}\ \bibnamefont {{Schaffer}}}, \bibinfo
  {author} {\bibfnamefont {L.}~\bibnamefont {{Shaw}}}, \bibinfo {author}
  {\bibfnamefont {E.}~\bibnamefont {{Shirokoff}}}, \bibinfo {author}
  {\bibfnamefont {H.~G.}\ \bibnamefont {{Spieler}}}, \bibinfo {author}
  {\bibfnamefont {B.}~\bibnamefont {{Stalder}}}, \bibinfo {author}
  {\bibfnamefont {Z.}~\bibnamefont {{Staniszewski}}}, \bibinfo {author}
  {\bibfnamefont {A.~A.}\ \bibnamefont {{Stark}}}, \bibinfo {author}
  {\bibfnamefont {K.}~\bibnamefont {{Vanderlinde}}}, \bibinfo {author}
  {\bibfnamefont {J.~D.}\ \bibnamefont {{Vieira}}}, \bibinfo {author}
  {\bibfnamefont {R.}~\bibnamefont {{Williamson}}},\ and\ \bibinfo {author}
  {\bibfnamefont {O.}~\bibnamefont {{Zahn}}},\ }\bibfield  {title} {\bibinfo
  {title} {{Sunyaev-Zel'dovich Cluster Profiles Measured with the South Pole
  Telescope}},\ }\href {https://doi.org/10.1088/0004-637X/716/2/1118}
  {\bibfield  {journal} {\bibinfo  {journal} {\apj}\ }\textbf {\bibinfo
  {volume} {716}},\ \bibinfo {pages} {1118} (\bibinfo {year} {2010})},\ \Eprint
  {https://arxiv.org/abs/0911.2444} {arXiv:0911.2444 [astro-ph.CO]}
  \BibitemShut {NoStop}%
\bibitem [{\citenamefont {{Hand}}\ \emph {et~al.}(2011)\citenamefont {{Hand}},
  \citenamefont {{Appel}}, \citenamefont {{Battaglia}}, \citenamefont {{Bond}},
  \citenamefont {{Das}}, \citenamefont {{Devlin}}, \citenamefont {{Dunkley}},
  \citenamefont {{D{\"u}nner}}, \citenamefont {{Essinger-Hileman}},
  \citenamefont {{Fowler}}, \citenamefont {{Hajian}}, \citenamefont
  {{Halpern}}, \citenamefont {{Hasselfield}}, \citenamefont {{Hilton}},
  \citenamefont {{Hincks}}, \citenamefont {{Hlozek}}, \citenamefont {{Hughes}},
  \citenamefont {{Irwin}}, \citenamefont {{Klein}}, \citenamefont {{Kosowsky}},
  \citenamefont {{Lin}}, \citenamefont {{Marriage}}, \citenamefont {{Marsden}},
  \citenamefont {{McLaren}}, \citenamefont {{Menanteau}}, \citenamefont
  {{Moodley}}, \citenamefont {{Niemack}}, \citenamefont {{Nolta}},
  \citenamefont {{Page}}, \citenamefont {{Parker}}, \citenamefont
  {{Partridge}}, \citenamefont {{Plimpton}}, \citenamefont {{Reese}},
  \citenamefont {{Rojas}}, \citenamefont {{Sehgal}}, \citenamefont {{Sherwin}},
  \citenamefont {{Sievers}}, \citenamefont {{Spergel}}, \citenamefont
  {{Staggs}}, \citenamefont {{Swetz}}, \citenamefont {{Switzer}}, \citenamefont
  {{Thornton}}, \citenamefont {{Trac}}, \citenamefont {{Visnjic}},\ and\
  \citenamefont {{Wollack}}}]{2011ApJ...736...39H}%
  \BibitemOpen
  \bibfield  {author} {\bibinfo {author} {\bibfnamefont {N.}~\bibnamefont
  {{Hand}}}, \bibinfo {author} {\bibfnamefont {J.~W.}\ \bibnamefont {{Appel}}},
  \bibinfo {author} {\bibfnamefont {N.}~\bibnamefont {{Battaglia}}}, \bibinfo
  {author} {\bibfnamefont {J.~R.}\ \bibnamefont {{Bond}}}, \bibinfo {author}
  {\bibfnamefont {S.}~\bibnamefont {{Das}}}, \bibinfo {author} {\bibfnamefont
  {M.~J.}\ \bibnamefont {{Devlin}}}, \bibinfo {author} {\bibfnamefont
  {J.}~\bibnamefont {{Dunkley}}}, \bibinfo {author} {\bibfnamefont
  {R.}~\bibnamefont {{D{\"u}nner}}}, \bibinfo {author} {\bibfnamefont
  {T.}~\bibnamefont {{Essinger-Hileman}}}, \bibinfo {author} {\bibfnamefont
  {J.~W.}\ \bibnamefont {{Fowler}}}, \bibinfo {author} {\bibfnamefont
  {A.}~\bibnamefont {{Hajian}}}, \bibinfo {author} {\bibfnamefont
  {M.}~\bibnamefont {{Halpern}}}, \bibinfo {author} {\bibfnamefont
  {M.}~\bibnamefont {{Hasselfield}}}, \bibinfo {author} {\bibfnamefont
  {M.}~\bibnamefont {{Hilton}}}, \bibinfo {author} {\bibfnamefont {A.~D.}\
  \bibnamefont {{Hincks}}}, \bibinfo {author} {\bibfnamefont {R.}~\bibnamefont
  {{Hlozek}}}, \bibinfo {author} {\bibfnamefont {J.~P.}\ \bibnamefont
  {{Hughes}}}, \bibinfo {author} {\bibfnamefont {K.~D.}\ \bibnamefont
  {{Irwin}}}, \bibinfo {author} {\bibfnamefont {J.}~\bibnamefont {{Klein}}},
  \bibinfo {author} {\bibfnamefont {A.}~\bibnamefont {{Kosowsky}}}, \bibinfo
  {author} {\bibfnamefont {Y.-T.}\ \bibnamefont {{Lin}}}, \bibinfo {author}
  {\bibfnamefont {T.~A.}\ \bibnamefont {{Marriage}}}, \bibinfo {author}
  {\bibfnamefont {D.}~\bibnamefont {{Marsden}}}, \bibinfo {author}
  {\bibfnamefont {M.}~\bibnamefont {{McLaren}}}, \bibinfo {author}
  {\bibfnamefont {F.}~\bibnamefont {{Menanteau}}}, \bibinfo {author}
  {\bibfnamefont {K.}~\bibnamefont {{Moodley}}}, \bibinfo {author}
  {\bibfnamefont {M.~D.}\ \bibnamefont {{Niemack}}}, \bibinfo {author}
  {\bibfnamefont {M.~R.}\ \bibnamefont {{Nolta}}}, \bibinfo {author}
  {\bibfnamefont {L.~A.}\ \bibnamefont {{Page}}}, \bibinfo {author}
  {\bibfnamefont {L.}~\bibnamefont {{Parker}}}, \bibinfo {author}
  {\bibfnamefont {B.}~\bibnamefont {{Partridge}}}, \bibinfo {author}
  {\bibfnamefont {R.}~\bibnamefont {{Plimpton}}}, \bibinfo {author}
  {\bibfnamefont {E.~D.}\ \bibnamefont {{Reese}}}, \bibinfo {author}
  {\bibfnamefont {F.}~\bibnamefont {{Rojas}}}, \bibinfo {author} {\bibfnamefont
  {N.}~\bibnamefont {{Sehgal}}}, \bibinfo {author} {\bibfnamefont {B.~D.}\
  \bibnamefont {{Sherwin}}}, \bibinfo {author} {\bibfnamefont {J.~L.}\
  \bibnamefont {{Sievers}}}, \bibinfo {author} {\bibfnamefont {D.~N.}\
  \bibnamefont {{Spergel}}}, \bibinfo {author} {\bibfnamefont {S.~T.}\
  \bibnamefont {{Staggs}}}, \bibinfo {author} {\bibfnamefont {D.~S.}\
  \bibnamefont {{Swetz}}}, \bibinfo {author} {\bibfnamefont {E.~R.}\
  \bibnamefont {{Switzer}}}, \bibinfo {author} {\bibfnamefont {R.}~\bibnamefont
  {{Thornton}}}, \bibinfo {author} {\bibfnamefont {H.}~\bibnamefont {{Trac}}},
  \bibinfo {author} {\bibfnamefont {K.}~\bibnamefont {{Visnjic}}},\ and\
  \bibinfo {author} {\bibfnamefont {E.}~\bibnamefont {{Wollack}}},\ }\bibfield
  {title} {\bibinfo {title} {{The Atacama Cosmology Telescope: Detection of
  Sunyaev-Zel'Dovich Decrement in Groups and Clusters Associated with Luminous
  Red Galaxies}},\ }\href {https://doi.org/10.1088/0004-637X/736/1/39}
  {\bibfield  {journal} {\bibinfo  {journal} {\apj}\ }\textbf {\bibinfo
  {volume} {736}},\ \bibinfo {eid} {39} (\bibinfo {year} {2011})},\ \Eprint
  {https://arxiv.org/abs/1101.1951} {arXiv:1101.1951 [astro-ph.CO]}
  \BibitemShut {NoStop}%
\bibitem [{\citenamefont {{Sehgal}}\ \emph {et~al.}(2011)\citenamefont
  {{Sehgal}}, \citenamefont {{Trac}}, \citenamefont {{Acquaviva}},
  \citenamefont {{Ade}}, \citenamefont {{Aguirre}}, \citenamefont {{Amiri}},
  \citenamefont {{Appel}}, \citenamefont {{Barrientos}}, \citenamefont
  {{Battistelli}}, \citenamefont {{Bond}}, \citenamefont {{Brown}},
  \citenamefont {{Burger}}, \citenamefont {{Chervenak}}, \citenamefont {{Das}},
  \citenamefont {{Devlin}}, \citenamefont {{Dicker}}, \citenamefont {{Bertrand
  Doriese}}, \citenamefont {{Dunkley}}, \citenamefont {{D{\"u}nner}},
  \citenamefont {{Essinger-Hileman}}, \citenamefont {{Fisher}}, \citenamefont
  {{Fowler}}, \citenamefont {{Hajian}}, \citenamefont {{Halpern}},
  \citenamefont {{Hasselfield}}, \citenamefont {{Hern{\'a}ndez-Monteagudo}},
  \citenamefont {{Hilton}}, \citenamefont {{Hilton}}, \citenamefont {{Hincks}},
  \citenamefont {{Hlozek}}, \citenamefont {{Holtz}}, \citenamefont
  {{Huffenberger}}, \citenamefont {{Hughes}}, \citenamefont {{Hughes}},
  \citenamefont {{Infante}}, \citenamefont {{Irwin}}, \citenamefont {{Jones}},
  \citenamefont {{Baptiste Juin}}, \citenamefont {{Klein}}, \citenamefont
  {{Kosowsky}}, \citenamefont {{Lau}}, \citenamefont {{Limon}}, \citenamefont
  {{Lin}}, \citenamefont {{Lupton}}, \citenamefont {{Marriage}}, \citenamefont
  {{Marsden}}, \citenamefont {{Martocci}}, \citenamefont {{Mauskopf}},
  \citenamefont {{Menanteau}}, \citenamefont {{Moodley}}, \citenamefont
  {{Moseley}}, \citenamefont {{Netterfield}}, \citenamefont {{Niemack}},
  \citenamefont {{Nolta}}, \citenamefont {{Page}}, \citenamefont {{Parker}},
  \citenamefont {{Partridge}}, \citenamefont {{Reid}}, \citenamefont
  {{Sherwin}}, \citenamefont {{Sievers}}, \citenamefont {{Spergel}},
  \citenamefont {{Staggs}}, \citenamefont {{Swetz}}, \citenamefont {{Switzer}},
  \citenamefont {{Thornton}}, \citenamefont {{Tucker}}, \citenamefont
  {{Warne}}, \citenamefont {{Wollack}},\ and\ \citenamefont
  {{Zhao}}}]{2011ApJ...732...44S}%
  \BibitemOpen
  \bibfield  {author} {\bibinfo {author} {\bibfnamefont {N.}~\bibnamefont
  {{Sehgal}}}, \bibinfo {author} {\bibfnamefont {H.}~\bibnamefont {{Trac}}},
  \bibinfo {author} {\bibfnamefont {V.}~\bibnamefont {{Acquaviva}}}, \bibinfo
  {author} {\bibfnamefont {P.~A.~R.}\ \bibnamefont {{Ade}}}, \bibinfo {author}
  {\bibfnamefont {P.}~\bibnamefont {{Aguirre}}}, \bibinfo {author}
  {\bibfnamefont {M.}~\bibnamefont {{Amiri}}}, \bibinfo {author} {\bibfnamefont
  {J.~W.}\ \bibnamefont {{Appel}}}, \bibinfo {author} {\bibfnamefont {L.~F.}\
  \bibnamefont {{Barrientos}}}, \bibinfo {author} {\bibfnamefont {E.~S.}\
  \bibnamefont {{Battistelli}}}, \bibinfo {author} {\bibfnamefont {J.~R.}\
  \bibnamefont {{Bond}}}, \bibinfo {author} {\bibfnamefont {B.}~\bibnamefont
  {{Brown}}}, \bibinfo {author} {\bibfnamefont {B.}~\bibnamefont {{Burger}}},
  \bibinfo {author} {\bibfnamefont {J.}~\bibnamefont {{Chervenak}}}, \bibinfo
  {author} {\bibfnamefont {S.}~\bibnamefont {{Das}}}, \bibinfo {author}
  {\bibfnamefont {M.~J.}\ \bibnamefont {{Devlin}}}, \bibinfo {author}
  {\bibfnamefont {S.~R.}\ \bibnamefont {{Dicker}}}, \bibinfo {author}
  {\bibfnamefont {W.}~\bibnamefont {{Bertrand Doriese}}}, \bibinfo {author}
  {\bibfnamefont {J.}~\bibnamefont {{Dunkley}}}, \bibinfo {author}
  {\bibfnamefont {R.}~\bibnamefont {{D{\"u}nner}}}, \bibinfo {author}
  {\bibfnamefont {T.}~\bibnamefont {{Essinger-Hileman}}}, \bibinfo {author}
  {\bibfnamefont {R.~P.}\ \bibnamefont {{Fisher}}}, \bibinfo {author}
  {\bibfnamefont {J.~W.}\ \bibnamefont {{Fowler}}}, \bibinfo {author}
  {\bibfnamefont {A.}~\bibnamefont {{Hajian}}}, \bibinfo {author}
  {\bibfnamefont {M.}~\bibnamefont {{Halpern}}}, \bibinfo {author}
  {\bibfnamefont {M.}~\bibnamefont {{Hasselfield}}}, \bibinfo {author}
  {\bibfnamefont {C.}~\bibnamefont {{Hern{\'a}ndez-Monteagudo}}}, \bibinfo
  {author} {\bibfnamefont {G.~C.}\ \bibnamefont {{Hilton}}}, \bibinfo {author}
  {\bibfnamefont {M.}~\bibnamefont {{Hilton}}}, \bibinfo {author}
  {\bibfnamefont {A.~D.}\ \bibnamefont {{Hincks}}}, \bibinfo {author}
  {\bibfnamefont {R.}~\bibnamefont {{Hlozek}}}, \bibinfo {author}
  {\bibfnamefont {D.}~\bibnamefont {{Holtz}}}, \bibinfo {author} {\bibfnamefont
  {K.~M.}\ \bibnamefont {{Huffenberger}}}, \bibinfo {author} {\bibfnamefont
  {D.~H.}\ \bibnamefont {{Hughes}}}, \bibinfo {author} {\bibfnamefont {J.~P.}\
  \bibnamefont {{Hughes}}}, \bibinfo {author} {\bibfnamefont {L.}~\bibnamefont
  {{Infante}}}, \bibinfo {author} {\bibfnamefont {K.~D.}\ \bibnamefont
  {{Irwin}}}, \bibinfo {author} {\bibfnamefont {A.}~\bibnamefont {{Jones}}},
  \bibinfo {author} {\bibfnamefont {J.}~\bibnamefont {{Baptiste Juin}}},
  \bibinfo {author} {\bibfnamefont {J.}~\bibnamefont {{Klein}}}, \bibinfo
  {author} {\bibfnamefont {A.}~\bibnamefont {{Kosowsky}}}, \bibinfo {author}
  {\bibfnamefont {J.~M.}\ \bibnamefont {{Lau}}}, \bibinfo {author}
  {\bibfnamefont {M.}~\bibnamefont {{Limon}}}, \bibinfo {author} {\bibfnamefont
  {Y.-T.}\ \bibnamefont {{Lin}}}, \bibinfo {author} {\bibfnamefont {R.~H.}\
  \bibnamefont {{Lupton}}}, \bibinfo {author} {\bibfnamefont {T.~A.}\
  \bibnamefont {{Marriage}}}, \bibinfo {author} {\bibfnamefont
  {D.}~\bibnamefont {{Marsden}}}, \bibinfo {author} {\bibfnamefont
  {K.}~\bibnamefont {{Martocci}}}, \bibinfo {author} {\bibfnamefont
  {P.}~\bibnamefont {{Mauskopf}}}, \bibinfo {author} {\bibfnamefont
  {F.}~\bibnamefont {{Menanteau}}}, \bibinfo {author} {\bibfnamefont
  {K.}~\bibnamefont {{Moodley}}}, \bibinfo {author} {\bibfnamefont
  {H.}~\bibnamefont {{Moseley}}}, \bibinfo {author} {\bibfnamefont {C.~B.}\
  \bibnamefont {{Netterfield}}}, \bibinfo {author} {\bibfnamefont {M.~D.}\
  \bibnamefont {{Niemack}}}, \bibinfo {author} {\bibfnamefont {M.~R.}\
  \bibnamefont {{Nolta}}}, \bibinfo {author} {\bibfnamefont {L.~A.}\
  \bibnamefont {{Page}}}, \bibinfo {author} {\bibfnamefont {L.}~\bibnamefont
  {{Parker}}}, \bibinfo {author} {\bibfnamefont {B.}~\bibnamefont
  {{Partridge}}}, \bibinfo {author} {\bibfnamefont {B.}~\bibnamefont {{Reid}}},
  \bibinfo {author} {\bibfnamefont {B.~D.}\ \bibnamefont {{Sherwin}}}, \bibinfo
  {author} {\bibfnamefont {J.}~\bibnamefont {{Sievers}}}, \bibinfo {author}
  {\bibfnamefont {D.~N.}\ \bibnamefont {{Spergel}}}, \bibinfo {author}
  {\bibfnamefont {S.~T.}\ \bibnamefont {{Staggs}}}, \bibinfo {author}
  {\bibfnamefont {D.~S.}\ \bibnamefont {{Swetz}}}, \bibinfo {author}
  {\bibfnamefont {E.~R.}\ \bibnamefont {{Switzer}}}, \bibinfo {author}
  {\bibfnamefont {R.}~\bibnamefont {{Thornton}}}, \bibinfo {author}
  {\bibfnamefont {C.}~\bibnamefont {{Tucker}}}, \bibinfo {author}
  {\bibfnamefont {R.}~\bibnamefont {{Warne}}}, \bibinfo {author} {\bibfnamefont
  {E.}~\bibnamefont {{Wollack}}},\ and\ \bibinfo {author} {\bibfnamefont
  {Y.}~\bibnamefont {{Zhao}}},\ }\bibfield  {title} {\bibinfo {title} {{The
  Atacama Cosmology Telescope: Cosmology from Galaxy Clusters Detected via the
  Sunyaev-Zel'dovich Effect}},\ }\href
  {https://doi.org/10.1088/0004-637X/732/1/44} {\bibfield  {journal} {\bibinfo
  {journal} {\apj}\ }\textbf {\bibinfo {volume} {732}},\ \bibinfo {eid} {44}
  (\bibinfo {year} {2011})},\ \Eprint {https://arxiv.org/abs/1010.1025}
  {arXiv:1010.1025 [astro-ph.CO]} \BibitemShut {NoStop}%
\bibitem [{\citenamefont {{Sehgal}}\ \emph {et~al.}(2013)\citenamefont
  {{Sehgal}}, \citenamefont {{Addison}}, \citenamefont {{Battaglia}},
  \citenamefont {{Battistelli}}, \citenamefont {{Bond}}, \citenamefont {{Das}},
  \citenamefont {{Devlin}}, \citenamefont {{Dunkley}}, \citenamefont
  {{D{\"u}nner}}, \citenamefont {{Gralla}}, \citenamefont {{Hajian}},
  \citenamefont {{Halpern}}, \citenamefont {{Hasselfield}}, \citenamefont
  {{Hilton}}, \citenamefont {{Hincks}}, \citenamefont {{Hlozek}}, \citenamefont
  {{Hughes}}, \citenamefont {{Kosowsky}}, \citenamefont {{Lin}}, \citenamefont
  {{Louis}}, \citenamefont {{Marriage}}, \citenamefont {{Marsden}},
  \citenamefont {{Menanteau}}, \citenamefont {{Moodley}}, \citenamefont
  {{Niemack}}, \citenamefont {{Page}}, \citenamefont {{Partridge}},
  \citenamefont {{Reese}}, \citenamefont {{Sherwin}}, \citenamefont
  {{Sievers}}, \citenamefont {{Sif{\'o}n}}, \citenamefont {{Spergel}},
  \citenamefont {{Staggs}}, \citenamefont {{Swetz}}, \citenamefont
  {{Switzer}},\ and\ \citenamefont {{Wollack}}}]{2013ApJ...767...38S}%
  \BibitemOpen
  \bibfield  {author} {\bibinfo {author} {\bibfnamefont {N.}~\bibnamefont
  {{Sehgal}}}, \bibinfo {author} {\bibfnamefont {G.}~\bibnamefont {{Addison}}},
  \bibinfo {author} {\bibfnamefont {N.}~\bibnamefont {{Battaglia}}}, \bibinfo
  {author} {\bibfnamefont {E.~S.}\ \bibnamefont {{Battistelli}}}, \bibinfo
  {author} {\bibfnamefont {J.~R.}\ \bibnamefont {{Bond}}}, \bibinfo {author}
  {\bibfnamefont {S.}~\bibnamefont {{Das}}}, \bibinfo {author} {\bibfnamefont
  {M.~J.}\ \bibnamefont {{Devlin}}}, \bibinfo {author} {\bibfnamefont
  {J.}~\bibnamefont {{Dunkley}}}, \bibinfo {author} {\bibfnamefont
  {R.}~\bibnamefont {{D{\"u}nner}}}, \bibinfo {author} {\bibfnamefont
  {M.}~\bibnamefont {{Gralla}}}, \bibinfo {author} {\bibfnamefont
  {A.}~\bibnamefont {{Hajian}}}, \bibinfo {author} {\bibfnamefont
  {M.}~\bibnamefont {{Halpern}}}, \bibinfo {author} {\bibfnamefont
  {M.}~\bibnamefont {{Hasselfield}}}, \bibinfo {author} {\bibfnamefont
  {M.}~\bibnamefont {{Hilton}}}, \bibinfo {author} {\bibfnamefont {A.~D.}\
  \bibnamefont {{Hincks}}}, \bibinfo {author} {\bibfnamefont {R.}~\bibnamefont
  {{Hlozek}}}, \bibinfo {author} {\bibfnamefont {J.~P.}\ \bibnamefont
  {{Hughes}}}, \bibinfo {author} {\bibfnamefont {A.}~\bibnamefont
  {{Kosowsky}}}, \bibinfo {author} {\bibfnamefont {Y.-T.}\ \bibnamefont
  {{Lin}}}, \bibinfo {author} {\bibfnamefont {T.}~\bibnamefont {{Louis}}},
  \bibinfo {author} {\bibfnamefont {T.~A.}\ \bibnamefont {{Marriage}}},
  \bibinfo {author} {\bibfnamefont {D.}~\bibnamefont {{Marsden}}}, \bibinfo
  {author} {\bibfnamefont {F.}~\bibnamefont {{Menanteau}}}, \bibinfo {author}
  {\bibfnamefont {K.}~\bibnamefont {{Moodley}}}, \bibinfo {author}
  {\bibfnamefont {M.~D.}\ \bibnamefont {{Niemack}}}, \bibinfo {author}
  {\bibfnamefont {L.~A.}\ \bibnamefont {{Page}}}, \bibinfo {author}
  {\bibfnamefont {B.}~\bibnamefont {{Partridge}}}, \bibinfo {author}
  {\bibfnamefont {E.~D.}\ \bibnamefont {{Reese}}}, \bibinfo {author}
  {\bibfnamefont {B.~D.}\ \bibnamefont {{Sherwin}}}, \bibinfo {author}
  {\bibfnamefont {J.}~\bibnamefont {{Sievers}}}, \bibinfo {author}
  {\bibfnamefont {C.}~\bibnamefont {{Sif{\'o}n}}}, \bibinfo {author}
  {\bibfnamefont {D.~N.}\ \bibnamefont {{Spergel}}}, \bibinfo {author}
  {\bibfnamefont {S.~T.}\ \bibnamefont {{Staggs}}}, \bibinfo {author}
  {\bibfnamefont {D.~S.}\ \bibnamefont {{Swetz}}}, \bibinfo {author}
  {\bibfnamefont {E.~R.}\ \bibnamefont {{Switzer}}},\ and\ \bibinfo {author}
  {\bibfnamefont {E.}~\bibnamefont {{Wollack}}},\ }\bibfield  {title} {\bibinfo
  {title} {{The Atacama Cosmology Telescope: Relation between Galaxy Cluster
  Optical Richness and Sunyaev-Zel'dovich Effect}},\ }\href
  {https://doi.org/10.1088/0004-637X/767/1/38} {\bibfield  {journal} {\bibinfo
  {journal} {\apj}\ }\textbf {\bibinfo {volume} {767}},\ \bibinfo {eid} {38}
  (\bibinfo {year} {2013})},\ \Eprint {https://arxiv.org/abs/1205.2369}
  {arXiv:1205.2369 [astro-ph.CO]} \BibitemShut {NoStop}%
\bibitem [{\citenamefont {{Planck Collaboration}}\ \emph
  {et~al.}(2013)\citenamefont {{Planck Collaboration}}, \citenamefont {{Ade}},
  \citenamefont {{Aghanim}}, \citenamefont {{Arnaud}}, \citenamefont
  {{Ashdown}}, \citenamefont {{Atrio-Barandela}}, \citenamefont {{Aumont}},
  \citenamefont {{Baccigalupi}}, \citenamefont {{Balbi}}, \citenamefont
  {{Banday}}, \citenamefont {{Barreiro}}, \citenamefont {{Bartlett}},
  \citenamefont {{Battaner}}, \citenamefont {{Benabed}}, \citenamefont
  {{Beno{\^\i}t}}, \citenamefont {{Bernard}}, \citenamefont {{Bersanelli}},
  \citenamefont {{Bhatia}}, \citenamefont {{Bikmaev}}, \citenamefont {{Bobin}},
  \citenamefont {{B{\"o}hringer}}, \citenamefont {{Bonaldi}}, \citenamefont
  {{Bond}}, \citenamefont {{Borgani}}, \citenamefont {{Borrill}}, \citenamefont
  {{Bouchet}}, \citenamefont {{Bourdin}}, \citenamefont {{Brown}},
  \citenamefont {{Burenin}}, \citenamefont {{Burigana}}, \citenamefont
  {{Cabella}}, \citenamefont {{Cardoso}}, \citenamefont {{Carvalho}},
  \citenamefont {{Castex}}, \citenamefont {{Catalano}}, \citenamefont
  {{Cay{\'o}n}}, \citenamefont {{Chamballu}}, \citenamefont {{Chiang}},
  \citenamefont {{Chon}}, \citenamefont {{Christensen}}, \citenamefont
  {{Churazov}}, \citenamefont {{Clements}}, \citenamefont {{Colafrancesco}},
  \citenamefont {{Colombi}}, \citenamefont {{Colombo}}, \citenamefont
  {{Comis}}, \citenamefont {{Coulais}}, \citenamefont {{Crill}}, \citenamefont
  {{Cuttaia}}, \citenamefont {{Da Silva}}, \citenamefont {{Dahle}},
  \citenamefont {{Danese}}, \citenamefont {{Davis}}, \citenamefont {{de
  Bernardis}}, \citenamefont {{de Gasperis}}, \citenamefont {{de Zotti}},
  \citenamefont {{Delabrouille}}, \citenamefont {{D{\'e}mocl{\`e}s}},
  \citenamefont {{D{\'e}sert}}, \citenamefont {{Diego}}, \citenamefont
  {{Dolag}}, \citenamefont {{Dole}}, \citenamefont {{Donzelli}}, \citenamefont
  {{Dor{\'e}}}, \citenamefont {{D{\"o}rl}}, \citenamefont {{Douspis}},
  \citenamefont {{Dupac}}, \citenamefont {{Efstathiou}}, \citenamefont
  {{En{\ss}lin}}, \citenamefont {{Eriksen}}, \citenamefont {{Finelli}},
  \citenamefont {{Flores-Cacho}}, \citenamefont {{Forni}}, \citenamefont
  {{Fosalba}}, \citenamefont {{Frailis}}, \citenamefont {{Franceschi}},
  \citenamefont {{Frommert}}, \citenamefont {{Galeotta}}, \citenamefont
  {{Ganga}}, \citenamefont {{G{\'e}nova-Santos}}, \citenamefont {{Giard}},
  \citenamefont {{Giraud-H{\'e}raud}}, \citenamefont {{Gonz{\'a}lez-Nuevo}},
  \citenamefont {{G{\'o}rski}}, \citenamefont {{Gregorio}}, \citenamefont
  {{Gruppuso}}, \citenamefont {{Hansen}}, \citenamefont {{Harrison}},
  \citenamefont {{Hempel}}, \citenamefont {{Henrot-Versill{\'e}}},
  \citenamefont {{Hern{\'a}ndez-Monteagudo}}, \citenamefont {{Herranz}},
  \citenamefont {{Hildebrandt}}, \citenamefont {{Hivon}}, \citenamefont
  {{Hobson}}, \citenamefont {{Holmes}}, \citenamefont {{Hurier}}, \citenamefont
  {{Jaffe}}, \citenamefont {{Jaffe}}, \citenamefont {{Jagemann}}, \citenamefont
  {{Jones}}, \citenamefont {{Juvela}}, \citenamefont {{Keih{\"a}nen}},
  \citenamefont {{Khamitov}}, \citenamefont {{Kisner}}, \citenamefont
  {{Kneissl}}, \citenamefont {{Knoche}}, \citenamefont {{Knox}}, \citenamefont
  {{Kunz}}, \citenamefont {{Kurki-Suonio}}, \citenamefont {{Lagache}},
  \citenamefont {{L{\"a}hteenm{\"a}ki}}, \citenamefont {{Lamarre}},
  \citenamefont {{Lasenby}}, \citenamefont {{Lawrence}}, \citenamefont {{Le
  Jeune}}, \citenamefont {{Leonardi}}, \citenamefont {{Liddle}}, \citenamefont
  {{Lilje}}, \citenamefont {{L{\'o}pez-Caniego}}, \citenamefont {{Luzzi}},
  \citenamefont {{Mac{\'\i}as-P{\'e}rez}}, \citenamefont {{Maino}},
  \citenamefont {{Mandolesi}}, \citenamefont {{Maris}}, \citenamefont
  {{Marleau}}, \citenamefont {{Marshall}}, \citenamefont
  {{Mart{\'\i}nez-Gonz{\'a}lez}}, \citenamefont {{Masi}}, \citenamefont
  {{Massardi}}, \citenamefont {{Matarrese}}, \citenamefont {{Mazzotta}},
  \citenamefont {{Mei}}, \citenamefont {{Melchiorri}}, \citenamefont {{Melin}},
  \citenamefont {{Mendes}}, \citenamefont {{Mennella}}, \citenamefont
  {{Mitra}}, \citenamefont {{Miville-Desch{\^e}nes}}, \citenamefont {{Moneti}},
  \citenamefont {{Montier}}, \citenamefont {{Morgante}}, \citenamefont
  {{Mortlock}}, \citenamefont {{Munshi}}, \citenamefont {{Murphy}},
  \citenamefont {{Naselsky}}, \citenamefont {{Nati}}, \citenamefont {{Natoli}},
  \citenamefont {{N{\o}rgaard-Nielsen}}, \citenamefont {{Noviello}},
  \citenamefont {{Novikov}}, \citenamefont {{Novikov}}, \citenamefont
  {{Osborne}}, \citenamefont {{Pajot}}, \citenamefont {{Paoletti}},
  \citenamefont {{Pasian}}, \citenamefont {{Patanchon}}, \citenamefont
  {{Perdereau}}, \citenamefont {{Perotto}}, \citenamefont {{Perrotta}},
  \citenamefont {{Piacentini}}, \citenamefont {{Piat}}, \citenamefont
  {{Pierpaoli}}, \citenamefont {{Piffaretti}}, \citenamefont {{Plaszczynski}},
  \citenamefont {{Pointecouteau}}, \citenamefont {{Polenta}}, \citenamefont
  {{Ponthieu}}, \citenamefont {{Popa}}, \citenamefont {{Poutanen}},
  \citenamefont {{Pratt}}, \citenamefont {{Prunet}}, \citenamefont {{Puget}},
  \citenamefont {{Rachen}}, \citenamefont {{Reach}}, \citenamefont {{Rebolo}},
  \citenamefont {{Reinecke}}, \citenamefont {{Remazeilles}}, \citenamefont
  {{Renault}}, \citenamefont {{Ricciardi}}, \citenamefont {{Riller}},
  \citenamefont {{Ristorcelli}}, \citenamefont {{Rocha}}, \citenamefont
  {{Roman}}, \citenamefont {{Rosset}}, \citenamefont {{Rossetti}},
  \citenamefont {{Rubi{\~n}o-Mart{\'\i}n}}, \citenamefont {{Rusholme}},
  \citenamefont {{Sandri}}, \citenamefont {{Savini}}, \citenamefont {{Scott}},
  \citenamefont {{Smoot}}, \citenamefont {{Starck}}, \citenamefont
  {{Sudiwala}}, \citenamefont {{Sunyaev}}, \citenamefont {{Sutton}},
  \citenamefont {{Suur-Uski}}, \citenamefont {{Sygnet}}, \citenamefont
  {{Tauber}}, \citenamefont {{Terenzi}}, \citenamefont {{Toffolatti}},
  \citenamefont {{Tomasi}}, \citenamefont {{Tristram}}, \citenamefont
  {{Tuovinen}}, \citenamefont {{Valenziano}}, \citenamefont {{Van Tent}},
  \citenamefont {{Varis}}, \citenamefont {{Vielva}}, \citenamefont {{Villa}},
  \citenamefont {{Vittorio}}, \citenamefont {{Wade}}, \citenamefont
  {{Wandelt}}, \citenamefont {{Welikala}}, \citenamefont {{White}},
  \citenamefont {{White}}, \citenamefont {{Yvon}}, \citenamefont {{Zacchei}},\
  and\ \citenamefont {{Zonca}}}]{2013A&A...550A.131P}%
  \BibitemOpen
  \bibfield  {author} {\bibinfo {author} {\bibnamefont {{Planck
  Collaboration}}}, \bibinfo {author} {\bibfnamefont {P.~A.~R.}\ \bibnamefont
  {{Ade}}}, \bibinfo {author} {\bibfnamefont {N.}~\bibnamefont {{Aghanim}}},
  \bibinfo {author} {\bibfnamefont {M.}~\bibnamefont {{Arnaud}}}, \bibinfo
  {author} {\bibfnamefont {M.}~\bibnamefont {{Ashdown}}}, \bibinfo {author}
  {\bibfnamefont {F.}~\bibnamefont {{Atrio-Barandela}}}, \bibinfo {author}
  {\bibfnamefont {J.}~\bibnamefont {{Aumont}}}, \bibinfo {author}
  {\bibfnamefont {C.}~\bibnamefont {{Baccigalupi}}}, \bibinfo {author}
  {\bibfnamefont {A.}~\bibnamefont {{Balbi}}}, \bibinfo {author} {\bibfnamefont
  {A.~J.}\ \bibnamefont {{Banday}}}, \bibinfo {author} {\bibfnamefont {R.~B.}\
  \bibnamefont {{Barreiro}}}, \bibinfo {author} {\bibfnamefont {J.~G.}\
  \bibnamefont {{Bartlett}}}, \bibinfo {author} {\bibfnamefont
  {E.}~\bibnamefont {{Battaner}}}, \bibinfo {author} {\bibfnamefont
  {K.}~\bibnamefont {{Benabed}}}, \bibinfo {author} {\bibfnamefont
  {A.}~\bibnamefont {{Beno{\^\i}t}}}, \bibinfo {author} {\bibfnamefont {J.~P.}\
  \bibnamefont {{Bernard}}}, \bibinfo {author} {\bibfnamefont {M.}~\bibnamefont
  {{Bersanelli}}}, \bibinfo {author} {\bibfnamefont {R.}~\bibnamefont
  {{Bhatia}}}, \bibinfo {author} {\bibfnamefont {I.}~\bibnamefont {{Bikmaev}}},
  \bibinfo {author} {\bibfnamefont {J.}~\bibnamefont {{Bobin}}}, \bibinfo
  {author} {\bibfnamefont {H.}~\bibnamefont {{B{\"o}hringer}}}, \bibinfo
  {author} {\bibfnamefont {A.}~\bibnamefont {{Bonaldi}}}, \bibinfo {author}
  {\bibfnamefont {J.~R.}\ \bibnamefont {{Bond}}}, \bibinfo {author}
  {\bibfnamefont {S.}~\bibnamefont {{Borgani}}}, \bibinfo {author}
  {\bibfnamefont {J.}~\bibnamefont {{Borrill}}}, \bibinfo {author}
  {\bibfnamefont {F.~R.}\ \bibnamefont {{Bouchet}}}, \bibinfo {author}
  {\bibfnamefont {H.}~\bibnamefont {{Bourdin}}}, \bibinfo {author}
  {\bibfnamefont {M.~L.}\ \bibnamefont {{Brown}}}, \bibinfo {author}
  {\bibfnamefont {R.}~\bibnamefont {{Burenin}}}, \bibinfo {author}
  {\bibfnamefont {C.}~\bibnamefont {{Burigana}}}, \bibinfo {author}
  {\bibfnamefont {P.}~\bibnamefont {{Cabella}}}, \bibinfo {author}
  {\bibfnamefont {J.~F.}\ \bibnamefont {{Cardoso}}}, \bibinfo {author}
  {\bibfnamefont {P.}~\bibnamefont {{Carvalho}}}, \bibinfo {author}
  {\bibfnamefont {G.}~\bibnamefont {{Castex}}}, \bibinfo {author}
  {\bibfnamefont {A.}~\bibnamefont {{Catalano}}}, \bibinfo {author}
  {\bibfnamefont {L.}~\bibnamefont {{Cay{\'o}n}}}, \bibinfo {author}
  {\bibfnamefont {A.}~\bibnamefont {{Chamballu}}}, \bibinfo {author}
  {\bibfnamefont {L.~Y.}\ \bibnamefont {{Chiang}}}, \bibinfo {author}
  {\bibfnamefont {G.}~\bibnamefont {{Chon}}}, \bibinfo {author} {\bibfnamefont
  {P.~R.}\ \bibnamefont {{Christensen}}}, \bibinfo {author} {\bibfnamefont
  {E.}~\bibnamefont {{Churazov}}}, \bibinfo {author} {\bibfnamefont {D.~L.}\
  \bibnamefont {{Clements}}}, \bibinfo {author} {\bibfnamefont
  {S.}~\bibnamefont {{Colafrancesco}}}, \bibinfo {author} {\bibfnamefont
  {S.}~\bibnamefont {{Colombi}}}, \bibinfo {author} {\bibfnamefont {L.~P.~L.}\
  \bibnamefont {{Colombo}}}, \bibinfo {author} {\bibfnamefont {B.}~\bibnamefont
  {{Comis}}}, \bibinfo {author} {\bibfnamefont {A.}~\bibnamefont {{Coulais}}},
  \bibinfo {author} {\bibfnamefont {B.~P.}\ \bibnamefont {{Crill}}}, \bibinfo
  {author} {\bibfnamefont {F.}~\bibnamefont {{Cuttaia}}}, \bibinfo {author}
  {\bibfnamefont {A.}~\bibnamefont {{Da Silva}}}, \bibinfo {author}
  {\bibfnamefont {H.}~\bibnamefont {{Dahle}}}, \bibinfo {author} {\bibfnamefont
  {L.}~\bibnamefont {{Danese}}}, \bibinfo {author} {\bibfnamefont {R.~J.}\
  \bibnamefont {{Davis}}}, \bibinfo {author} {\bibfnamefont {P.}~\bibnamefont
  {{de Bernardis}}}, \bibinfo {author} {\bibfnamefont {G.}~\bibnamefont {{de
  Gasperis}}}, \bibinfo {author} {\bibfnamefont {G.}~\bibnamefont {{de
  Zotti}}}, \bibinfo {author} {\bibfnamefont {J.}~\bibnamefont
  {{Delabrouille}}}, \bibinfo {author} {\bibfnamefont {J.}~\bibnamefont
  {{D{\'e}mocl{\`e}s}}}, \bibinfo {author} {\bibfnamefont {F.~X.}\ \bibnamefont
  {{D{\'e}sert}}}, \bibinfo {author} {\bibfnamefont {J.~M.}\ \bibnamefont
  {{Diego}}}, \bibinfo {author} {\bibfnamefont {K.}~\bibnamefont {{Dolag}}},
  \bibinfo {author} {\bibfnamefont {H.}~\bibnamefont {{Dole}}}, \bibinfo
  {author} {\bibfnamefont {S.}~\bibnamefont {{Donzelli}}}, \bibinfo {author}
  {\bibfnamefont {O.}~\bibnamefont {{Dor{\'e}}}}, \bibinfo {author}
  {\bibfnamefont {U.}~\bibnamefont {{D{\"o}rl}}}, \bibinfo {author}
  {\bibfnamefont {M.}~\bibnamefont {{Douspis}}}, \bibinfo {author}
  {\bibfnamefont {X.}~\bibnamefont {{Dupac}}}, \bibinfo {author} {\bibfnamefont
  {G.}~\bibnamefont {{Efstathiou}}}, \bibinfo {author} {\bibfnamefont {T.~A.}\
  \bibnamefont {{En{\ss}lin}}}, \bibinfo {author} {\bibfnamefont {H.~K.}\
  \bibnamefont {{Eriksen}}}, \bibinfo {author} {\bibfnamefont {F.}~\bibnamefont
  {{Finelli}}}, \bibinfo {author} {\bibfnamefont {I.}~\bibnamefont
  {{Flores-Cacho}}}, \bibinfo {author} {\bibfnamefont {O.}~\bibnamefont
  {{Forni}}}, \bibinfo {author} {\bibfnamefont {P.}~\bibnamefont {{Fosalba}}},
  \bibinfo {author} {\bibfnamefont {M.}~\bibnamefont {{Frailis}}}, \bibinfo
  {author} {\bibfnamefont {E.}~\bibnamefont {{Franceschi}}}, \bibinfo {author}
  {\bibfnamefont {M.}~\bibnamefont {{Frommert}}}, \bibinfo {author}
  {\bibfnamefont {S.}~\bibnamefont {{Galeotta}}}, \bibinfo {author}
  {\bibfnamefont {K.}~\bibnamefont {{Ganga}}}, \bibinfo {author} {\bibfnamefont
  {R.~T.}\ \bibnamefont {{G{\'e}nova-Santos}}}, \bibinfo {author}
  {\bibfnamefont {M.}~\bibnamefont {{Giard}}}, \bibinfo {author} {\bibfnamefont
  {Y.}~\bibnamefont {{Giraud-H{\'e}raud}}}, \bibinfo {author} {\bibfnamefont
  {J.}~\bibnamefont {{Gonz{\'a}lez-Nuevo}}}, \bibinfo {author} {\bibfnamefont
  {K.~M.}\ \bibnamefont {{G{\'o}rski}}}, \bibinfo {author} {\bibfnamefont
  {A.}~\bibnamefont {{Gregorio}}}, \bibinfo {author} {\bibfnamefont
  {A.}~\bibnamefont {{Gruppuso}}}, \bibinfo {author} {\bibfnamefont {F.~K.}\
  \bibnamefont {{Hansen}}}, \bibinfo {author} {\bibfnamefont {D.}~\bibnamefont
  {{Harrison}}}, \bibinfo {author} {\bibfnamefont {A.}~\bibnamefont
  {{Hempel}}}, \bibinfo {author} {\bibfnamefont {S.}~\bibnamefont
  {{Henrot-Versill{\'e}}}}, \bibinfo {author} {\bibfnamefont {C.}~\bibnamefont
  {{Hern{\'a}ndez-Monteagudo}}}, \bibinfo {author} {\bibfnamefont
  {D.}~\bibnamefont {{Herranz}}}, \bibinfo {author} {\bibfnamefont {S.~R.}\
  \bibnamefont {{Hildebrandt}}}, \bibinfo {author} {\bibfnamefont
  {E.}~\bibnamefont {{Hivon}}}, \bibinfo {author} {\bibfnamefont
  {M.}~\bibnamefont {{Hobson}}}, \bibinfo {author} {\bibfnamefont {W.~A.}\
  \bibnamefont {{Holmes}}}, \bibinfo {author} {\bibfnamefont {G.}~\bibnamefont
  {{Hurier}}}, \bibinfo {author} {\bibfnamefont {T.~R.}\ \bibnamefont
  {{Jaffe}}}, \bibinfo {author} {\bibfnamefont {A.~H.}\ \bibnamefont
  {{Jaffe}}}, \bibinfo {author} {\bibfnamefont {T.}~\bibnamefont {{Jagemann}}},
  \bibinfo {author} {\bibfnamefont {W.~C.}\ \bibnamefont {{Jones}}}, \bibinfo
  {author} {\bibfnamefont {M.}~\bibnamefont {{Juvela}}}, \bibinfo {author}
  {\bibfnamefont {E.}~\bibnamefont {{Keih{\"a}nen}}}, \bibinfo {author}
  {\bibfnamefont {I.}~\bibnamefont {{Khamitov}}}, \bibinfo {author}
  {\bibfnamefont {T.~S.}\ \bibnamefont {{Kisner}}}, \bibinfo {author}
  {\bibfnamefont {R.}~\bibnamefont {{Kneissl}}}, \bibinfo {author}
  {\bibfnamefont {J.}~\bibnamefont {{Knoche}}}, \bibinfo {author}
  {\bibfnamefont {L.}~\bibnamefont {{Knox}}}, \bibinfo {author} {\bibfnamefont
  {M.}~\bibnamefont {{Kunz}}}, \bibinfo {author} {\bibfnamefont
  {H.}~\bibnamefont {{Kurki-Suonio}}}, \bibinfo {author} {\bibfnamefont
  {G.}~\bibnamefont {{Lagache}}}, \bibinfo {author} {\bibfnamefont
  {A.}~\bibnamefont {{L{\"a}hteenm{\"a}ki}}}, \bibinfo {author} {\bibfnamefont
  {J.~M.}\ \bibnamefont {{Lamarre}}}, \bibinfo {author} {\bibfnamefont
  {A.}~\bibnamefont {{Lasenby}}}, \bibinfo {author} {\bibfnamefont {C.~R.}\
  \bibnamefont {{Lawrence}}}, \bibinfo {author} {\bibfnamefont
  {M.}~\bibnamefont {{Le Jeune}}}, \bibinfo {author} {\bibfnamefont
  {R.}~\bibnamefont {{Leonardi}}}, \bibinfo {author} {\bibfnamefont
  {A.}~\bibnamefont {{Liddle}}}, \bibinfo {author} {\bibfnamefont {P.~B.}\
  \bibnamefont {{Lilje}}}, \bibinfo {author} {\bibfnamefont {M.}~\bibnamefont
  {{L{\'o}pez-Caniego}}}, \bibinfo {author} {\bibfnamefont {G.}~\bibnamefont
  {{Luzzi}}}, \bibinfo {author} {\bibfnamefont {J.~F.}\ \bibnamefont
  {{Mac{\'\i}as-P{\'e}rez}}}, \bibinfo {author} {\bibfnamefont
  {D.}~\bibnamefont {{Maino}}}, \bibinfo {author} {\bibfnamefont
  {N.}~\bibnamefont {{Mandolesi}}}, \bibinfo {author} {\bibfnamefont
  {M.}~\bibnamefont {{Maris}}}, \bibinfo {author} {\bibfnamefont
  {F.}~\bibnamefont {{Marleau}}}, \bibinfo {author} {\bibfnamefont {D.~J.}\
  \bibnamefont {{Marshall}}}, \bibinfo {author} {\bibfnamefont
  {E.}~\bibnamefont {{Mart{\'\i}nez-Gonz{\'a}lez}}}, \bibinfo {author}
  {\bibfnamefont {S.}~\bibnamefont {{Masi}}}, \bibinfo {author} {\bibfnamefont
  {M.}~\bibnamefont {{Massardi}}}, \bibinfo {author} {\bibfnamefont
  {S.}~\bibnamefont {{Matarrese}}}, \bibinfo {author} {\bibfnamefont
  {P.}~\bibnamefont {{Mazzotta}}}, \bibinfo {author} {\bibfnamefont
  {S.}~\bibnamefont {{Mei}}}, \bibinfo {author} {\bibfnamefont
  {A.}~\bibnamefont {{Melchiorri}}}, \bibinfo {author} {\bibfnamefont {J.~B.}\
  \bibnamefont {{Melin}}}, \bibinfo {author} {\bibfnamefont {L.}~\bibnamefont
  {{Mendes}}}, \bibinfo {author} {\bibfnamefont {A.}~\bibnamefont
  {{Mennella}}}, \bibinfo {author} {\bibfnamefont {S.}~\bibnamefont {{Mitra}}},
  \bibinfo {author} {\bibfnamefont {M.~A.}\ \bibnamefont
  {{Miville-Desch{\^e}nes}}}, \bibinfo {author} {\bibfnamefont
  {A.}~\bibnamefont {{Moneti}}}, \bibinfo {author} {\bibfnamefont
  {L.}~\bibnamefont {{Montier}}}, \bibinfo {author} {\bibfnamefont
  {G.}~\bibnamefont {{Morgante}}}, \bibinfo {author} {\bibfnamefont
  {D.}~\bibnamefont {{Mortlock}}}, \bibinfo {author} {\bibfnamefont
  {D.}~\bibnamefont {{Munshi}}}, \bibinfo {author} {\bibfnamefont {J.~A.}\
  \bibnamefont {{Murphy}}}, \bibinfo {author} {\bibfnamefont {P.}~\bibnamefont
  {{Naselsky}}}, \bibinfo {author} {\bibfnamefont {F.}~\bibnamefont {{Nati}}},
  \bibinfo {author} {\bibfnamefont {P.}~\bibnamefont {{Natoli}}}, \bibinfo
  {author} {\bibfnamefont {H.~U.}\ \bibnamefont {{N{\o}rgaard-Nielsen}}},
  \bibinfo {author} {\bibfnamefont {F.}~\bibnamefont {{Noviello}}}, \bibinfo
  {author} {\bibfnamefont {D.}~\bibnamefont {{Novikov}}}, \bibinfo {author}
  {\bibfnamefont {I.}~\bibnamefont {{Novikov}}}, \bibinfo {author}
  {\bibfnamefont {S.}~\bibnamefont {{Osborne}}}, \bibinfo {author}
  {\bibfnamefont {F.}~\bibnamefont {{Pajot}}}, \bibinfo {author} {\bibfnamefont
  {D.}~\bibnamefont {{Paoletti}}}, \bibinfo {author} {\bibfnamefont
  {F.}~\bibnamefont {{Pasian}}}, \bibinfo {author} {\bibfnamefont
  {G.}~\bibnamefont {{Patanchon}}}, \bibinfo {author} {\bibfnamefont
  {O.}~\bibnamefont {{Perdereau}}}, \bibinfo {author} {\bibfnamefont
  {L.}~\bibnamefont {{Perotto}}}, \bibinfo {author} {\bibfnamefont
  {F.}~\bibnamefont {{Perrotta}}}, \bibinfo {author} {\bibfnamefont
  {F.}~\bibnamefont {{Piacentini}}}, \bibinfo {author} {\bibfnamefont
  {M.}~\bibnamefont {{Piat}}}, \bibinfo {author} {\bibfnamefont
  {E.}~\bibnamefont {{Pierpaoli}}}, \bibinfo {author} {\bibfnamefont
  {R.}~\bibnamefont {{Piffaretti}}}, \bibinfo {author} {\bibfnamefont
  {S.}~\bibnamefont {{Plaszczynski}}}, \bibinfo {author} {\bibfnamefont
  {E.}~\bibnamefont {{Pointecouteau}}}, \bibinfo {author} {\bibfnamefont
  {G.}~\bibnamefont {{Polenta}}}, \bibinfo {author} {\bibfnamefont
  {N.}~\bibnamefont {{Ponthieu}}}, \bibinfo {author} {\bibfnamefont
  {L.}~\bibnamefont {{Popa}}}, \bibinfo {author} {\bibfnamefont
  {T.}~\bibnamefont {{Poutanen}}}, \bibinfo {author} {\bibfnamefont {G.~W.}\
  \bibnamefont {{Pratt}}}, \bibinfo {author} {\bibfnamefont {S.}~\bibnamefont
  {{Prunet}}}, \bibinfo {author} {\bibfnamefont {J.~L.}\ \bibnamefont
  {{Puget}}}, \bibinfo {author} {\bibfnamefont {J.~P.}\ \bibnamefont
  {{Rachen}}}, \bibinfo {author} {\bibfnamefont {W.~T.}\ \bibnamefont
  {{Reach}}}, \bibinfo {author} {\bibfnamefont {R.}~\bibnamefont {{Rebolo}}},
  \bibinfo {author} {\bibfnamefont {M.}~\bibnamefont {{Reinecke}}}, \bibinfo
  {author} {\bibfnamefont {M.}~\bibnamefont {{Remazeilles}}}, \bibinfo {author}
  {\bibfnamefont {C.}~\bibnamefont {{Renault}}}, \bibinfo {author}
  {\bibfnamefont {S.}~\bibnamefont {{Ricciardi}}}, \bibinfo {author}
  {\bibfnamefont {T.}~\bibnamefont {{Riller}}}, \bibinfo {author}
  {\bibfnamefont {I.}~\bibnamefont {{Ristorcelli}}}, \bibinfo {author}
  {\bibfnamefont {G.}~\bibnamefont {{Rocha}}}, \bibinfo {author} {\bibfnamefont
  {M.}~\bibnamefont {{Roman}}}, \bibinfo {author} {\bibfnamefont
  {C.}~\bibnamefont {{Rosset}}}, \bibinfo {author} {\bibfnamefont
  {M.}~\bibnamefont {{Rossetti}}}, \bibinfo {author} {\bibfnamefont {J.~A.}\
  \bibnamefont {{Rubi{\~n}o-Mart{\'\i}n}}}, \bibinfo {author} {\bibfnamefont
  {B.}~\bibnamefont {{Rusholme}}}, \bibinfo {author} {\bibfnamefont
  {M.}~\bibnamefont {{Sandri}}}, \bibinfo {author} {\bibfnamefont
  {G.}~\bibnamefont {{Savini}}}, \bibinfo {author} {\bibfnamefont
  {D.}~\bibnamefont {{Scott}}}, \bibinfo {author} {\bibfnamefont {G.~F.}\
  \bibnamefont {{Smoot}}}, \bibinfo {author} {\bibfnamefont {J.~L.}\
  \bibnamefont {{Starck}}}, \bibinfo {author} {\bibfnamefont {R.}~\bibnamefont
  {{Sudiwala}}}, \bibinfo {author} {\bibfnamefont {R.}~\bibnamefont
  {{Sunyaev}}}, \bibinfo {author} {\bibfnamefont {D.}~\bibnamefont {{Sutton}}},
  \bibinfo {author} {\bibfnamefont {A.~S.}\ \bibnamefont {{Suur-Uski}}},
  \bibinfo {author} {\bibfnamefont {J.~F.}\ \bibnamefont {{Sygnet}}}, \bibinfo
  {author} {\bibfnamefont {J.~A.}\ \bibnamefont {{Tauber}}}, \bibinfo {author}
  {\bibfnamefont {L.}~\bibnamefont {{Terenzi}}}, \bibinfo {author}
  {\bibfnamefont {L.}~\bibnamefont {{Toffolatti}}}, \bibinfo {author}
  {\bibfnamefont {M.}~\bibnamefont {{Tomasi}}}, \bibinfo {author}
  {\bibfnamefont {M.}~\bibnamefont {{Tristram}}}, \bibinfo {author}
  {\bibfnamefont {J.}~\bibnamefont {{Tuovinen}}}, \bibinfo {author}
  {\bibfnamefont {L.}~\bibnamefont {{Valenziano}}}, \bibinfo {author}
  {\bibfnamefont {B.}~\bibnamefont {{Van Tent}}}, \bibinfo {author}
  {\bibfnamefont {J.}~\bibnamefont {{Varis}}}, \bibinfo {author} {\bibfnamefont
  {P.}~\bibnamefont {{Vielva}}}, \bibinfo {author} {\bibfnamefont
  {F.}~\bibnamefont {{Villa}}}, \bibinfo {author} {\bibfnamefont
  {N.}~\bibnamefont {{Vittorio}}}, \bibinfo {author} {\bibfnamefont {L.~A.}\
  \bibnamefont {{Wade}}}, \bibinfo {author} {\bibfnamefont {B.~D.}\
  \bibnamefont {{Wandelt}}}, \bibinfo {author} {\bibfnamefont {N.}~\bibnamefont
  {{Welikala}}}, \bibinfo {author} {\bibfnamefont {S.~D.~M.}\ \bibnamefont
  {{White}}}, \bibinfo {author} {\bibfnamefont {M.}~\bibnamefont {{White}}},
  \bibinfo {author} {\bibfnamefont {D.}~\bibnamefont {{Yvon}}}, \bibinfo
  {author} {\bibfnamefont {A.}~\bibnamefont {{Zacchei}}},\ and\ \bibinfo
  {author} {\bibfnamefont {A.}~\bibnamefont {{Zonca}}},\ }\bibfield  {title}
  {\bibinfo {title} {{Planck intermediate results. V. Pressure profiles of
  galaxy clusters from the Sunyaev-Zeldovich effect}},\ }\href
  {https://doi.org/10.1051/0004-6361/201220040} {\bibfield  {journal} {\bibinfo
   {journal} {\aap}\ }\textbf {\bibinfo {volume} {550}},\ \bibinfo {eid} {A131}
  (\bibinfo {year} {2013})},\ \Eprint {https://arxiv.org/abs/1207.4061}
  {arXiv:1207.4061 [astro-ph.CO]} \BibitemShut {NoStop}%
\bibitem [{\citenamefont {{Amodeo}}\ \emph {et~al.}(2021)\citenamefont
  {{Amodeo}}, \citenamefont {{Battaglia}}, \citenamefont {{Schaan}},
  \citenamefont {{Ferraro}}, \citenamefont {{Moser}}, \citenamefont {{Aiola}},
  \citenamefont {{Austermann}}, \citenamefont {{Beall}}, \citenamefont
  {{Bean}}, \citenamefont {{Becker}}, \citenamefont {{Bond}}, \citenamefont
  {{Calabrese}}, \citenamefont {{Calafut}}, \citenamefont {{Choi}},
  \citenamefont {{Denison}}, \citenamefont {{Devlin}}, \citenamefont {{Duff}},
  \citenamefont {{Duivenvoorden}}, \citenamefont {{Dunkley}}, \citenamefont
  {{D{\"u}nner}}, \citenamefont {{Gallardo}}, \citenamefont {{Hall}},
  \citenamefont {{Han}}, \citenamefont {{Hill}}, \citenamefont {{Hilton}},
  \citenamefont {{Hilton}}, \citenamefont {{Hlo{\v{z}}ek}}, \citenamefont
  {{Hubmayr}}, \citenamefont {{Huffenberger}}, \citenamefont {{Hughes}},
  \citenamefont {{Koopman}}, \citenamefont {{MacInnis}}, \citenamefont
  {{McMahon}}, \citenamefont {{Madhavacheril}}, \citenamefont {{Moodley}},
  \citenamefont {{Mroczkowski}}, \citenamefont {{Naess}}, \citenamefont
  {{Nati}}, \citenamefont {{Newburgh}}, \citenamefont {{Niemack}},
  \citenamefont {{Page}}, \citenamefont {{Partridge}}, \citenamefont
  {{Schillaci}}, \citenamefont {{Sehgal}}, \citenamefont {{Sif{\'o}n}},
  \citenamefont {{Spergel}}, \citenamefont {{Staggs}}, \citenamefont
  {{Storer}}, \citenamefont {{Ullom}}, \citenamefont {{Vale}}, \citenamefont
  {{van Engelen}}, \citenamefont {{Van Lanen}}, \citenamefont {{Vavagiakis}},
  \citenamefont {{Wollack}},\ and\ \citenamefont {{Xu}}}]{2021PhRvD.103f3514A}%
  \BibitemOpen
  \bibfield  {author} {\bibinfo {author} {\bibfnamefont {S.}~\bibnamefont
  {{Amodeo}}}, \bibinfo {author} {\bibfnamefont {N.}~\bibnamefont
  {{Battaglia}}}, \bibinfo {author} {\bibfnamefont {E.}~\bibnamefont
  {{Schaan}}}, \bibinfo {author} {\bibfnamefont {S.}~\bibnamefont {{Ferraro}}},
  \bibinfo {author} {\bibfnamefont {E.}~\bibnamefont {{Moser}}}, \bibinfo
  {author} {\bibfnamefont {S.}~\bibnamefont {{Aiola}}}, \bibinfo {author}
  {\bibfnamefont {J.~E.}\ \bibnamefont {{Austermann}}}, \bibinfo {author}
  {\bibfnamefont {J.~A.}\ \bibnamefont {{Beall}}}, \bibinfo {author}
  {\bibfnamefont {R.}~\bibnamefont {{Bean}}}, \bibinfo {author} {\bibfnamefont
  {D.~T.}\ \bibnamefont {{Becker}}}, \bibinfo {author} {\bibfnamefont {R.~J.}\
  \bibnamefont {{Bond}}}, \bibinfo {author} {\bibfnamefont {E.}~\bibnamefont
  {{Calabrese}}}, \bibinfo {author} {\bibfnamefont {V.}~\bibnamefont
  {{Calafut}}}, \bibinfo {author} {\bibfnamefont {S.~K.}\ \bibnamefont
  {{Choi}}}, \bibinfo {author} {\bibfnamefont {E.~V.}\ \bibnamefont
  {{Denison}}}, \bibinfo {author} {\bibfnamefont {M.}~\bibnamefont {{Devlin}}},
  \bibinfo {author} {\bibfnamefont {S.~M.}\ \bibnamefont {{Duff}}}, \bibinfo
  {author} {\bibfnamefont {A.~J.}\ \bibnamefont {{Duivenvoorden}}}, \bibinfo
  {author} {\bibfnamefont {J.}~\bibnamefont {{Dunkley}}}, \bibinfo {author}
  {\bibfnamefont {R.}~\bibnamefont {{D{\"u}nner}}}, \bibinfo {author}
  {\bibfnamefont {P.~A.}\ \bibnamefont {{Gallardo}}}, \bibinfo {author}
  {\bibfnamefont {K.~R.}\ \bibnamefont {{Hall}}}, \bibinfo {author}
  {\bibfnamefont {D.}~\bibnamefont {{Han}}}, \bibinfo {author} {\bibfnamefont
  {J.~C.}\ \bibnamefont {{Hill}}}, \bibinfo {author} {\bibfnamefont {G.~C.}\
  \bibnamefont {{Hilton}}}, \bibinfo {author} {\bibfnamefont {M.}~\bibnamefont
  {{Hilton}}}, \bibinfo {author} {\bibfnamefont {R.}~\bibnamefont
  {{Hlo{\v{z}}ek}}}, \bibinfo {author} {\bibfnamefont {J.}~\bibnamefont
  {{Hubmayr}}}, \bibinfo {author} {\bibfnamefont {K.~M.}\ \bibnamefont
  {{Huffenberger}}}, \bibinfo {author} {\bibfnamefont {J.~P.}\ \bibnamefont
  {{Hughes}}}, \bibinfo {author} {\bibfnamefont {B.~J.}\ \bibnamefont
  {{Koopman}}}, \bibinfo {author} {\bibfnamefont {A.}~\bibnamefont
  {{MacInnis}}}, \bibinfo {author} {\bibfnamefont {J.}~\bibnamefont
  {{McMahon}}}, \bibinfo {author} {\bibfnamefont {M.~S.}\ \bibnamefont
  {{Madhavacheril}}}, \bibinfo {author} {\bibfnamefont {K.}~\bibnamefont
  {{Moodley}}}, \bibinfo {author} {\bibfnamefont {T.}~\bibnamefont
  {{Mroczkowski}}}, \bibinfo {author} {\bibfnamefont {S.}~\bibnamefont
  {{Naess}}}, \bibinfo {author} {\bibfnamefont {F.}~\bibnamefont {{Nati}}},
  \bibinfo {author} {\bibfnamefont {L.~B.}\ \bibnamefont {{Newburgh}}},
  \bibinfo {author} {\bibfnamefont {M.~D.}\ \bibnamefont {{Niemack}}}, \bibinfo
  {author} {\bibfnamefont {L.~A.}\ \bibnamefont {{Page}}}, \bibinfo {author}
  {\bibfnamefont {B.}~\bibnamefont {{Partridge}}}, \bibinfo {author}
  {\bibfnamefont {A.}~\bibnamefont {{Schillaci}}}, \bibinfo {author}
  {\bibfnamefont {N.}~\bibnamefont {{Sehgal}}}, \bibinfo {author}
  {\bibfnamefont {C.}~\bibnamefont {{Sif{\'o}n}}}, \bibinfo {author}
  {\bibfnamefont {D.~N.}\ \bibnamefont {{Spergel}}}, \bibinfo {author}
  {\bibfnamefont {S.}~\bibnamefont {{Staggs}}}, \bibinfo {author}
  {\bibfnamefont {E.~R.}\ \bibnamefont {{Storer}}}, \bibinfo {author}
  {\bibfnamefont {J.~N.}\ \bibnamefont {{Ullom}}}, \bibinfo {author}
  {\bibfnamefont {L.~R.}\ \bibnamefont {{Vale}}}, \bibinfo {author}
  {\bibfnamefont {A.}~\bibnamefont {{van Engelen}}}, \bibinfo {author}
  {\bibfnamefont {J.}~\bibnamefont {{Van Lanen}}}, \bibinfo {author}
  {\bibfnamefont {E.~M.}\ \bibnamefont {{Vavagiakis}}}, \bibinfo {author}
  {\bibfnamefont {E.~J.}\ \bibnamefont {{Wollack}}},\ and\ \bibinfo {author}
  {\bibfnamefont {Z.}~\bibnamefont {{Xu}}},\ }\bibfield  {title} {\bibinfo
  {title} {{Atacama Cosmology Telescope: Modeling the gas thermodynamics in
  BOSS CMASS galaxies from kinematic and thermal Sunyaev-Zel'dovich
  measurements}},\ }\href {https://doi.org/10.1103/PhysRevD.103.063514}
  {\bibfield  {journal} {\bibinfo  {journal} {\prd}\ }\textbf {\bibinfo
  {volume} {103}},\ \bibinfo {eid} {063514} (\bibinfo {year} {2021})},\ \Eprint
  {https://arxiv.org/abs/2009.05558} {arXiv:2009.05558 [astro-ph.CO]}
  \BibitemShut {NoStop}%
\bibitem [{\citenamefont {{Bond}}\ \emph {et~al.}(1996)\citenamefont {{Bond}},
  \citenamefont {{Kofman}},\ and\ \citenamefont
  {{Pogosyan}}}]{1996Natur.380..603B}%
  \BibitemOpen
  \bibfield  {author} {\bibinfo {author} {\bibfnamefont {J.~R.}\ \bibnamefont
  {{Bond}}}, \bibinfo {author} {\bibfnamefont {L.}~\bibnamefont {{Kofman}}},\
  and\ \bibinfo {author} {\bibfnamefont {D.}~\bibnamefont {{Pogosyan}}},\
  }\bibfield  {title} {\bibinfo {title} {{How filaments of galaxies are woven
  into the cosmic web}},\ }\href {https://doi.org/10.1038/380603a0} {\bibfield
  {journal} {\bibinfo  {journal} {\nat}\ }\textbf {\bibinfo {volume} {380}},\
  \bibinfo {pages} {603} (\bibinfo {year} {1996})},\ \Eprint
  {https://arxiv.org/abs/astro-ph/9512141} {arXiv:astro-ph/9512141 [astro-ph]}
  \BibitemShut {NoStop}%
\bibitem [{\citenamefont {{Lokken}}\ \emph {et~al.}(2022)\citenamefont
  {{Lokken}}, \citenamefont {{Hlo{\v{z}}ek}}, \citenamefont {{van Engelen}},
  \citenamefont {{Madhavacheril}}, \citenamefont {{Baxter}}, \citenamefont
  {{DeRose}}, \citenamefont {{Doux}}, \citenamefont {{Pandey}}, \citenamefont
  {{Rykoff}}, \citenamefont {{Stein}}, \citenamefont {{To}}, \citenamefont
  {{Abbott}}, \citenamefont {{Adhikari}}, \citenamefont {{Aguena}},
  \citenamefont {{Allam}}, \citenamefont {{Andrade-Oliveira}}, \citenamefont
  {{Annis}}, \citenamefont {{Battaglia}}, \citenamefont {{Bernstein}},
  \citenamefont {{Bertin}}, \citenamefont {{Bond}}, \citenamefont {{Brooks}},
  \citenamefont {{Calabrese}}, \citenamefont {{Carnero Rosell}}, \citenamefont
  {{Carrasco Kind}}, \citenamefont {{Carretero}}, \citenamefont {{Cawthon}},
  \citenamefont {{Choi}}, \citenamefont {{Costanzi}}, \citenamefont {{Crocce}},
  \citenamefont {{da Costa}}, \citenamefont {{da Silva Pereira}}, \citenamefont
  {{De Vicente}}, \citenamefont {{Desai}}, \citenamefont {{Dietrich}},
  \citenamefont {{Doel}}, \citenamefont {{Dunkley}}, \citenamefont {{Everett}},
  \citenamefont {{Evrard}}, \citenamefont {{Ferraro}}, \citenamefont
  {{Flaugher}}, \citenamefont {{Fosalba}}, \citenamefont {{Frieman}},
  \citenamefont {{Gallardo}}, \citenamefont {{Garc{\'\i}a-Bellido}},
  \citenamefont {{Gaztanaga}}, \citenamefont {{Gerdes}}, \citenamefont
  {{Giannantonio}}, \citenamefont {{Gruen}}, \citenamefont {{Gruendl}},
  \citenamefont {{Gschwend}}, \citenamefont {{Gutierrez}}, \citenamefont
  {{Hill}}, \citenamefont {{Hilton}}, \citenamefont {{Hincks}}, \citenamefont
  {{Hinton}}, \citenamefont {{Hollowood}}, \citenamefont {{Honscheid}},
  \citenamefont {{Hoyle}}, \citenamefont {{Huang}}, \citenamefont {{Hughes}},
  \citenamefont {{Huterer}}, \citenamefont {{Jain}}, \citenamefont {{James}},
  \citenamefont {{Jeltema}}, \citenamefont {{Kuehn}}, \citenamefont {{Lima}},
  \citenamefont {{Maia}}, \citenamefont {{Marshall}}, \citenamefont
  {{McMahon}}, \citenamefont {{Melchior}}, \citenamefont {{Menanteau}},
  \citenamefont {{Miquel}}, \citenamefont {{Mohr}}, \citenamefont {{Moodley}},
  \citenamefont {{Morgan}}, \citenamefont {{Nati}}, \citenamefont {{Page}},
  \citenamefont {{Ogando}}, \citenamefont {{Palmese}}, \citenamefont
  {{Paz-Chinch{\'o}n}}, \citenamefont {{Plazas Malag{\'o}n}}, \citenamefont
  {{Pieres}}, \citenamefont {{Romer}}, \citenamefont {{Rozo}}, \citenamefont
  {{Sanchez}}, \citenamefont {{Scarpine}}, \citenamefont {{Schillaci}},
  \citenamefont {{Schubnell}}, \citenamefont {{Serrano}}, \citenamefont
  {{Sevilla-Noarbe}}, \citenamefont {{Sheldon}}, \citenamefont {{Shin}},
  \citenamefont {{Sif{\'o}n}}, \citenamefont {{Smith}}, \citenamefont
  {{Soares-Santos}}, \citenamefont {{Suchyta}}, \citenamefont {{Swanson}},
  \citenamefont {{Tarle}}, \citenamefont {{Thomas}}, \citenamefont {{Tucker}},
  \citenamefont {{Varga}}, \citenamefont {{Weller}}, \citenamefont
  {{Wechsler}}, \citenamefont {{Wilkinson}}, \citenamefont {{Wollack}},\ and\
  \citenamefont {{Xu}}}]{2022ApJ...933..134L}%
  \BibitemOpen
  \bibfield  {author} {\bibinfo {author} {\bibfnamefont {M.}~\bibnamefont
  {{Lokken}}}, \bibinfo {author} {\bibfnamefont {R.}~\bibnamefont
  {{Hlo{\v{z}}ek}}}, \bibinfo {author} {\bibfnamefont {A.}~\bibnamefont {{van
  Engelen}}}, \bibinfo {author} {\bibfnamefont {M.}~\bibnamefont
  {{Madhavacheril}}}, \bibinfo {author} {\bibfnamefont {E.}~\bibnamefont
  {{Baxter}}}, \bibinfo {author} {\bibfnamefont {J.}~\bibnamefont {{DeRose}}},
  \bibinfo {author} {\bibfnamefont {C.}~\bibnamefont {{Doux}}}, \bibinfo
  {author} {\bibfnamefont {S.}~\bibnamefont {{Pandey}}}, \bibinfo {author}
  {\bibfnamefont {E.~S.}\ \bibnamefont {{Rykoff}}}, \bibinfo {author}
  {\bibfnamefont {G.}~\bibnamefont {{Stein}}}, \bibinfo {author} {\bibfnamefont
  {C.}~\bibnamefont {{To}}}, \bibinfo {author} {\bibfnamefont {T.~M.~C.}\
  \bibnamefont {{Abbott}}}, \bibinfo {author} {\bibfnamefont {S.}~\bibnamefont
  {{Adhikari}}}, \bibinfo {author} {\bibfnamefont {M.}~\bibnamefont
  {{Aguena}}}, \bibinfo {author} {\bibfnamefont {S.}~\bibnamefont {{Allam}}},
  \bibinfo {author} {\bibfnamefont {F.}~\bibnamefont {{Andrade-Oliveira}}},
  \bibinfo {author} {\bibfnamefont {J.}~\bibnamefont {{Annis}}}, \bibinfo
  {author} {\bibfnamefont {N.}~\bibnamefont {{Battaglia}}}, \bibinfo {author}
  {\bibfnamefont {G.~M.}\ \bibnamefont {{Bernstein}}}, \bibinfo {author}
  {\bibfnamefont {E.}~\bibnamefont {{Bertin}}}, \bibinfo {author}
  {\bibfnamefont {J.~R.}\ \bibnamefont {{Bond}}}, \bibinfo {author}
  {\bibfnamefont {D.}~\bibnamefont {{Brooks}}}, \bibinfo {author}
  {\bibfnamefont {E.}~\bibnamefont {{Calabrese}}}, \bibinfo {author}
  {\bibfnamefont {A.}~\bibnamefont {{Carnero Rosell}}}, \bibinfo {author}
  {\bibfnamefont {M.}~\bibnamefont {{Carrasco Kind}}}, \bibinfo {author}
  {\bibfnamefont {J.}~\bibnamefont {{Carretero}}}, \bibinfo {author}
  {\bibfnamefont {R.}~\bibnamefont {{Cawthon}}}, \bibinfo {author}
  {\bibfnamefont {A.}~\bibnamefont {{Choi}}}, \bibinfo {author} {\bibfnamefont
  {M.}~\bibnamefont {{Costanzi}}}, \bibinfo {author} {\bibfnamefont
  {M.}~\bibnamefont {{Crocce}}}, \bibinfo {author} {\bibfnamefont {L.~N.}\
  \bibnamefont {{da Costa}}}, \bibinfo {author} {\bibfnamefont {M.~E.}\
  \bibnamefont {{da Silva Pereira}}}, \bibinfo {author} {\bibfnamefont
  {J.}~\bibnamefont {{De Vicente}}}, \bibinfo {author} {\bibfnamefont
  {S.}~\bibnamefont {{Desai}}}, \bibinfo {author} {\bibfnamefont {J.~P.}\
  \bibnamefont {{Dietrich}}}, \bibinfo {author} {\bibfnamefont
  {P.}~\bibnamefont {{Doel}}}, \bibinfo {author} {\bibfnamefont
  {J.}~\bibnamefont {{Dunkley}}}, \bibinfo {author} {\bibfnamefont
  {S.}~\bibnamefont {{Everett}}}, \bibinfo {author} {\bibfnamefont {A.~E.}\
  \bibnamefont {{Evrard}}}, \bibinfo {author} {\bibfnamefont {S.}~\bibnamefont
  {{Ferraro}}}, \bibinfo {author} {\bibfnamefont {B.}~\bibnamefont
  {{Flaugher}}}, \bibinfo {author} {\bibfnamefont {P.}~\bibnamefont
  {{Fosalba}}}, \bibinfo {author} {\bibfnamefont {J.}~\bibnamefont
  {{Frieman}}}, \bibinfo {author} {\bibfnamefont {P.~A.}\ \bibnamefont
  {{Gallardo}}}, \bibinfo {author} {\bibfnamefont {J.}~\bibnamefont
  {{Garc{\'\i}a-Bellido}}}, \bibinfo {author} {\bibfnamefont {E.}~\bibnamefont
  {{Gaztanaga}}}, \bibinfo {author} {\bibfnamefont {D.~W.}\ \bibnamefont
  {{Gerdes}}}, \bibinfo {author} {\bibfnamefont {T.}~\bibnamefont
  {{Giannantonio}}}, \bibinfo {author} {\bibfnamefont {D.}~\bibnamefont
  {{Gruen}}}, \bibinfo {author} {\bibfnamefont {R.~A.}\ \bibnamefont
  {{Gruendl}}}, \bibinfo {author} {\bibfnamefont {J.}~\bibnamefont
  {{Gschwend}}}, \bibinfo {author} {\bibfnamefont {G.}~\bibnamefont
  {{Gutierrez}}}, \bibinfo {author} {\bibfnamefont {J.~C.}\ \bibnamefont
  {{Hill}}}, \bibinfo {author} {\bibfnamefont {M.}~\bibnamefont {{Hilton}}},
  \bibinfo {author} {\bibfnamefont {A.~D.}\ \bibnamefont {{Hincks}}}, \bibinfo
  {author} {\bibfnamefont {S.~R.}\ \bibnamefont {{Hinton}}}, \bibinfo {author}
  {\bibfnamefont {D.~L.}\ \bibnamefont {{Hollowood}}}, \bibinfo {author}
  {\bibfnamefont {K.}~\bibnamefont {{Honscheid}}}, \bibinfo {author}
  {\bibfnamefont {B.}~\bibnamefont {{Hoyle}}}, \bibinfo {author} {\bibfnamefont
  {Z.}~\bibnamefont {{Huang}}}, \bibinfo {author} {\bibfnamefont {J.~P.}\
  \bibnamefont {{Hughes}}}, \bibinfo {author} {\bibfnamefont {D.}~\bibnamefont
  {{Huterer}}}, \bibinfo {author} {\bibfnamefont {B.}~\bibnamefont {{Jain}}},
  \bibinfo {author} {\bibfnamefont {D.~J.}\ \bibnamefont {{James}}}, \bibinfo
  {author} {\bibfnamefont {T.}~\bibnamefont {{Jeltema}}}, \bibinfo {author}
  {\bibfnamefont {K.}~\bibnamefont {{Kuehn}}}, \bibinfo {author} {\bibfnamefont
  {M.}~\bibnamefont {{Lima}}}, \bibinfo {author} {\bibfnamefont {M.~A.~G.}\
  \bibnamefont {{Maia}}}, \bibinfo {author} {\bibfnamefont {J.~L.}\
  \bibnamefont {{Marshall}}}, \bibinfo {author} {\bibfnamefont
  {J.}~\bibnamefont {{McMahon}}}, \bibinfo {author} {\bibfnamefont
  {P.}~\bibnamefont {{Melchior}}}, \bibinfo {author} {\bibfnamefont
  {F.}~\bibnamefont {{Menanteau}}}, \bibinfo {author} {\bibfnamefont
  {R.}~\bibnamefont {{Miquel}}}, \bibinfo {author} {\bibfnamefont {J.~J.}\
  \bibnamefont {{Mohr}}}, \bibinfo {author} {\bibfnamefont {K.}~\bibnamefont
  {{Moodley}}}, \bibinfo {author} {\bibfnamefont {R.}~\bibnamefont {{Morgan}}},
  \bibinfo {author} {\bibfnamefont {F.}~\bibnamefont {{Nati}}}, \bibinfo
  {author} {\bibfnamefont {L.}~\bibnamefont {{Page}}}, \bibinfo {author}
  {\bibfnamefont {R.~L.~C.}\ \bibnamefont {{Ogando}}}, \bibinfo {author}
  {\bibfnamefont {A.}~\bibnamefont {{Palmese}}}, \bibinfo {author}
  {\bibfnamefont {F.}~\bibnamefont {{Paz-Chinch{\'o}n}}}, \bibinfo {author}
  {\bibfnamefont {A.~A.}\ \bibnamefont {{Plazas Malag{\'o}n}}}, \bibinfo
  {author} {\bibfnamefont {A.}~\bibnamefont {{Pieres}}}, \bibinfo {author}
  {\bibfnamefont {A.~K.}\ \bibnamefont {{Romer}}}, \bibinfo {author}
  {\bibfnamefont {E.}~\bibnamefont {{Rozo}}}, \bibinfo {author} {\bibfnamefont
  {E.}~\bibnamefont {{Sanchez}}}, \bibinfo {author} {\bibfnamefont
  {V.}~\bibnamefont {{Scarpine}}}, \bibinfo {author} {\bibfnamefont
  {A.}~\bibnamefont {{Schillaci}}}, \bibinfo {author} {\bibfnamefont
  {M.}~\bibnamefont {{Schubnell}}}, \bibinfo {author} {\bibfnamefont
  {S.}~\bibnamefont {{Serrano}}}, \bibinfo {author} {\bibfnamefont
  {I.}~\bibnamefont {{Sevilla-Noarbe}}}, \bibinfo {author} {\bibfnamefont
  {E.}~\bibnamefont {{Sheldon}}}, \bibinfo {author} {\bibfnamefont
  {T.}~\bibnamefont {{Shin}}}, \bibinfo {author} {\bibfnamefont
  {C.}~\bibnamefont {{Sif{\'o}n}}}, \bibinfo {author} {\bibfnamefont
  {M.}~\bibnamefont {{Smith}}}, \bibinfo {author} {\bibfnamefont
  {M.}~\bibnamefont {{Soares-Santos}}}, \bibinfo {author} {\bibfnamefont
  {E.}~\bibnamefont {{Suchyta}}}, \bibinfo {author} {\bibfnamefont {M.~E.~C.}\
  \bibnamefont {{Swanson}}}, \bibinfo {author} {\bibfnamefont {G.}~\bibnamefont
  {{Tarle}}}, \bibinfo {author} {\bibfnamefont {D.}~\bibnamefont {{Thomas}}},
  \bibinfo {author} {\bibfnamefont {D.~L.}\ \bibnamefont {{Tucker}}}, \bibinfo
  {author} {\bibfnamefont {T.~N.}\ \bibnamefont {{Varga}}}, \bibinfo {author}
  {\bibfnamefont {J.}~\bibnamefont {{Weller}}}, \bibinfo {author}
  {\bibfnamefont {R.~H.}\ \bibnamefont {{Wechsler}}}, \bibinfo {author}
  {\bibfnamefont {R.~D.}\ \bibnamefont {{Wilkinson}}}, \bibinfo {author}
  {\bibfnamefont {E.~J.}\ \bibnamefont {{Wollack}}},\ and\ \bibinfo {author}
  {\bibfnamefont {Z.}~\bibnamefont {{Xu}}},\ }\bibfield  {title} {\bibinfo
  {title} {{Superclustering with the Atacama Cosmology Telescope and Dark
  Energy Survey. I. Evidence for Thermal Energy Anisotropy Using Oriented
  Stacking}},\ }\href {https://doi.org/10.3847/1538-4357/ac7043} {\bibfield
  {journal} {\bibinfo  {journal} {\apj}\ }\textbf {\bibinfo {volume} {933}},\
  \bibinfo {eid} {134} (\bibinfo {year} {2022})},\ \Eprint
  {https://arxiv.org/abs/2107.05523} {arXiv:2107.05523 [astro-ph.CO]}
  \BibitemShut {NoStop}%
\bibitem [{\citenamefont {{Lokken}}\ \emph {et~al.}(2023)\citenamefont
  {{Lokken}}, \citenamefont {{Cui}}, \citenamefont {{Bond}}, \citenamefont
  {{Hlo{\v{z}}ek}}, \citenamefont {{Murray}}, \citenamefont {{Dav{\'e}}},\ and\
  \citenamefont {{van Engelen}}}]{2023MNRAS.523.1346L}%
  \BibitemOpen
  \bibfield  {author} {\bibinfo {author} {\bibfnamefont {M.}~\bibnamefont
  {{Lokken}}}, \bibinfo {author} {\bibfnamefont {W.}~\bibnamefont {{Cui}}},
  \bibinfo {author} {\bibfnamefont {J.~R.}\ \bibnamefont {{Bond}}}, \bibinfo
  {author} {\bibfnamefont {R.}~\bibnamefont {{Hlo{\v{z}}ek}}}, \bibinfo
  {author} {\bibfnamefont {N.}~\bibnamefont {{Murray}}}, \bibinfo {author}
  {\bibfnamefont {R.}~\bibnamefont {{Dav{\'e}}}},\ and\ \bibinfo {author}
  {\bibfnamefont {A.}~\bibnamefont {{van Engelen}}},\ }\bibfield  {title}
  {\bibinfo {title} {{Boundless baryons: how diffuse gas contributes to
  anisotropic tSZ signal around simulated Three Hundred clusters}},\ }\href
  {https://doi.org/10.1093/mnras/stad1414} {\bibfield  {journal} {\bibinfo
  {journal} {\mnras}\ }\textbf {\bibinfo {volume} {523}},\ \bibinfo {pages}
  {1346} (\bibinfo {year} {2023})},\ \Eprint {https://arxiv.org/abs/2211.00242}
  {arXiv:2211.00242 [astro-ph.CO]} \BibitemShut {NoStop}%
\bibitem [{\citenamefont {{Schaan}}\ \emph {et~al.}(2016)\citenamefont
  {{Schaan}}, \citenamefont {{Ferraro}}, \citenamefont {{Vargas-Maga{\~n}a}},
  \citenamefont {{Smith}}, \citenamefont {{Ho}}, \citenamefont {{Aiola}},
  \citenamefont {{Battaglia}}, \citenamefont {{Bond}}, \citenamefont {{De
  Bernardis}}, \citenamefont {{Calabrese}}, \citenamefont {{Cho}},
  \citenamefont {{Devlin}}, \citenamefont {{Dunkley}}, \citenamefont
  {{Gallardo}}, \citenamefont {{Hasselfield}}, \citenamefont {{Henderson}},
  \citenamefont {{Hill}}, \citenamefont {{Hincks}}, \citenamefont {{Hlozek}},
  \citenamefont {{Hubmayr}}, \citenamefont {{Hughes}}, \citenamefont {{Irwin}},
  \citenamefont {{Koopman}}, \citenamefont {{Kosowsky}}, \citenamefont {{Li}},
  \citenamefont {{Louis}}, \citenamefont {{Lungu}}, \citenamefont
  {{Madhavacheril}}, \citenamefont {{Maurin}}, \citenamefont {{McMahon}},
  \citenamefont {{Moodley}}, \citenamefont {{Naess}}, \citenamefont {{Nati}},
  \citenamefont {{Newburgh}}, \citenamefont {{Niemack}}, \citenamefont
  {{Page}}, \citenamefont {{Pappas}}, \citenamefont {{Partridge}},
  \citenamefont {{Schmitt}}, \citenamefont {{Sehgal}}, \citenamefont
  {{Sherwin}}, \citenamefont {{Sievers}}, \citenamefont {{Spergel}},
  \citenamefont {{Staggs}}, \citenamefont {{van Engelen}}, \citenamefont
  {{Wollack}},\ and\ \citenamefont {{ACTPol
  Collaboration}}}]{2016PhRvD..93h2002S}%
  \BibitemOpen
  \bibfield  {author} {\bibinfo {author} {\bibfnamefont {E.}~\bibnamefont
  {{Schaan}}}, \bibinfo {author} {\bibfnamefont {S.}~\bibnamefont {{Ferraro}}},
  \bibinfo {author} {\bibfnamefont {M.}~\bibnamefont {{Vargas-Maga{\~n}a}}},
  \bibinfo {author} {\bibfnamefont {K.~M.}\ \bibnamefont {{Smith}}}, \bibinfo
  {author} {\bibfnamefont {S.}~\bibnamefont {{Ho}}}, \bibinfo {author}
  {\bibfnamefont {S.}~\bibnamefont {{Aiola}}}, \bibinfo {author} {\bibfnamefont
  {N.}~\bibnamefont {{Battaglia}}}, \bibinfo {author} {\bibfnamefont {J.~R.}\
  \bibnamefont {{Bond}}}, \bibinfo {author} {\bibfnamefont {F.}~\bibnamefont
  {{De Bernardis}}}, \bibinfo {author} {\bibfnamefont {E.}~\bibnamefont
  {{Calabrese}}}, \bibinfo {author} {\bibfnamefont {H.-M.}\ \bibnamefont
  {{Cho}}}, \bibinfo {author} {\bibfnamefont {M.~J.}\ \bibnamefont {{Devlin}}},
  \bibinfo {author} {\bibfnamefont {J.}~\bibnamefont {{Dunkley}}}, \bibinfo
  {author} {\bibfnamefont {P.~A.}\ \bibnamefont {{Gallardo}}}, \bibinfo
  {author} {\bibfnamefont {M.}~\bibnamefont {{Hasselfield}}}, \bibinfo {author}
  {\bibfnamefont {S.}~\bibnamefont {{Henderson}}}, \bibinfo {author}
  {\bibfnamefont {J.~C.}\ \bibnamefont {{Hill}}}, \bibinfo {author}
  {\bibfnamefont {A.~D.}\ \bibnamefont {{Hincks}}}, \bibinfo {author}
  {\bibfnamefont {R.}~\bibnamefont {{Hlozek}}}, \bibinfo {author}
  {\bibfnamefont {J.}~\bibnamefont {{Hubmayr}}}, \bibinfo {author}
  {\bibfnamefont {J.~P.}\ \bibnamefont {{Hughes}}}, \bibinfo {author}
  {\bibfnamefont {K.~D.}\ \bibnamefont {{Irwin}}}, \bibinfo {author}
  {\bibfnamefont {B.}~\bibnamefont {{Koopman}}}, \bibinfo {author}
  {\bibfnamefont {A.}~\bibnamefont {{Kosowsky}}}, \bibinfo {author}
  {\bibfnamefont {D.}~\bibnamefont {{Li}}}, \bibinfo {author} {\bibfnamefont
  {T.}~\bibnamefont {{Louis}}}, \bibinfo {author} {\bibfnamefont
  {M.}~\bibnamefont {{Lungu}}}, \bibinfo {author} {\bibfnamefont
  {M.}~\bibnamefont {{Madhavacheril}}}, \bibinfo {author} {\bibfnamefont
  {L.}~\bibnamefont {{Maurin}}}, \bibinfo {author} {\bibfnamefont {J.~J.}\
  \bibnamefont {{McMahon}}}, \bibinfo {author} {\bibfnamefont {K.}~\bibnamefont
  {{Moodley}}}, \bibinfo {author} {\bibfnamefont {S.}~\bibnamefont {{Naess}}},
  \bibinfo {author} {\bibfnamefont {F.}~\bibnamefont {{Nati}}}, \bibinfo
  {author} {\bibfnamefont {L.}~\bibnamefont {{Newburgh}}}, \bibinfo {author}
  {\bibfnamefont {M.~D.}\ \bibnamefont {{Niemack}}}, \bibinfo {author}
  {\bibfnamefont {L.~A.}\ \bibnamefont {{Page}}}, \bibinfo {author}
  {\bibfnamefont {C.~G.}\ \bibnamefont {{Pappas}}}, \bibinfo {author}
  {\bibfnamefont {B.}~\bibnamefont {{Partridge}}}, \bibinfo {author}
  {\bibfnamefont {B.~L.}\ \bibnamefont {{Schmitt}}}, \bibinfo {author}
  {\bibfnamefont {N.}~\bibnamefont {{Sehgal}}}, \bibinfo {author}
  {\bibfnamefont {B.~D.}\ \bibnamefont {{Sherwin}}}, \bibinfo {author}
  {\bibfnamefont {J.~L.}\ \bibnamefont {{Sievers}}}, \bibinfo {author}
  {\bibfnamefont {D.~N.}\ \bibnamefont {{Spergel}}}, \bibinfo {author}
  {\bibfnamefont {S.~T.}\ \bibnamefont {{Staggs}}}, \bibinfo {author}
  {\bibfnamefont {A.}~\bibnamefont {{van Engelen}}}, \bibinfo {author}
  {\bibfnamefont {E.~J.}\ \bibnamefont {{Wollack}}},\ and\ \bibinfo {author}
  {\bibnamefont {{ACTPol Collaboration}}},\ }\bibfield  {title} {\bibinfo
  {title} {{Evidence for the kinematic Sunyaev-Zel'dovich effect with the
  Atacama Cosmology Telescope and velocity reconstruction from the Baryon
  Oscillation Spectroscopic Survey}},\ }\href
  {https://doi.org/10.1103/PhysRevD.93.082002} {\bibfield  {journal} {\bibinfo
  {journal} {\prd}\ }\textbf {\bibinfo {volume} {93}},\ \bibinfo {eid} {082002}
  (\bibinfo {year} {2016})},\ \Eprint {https://arxiv.org/abs/1510.06442}
  {arXiv:1510.06442 [astro-ph.CO]} \BibitemShut {NoStop}%
\bibitem [{\citenamefont {{DESI Collaboration}}\ \emph
  {et~al.}(2022)\citenamefont {{DESI Collaboration}}, \citenamefont
  {{Abareshi}}, \citenamefont {{Aguilar}}, \citenamefont {{Ahlen}},
  \citenamefont {{Alam}}, \citenamefont {{Alexander}}, \citenamefont
  {{Alfarsy}}, \citenamefont {{Allen}}, \citenamefont {{Allende Prieto}},
  \citenamefont {{Alves}}, \citenamefont {{Ameel}}, \citenamefont
  {{Armengaud}}, \citenamefont {{Asorey}}, \citenamefont {{Aviles}},
  \citenamefont {{Bailey}}, \citenamefont {{Balaguera-Antol{\'\i}nez}},
  \citenamefont {{Ballester}}, \citenamefont {{Baltay}}, \citenamefont
  {{Bault}}, \citenamefont {{Beltran}}, \citenamefont {{Benavides}},
  \citenamefont {{BenZvi}}, \citenamefont {{Berti}}, \citenamefont {{Besuner}},
  \citenamefont {{Beutler}}, \citenamefont {{Bianchi}}, \citenamefont
  {{Blake}}, \citenamefont {{Blanc}}, \citenamefont {{Blum}}, \citenamefont
  {{Bolton}}, \citenamefont {{Bose}}, \citenamefont {{Bramall}}, \citenamefont
  {{Brieden}}, \citenamefont {{Brodzeller}}, \citenamefont {{Brooks}},
  \citenamefont {{Brownewell}}, \citenamefont {{Buckley-Geer}}, \citenamefont
  {{Cahn}}, \citenamefont {{Cai}}, \citenamefont {{Canning}}, \citenamefont
  {{Capasso}}, \citenamefont {{Carnero Rosell}}, \citenamefont {{Carton}},
  \citenamefont {{Casas}}, \citenamefont {{Castander}}, \citenamefont
  {{Cervantes-Cota}}, \citenamefont {{Chabanier}}, \citenamefont
  {{Chaussidon}}, \citenamefont {{Chuang}}, \citenamefont {{Circosta}},
  \citenamefont {{Cole}}, \citenamefont {{Cooper}}, \citenamefont {{da Costa}},
  \citenamefont {{Cousinou}}, \citenamefont {{Cuceu}}, \citenamefont {{Davis}},
  \citenamefont {{Dawson}}, \citenamefont {{de la Cruz-Noriega}}, \citenamefont
  {{de la Macorra}}, \citenamefont {{de Mattia}}, \citenamefont {{Della
  Costa}}, \citenamefont {{Demmer}}, \citenamefont {{Derwent}}, \citenamefont
  {{Dey}}, \citenamefont {{Dey}}, \citenamefont {{Dhungana}}, \citenamefont
  {{Ding}}, \citenamefont {{Dobson}}, \citenamefont {{Doel}}, \citenamefont
  {{Donald-McCann}}, \citenamefont {{Donaldson}}, \citenamefont {{Douglass}},
  \citenamefont {{Duan}}, \citenamefont {{Dunlop}}, \citenamefont
  {{Edelstein}}, \citenamefont {{Eftekharzadeh}}, \citenamefont {{Eisenstein}},
  \citenamefont {{Enriquez-Vargas}}, \citenamefont {{Escoffier}}, \citenamefont
  {{Evatt}}, \citenamefont {{Fagrelius}}, \citenamefont {{Fan}}, \citenamefont
  {{Fanning}}, \citenamefont {{Fawcett}}, \citenamefont {{Ferraro}},
  \citenamefont {{Ereza}}, \citenamefont {{Flaugher}}, \citenamefont
  {{Font-Ribera}}, \citenamefont {{Forero-Romero}}, \citenamefont {{Frenk}},
  \citenamefont {{Fromenteau}}, \citenamefont {{G{\"a}nsicke}}, \citenamefont
  {{Garcia-Quintero}}, \citenamefont {{Garrison}}, \citenamefont
  {{Gazta{\~n}aga}}, \citenamefont {{Gerardi}}, \citenamefont
  {{Gil-Mar{\'\i}n}}, \citenamefont {{Gontcho a Gontcho}}, \citenamefont
  {{Gonzalez-Morales}}, \citenamefont {{Gonzalez-de-Rivera}}, \citenamefont
  {{Gonzalez-Perez}}, \citenamefont {{Gordon}}, \citenamefont {{Graur}},
  \citenamefont {{Green}}, \citenamefont {{Grove}}, \citenamefont {{Gruen}},
  \citenamefont {{Gutierrez}}, \citenamefont {{Guy}}, \citenamefont {{Hahn}},
  \citenamefont {{Harris}}, \citenamefont {{Herrera}}, \citenamefont
  {{Herrera-Alcantar}}, \citenamefont {{Honscheid}}, \citenamefont {{Howlett}},
  \citenamefont {{Huterer}}, \citenamefont {{Ir{\v{s}}i{\v{c}}}}, \citenamefont
  {{Ishak}}, \citenamefont {{Jelinsky}}, \citenamefont {{Jiang}}, \citenamefont
  {{Jimenez}}, \citenamefont {{Jing}}, \citenamefont {{Joyce}}, \citenamefont
  {{Jullo}}, \citenamefont {{Juneau}}, \citenamefont {{Kara{\c{c}}ayl{\i}}},
  \citenamefont {{Karamanis}}, \citenamefont {{Karcher}}, \citenamefont
  {{Karim}}, \citenamefont {{Kehoe}}, \citenamefont {{Kent}}, \citenamefont
  {{Kirkby}}, \citenamefont {{Kisner}}, \citenamefont {{Kitaura}},
  \citenamefont {{Koposov}}, \citenamefont {{Kov{\'a}cs}}, \citenamefont
  {{Kremin}}, \citenamefont {{Krolewski}}, \citenamefont {{L'Huillier}},
  \citenamefont {{Lahav}}, \citenamefont {{Lambert}}, \citenamefont {{Lamman}},
  \citenamefont {{Lan}}, \citenamefont {{Landriau}}, \citenamefont {{Lane}},
  \citenamefont {{Lang}}, \citenamefont {{Lange}}, \citenamefont {{Lasker}},
  \citenamefont {{Le Guillou}}, \citenamefont {{Leauthaud}}, \citenamefont {{Le
  Van Suu}}, \citenamefont {{Levi}}, \citenamefont {{Li}}, \citenamefont
  {{Magneville}}, \citenamefont {{Manera}}, \citenamefont {{Manser}},
  \citenamefont {{Marshall}}, \citenamefont {{Martini}}, \citenamefont
  {{McCollam}}, \citenamefont {{McDonald}}, \citenamefont {{Meisner}},
  \citenamefont {{Mena-Fern{\'a}ndez}}, \citenamefont {{Meneses-Rizo}},
  \citenamefont {{Mezcua}}, \citenamefont {{Miller}}, \citenamefont {{Miquel}},
  \citenamefont {{Montero-Camacho}}, \citenamefont {{Moon}}, \citenamefont
  {{Moustakas}}, \citenamefont {{Mueller}}, \citenamefont
  {{Mu{\~n}oz-Guti{\'e}rrez}}, \citenamefont {{Myers}}, \citenamefont
  {{Nadathur}}, \citenamefont {{Najita}}, \citenamefont {{Napolitano}},
  \citenamefont {{Neilsen}}, \citenamefont {{Newman}}, \citenamefont {{Nie}},
  \citenamefont {{Ning}}, \citenamefont {{Niz}}, \citenamefont {{Norberg}},
  \citenamefont {{Noriega}}, \citenamefont {{O'Brien}}, \citenamefont
  {{Obuljen}}, \citenamefont {{Palanque-Delabrouille}}, \citenamefont
  {{Palmese}}, \citenamefont {{Zhiwei}}, \citenamefont {{Pappalardo}},
  \citenamefont {{PENG}}, \citenamefont {{Percival}}, \citenamefont
  {{Perruchot}}, \citenamefont {{Pogge}}, \citenamefont {{Poppett}},
  \citenamefont {{Porredon}}, \citenamefont {{Prada}}, \citenamefont
  {{Prochaska}}, \citenamefont {{Pucha}}, \citenamefont
  {{P{\'e}rez-Fern{\'a}ndez}}, \citenamefont {{P{\'e}rez-R{\`a}fols}},
  \citenamefont {{Rabinowitz}}, \citenamefont {{Raichoor}}, \citenamefont
  {{Ramirez-Solano}}, \citenamefont {{Ram{\'\i}rez-P{\'e}rez}}, \citenamefont
  {{Ravoux}}, \citenamefont {{Reil}}, \citenamefont {{Rezaie}}, \citenamefont
  {{Rocher}}, \citenamefont {{Rockosi}}, \citenamefont {{Roe}}, \citenamefont
  {{Roodman}}, \citenamefont {{Ross}}, \citenamefont {{Rossi}}, \citenamefont
  {{Ruggeri}}, \citenamefont {{Ruhlmann-Kleider}}, \citenamefont {{Sabiu}},
  \citenamefont {{Safonova}}, \citenamefont {{Said}}, \citenamefont
  {{Saintonge}}, \citenamefont {{Salas Catonga}}, \citenamefont {{Samushia}},
  \citenamefont {{Sanchez}}, \citenamefont {{Saulder}}, \citenamefont
  {{Schaan}}, \citenamefont {{Schlafly}}, \citenamefont {{Schlegel}},
  \citenamefont {{Schmoll}}, \citenamefont {{Scholte}}, \citenamefont
  {{Schubnell}}, \citenamefont {{Secroun}}, \citenamefont {{Seo}},
  \citenamefont {{Serrano}}, \citenamefont {{Sharples}}, \citenamefont
  {{Sholl}}, \citenamefont {{Silber}}, \citenamefont {{Silva}}, \citenamefont
  {{Sirk}}, \citenamefont {{Siudek}}, \citenamefont {{Smith}}, \citenamefont
  {{Sprayberry}}, \citenamefont {{Staten}}, \citenamefont {{Stupak}},
  \citenamefont {{Tan}}, \citenamefont {{Tarl{\'e}}}, \citenamefont {{Tie}},
  \citenamefont {{Tojeiro}}, \citenamefont {{Ure{\~n}a-L{\'o}pez}},
  \citenamefont {{Valdes}}, \citenamefont {{Valenzuela}}, \citenamefont
  {{Valluri}}, \citenamefont {{Vargas-Maga{\~n}a}}, \citenamefont {{Verde}},
  \citenamefont {{Walther}}, \citenamefont {{Wang}}, \citenamefont {{Wang}},
  \citenamefont {{Weaver}}, \citenamefont {{Weaverdyck}}, \citenamefont
  {{Wechsler}}, \citenamefont {{Wilson}}, \citenamefont {{Yang}}, \citenamefont
  {{Yu}}, \citenamefont {{Yuan}}, \citenamefont {{Y{\`e}che}}, \citenamefont
  {{Zhang}}, \citenamefont {{Zhang}}, \citenamefont {{Zhao}}, \citenamefont
  {{Zhou}}, \citenamefont {{Zhou}}, \citenamefont {{Zou}}, \citenamefont
  {{Zou}}, \citenamefont {{Zou}}, \citenamefont {{Zu}},\ and\ \citenamefont
  {{DESI Collaboration}}}]{2022AJ....164..207D}%
  \BibitemOpen
  \bibfield  {author} {\bibinfo {author} {\bibnamefont {{DESI Collaboration}}},
  \bibinfo {author} {\bibfnamefont {B.}~\bibnamefont {{Abareshi}}}, \bibinfo
  {author} {\bibfnamefont {J.}~\bibnamefont {{Aguilar}}}, \bibinfo {author}
  {\bibfnamefont {S.}~\bibnamefont {{Ahlen}}}, \bibinfo {author} {\bibfnamefont
  {S.}~\bibnamefont {{Alam}}}, \bibinfo {author} {\bibfnamefont {D.~M.}\
  \bibnamefont {{Alexander}}}, \bibinfo {author} {\bibfnamefont
  {R.}~\bibnamefont {{Alfarsy}}}, \bibinfo {author} {\bibfnamefont
  {L.}~\bibnamefont {{Allen}}}, \bibinfo {author} {\bibfnamefont
  {C.}~\bibnamefont {{Allende Prieto}}}, \bibinfo {author} {\bibfnamefont
  {O.}~\bibnamefont {{Alves}}}, \bibinfo {author} {\bibfnamefont
  {J.}~\bibnamefont {{Ameel}}}, \bibinfo {author} {\bibfnamefont
  {E.}~\bibnamefont {{Armengaud}}}, \bibinfo {author} {\bibfnamefont
  {J.}~\bibnamefont {{Asorey}}}, \bibinfo {author} {\bibfnamefont
  {A.}~\bibnamefont {{Aviles}}}, \bibinfo {author} {\bibfnamefont
  {S.}~\bibnamefont {{Bailey}}}, \bibinfo {author} {\bibfnamefont
  {A.}~\bibnamefont {{Balaguera-Antol{\'\i}nez}}}, \bibinfo {author}
  {\bibfnamefont {O.}~\bibnamefont {{Ballester}}}, \bibinfo {author}
  {\bibfnamefont {C.}~\bibnamefont {{Baltay}}}, \bibinfo {author}
  {\bibfnamefont {A.}~\bibnamefont {{Bault}}}, \bibinfo {author} {\bibfnamefont
  {S.~F.}\ \bibnamefont {{Beltran}}}, \bibinfo {author} {\bibfnamefont
  {B.}~\bibnamefont {{Benavides}}}, \bibinfo {author} {\bibfnamefont
  {S.}~\bibnamefont {{BenZvi}}}, \bibinfo {author} {\bibfnamefont
  {A.}~\bibnamefont {{Berti}}}, \bibinfo {author} {\bibfnamefont
  {R.}~\bibnamefont {{Besuner}}}, \bibinfo {author} {\bibfnamefont
  {F.}~\bibnamefont {{Beutler}}}, \bibinfo {author} {\bibfnamefont
  {D.}~\bibnamefont {{Bianchi}}}, \bibinfo {author} {\bibfnamefont
  {C.}~\bibnamefont {{Blake}}}, \bibinfo {author} {\bibfnamefont
  {P.}~\bibnamefont {{Blanc}}}, \bibinfo {author} {\bibfnamefont
  {R.}~\bibnamefont {{Blum}}}, \bibinfo {author} {\bibfnamefont
  {A.}~\bibnamefont {{Bolton}}}, \bibinfo {author} {\bibfnamefont
  {S.}~\bibnamefont {{Bose}}}, \bibinfo {author} {\bibfnamefont
  {D.}~\bibnamefont {{Bramall}}}, \bibinfo {author} {\bibfnamefont
  {S.}~\bibnamefont {{Brieden}}}, \bibinfo {author} {\bibfnamefont
  {A.}~\bibnamefont {{Brodzeller}}}, \bibinfo {author} {\bibfnamefont
  {D.}~\bibnamefont {{Brooks}}}, \bibinfo {author} {\bibfnamefont
  {C.}~\bibnamefont {{Brownewell}}}, \bibinfo {author} {\bibfnamefont
  {E.}~\bibnamefont {{Buckley-Geer}}}, \bibinfo {author} {\bibfnamefont
  {R.~N.}\ \bibnamefont {{Cahn}}}, \bibinfo {author} {\bibfnamefont
  {Z.}~\bibnamefont {{Cai}}}, \bibinfo {author} {\bibfnamefont
  {R.}~\bibnamefont {{Canning}}}, \bibinfo {author} {\bibfnamefont
  {R.}~\bibnamefont {{Capasso}}}, \bibinfo {author} {\bibfnamefont
  {A.}~\bibnamefont {{Carnero Rosell}}}, \bibinfo {author} {\bibfnamefont
  {P.}~\bibnamefont {{Carton}}}, \bibinfo {author} {\bibfnamefont
  {R.}~\bibnamefont {{Casas}}}, \bibinfo {author} {\bibfnamefont {F.~J.}\
  \bibnamefont {{Castander}}}, \bibinfo {author} {\bibfnamefont {J.~L.}\
  \bibnamefont {{Cervantes-Cota}}}, \bibinfo {author} {\bibfnamefont
  {S.}~\bibnamefont {{Chabanier}}}, \bibinfo {author} {\bibfnamefont
  {E.}~\bibnamefont {{Chaussidon}}}, \bibinfo {author} {\bibfnamefont
  {C.}~\bibnamefont {{Chuang}}}, \bibinfo {author} {\bibfnamefont
  {C.}~\bibnamefont {{Circosta}}}, \bibinfo {author} {\bibfnamefont
  {S.}~\bibnamefont {{Cole}}}, \bibinfo {author} {\bibfnamefont {A.~P.}\
  \bibnamefont {{Cooper}}}, \bibinfo {author} {\bibfnamefont {L.}~\bibnamefont
  {{da Costa}}}, \bibinfo {author} {\bibfnamefont {M.~C.}\ \bibnamefont
  {{Cousinou}}}, \bibinfo {author} {\bibfnamefont {A.}~\bibnamefont {{Cuceu}}},
  \bibinfo {author} {\bibfnamefont {T.~M.}\ \bibnamefont {{Davis}}}, \bibinfo
  {author} {\bibfnamefont {K.}~\bibnamefont {{Dawson}}}, \bibinfo {author}
  {\bibfnamefont {R.}~\bibnamefont {{de la Cruz-Noriega}}}, \bibinfo {author}
  {\bibfnamefont {A.}~\bibnamefont {{de la Macorra}}}, \bibinfo {author}
  {\bibfnamefont {A.}~\bibnamefont {{de Mattia}}}, \bibinfo {author}
  {\bibfnamefont {J.}~\bibnamefont {{Della Costa}}}, \bibinfo {author}
  {\bibfnamefont {P.}~\bibnamefont {{Demmer}}}, \bibinfo {author}
  {\bibfnamefont {M.}~\bibnamefont {{Derwent}}}, \bibinfo {author}
  {\bibfnamefont {A.}~\bibnamefont {{Dey}}}, \bibinfo {author} {\bibfnamefont
  {B.}~\bibnamefont {{Dey}}}, \bibinfo {author} {\bibfnamefont
  {G.}~\bibnamefont {{Dhungana}}}, \bibinfo {author} {\bibfnamefont
  {Z.}~\bibnamefont {{Ding}}}, \bibinfo {author} {\bibfnamefont
  {C.}~\bibnamefont {{Dobson}}}, \bibinfo {author} {\bibfnamefont
  {P.}~\bibnamefont {{Doel}}}, \bibinfo {author} {\bibfnamefont
  {J.}~\bibnamefont {{Donald-McCann}}}, \bibinfo {author} {\bibfnamefont
  {J.}~\bibnamefont {{Donaldson}}}, \bibinfo {author} {\bibfnamefont
  {K.}~\bibnamefont {{Douglass}}}, \bibinfo {author} {\bibfnamefont
  {Y.}~\bibnamefont {{Duan}}}, \bibinfo {author} {\bibfnamefont
  {P.}~\bibnamefont {{Dunlop}}}, \bibinfo {author} {\bibfnamefont
  {J.}~\bibnamefont {{Edelstein}}}, \bibinfo {author} {\bibfnamefont
  {S.}~\bibnamefont {{Eftekharzadeh}}}, \bibinfo {author} {\bibfnamefont
  {D.~J.}\ \bibnamefont {{Eisenstein}}}, \bibinfo {author} {\bibfnamefont
  {M.}~\bibnamefont {{Enriquez-Vargas}}}, \bibinfo {author} {\bibfnamefont
  {S.}~\bibnamefont {{Escoffier}}}, \bibinfo {author} {\bibfnamefont
  {M.}~\bibnamefont {{Evatt}}}, \bibinfo {author} {\bibfnamefont
  {P.}~\bibnamefont {{Fagrelius}}}, \bibinfo {author} {\bibfnamefont
  {X.}~\bibnamefont {{Fan}}}, \bibinfo {author} {\bibfnamefont
  {K.}~\bibnamefont {{Fanning}}}, \bibinfo {author} {\bibfnamefont {V.~A.}\
  \bibnamefont {{Fawcett}}}, \bibinfo {author} {\bibfnamefont {S.}~\bibnamefont
  {{Ferraro}}}, \bibinfo {author} {\bibfnamefont {J.}~\bibnamefont {{Ereza}}},
  \bibinfo {author} {\bibfnamefont {B.}~\bibnamefont {{Flaugher}}}, \bibinfo
  {author} {\bibfnamefont {A.}~\bibnamefont {{Font-Ribera}}}, \bibinfo {author}
  {\bibfnamefont {J.~E.}\ \bibnamefont {{Forero-Romero}}}, \bibinfo {author}
  {\bibfnamefont {C.~S.}\ \bibnamefont {{Frenk}}}, \bibinfo {author}
  {\bibfnamefont {S.}~\bibnamefont {{Fromenteau}}}, \bibinfo {author}
  {\bibfnamefont {B.~T.}\ \bibnamefont {{G{\"a}nsicke}}}, \bibinfo {author}
  {\bibfnamefont {C.}~\bibnamefont {{Garcia-Quintero}}}, \bibinfo {author}
  {\bibfnamefont {L.}~\bibnamefont {{Garrison}}}, \bibinfo {author}
  {\bibfnamefont {E.}~\bibnamefont {{Gazta{\~n}aga}}}, \bibinfo {author}
  {\bibfnamefont {F.}~\bibnamefont {{Gerardi}}}, \bibinfo {author}
  {\bibfnamefont {H.}~\bibnamefont {{Gil-Mar{\'\i}n}}}, \bibinfo {author}
  {\bibfnamefont {S.}~\bibnamefont {{Gontcho a Gontcho}}}, \bibinfo {author}
  {\bibfnamefont {A.~X.}\ \bibnamefont {{Gonzalez-Morales}}}, \bibinfo {author}
  {\bibfnamefont {G.}~\bibnamefont {{Gonzalez-de-Rivera}}}, \bibinfo {author}
  {\bibfnamefont {V.}~\bibnamefont {{Gonzalez-Perez}}}, \bibinfo {author}
  {\bibfnamefont {C.}~\bibnamefont {{Gordon}}}, \bibinfo {author}
  {\bibfnamefont {O.}~\bibnamefont {{Graur}}}, \bibinfo {author} {\bibfnamefont
  {D.}~\bibnamefont {{Green}}}, \bibinfo {author} {\bibfnamefont
  {C.}~\bibnamefont {{Grove}}}, \bibinfo {author} {\bibfnamefont
  {D.}~\bibnamefont {{Gruen}}}, \bibinfo {author} {\bibfnamefont
  {G.}~\bibnamefont {{Gutierrez}}}, \bibinfo {author} {\bibfnamefont
  {J.}~\bibnamefont {{Guy}}}, \bibinfo {author} {\bibfnamefont
  {C.}~\bibnamefont {{Hahn}}}, \bibinfo {author} {\bibfnamefont
  {S.}~\bibnamefont {{Harris}}}, \bibinfo {author} {\bibfnamefont
  {D.}~\bibnamefont {{Herrera}}}, \bibinfo {author} {\bibfnamefont {H.~K.}\
  \bibnamefont {{Herrera-Alcantar}}}, \bibinfo {author} {\bibfnamefont
  {K.}~\bibnamefont {{Honscheid}}}, \bibinfo {author} {\bibfnamefont
  {C.}~\bibnamefont {{Howlett}}}, \bibinfo {author} {\bibfnamefont
  {D.}~\bibnamefont {{Huterer}}}, \bibinfo {author} {\bibfnamefont
  {V.}~\bibnamefont {{Ir{\v{s}}i{\v{c}}}}}, \bibinfo {author} {\bibfnamefont
  {M.}~\bibnamefont {{Ishak}}}, \bibinfo {author} {\bibfnamefont
  {P.}~\bibnamefont {{Jelinsky}}}, \bibinfo {author} {\bibfnamefont
  {L.}~\bibnamefont {{Jiang}}}, \bibinfo {author} {\bibfnamefont
  {J.}~\bibnamefont {{Jimenez}}}, \bibinfo {author} {\bibfnamefont {Y.~P.}\
  \bibnamefont {{Jing}}}, \bibinfo {author} {\bibfnamefont {R.}~\bibnamefont
  {{Joyce}}}, \bibinfo {author} {\bibfnamefont {E.}~\bibnamefont {{Jullo}}},
  \bibinfo {author} {\bibfnamefont {S.}~\bibnamefont {{Juneau}}}, \bibinfo
  {author} {\bibfnamefont {N.~G.}\ \bibnamefont {{Kara{\c{c}}ayl{\i}}}},
  \bibinfo {author} {\bibfnamefont {M.}~\bibnamefont {{Karamanis}}}, \bibinfo
  {author} {\bibfnamefont {A.}~\bibnamefont {{Karcher}}}, \bibinfo {author}
  {\bibfnamefont {T.}~\bibnamefont {{Karim}}}, \bibinfo {author} {\bibfnamefont
  {R.}~\bibnamefont {{Kehoe}}}, \bibinfo {author} {\bibfnamefont
  {S.}~\bibnamefont {{Kent}}}, \bibinfo {author} {\bibfnamefont
  {D.}~\bibnamefont {{Kirkby}}}, \bibinfo {author} {\bibfnamefont
  {T.}~\bibnamefont {{Kisner}}}, \bibinfo {author} {\bibfnamefont
  {F.}~\bibnamefont {{Kitaura}}}, \bibinfo {author} {\bibfnamefont {S.~E.}\
  \bibnamefont {{Koposov}}}, \bibinfo {author} {\bibfnamefont {A.}~\bibnamefont
  {{Kov{\'a}cs}}}, \bibinfo {author} {\bibfnamefont {A.}~\bibnamefont
  {{Kremin}}}, \bibinfo {author} {\bibfnamefont {A.}~\bibnamefont
  {{Krolewski}}}, \bibinfo {author} {\bibfnamefont {B.}~\bibnamefont
  {{L'Huillier}}}, \bibinfo {author} {\bibfnamefont {O.}~\bibnamefont
  {{Lahav}}}, \bibinfo {author} {\bibfnamefont {A.}~\bibnamefont {{Lambert}}},
  \bibinfo {author} {\bibfnamefont {C.}~\bibnamefont {{Lamman}}}, \bibinfo
  {author} {\bibfnamefont {T.-W.}\ \bibnamefont {{Lan}}}, \bibinfo {author}
  {\bibfnamefont {M.}~\bibnamefont {{Landriau}}}, \bibinfo {author}
  {\bibfnamefont {S.}~\bibnamefont {{Lane}}}, \bibinfo {author} {\bibfnamefont
  {D.}~\bibnamefont {{Lang}}}, \bibinfo {author} {\bibfnamefont {J.~U.}\
  \bibnamefont {{Lange}}}, \bibinfo {author} {\bibfnamefont {J.}~\bibnamefont
  {{Lasker}}}, \bibinfo {author} {\bibfnamefont {L.}~\bibnamefont {{Le
  Guillou}}}, \bibinfo {author} {\bibfnamefont {A.}~\bibnamefont
  {{Leauthaud}}}, \bibinfo {author} {\bibfnamefont {A.}~\bibnamefont {{Le Van
  Suu}}}, \bibinfo {author} {\bibfnamefont {M.~E.}\ \bibnamefont {{Levi}}},
  \bibinfo {author} {\bibfnamefont {T.~S.}\ \bibnamefont {{Li}}}, \bibinfo
  {author} {\bibfnamefont {C.}~\bibnamefont {{Magneville}}}, \bibinfo {author}
  {\bibfnamefont {M.}~\bibnamefont {{Manera}}}, \bibinfo {author}
  {\bibfnamefont {C.~J.}\ \bibnamefont {{Manser}}}, \bibinfo {author}
  {\bibfnamefont {B.}~\bibnamefont {{Marshall}}}, \bibinfo {author}
  {\bibfnamefont {P.}~\bibnamefont {{Martini}}}, \bibinfo {author}
  {\bibfnamefont {W.}~\bibnamefont {{McCollam}}}, \bibinfo {author}
  {\bibfnamefont {P.}~\bibnamefont {{McDonald}}}, \bibinfo {author}
  {\bibfnamefont {A.~M.}\ \bibnamefont {{Meisner}}}, \bibinfo {author}
  {\bibfnamefont {J.}~\bibnamefont {{Mena-Fern{\'a}ndez}}}, \bibinfo {author}
  {\bibfnamefont {J.}~\bibnamefont {{Meneses-Rizo}}}, \bibinfo {author}
  {\bibfnamefont {M.}~\bibnamefont {{Mezcua}}}, \bibinfo {author}
  {\bibfnamefont {T.}~\bibnamefont {{Miller}}}, \bibinfo {author}
  {\bibfnamefont {R.}~\bibnamefont {{Miquel}}}, \bibinfo {author}
  {\bibfnamefont {P.}~\bibnamefont {{Montero-Camacho}}}, \bibinfo {author}
  {\bibfnamefont {J.}~\bibnamefont {{Moon}}}, \bibinfo {author} {\bibfnamefont
  {J.}~\bibnamefont {{Moustakas}}}, \bibinfo {author} {\bibfnamefont
  {E.}~\bibnamefont {{Mueller}}}, \bibinfo {author} {\bibfnamefont
  {A.}~\bibnamefont {{Mu{\~n}oz-Guti{\'e}rrez}}}, \bibinfo {author}
  {\bibfnamefont {A.~D.}\ \bibnamefont {{Myers}}}, \bibinfo {author}
  {\bibfnamefont {S.}~\bibnamefont {{Nadathur}}}, \bibinfo {author}
  {\bibfnamefont {J.}~\bibnamefont {{Najita}}}, \bibinfo {author}
  {\bibfnamefont {L.}~\bibnamefont {{Napolitano}}}, \bibinfo {author}
  {\bibfnamefont {E.}~\bibnamefont {{Neilsen}}}, \bibinfo {author}
  {\bibfnamefont {J.~A.}\ \bibnamefont {{Newman}}}, \bibinfo {author}
  {\bibfnamefont {J.~D.}\ \bibnamefont {{Nie}}}, \bibinfo {author}
  {\bibfnamefont {Y.}~\bibnamefont {{Ning}}}, \bibinfo {author} {\bibfnamefont
  {G.}~\bibnamefont {{Niz}}}, \bibinfo {author} {\bibfnamefont
  {P.}~\bibnamefont {{Norberg}}}, \bibinfo {author} {\bibfnamefont {H.~E.}\
  \bibnamefont {{Noriega}}}, \bibinfo {author} {\bibfnamefont {T.}~\bibnamefont
  {{O'Brien}}}, \bibinfo {author} {\bibfnamefont {A.}~\bibnamefont
  {{Obuljen}}}, \bibinfo {author} {\bibfnamefont {N.}~\bibnamefont
  {{Palanque-Delabrouille}}}, \bibinfo {author} {\bibfnamefont
  {A.}~\bibnamefont {{Palmese}}}, \bibinfo {author} {\bibfnamefont
  {P.}~\bibnamefont {{Zhiwei}}}, \bibinfo {author} {\bibfnamefont
  {D.}~\bibnamefont {{Pappalardo}}}, \bibinfo {author} {\bibfnamefont
  {X.}~\bibnamefont {{PENG}}}, \bibinfo {author} {\bibfnamefont {W.~J.}\
  \bibnamefont {{Percival}}}, \bibinfo {author} {\bibfnamefont
  {S.}~\bibnamefont {{Perruchot}}}, \bibinfo {author} {\bibfnamefont
  {R.}~\bibnamefont {{Pogge}}}, \bibinfo {author} {\bibfnamefont
  {C.}~\bibnamefont {{Poppett}}}, \bibinfo {author} {\bibfnamefont
  {A.}~\bibnamefont {{Porredon}}}, \bibinfo {author} {\bibfnamefont
  {F.}~\bibnamefont {{Prada}}}, \bibinfo {author} {\bibfnamefont
  {J.}~\bibnamefont {{Prochaska}}}, \bibinfo {author} {\bibfnamefont
  {R.}~\bibnamefont {{Pucha}}}, \bibinfo {author} {\bibfnamefont
  {A.}~\bibnamefont {{P{\'e}rez-Fern{\'a}ndez}}}, \bibinfo {author}
  {\bibfnamefont {I.}~\bibnamefont {{P{\'e}rez-R{\`a}fols}}}, \bibinfo {author}
  {\bibfnamefont {D.}~\bibnamefont {{Rabinowitz}}}, \bibinfo {author}
  {\bibfnamefont {A.}~\bibnamefont {{Raichoor}}}, \bibinfo {author}
  {\bibfnamefont {S.}~\bibnamefont {{Ramirez-Solano}}}, \bibinfo {author}
  {\bibfnamefont {C.}~\bibnamefont {{Ram{\'\i}rez-P{\'e}rez}}}, \bibinfo
  {author} {\bibfnamefont {C.}~\bibnamefont {{Ravoux}}}, \bibinfo {author}
  {\bibfnamefont {K.}~\bibnamefont {{Reil}}}, \bibinfo {author} {\bibfnamefont
  {M.}~\bibnamefont {{Rezaie}}}, \bibinfo {author} {\bibfnamefont
  {A.}~\bibnamefont {{Rocher}}}, \bibinfo {author} {\bibfnamefont
  {C.}~\bibnamefont {{Rockosi}}}, \bibinfo {author} {\bibfnamefont {N.~A.}\
  \bibnamefont {{Roe}}}, \bibinfo {author} {\bibfnamefont {A.}~\bibnamefont
  {{Roodman}}}, \bibinfo {author} {\bibfnamefont {A.~J.}\ \bibnamefont
  {{Ross}}}, \bibinfo {author} {\bibfnamefont {G.}~\bibnamefont {{Rossi}}},
  \bibinfo {author} {\bibfnamefont {R.}~\bibnamefont {{Ruggeri}}}, \bibinfo
  {author} {\bibfnamefont {V.}~\bibnamefont {{Ruhlmann-Kleider}}}, \bibinfo
  {author} {\bibfnamefont {C.~G.}\ \bibnamefont {{Sabiu}}}, \bibinfo {author}
  {\bibfnamefont {S.}~\bibnamefont {{Safonova}}}, \bibinfo {author}
  {\bibfnamefont {K.}~\bibnamefont {{Said}}}, \bibinfo {author} {\bibfnamefont
  {A.}~\bibnamefont {{Saintonge}}}, \bibinfo {author} {\bibfnamefont
  {J.}~\bibnamefont {{Salas Catonga}}}, \bibinfo {author} {\bibfnamefont
  {L.}~\bibnamefont {{Samushia}}}, \bibinfo {author} {\bibfnamefont
  {E.}~\bibnamefont {{Sanchez}}}, \bibinfo {author} {\bibfnamefont
  {C.}~\bibnamefont {{Saulder}}}, \bibinfo {author} {\bibfnamefont
  {E.}~\bibnamefont {{Schaan}}}, \bibinfo {author} {\bibfnamefont
  {E.}~\bibnamefont {{Schlafly}}}, \bibinfo {author} {\bibfnamefont
  {D.}~\bibnamefont {{Schlegel}}}, \bibinfo {author} {\bibfnamefont
  {J.}~\bibnamefont {{Schmoll}}}, \bibinfo {author} {\bibfnamefont
  {D.}~\bibnamefont {{Scholte}}}, \bibinfo {author} {\bibfnamefont
  {M.}~\bibnamefont {{Schubnell}}}, \bibinfo {author} {\bibfnamefont
  {A.}~\bibnamefont {{Secroun}}}, \bibinfo {author} {\bibfnamefont
  {H.}~\bibnamefont {{Seo}}}, \bibinfo {author} {\bibfnamefont
  {S.}~\bibnamefont {{Serrano}}}, \bibinfo {author} {\bibfnamefont {R.~M.}\
  \bibnamefont {{Sharples}}}, \bibinfo {author} {\bibfnamefont {M.~J.}\
  \bibnamefont {{Sholl}}}, \bibinfo {author} {\bibfnamefont {J.~H.}\
  \bibnamefont {{Silber}}}, \bibinfo {author} {\bibfnamefont {D.~R.}\
  \bibnamefont {{Silva}}}, \bibinfo {author} {\bibfnamefont {M.}~\bibnamefont
  {{Sirk}}}, \bibinfo {author} {\bibfnamefont {M.}~\bibnamefont {{Siudek}}},
  \bibinfo {author} {\bibfnamefont {A.}~\bibnamefont {{Smith}}}, \bibinfo
  {author} {\bibfnamefont {D.}~\bibnamefont {{Sprayberry}}}, \bibinfo {author}
  {\bibfnamefont {R.}~\bibnamefont {{Staten}}}, \bibinfo {author}
  {\bibfnamefont {B.}~\bibnamefont {{Stupak}}}, \bibinfo {author}
  {\bibfnamefont {T.}~\bibnamefont {{Tan}}}, \bibinfo {author} {\bibfnamefont
  {G.}~\bibnamefont {{Tarl{\'e}}}}, \bibinfo {author} {\bibfnamefont {S.~S.}\
  \bibnamefont {{Tie}}}, \bibinfo {author} {\bibfnamefont {R.}~\bibnamefont
  {{Tojeiro}}}, \bibinfo {author} {\bibfnamefont {L.~A.}\ \bibnamefont
  {{Ure{\~n}a-L{\'o}pez}}}, \bibinfo {author} {\bibfnamefont {F.}~\bibnamefont
  {{Valdes}}}, \bibinfo {author} {\bibfnamefont {O.}~\bibnamefont
  {{Valenzuela}}}, \bibinfo {author} {\bibfnamefont {M.}~\bibnamefont
  {{Valluri}}}, \bibinfo {author} {\bibfnamefont {M.}~\bibnamefont
  {{Vargas-Maga{\~n}a}}}, \bibinfo {author} {\bibfnamefont {L.}~\bibnamefont
  {{Verde}}}, \bibinfo {author} {\bibfnamefont {M.}~\bibnamefont {{Walther}}},
  \bibinfo {author} {\bibfnamefont {B.}~\bibnamefont {{Wang}}}, \bibinfo
  {author} {\bibfnamefont {M.~S.}\ \bibnamefont {{Wang}}}, \bibinfo {author}
  {\bibfnamefont {B.~A.}\ \bibnamefont {{Weaver}}}, \bibinfo {author}
  {\bibfnamefont {C.}~\bibnamefont {{Weaverdyck}}}, \bibinfo {author}
  {\bibfnamefont {R.}~\bibnamefont {{Wechsler}}}, \bibinfo {author}
  {\bibfnamefont {M.~J.}\ \bibnamefont {{Wilson}}}, \bibinfo {author}
  {\bibfnamefont {J.}~\bibnamefont {{Yang}}}, \bibinfo {author} {\bibfnamefont
  {Y.}~\bibnamefont {{Yu}}}, \bibinfo {author} {\bibfnamefont {S.}~\bibnamefont
  {{Yuan}}}, \bibinfo {author} {\bibfnamefont {C.}~\bibnamefont {{Y{\`e}che}}},
  \bibinfo {author} {\bibfnamefont {H.}~\bibnamefont {{Zhang}}}, \bibinfo
  {author} {\bibfnamefont {K.}~\bibnamefont {{Zhang}}}, \bibinfo {author}
  {\bibfnamefont {C.}~\bibnamefont {{Zhao}}}, \bibinfo {author} {\bibfnamefont
  {R.}~\bibnamefont {{Zhou}}}, \bibinfo {author} {\bibfnamefont
  {Z.}~\bibnamefont {{Zhou}}}, \bibinfo {author} {\bibfnamefont
  {H.}~\bibnamefont {{Zou}}}, \bibinfo {author} {\bibfnamefont
  {J.}~\bibnamefont {{Zou}}}, \bibinfo {author} {\bibfnamefont
  {S.}~\bibnamefont {{Zou}}}, \bibinfo {author} {\bibfnamefont
  {Y.}~\bibnamefont {{Zu}}},\ and\ \bibinfo {author} {\bibnamefont {{DESI
  Collaboration}}},\ }\bibfield  {title} {\bibinfo {title} {{Overview of the
  Instrumentation for the Dark Energy Spectroscopic Instrument}},\ }\href
  {https://doi.org/10.3847/1538-3881/ac882b} {\bibfield  {journal} {\bibinfo
  {journal} {\aj}\ }\textbf {\bibinfo {volume} {164}},\ \bibinfo {eid} {207}
  (\bibinfo {year} {2022})},\ \Eprint {https://arxiv.org/abs/2205.10939}
  {arXiv:2205.10939 [astro-ph.IM]} \BibitemShut {NoStop}%
\bibitem [{\citenamefont {{DESI Collaboration}}\ \emph
  {et~al.}(2016{\natexlab{a}})\citenamefont {{DESI Collaboration}},
  \citenamefont {{Aghamousa}}, \citenamefont {{Aguilar}}, \citenamefont
  {{Ahlen}}, \citenamefont {{Alam}}, \citenamefont {{Allen}}, \citenamefont
  {{Allende Prieto}}, \citenamefont {{Annis}}, \citenamefont {{Bailey}},
  \citenamefont {{Balland}}, \citenamefont {{Ballester}}, \citenamefont
  {{Baltay}}, \citenamefont {{Beaufore}}, \citenamefont {{Bebek}},
  \citenamefont {{Beers}}, \citenamefont {{Bell}}, \citenamefont {{Bernal}},
  \citenamefont {{Besuner}}, \citenamefont {{Beutler}}, \citenamefont
  {{Blake}}, \citenamefont {{Bleuler}}, \citenamefont {{Blomqvist}},
  \citenamefont {{Blum}}, \citenamefont {{Bolton}}, \citenamefont {{Briceno}},
  \citenamefont {{Brooks}}, \citenamefont {{Brownstein}}, \citenamefont
  {{Buckley-Geer}}, \citenamefont {{Burden}}, \citenamefont {{Burtin}},
  \citenamefont {{Busca}}, \citenamefont {{Cahn}}, \citenamefont {{Cai}},
  \citenamefont {{Cardiel-Sas}}, \citenamefont {{Carlberg}}, \citenamefont
  {{Carton}}, \citenamefont {{Casas}}, \citenamefont {{Castander}},
  \citenamefont {{Cervantes-Cota}}, \citenamefont {{Claybaugh}}, \citenamefont
  {{Close}}, \citenamefont {{Coker}}, \citenamefont {{Cole}}, \citenamefont
  {{Comparat}}, \citenamefont {{Cooper}}, \citenamefont {{Cousinou}},
  \citenamefont {{Crocce}}, \citenamefont {{Cuby}}, \citenamefont
  {{Cunningham}}, \citenamefont {{Davis}}, \citenamefont {{Dawson}},
  \citenamefont {{de la Macorra}}, \citenamefont {{De Vicente}}, \citenamefont
  {{Delubac}}, \citenamefont {{Derwent}}, \citenamefont {{Dey}}, \citenamefont
  {{Dhungana}}, \citenamefont {{Ding}}, \citenamefont {{Doel}}, \citenamefont
  {{Duan}}, \citenamefont {{Ealet}}, \citenamefont {{Edelstein}}, \citenamefont
  {{Eftekharzadeh}}, \citenamefont {{Eisenstein}}, \citenamefont {{Elliott}},
  \citenamefont {{Escoffier}}, \citenamefont {{Evatt}}, \citenamefont
  {{Fagrelius}}, \citenamefont {{Fan}}, \citenamefont {{Fanning}},
  \citenamefont {{Farahi}}, \citenamefont {{Farihi}}, \citenamefont {{Favole}},
  \citenamefont {{Feng}}, \citenamefont {{Fernandez}}, \citenamefont
  {{Findlay}}, \citenamefont {{Finkbeiner}}, \citenamefont {{Fitzpatrick}},
  \citenamefont {{Flaugher}}, \citenamefont {{Flender}}, \citenamefont
  {{Font-Ribera}}, \citenamefont {{Forero-Romero}}, \citenamefont {{Fosalba}},
  \citenamefont {{Frenk}}, \citenamefont {{Fumagalli}}, \citenamefont
  {{Gaensicke}}, \citenamefont {{Gallo}}, \citenamefont {{Garcia-Bellido}},
  \citenamefont {{Gaztanaga}}, \citenamefont {{Pietro Gentile Fusillo}},
  \citenamefont {{Gerard}}, \citenamefont {{Gershkovich}}, \citenamefont
  {{Giannantonio}}, \citenamefont {{Gillet}}, \citenamefont
  {{Gonzalez-de-Rivera}}, \citenamefont {{Gonzalez-Perez}}, \citenamefont
  {{Gott}}, \citenamefont {{Graur}}, \citenamefont {{Gutierrez}}, \citenamefont
  {{Guy}}, \citenamefont {{Habib}}, \citenamefont {{Heetderks}}, \citenamefont
  {{Heetderks}}, \citenamefont {{Heitmann}}, \citenamefont {{Hellwing}},
  \citenamefont {{Herrera}}, \citenamefont {{Ho}}, \citenamefont {{Holland}},
  \citenamefont {{Honscheid}}, \citenamefont {{Huff}}, \citenamefont
  {{Hutchinson}}, \citenamefont {{Huterer}}, \citenamefont {{Hwang}},
  \citenamefont {{Illa Laguna}}, \citenamefont {{Ishikawa}}, \citenamefont
  {{Jacobs}}, \citenamefont {{Jeffrey}}, \citenamefont {{Jelinsky}},
  \citenamefont {{Jennings}}, \citenamefont {{Jiang}}, \citenamefont
  {{Jimenez}}, \citenamefont {{Johnson}}, \citenamefont {{Joyce}},
  \citenamefont {{Jullo}}, \citenamefont {{Juneau}}, \citenamefont {{Kama}},
  \citenamefont {{Karcher}}, \citenamefont {{Karkar}}, \citenamefont {{Kehoe}},
  \citenamefont {{Kennamer}}, \citenamefont {{Kent}}, \citenamefont
  {{Kilbinger}}, \citenamefont {{Kim}}, \citenamefont {{Kirkby}}, \citenamefont
  {{Kisner}}, \citenamefont {{Kitanidis}}, \citenamefont {{Kneib}},
  \citenamefont {{Koposov}}, \citenamefont {{Kovacs}}, \citenamefont
  {{Koyama}}, \citenamefont {{Kremin}}, \citenamefont {{Kron}}, \citenamefont
  {{Kronig}}, \citenamefont {{Kueter-Young}}, \citenamefont {{Lacey}},
  \citenamefont {{Lafever}}, \citenamefont {{Lahav}}, \citenamefont
  {{Lambert}}, \citenamefont {{Lampton}}, \citenamefont {{Landriau}},
  \citenamefont {{Lang}}, \citenamefont {{Lauer}}, \citenamefont {{Le Goff}},
  \citenamefont {{Le Guillou}}, \citenamefont {{Le Van Suu}}, \citenamefont
  {{Lee}}, \citenamefont {{Lee}}, \citenamefont {{Leitner}}, \citenamefont
  {{Lesser}}, \citenamefont {{Levi}}, \citenamefont {{L'Huillier}},
  \citenamefont {{Li}}, \citenamefont {{Liang}}, \citenamefont {{Lin}},
  \citenamefont {{Linder}}, \citenamefont {{Loebman}}, \citenamefont
  {{Luki{\'c}}}, \citenamefont {{Ma}}, \citenamefont {{MacCrann}},
  \citenamefont {{Magneville}}, \citenamefont {{Makarem}}, \citenamefont
  {{Manera}}, \citenamefont {{Manser}}, \citenamefont {{Marshall}},
  \citenamefont {{Martini}}, \citenamefont {{Massey}}, \citenamefont
  {{Matheson}}, \citenamefont {{McCauley}}, \citenamefont {{McDonald}},
  \citenamefont {{McGreer}}, \citenamefont {{Meisner}}, \citenamefont
  {{Metcalfe}}, \citenamefont {{Miller}}, \citenamefont {{Miquel}},
  \citenamefont {{Moustakas}}, \citenamefont {{Myers}}, \citenamefont {{Naik}},
  \citenamefont {{Newman}}, \citenamefont {{Nichol}}, \citenamefont {{Nicola}},
  \citenamefont {{Nicolati da Costa}}, \citenamefont {{Nie}}, \citenamefont
  {{Niz}}, \citenamefont {{Norberg}}, \citenamefont {{Nord}}, \citenamefont
  {{Norman}}, \citenamefont {{Nugent}}, \citenamefont {{O'Brien}},
  \citenamefont {{Oh}}, \citenamefont {{Olsen}}, \citenamefont {{Padilla}},
  \citenamefont {{Padmanabhan}}, \citenamefont {{Padmanabhan}}, \citenamefont
  {{Palanque-Delabrouille}}, \citenamefont {{Palmese}}, \citenamefont
  {{Pappalardo}}, \citenamefont {{P{\^a}ris}}, \citenamefont {{Park}},
  \citenamefont {{Patej}}, \citenamefont {{Peacock}}, \citenamefont {{Peiris}},
  \citenamefont {{Peng}}, \citenamefont {{Percival}}, \citenamefont
  {{Perruchot}}, \citenamefont {{Pieri}}, \citenamefont {{Pogge}},
  \citenamefont {{Pollack}}, \citenamefont {{Poppett}}, \citenamefont
  {{Prada}}, \citenamefont {{Prakash}}, \citenamefont {{Probst}}, \citenamefont
  {{Rabinowitz}}, \citenamefont {{Raichoor}}, \citenamefont {{Ree}},
  \citenamefont {{Refregier}}, \citenamefont {{Regal}}, \citenamefont {{Reid}},
  \citenamefont {{Reil}}, \citenamefont {{Rezaie}}, \citenamefont {{Rockosi}},
  \citenamefont {{Roe}}, \citenamefont {{Ronayette}}, \citenamefont
  {{Roodman}}, \citenamefont {{Ross}}, \citenamefont {{Ross}}, \citenamefont
  {{Rossi}}, \citenamefont {{Rozo}}, \citenamefont {{Ruhlmann-Kleider}},
  \citenamefont {{Rykoff}}, \citenamefont {{Sabiu}}, \citenamefont
  {{Samushia}}, \citenamefont {{Sanchez}}, \citenamefont {{Sanchez}},
  \citenamefont {{Schlegel}}, \citenamefont {{Schneider}}, \citenamefont
  {{Schubnell}}, \citenamefont {{Secroun}}, \citenamefont {{Seljak}},
  \citenamefont {{Seo}}, \citenamefont {{Serrano}}, \citenamefont
  {{Shafieloo}}, \citenamefont {{Shan}}, \citenamefont {{Sharples}},
  \citenamefont {{Sholl}}, \citenamefont {{Shourt}}, \citenamefont {{Silber}},
  \citenamefont {{Silva}}, \citenamefont {{Sirk}}, \citenamefont {{Slosar}},
  \citenamefont {{Smith}}, \citenamefont {{Smoot}}, \citenamefont {{Som}},
  \citenamefont {{Song}}, \citenamefont {{Sprayberry}}, \citenamefont
  {{Staten}}, \citenamefont {{Stefanik}}, \citenamefont {{Tarle}},
  \citenamefont {{Sien Tie}}, \citenamefont {{Tinker}}, \citenamefont
  {{Tojeiro}}, \citenamefont {{Valdes}}, \citenamefont {{Valenzuela}},
  \citenamefont {{Valluri}}, \citenamefont {{Vargas-Magana}}, \citenamefont
  {{Verde}}, \citenamefont {{Walker}}, \citenamefont {{Wang}}, \citenamefont
  {{Wang}}, \citenamefont {{Weaver}}, \citenamefont {{Weaverdyck}},
  \citenamefont {{Wechsler}}, \citenamefont {{Weinberg}}, \citenamefont
  {{White}}, \citenamefont {{Yang}}, \citenamefont {{Yeche}}, \citenamefont
  {{Zhang}}, \citenamefont {{Zhao}}, \citenamefont {{Zheng}}, \citenamefont
  {{Zhou}}, \citenamefont {{Zhou}}, \citenamefont {{Zhu}}, \citenamefont
  {{Zou}},\ and\ \citenamefont {{Zu}}}]{2016arXiv161100037D}%
  \BibitemOpen
  \bibfield  {author} {\bibinfo {author} {\bibnamefont {{DESI Collaboration}}},
  \bibinfo {author} {\bibfnamefont {A.}~\bibnamefont {{Aghamousa}}}, \bibinfo
  {author} {\bibfnamefont {J.}~\bibnamefont {{Aguilar}}}, \bibinfo {author}
  {\bibfnamefont {S.}~\bibnamefont {{Ahlen}}}, \bibinfo {author} {\bibfnamefont
  {S.}~\bibnamefont {{Alam}}}, \bibinfo {author} {\bibfnamefont {L.~E.}\
  \bibnamefont {{Allen}}}, \bibinfo {author} {\bibfnamefont {C.}~\bibnamefont
  {{Allende Prieto}}}, \bibinfo {author} {\bibfnamefont {J.}~\bibnamefont
  {{Annis}}}, \bibinfo {author} {\bibfnamefont {S.}~\bibnamefont {{Bailey}}},
  \bibinfo {author} {\bibfnamefont {C.}~\bibnamefont {{Balland}}}, \bibinfo
  {author} {\bibfnamefont {O.}~\bibnamefont {{Ballester}}}, \bibinfo {author}
  {\bibfnamefont {C.}~\bibnamefont {{Baltay}}}, \bibinfo {author}
  {\bibfnamefont {L.}~\bibnamefont {{Beaufore}}}, \bibinfo {author}
  {\bibfnamefont {C.}~\bibnamefont {{Bebek}}}, \bibinfo {author} {\bibfnamefont
  {T.~C.}\ \bibnamefont {{Beers}}}, \bibinfo {author} {\bibfnamefont {E.~F.}\
  \bibnamefont {{Bell}}}, \bibinfo {author} {\bibfnamefont {J.~L.}\
  \bibnamefont {{Bernal}}}, \bibinfo {author} {\bibfnamefont {R.}~\bibnamefont
  {{Besuner}}}, \bibinfo {author} {\bibfnamefont {F.}~\bibnamefont
  {{Beutler}}}, \bibinfo {author} {\bibfnamefont {C.}~\bibnamefont {{Blake}}},
  \bibinfo {author} {\bibfnamefont {H.}~\bibnamefont {{Bleuler}}}, \bibinfo
  {author} {\bibfnamefont {M.}~\bibnamefont {{Blomqvist}}}, \bibinfo {author}
  {\bibfnamefont {R.}~\bibnamefont {{Blum}}}, \bibinfo {author} {\bibfnamefont
  {A.~S.}\ \bibnamefont {{Bolton}}}, \bibinfo {author} {\bibfnamefont
  {C.}~\bibnamefont {{Briceno}}}, \bibinfo {author} {\bibfnamefont
  {D.}~\bibnamefont {{Brooks}}}, \bibinfo {author} {\bibfnamefont {J.~R.}\
  \bibnamefont {{Brownstein}}}, \bibinfo {author} {\bibfnamefont
  {E.}~\bibnamefont {{Buckley-Geer}}}, \bibinfo {author} {\bibfnamefont
  {A.}~\bibnamefont {{Burden}}}, \bibinfo {author} {\bibfnamefont
  {E.}~\bibnamefont {{Burtin}}}, \bibinfo {author} {\bibfnamefont {N.~G.}\
  \bibnamefont {{Busca}}}, \bibinfo {author} {\bibfnamefont {R.~N.}\
  \bibnamefont {{Cahn}}}, \bibinfo {author} {\bibfnamefont {Y.-C.}\
  \bibnamefont {{Cai}}}, \bibinfo {author} {\bibfnamefont {L.}~\bibnamefont
  {{Cardiel-Sas}}}, \bibinfo {author} {\bibfnamefont {R.~G.}\ \bibnamefont
  {{Carlberg}}}, \bibinfo {author} {\bibfnamefont {P.-H.}\ \bibnamefont
  {{Carton}}}, \bibinfo {author} {\bibfnamefont {R.}~\bibnamefont {{Casas}}},
  \bibinfo {author} {\bibfnamefont {F.~J.}\ \bibnamefont {{Castander}}},
  \bibinfo {author} {\bibfnamefont {J.~L.}\ \bibnamefont {{Cervantes-Cota}}},
  \bibinfo {author} {\bibfnamefont {T.~M.}\ \bibnamefont {{Claybaugh}}},
  \bibinfo {author} {\bibfnamefont {M.}~\bibnamefont {{Close}}}, \bibinfo
  {author} {\bibfnamefont {C.~T.}\ \bibnamefont {{Coker}}}, \bibinfo {author}
  {\bibfnamefont {S.}~\bibnamefont {{Cole}}}, \bibinfo {author} {\bibfnamefont
  {J.}~\bibnamefont {{Comparat}}}, \bibinfo {author} {\bibfnamefont {A.~P.}\
  \bibnamefont {{Cooper}}}, \bibinfo {author} {\bibfnamefont {M.~C.}\
  \bibnamefont {{Cousinou}}}, \bibinfo {author} {\bibfnamefont
  {M.}~\bibnamefont {{Crocce}}}, \bibinfo {author} {\bibfnamefont {J.-G.}\
  \bibnamefont {{Cuby}}}, \bibinfo {author} {\bibfnamefont {D.~P.}\
  \bibnamefont {{Cunningham}}}, \bibinfo {author} {\bibfnamefont {T.~M.}\
  \bibnamefont {{Davis}}}, \bibinfo {author} {\bibfnamefont {K.~S.}\
  \bibnamefont {{Dawson}}}, \bibinfo {author} {\bibfnamefont {A.}~\bibnamefont
  {{de la Macorra}}}, \bibinfo {author} {\bibfnamefont {J.}~\bibnamefont {{De
  Vicente}}}, \bibinfo {author} {\bibfnamefont {T.}~\bibnamefont {{Delubac}}},
  \bibinfo {author} {\bibfnamefont {M.}~\bibnamefont {{Derwent}}}, \bibinfo
  {author} {\bibfnamefont {A.}~\bibnamefont {{Dey}}}, \bibinfo {author}
  {\bibfnamefont {G.}~\bibnamefont {{Dhungana}}}, \bibinfo {author}
  {\bibfnamefont {Z.}~\bibnamefont {{Ding}}}, \bibinfo {author} {\bibfnamefont
  {P.}~\bibnamefont {{Doel}}}, \bibinfo {author} {\bibfnamefont {Y.~T.}\
  \bibnamefont {{Duan}}}, \bibinfo {author} {\bibfnamefont {A.}~\bibnamefont
  {{Ealet}}}, \bibinfo {author} {\bibfnamefont {J.}~\bibnamefont
  {{Edelstein}}}, \bibinfo {author} {\bibfnamefont {S.}~\bibnamefont
  {{Eftekharzadeh}}}, \bibinfo {author} {\bibfnamefont {D.~J.}\ \bibnamefont
  {{Eisenstein}}}, \bibinfo {author} {\bibfnamefont {A.}~\bibnamefont
  {{Elliott}}}, \bibinfo {author} {\bibfnamefont {S.}~\bibnamefont
  {{Escoffier}}}, \bibinfo {author} {\bibfnamefont {M.}~\bibnamefont
  {{Evatt}}}, \bibinfo {author} {\bibfnamefont {P.}~\bibnamefont
  {{Fagrelius}}}, \bibinfo {author} {\bibfnamefont {X.}~\bibnamefont {{Fan}}},
  \bibinfo {author} {\bibfnamefont {K.}~\bibnamefont {{Fanning}}}, \bibinfo
  {author} {\bibfnamefont {A.}~\bibnamefont {{Farahi}}}, \bibinfo {author}
  {\bibfnamefont {J.}~\bibnamefont {{Farihi}}}, \bibinfo {author}
  {\bibfnamefont {G.}~\bibnamefont {{Favole}}}, \bibinfo {author}
  {\bibfnamefont {Y.}~\bibnamefont {{Feng}}}, \bibinfo {author} {\bibfnamefont
  {E.}~\bibnamefont {{Fernandez}}}, \bibinfo {author} {\bibfnamefont {J.~R.}\
  \bibnamefont {{Findlay}}}, \bibinfo {author} {\bibfnamefont {D.~P.}\
  \bibnamefont {{Finkbeiner}}}, \bibinfo {author} {\bibfnamefont {M.~J.}\
  \bibnamefont {{Fitzpatrick}}}, \bibinfo {author} {\bibfnamefont
  {B.}~\bibnamefont {{Flaugher}}}, \bibinfo {author} {\bibfnamefont
  {S.}~\bibnamefont {{Flender}}}, \bibinfo {author} {\bibfnamefont
  {A.}~\bibnamefont {{Font-Ribera}}}, \bibinfo {author} {\bibfnamefont {J.~E.}\
  \bibnamefont {{Forero-Romero}}}, \bibinfo {author} {\bibfnamefont
  {P.}~\bibnamefont {{Fosalba}}}, \bibinfo {author} {\bibfnamefont {C.~S.}\
  \bibnamefont {{Frenk}}}, \bibinfo {author} {\bibfnamefont {M.}~\bibnamefont
  {{Fumagalli}}}, \bibinfo {author} {\bibfnamefont {B.~T.}\ \bibnamefont
  {{Gaensicke}}}, \bibinfo {author} {\bibfnamefont {G.}~\bibnamefont
  {{Gallo}}}, \bibinfo {author} {\bibfnamefont {J.}~\bibnamefont
  {{Garcia-Bellido}}}, \bibinfo {author} {\bibfnamefont {E.}~\bibnamefont
  {{Gaztanaga}}}, \bibinfo {author} {\bibfnamefont {N.}~\bibnamefont {{Pietro
  Gentile Fusillo}}}, \bibinfo {author} {\bibfnamefont {T.}~\bibnamefont
  {{Gerard}}}, \bibinfo {author} {\bibfnamefont {I.}~\bibnamefont
  {{Gershkovich}}}, \bibinfo {author} {\bibfnamefont {T.}~\bibnamefont
  {{Giannantonio}}}, \bibinfo {author} {\bibfnamefont {D.}~\bibnamefont
  {{Gillet}}}, \bibinfo {author} {\bibfnamefont {G.}~\bibnamefont
  {{Gonzalez-de-Rivera}}}, \bibinfo {author} {\bibfnamefont {V.}~\bibnamefont
  {{Gonzalez-Perez}}}, \bibinfo {author} {\bibfnamefont {S.}~\bibnamefont
  {{Gott}}}, \bibinfo {author} {\bibfnamefont {O.}~\bibnamefont {{Graur}}},
  \bibinfo {author} {\bibfnamefont {G.}~\bibnamefont {{Gutierrez}}}, \bibinfo
  {author} {\bibfnamefont {J.}~\bibnamefont {{Guy}}}, \bibinfo {author}
  {\bibfnamefont {S.}~\bibnamefont {{Habib}}}, \bibinfo {author} {\bibfnamefont
  {H.}~\bibnamefont {{Heetderks}}}, \bibinfo {author} {\bibfnamefont
  {I.}~\bibnamefont {{Heetderks}}}, \bibinfo {author} {\bibfnamefont
  {K.}~\bibnamefont {{Heitmann}}}, \bibinfo {author} {\bibfnamefont {W.~A.}\
  \bibnamefont {{Hellwing}}}, \bibinfo {author} {\bibfnamefont {D.~A.}\
  \bibnamefont {{Herrera}}}, \bibinfo {author} {\bibfnamefont {S.}~\bibnamefont
  {{Ho}}}, \bibinfo {author} {\bibfnamefont {S.}~\bibnamefont {{Holland}}},
  \bibinfo {author} {\bibfnamefont {K.}~\bibnamefont {{Honscheid}}}, \bibinfo
  {author} {\bibfnamefont {E.}~\bibnamefont {{Huff}}}, \bibinfo {author}
  {\bibfnamefont {T.~A.}\ \bibnamefont {{Hutchinson}}}, \bibinfo {author}
  {\bibfnamefont {D.}~\bibnamefont {{Huterer}}}, \bibinfo {author}
  {\bibfnamefont {H.~S.}\ \bibnamefont {{Hwang}}}, \bibinfo {author}
  {\bibfnamefont {J.~M.}\ \bibnamefont {{Illa Laguna}}}, \bibinfo {author}
  {\bibfnamefont {Y.}~\bibnamefont {{Ishikawa}}}, \bibinfo {author}
  {\bibfnamefont {D.}~\bibnamefont {{Jacobs}}}, \bibinfo {author}
  {\bibfnamefont {N.}~\bibnamefont {{Jeffrey}}}, \bibinfo {author}
  {\bibfnamefont {P.}~\bibnamefont {{Jelinsky}}}, \bibinfo {author}
  {\bibfnamefont {E.}~\bibnamefont {{Jennings}}}, \bibinfo {author}
  {\bibfnamefont {L.}~\bibnamefont {{Jiang}}}, \bibinfo {author} {\bibfnamefont
  {J.}~\bibnamefont {{Jimenez}}}, \bibinfo {author} {\bibfnamefont
  {J.}~\bibnamefont {{Johnson}}}, \bibinfo {author} {\bibfnamefont
  {R.}~\bibnamefont {{Joyce}}}, \bibinfo {author} {\bibfnamefont
  {E.}~\bibnamefont {{Jullo}}}, \bibinfo {author} {\bibfnamefont
  {S.}~\bibnamefont {{Juneau}}}, \bibinfo {author} {\bibfnamefont
  {S.}~\bibnamefont {{Kama}}}, \bibinfo {author} {\bibfnamefont
  {A.}~\bibnamefont {{Karcher}}}, \bibinfo {author} {\bibfnamefont
  {S.}~\bibnamefont {{Karkar}}}, \bibinfo {author} {\bibfnamefont
  {R.}~\bibnamefont {{Kehoe}}}, \bibinfo {author} {\bibfnamefont
  {N.}~\bibnamefont {{Kennamer}}}, \bibinfo {author} {\bibfnamefont
  {S.}~\bibnamefont {{Kent}}}, \bibinfo {author} {\bibfnamefont
  {M.}~\bibnamefont {{Kilbinger}}}, \bibinfo {author} {\bibfnamefont {A.~G.}\
  \bibnamefont {{Kim}}}, \bibinfo {author} {\bibfnamefont {D.}~\bibnamefont
  {{Kirkby}}}, \bibinfo {author} {\bibfnamefont {T.}~\bibnamefont {{Kisner}}},
  \bibinfo {author} {\bibfnamefont {E.}~\bibnamefont {{Kitanidis}}}, \bibinfo
  {author} {\bibfnamefont {J.-P.}\ \bibnamefont {{Kneib}}}, \bibinfo {author}
  {\bibfnamefont {S.}~\bibnamefont {{Koposov}}}, \bibinfo {author}
  {\bibfnamefont {E.}~\bibnamefont {{Kovacs}}}, \bibinfo {author}
  {\bibfnamefont {K.}~\bibnamefont {{Koyama}}}, \bibinfo {author}
  {\bibfnamefont {A.}~\bibnamefont {{Kremin}}}, \bibinfo {author}
  {\bibfnamefont {R.}~\bibnamefont {{Kron}}}, \bibinfo {author} {\bibfnamefont
  {L.}~\bibnamefont {{Kronig}}}, \bibinfo {author} {\bibfnamefont
  {A.}~\bibnamefont {{Kueter-Young}}}, \bibinfo {author} {\bibfnamefont
  {C.~G.}\ \bibnamefont {{Lacey}}}, \bibinfo {author} {\bibfnamefont
  {R.}~\bibnamefont {{Lafever}}}, \bibinfo {author} {\bibfnamefont
  {O.}~\bibnamefont {{Lahav}}}, \bibinfo {author} {\bibfnamefont
  {A.}~\bibnamefont {{Lambert}}}, \bibinfo {author} {\bibfnamefont
  {M.}~\bibnamefont {{Lampton}}}, \bibinfo {author} {\bibfnamefont
  {M.}~\bibnamefont {{Landriau}}}, \bibinfo {author} {\bibfnamefont
  {D.}~\bibnamefont {{Lang}}}, \bibinfo {author} {\bibfnamefont {T.~R.}\
  \bibnamefont {{Lauer}}}, \bibinfo {author} {\bibfnamefont {J.-M.}\
  \bibnamefont {{Le Goff}}}, \bibinfo {author} {\bibfnamefont {L.}~\bibnamefont
  {{Le Guillou}}}, \bibinfo {author} {\bibfnamefont {A.}~\bibnamefont {{Le Van
  Suu}}}, \bibinfo {author} {\bibfnamefont {J.~H.}\ \bibnamefont {{Lee}}},
  \bibinfo {author} {\bibfnamefont {S.-J.}\ \bibnamefont {{Lee}}}, \bibinfo
  {author} {\bibfnamefont {D.}~\bibnamefont {{Leitner}}}, \bibinfo {author}
  {\bibfnamefont {M.}~\bibnamefont {{Lesser}}}, \bibinfo {author}
  {\bibfnamefont {M.~E.}\ \bibnamefont {{Levi}}}, \bibinfo {author}
  {\bibfnamefont {B.}~\bibnamefont {{L'Huillier}}}, \bibinfo {author}
  {\bibfnamefont {B.}~\bibnamefont {{Li}}}, \bibinfo {author} {\bibfnamefont
  {M.}~\bibnamefont {{Liang}}}, \bibinfo {author} {\bibfnamefont
  {H.}~\bibnamefont {{Lin}}}, \bibinfo {author} {\bibfnamefont
  {E.}~\bibnamefont {{Linder}}}, \bibinfo {author} {\bibfnamefont {S.~R.}\
  \bibnamefont {{Loebman}}}, \bibinfo {author} {\bibfnamefont {Z.}~\bibnamefont
  {{Luki{\'c}}}}, \bibinfo {author} {\bibfnamefont {J.}~\bibnamefont {{Ma}}},
  \bibinfo {author} {\bibfnamefont {N.}~\bibnamefont {{MacCrann}}}, \bibinfo
  {author} {\bibfnamefont {C.}~\bibnamefont {{Magneville}}}, \bibinfo {author}
  {\bibfnamefont {L.}~\bibnamefont {{Makarem}}}, \bibinfo {author}
  {\bibfnamefont {M.}~\bibnamefont {{Manera}}}, \bibinfo {author}
  {\bibfnamefont {C.~J.}\ \bibnamefont {{Manser}}}, \bibinfo {author}
  {\bibfnamefont {R.}~\bibnamefont {{Marshall}}}, \bibinfo {author}
  {\bibfnamefont {P.}~\bibnamefont {{Martini}}}, \bibinfo {author}
  {\bibfnamefont {R.}~\bibnamefont {{Massey}}}, \bibinfo {author}
  {\bibfnamefont {T.}~\bibnamefont {{Matheson}}}, \bibinfo {author}
  {\bibfnamefont {J.}~\bibnamefont {{McCauley}}}, \bibinfo {author}
  {\bibfnamefont {P.}~\bibnamefont {{McDonald}}}, \bibinfo {author}
  {\bibfnamefont {I.~D.}\ \bibnamefont {{McGreer}}}, \bibinfo {author}
  {\bibfnamefont {A.}~\bibnamefont {{Meisner}}}, \bibinfo {author}
  {\bibfnamefont {N.}~\bibnamefont {{Metcalfe}}}, \bibinfo {author}
  {\bibfnamefont {T.~N.}\ \bibnamefont {{Miller}}}, \bibinfo {author}
  {\bibfnamefont {R.}~\bibnamefont {{Miquel}}}, \bibinfo {author}
  {\bibfnamefont {J.}~\bibnamefont {{Moustakas}}}, \bibinfo {author}
  {\bibfnamefont {A.}~\bibnamefont {{Myers}}}, \bibinfo {author} {\bibfnamefont
  {M.}~\bibnamefont {{Naik}}}, \bibinfo {author} {\bibfnamefont {J.~A.}\
  \bibnamefont {{Newman}}}, \bibinfo {author} {\bibfnamefont {R.~C.}\
  \bibnamefont {{Nichol}}}, \bibinfo {author} {\bibfnamefont {A.}~\bibnamefont
  {{Nicola}}}, \bibinfo {author} {\bibfnamefont {L.}~\bibnamefont {{Nicolati da
  Costa}}}, \bibinfo {author} {\bibfnamefont {J.}~\bibnamefont {{Nie}}},
  \bibinfo {author} {\bibfnamefont {G.}~\bibnamefont {{Niz}}}, \bibinfo
  {author} {\bibfnamefont {P.}~\bibnamefont {{Norberg}}}, \bibinfo {author}
  {\bibfnamefont {B.}~\bibnamefont {{Nord}}}, \bibinfo {author} {\bibfnamefont
  {D.}~\bibnamefont {{Norman}}}, \bibinfo {author} {\bibfnamefont
  {P.}~\bibnamefont {{Nugent}}}, \bibinfo {author} {\bibfnamefont
  {T.}~\bibnamefont {{O'Brien}}}, \bibinfo {author} {\bibfnamefont
  {M.}~\bibnamefont {{Oh}}}, \bibinfo {author} {\bibfnamefont {K.~A.~G.}\
  \bibnamefont {{Olsen}}}, \bibinfo {author} {\bibfnamefont {C.}~\bibnamefont
  {{Padilla}}}, \bibinfo {author} {\bibfnamefont {H.}~\bibnamefont
  {{Padmanabhan}}}, \bibinfo {author} {\bibfnamefont {N.}~\bibnamefont
  {{Padmanabhan}}}, \bibinfo {author} {\bibfnamefont {N.}~\bibnamefont
  {{Palanque-Delabrouille}}}, \bibinfo {author} {\bibfnamefont
  {A.}~\bibnamefont {{Palmese}}}, \bibinfo {author} {\bibfnamefont
  {D.}~\bibnamefont {{Pappalardo}}}, \bibinfo {author} {\bibfnamefont
  {I.}~\bibnamefont {{P{\^a}ris}}}, \bibinfo {author} {\bibfnamefont
  {C.}~\bibnamefont {{Park}}}, \bibinfo {author} {\bibfnamefont
  {A.}~\bibnamefont {{Patej}}}, \bibinfo {author} {\bibfnamefont {J.~A.}\
  \bibnamefont {{Peacock}}}, \bibinfo {author} {\bibfnamefont {H.~V.}\
  \bibnamefont {{Peiris}}}, \bibinfo {author} {\bibfnamefont {X.}~\bibnamefont
  {{Peng}}}, \bibinfo {author} {\bibfnamefont {W.~J.}\ \bibnamefont
  {{Percival}}}, \bibinfo {author} {\bibfnamefont {S.}~\bibnamefont
  {{Perruchot}}}, \bibinfo {author} {\bibfnamefont {M.~M.}\ \bibnamefont
  {{Pieri}}}, \bibinfo {author} {\bibfnamefont {R.}~\bibnamefont {{Pogge}}},
  \bibinfo {author} {\bibfnamefont {J.~E.}\ \bibnamefont {{Pollack}}}, \bibinfo
  {author} {\bibfnamefont {C.}~\bibnamefont {{Poppett}}}, \bibinfo {author}
  {\bibfnamefont {F.}~\bibnamefont {{Prada}}}, \bibinfo {author} {\bibfnamefont
  {A.}~\bibnamefont {{Prakash}}}, \bibinfo {author} {\bibfnamefont {R.~G.}\
  \bibnamefont {{Probst}}}, \bibinfo {author} {\bibfnamefont {D.}~\bibnamefont
  {{Rabinowitz}}}, \bibinfo {author} {\bibfnamefont {A.}~\bibnamefont
  {{Raichoor}}}, \bibinfo {author} {\bibfnamefont {C.~H.}\ \bibnamefont
  {{Ree}}}, \bibinfo {author} {\bibfnamefont {A.}~\bibnamefont {{Refregier}}},
  \bibinfo {author} {\bibfnamefont {X.}~\bibnamefont {{Regal}}}, \bibinfo
  {author} {\bibfnamefont {B.}~\bibnamefont {{Reid}}}, \bibinfo {author}
  {\bibfnamefont {K.}~\bibnamefont {{Reil}}}, \bibinfo {author} {\bibfnamefont
  {M.}~\bibnamefont {{Rezaie}}}, \bibinfo {author} {\bibfnamefont {C.~M.}\
  \bibnamefont {{Rockosi}}}, \bibinfo {author} {\bibfnamefont {N.}~\bibnamefont
  {{Roe}}}, \bibinfo {author} {\bibfnamefont {S.}~\bibnamefont {{Ronayette}}},
  \bibinfo {author} {\bibfnamefont {A.}~\bibnamefont {{Roodman}}}, \bibinfo
  {author} {\bibfnamefont {A.~J.}\ \bibnamefont {{Ross}}}, \bibinfo {author}
  {\bibfnamefont {N.~P.}\ \bibnamefont {{Ross}}}, \bibinfo {author}
  {\bibfnamefont {G.}~\bibnamefont {{Rossi}}}, \bibinfo {author} {\bibfnamefont
  {E.}~\bibnamefont {{Rozo}}}, \bibinfo {author} {\bibfnamefont
  {V.}~\bibnamefont {{Ruhlmann-Kleider}}}, \bibinfo {author} {\bibfnamefont
  {E.~S.}\ \bibnamefont {{Rykoff}}}, \bibinfo {author} {\bibfnamefont
  {C.}~\bibnamefont {{Sabiu}}}, \bibinfo {author} {\bibfnamefont
  {L.}~\bibnamefont {{Samushia}}}, \bibinfo {author} {\bibfnamefont
  {E.}~\bibnamefont {{Sanchez}}}, \bibinfo {author} {\bibfnamefont
  {J.}~\bibnamefont {{Sanchez}}}, \bibinfo {author} {\bibfnamefont {D.~J.}\
  \bibnamefont {{Schlegel}}}, \bibinfo {author} {\bibfnamefont
  {M.}~\bibnamefont {{Schneider}}}, \bibinfo {author} {\bibfnamefont
  {M.}~\bibnamefont {{Schubnell}}}, \bibinfo {author} {\bibfnamefont
  {A.}~\bibnamefont {{Secroun}}}, \bibinfo {author} {\bibfnamefont
  {U.}~\bibnamefont {{Seljak}}}, \bibinfo {author} {\bibfnamefont {H.-J.}\
  \bibnamefont {{Seo}}}, \bibinfo {author} {\bibfnamefont {S.}~\bibnamefont
  {{Serrano}}}, \bibinfo {author} {\bibfnamefont {A.}~\bibnamefont
  {{Shafieloo}}}, \bibinfo {author} {\bibfnamefont {H.}~\bibnamefont {{Shan}}},
  \bibinfo {author} {\bibfnamefont {R.}~\bibnamefont {{Sharples}}}, \bibinfo
  {author} {\bibfnamefont {M.~J.}\ \bibnamefont {{Sholl}}}, \bibinfo {author}
  {\bibfnamefont {W.~V.}\ \bibnamefont {{Shourt}}}, \bibinfo {author}
  {\bibfnamefont {J.~H.}\ \bibnamefont {{Silber}}}, \bibinfo {author}
  {\bibfnamefont {D.~R.}\ \bibnamefont {{Silva}}}, \bibinfo {author}
  {\bibfnamefont {M.~M.}\ \bibnamefont {{Sirk}}}, \bibinfo {author}
  {\bibfnamefont {A.}~\bibnamefont {{Slosar}}}, \bibinfo {author}
  {\bibfnamefont {A.}~\bibnamefont {{Smith}}}, \bibinfo {author} {\bibfnamefont
  {G.~F.}\ \bibnamefont {{Smoot}}}, \bibinfo {author} {\bibfnamefont
  {D.}~\bibnamefont {{Som}}}, \bibinfo {author} {\bibfnamefont {Y.-S.}\
  \bibnamefont {{Song}}}, \bibinfo {author} {\bibfnamefont {D.}~\bibnamefont
  {{Sprayberry}}}, \bibinfo {author} {\bibfnamefont {R.}~\bibnamefont
  {{Staten}}}, \bibinfo {author} {\bibfnamefont {A.}~\bibnamefont
  {{Stefanik}}}, \bibinfo {author} {\bibfnamefont {G.}~\bibnamefont {{Tarle}}},
  \bibinfo {author} {\bibfnamefont {S.}~\bibnamefont {{Sien Tie}}}, \bibinfo
  {author} {\bibfnamefont {J.~L.}\ \bibnamefont {{Tinker}}}, \bibinfo {author}
  {\bibfnamefont {R.}~\bibnamefont {{Tojeiro}}}, \bibinfo {author}
  {\bibfnamefont {F.}~\bibnamefont {{Valdes}}}, \bibinfo {author}
  {\bibfnamefont {O.}~\bibnamefont {{Valenzuela}}}, \bibinfo {author}
  {\bibfnamefont {M.}~\bibnamefont {{Valluri}}}, \bibinfo {author}
  {\bibfnamefont {M.}~\bibnamefont {{Vargas-Magana}}}, \bibinfo {author}
  {\bibfnamefont {L.}~\bibnamefont {{Verde}}}, \bibinfo {author} {\bibfnamefont
  {A.~R.}\ \bibnamefont {{Walker}}}, \bibinfo {author} {\bibfnamefont
  {J.}~\bibnamefont {{Wang}}}, \bibinfo {author} {\bibfnamefont
  {Y.}~\bibnamefont {{Wang}}}, \bibinfo {author} {\bibfnamefont {B.~A.}\
  \bibnamefont {{Weaver}}}, \bibinfo {author} {\bibfnamefont {C.}~\bibnamefont
  {{Weaverdyck}}}, \bibinfo {author} {\bibfnamefont {R.~H.}\ \bibnamefont
  {{Wechsler}}}, \bibinfo {author} {\bibfnamefont {D.~H.}\ \bibnamefont
  {{Weinberg}}}, \bibinfo {author} {\bibfnamefont {M.}~\bibnamefont {{White}}},
  \bibinfo {author} {\bibfnamefont {Q.}~\bibnamefont {{Yang}}}, \bibinfo
  {author} {\bibfnamefont {C.}~\bibnamefont {{Yeche}}}, \bibinfo {author}
  {\bibfnamefont {T.}~\bibnamefont {{Zhang}}}, \bibinfo {author} {\bibfnamefont
  {G.-B.}\ \bibnamefont {{Zhao}}}, \bibinfo {author} {\bibfnamefont
  {Y.}~\bibnamefont {{Zheng}}}, \bibinfo {author} {\bibfnamefont
  {X.}~\bibnamefont {{Zhou}}}, \bibinfo {author} {\bibfnamefont
  {Z.}~\bibnamefont {{Zhou}}}, \bibinfo {author} {\bibfnamefont
  {Y.}~\bibnamefont {{Zhu}}}, \bibinfo {author} {\bibfnamefont
  {H.}~\bibnamefont {{Zou}}},\ and\ \bibinfo {author} {\bibfnamefont
  {Y.}~\bibnamefont {{Zu}}},\ }\bibfield  {title} {\bibinfo {title} {{The DESI
  Experiment Part II: Instrument Design}},\ }\href@noop {} {\bibfield
  {journal} {\bibinfo  {journal} {arXiv e-prints}\ ,\ \bibinfo {eid}
  {arXiv:1611.00037}} (\bibinfo {year} {2016}{\natexlab{a}})},\ \Eprint
  {https://arxiv.org/abs/1611.00037} {arXiv:1611.00037 [astro-ph.IM]}
  \BibitemShut {NoStop}%
\bibitem [{\citenamefont {{Silber}}\ \emph {et~al.}(2023)\citenamefont
  {{Silber}}, \citenamefont {{Fagrelius}}, \citenamefont {{Fanning}},
  \citenamefont {{Schubnell}}, \citenamefont {{Aguilar}}, \citenamefont
  {{Ahlen}}, \citenamefont {{Ameel}}, \citenamefont {{Ballester}},
  \citenamefont {{Baltay}}, \citenamefont {{Bebek}}, \citenamefont {{Benton
  Beard}}, \citenamefont {{Besuner}}, \citenamefont {{Cardiel-Sas}},
  \citenamefont {{Casas}}, \citenamefont {{Castander}}, \citenamefont
  {{Claybaugh}}, \citenamefont {{Dobson}}, \citenamefont {{Duan}},
  \citenamefont {{Dunlop}}, \citenamefont {{Edelstein}}, \citenamefont
  {{Emmet}}, \citenamefont {{Elliott}}, \citenamefont {{Evatt}}, \citenamefont
  {{Gershkovich}}, \citenamefont {{Guy}}, \citenamefont {{Harris}},
  \citenamefont {{Heetderks}}, \citenamefont {{Heetderks}}, \citenamefont
  {{Honscheid}}, \citenamefont {{Illa}}, \citenamefont {{Jelinsky}},
  \citenamefont {{Jelinsky}}, \citenamefont {{Jimenez}}, \citenamefont
  {{Karcher}}, \citenamefont {{Kent}}, \citenamefont {{Kirkby}}, \citenamefont
  {{Kneib}}, \citenamefont {{Lambert}}, \citenamefont {{Lampton}},
  \citenamefont {{Leitner}}, \citenamefont {{Levi}}, \citenamefont
  {{McCauley}}, \citenamefont {{Meisner}}, \citenamefont {{Miller}},
  \citenamefont {{Miquel}}, \citenamefont {{Mundet}}, \citenamefont
  {{Poppett}}, \citenamefont {{Rabinowitz}}, \citenamefont {{Reil}},
  \citenamefont {{Roman}}, \citenamefont {{Schlegel}}, \citenamefont
  {{Serrano}}, \citenamefont {{Van Shourt}}, \citenamefont {{Sprayberry}},
  \citenamefont {{Tarl{\'e}}}, \citenamefont {{Tie}}, \citenamefont
  {{Weaverdyck}}, \citenamefont {{Zhang}}, \citenamefont {{Azzaro}},
  \citenamefont {{Bailey}}, \citenamefont {{Becerril}}, \citenamefont
  {{Blackwell}}, \citenamefont {{Bouri}}, \citenamefont {{Brooks}},
  \citenamefont {{Buckley-Geer}}, \citenamefont {{Castro}}, \citenamefont
  {{Derwent}}, \citenamefont {{Dey}}, \citenamefont {{Dhungana}}, \citenamefont
  {{Doel}}, \citenamefont {{Eisenstein}}, \citenamefont {{Fahim}},
  \citenamefont {{Garcia-Bellido}}, \citenamefont {{Gazta{\~n}aga}},
  \citenamefont {{A Gontcho}}, \citenamefont {{Gutierrez}}, \citenamefont
  {{H{\"o}rler}}, \citenamefont {{Kehoe}}, \citenamefont {{Kisner}},
  \citenamefont {{Kremin}}, \citenamefont {{Kronig}}, \citenamefont
  {{Landriau}}, \citenamefont {{Le Guillou}}, \citenamefont {{Martini}},
  \citenamefont {{Moustakas}}, \citenamefont {{Palanque-Delabrouille}},
  \citenamefont {{Peng}}, \citenamefont {{Percival}}, \citenamefont {{Prada}},
  \citenamefont {{Allende Prieto}}, \citenamefont {{de Rivera}}, \citenamefont
  {{Sanchez}}, \citenamefont {{Sanchez}}, \citenamefont {{Sharples}},
  \citenamefont {{Soares-Santos}}, \citenamefont {{Schlafly}}, \citenamefont
  {{Weaver}}, \citenamefont {{Zhou}}, \citenamefont {{Zhu}}, \citenamefont
  {{Zou}},\ and\ \citenamefont {{DESI Collaboration}}}]{2023AJ....165....9S}%
  \BibitemOpen
  \bibfield  {author} {\bibinfo {author} {\bibfnamefont {J.~H.}\ \bibnamefont
  {{Silber}}}, \bibinfo {author} {\bibfnamefont {P.}~\bibnamefont
  {{Fagrelius}}}, \bibinfo {author} {\bibfnamefont {K.}~\bibnamefont
  {{Fanning}}}, \bibinfo {author} {\bibfnamefont {M.}~\bibnamefont
  {{Schubnell}}}, \bibinfo {author} {\bibfnamefont {J.~N.}\ \bibnamefont
  {{Aguilar}}}, \bibinfo {author} {\bibfnamefont {S.}~\bibnamefont {{Ahlen}}},
  \bibinfo {author} {\bibfnamefont {J.}~\bibnamefont {{Ameel}}}, \bibinfo
  {author} {\bibfnamefont {O.}~\bibnamefont {{Ballester}}}, \bibinfo {author}
  {\bibfnamefont {C.}~\bibnamefont {{Baltay}}}, \bibinfo {author}
  {\bibfnamefont {C.}~\bibnamefont {{Bebek}}}, \bibinfo {author} {\bibfnamefont
  {D.}~\bibnamefont {{Benton Beard}}}, \bibinfo {author} {\bibfnamefont
  {R.}~\bibnamefont {{Besuner}}}, \bibinfo {author} {\bibfnamefont
  {L.}~\bibnamefont {{Cardiel-Sas}}}, \bibinfo {author} {\bibfnamefont
  {R.}~\bibnamefont {{Casas}}}, \bibinfo {author} {\bibfnamefont {F.~J.}\
  \bibnamefont {{Castander}}}, \bibinfo {author} {\bibfnamefont
  {T.}~\bibnamefont {{Claybaugh}}}, \bibinfo {author} {\bibfnamefont
  {C.}~\bibnamefont {{Dobson}}}, \bibinfo {author} {\bibfnamefont
  {Y.}~\bibnamefont {{Duan}}}, \bibinfo {author} {\bibfnamefont
  {P.}~\bibnamefont {{Dunlop}}}, \bibinfo {author} {\bibfnamefont
  {J.}~\bibnamefont {{Edelstein}}}, \bibinfo {author} {\bibfnamefont {W.~T.}\
  \bibnamefont {{Emmet}}}, \bibinfo {author} {\bibfnamefont {A.}~\bibnamefont
  {{Elliott}}}, \bibinfo {author} {\bibfnamefont {M.}~\bibnamefont {{Evatt}}},
  \bibinfo {author} {\bibfnamefont {I.}~\bibnamefont {{Gershkovich}}}, \bibinfo
  {author} {\bibfnamefont {J.}~\bibnamefont {{Guy}}}, \bibinfo {author}
  {\bibfnamefont {S.}~\bibnamefont {{Harris}}}, \bibinfo {author}
  {\bibfnamefont {H.}~\bibnamefont {{Heetderks}}}, \bibinfo {author}
  {\bibfnamefont {I.}~\bibnamefont {{Heetderks}}}, \bibinfo {author}
  {\bibfnamefont {K.}~\bibnamefont {{Honscheid}}}, \bibinfo {author}
  {\bibfnamefont {J.~M.}\ \bibnamefont {{Illa}}}, \bibinfo {author}
  {\bibfnamefont {P.}~\bibnamefont {{Jelinsky}}}, \bibinfo {author}
  {\bibfnamefont {S.~R.}\ \bibnamefont {{Jelinsky}}}, \bibinfo {author}
  {\bibfnamefont {J.}~\bibnamefont {{Jimenez}}}, \bibinfo {author}
  {\bibfnamefont {A.}~\bibnamefont {{Karcher}}}, \bibinfo {author}
  {\bibfnamefont {S.}~\bibnamefont {{Kent}}}, \bibinfo {author} {\bibfnamefont
  {D.}~\bibnamefont {{Kirkby}}}, \bibinfo {author} {\bibfnamefont {J.-P.}\
  \bibnamefont {{Kneib}}}, \bibinfo {author} {\bibfnamefont {A.}~\bibnamefont
  {{Lambert}}}, \bibinfo {author} {\bibfnamefont {M.}~\bibnamefont
  {{Lampton}}}, \bibinfo {author} {\bibfnamefont {D.}~\bibnamefont
  {{Leitner}}}, \bibinfo {author} {\bibfnamefont {M.}~\bibnamefont {{Levi}}},
  \bibinfo {author} {\bibfnamefont {J.}~\bibnamefont {{McCauley}}}, \bibinfo
  {author} {\bibfnamefont {A.}~\bibnamefont {{Meisner}}}, \bibinfo {author}
  {\bibfnamefont {T.~N.}\ \bibnamefont {{Miller}}}, \bibinfo {author}
  {\bibfnamefont {R.}~\bibnamefont {{Miquel}}}, \bibinfo {author}
  {\bibfnamefont {J.}~\bibnamefont {{Mundet}}}, \bibinfo {author}
  {\bibfnamefont {C.}~\bibnamefont {{Poppett}}}, \bibinfo {author}
  {\bibfnamefont {D.}~\bibnamefont {{Rabinowitz}}}, \bibinfo {author}
  {\bibfnamefont {K.}~\bibnamefont {{Reil}}}, \bibinfo {author} {\bibfnamefont
  {D.}~\bibnamefont {{Roman}}}, \bibinfo {author} {\bibfnamefont
  {D.}~\bibnamefont {{Schlegel}}}, \bibinfo {author} {\bibfnamefont
  {S.}~\bibnamefont {{Serrano}}}, \bibinfo {author} {\bibfnamefont
  {W.}~\bibnamefont {{Van Shourt}}}, \bibinfo {author} {\bibfnamefont
  {D.}~\bibnamefont {{Sprayberry}}}, \bibinfo {author} {\bibfnamefont
  {G.}~\bibnamefont {{Tarl{\'e}}}}, \bibinfo {author} {\bibfnamefont {S.~S.}\
  \bibnamefont {{Tie}}}, \bibinfo {author} {\bibfnamefont {C.}~\bibnamefont
  {{Weaverdyck}}}, \bibinfo {author} {\bibfnamefont {K.}~\bibnamefont
  {{Zhang}}}, \bibinfo {author} {\bibfnamefont {M.}~\bibnamefont {{Azzaro}}},
  \bibinfo {author} {\bibfnamefont {S.}~\bibnamefont {{Bailey}}}, \bibinfo
  {author} {\bibfnamefont {S.}~\bibnamefont {{Becerril}}}, \bibinfo {author}
  {\bibfnamefont {T.}~\bibnamefont {{Blackwell}}}, \bibinfo {author}
  {\bibfnamefont {M.}~\bibnamefont {{Bouri}}}, \bibinfo {author} {\bibfnamefont
  {D.}~\bibnamefont {{Brooks}}}, \bibinfo {author} {\bibfnamefont
  {E.}~\bibnamefont {{Buckley-Geer}}}, \bibinfo {author} {\bibfnamefont
  {J.~P.}\ \bibnamefont {{Castro}}}, \bibinfo {author} {\bibfnamefont
  {M.}~\bibnamefont {{Derwent}}}, \bibinfo {author} {\bibfnamefont
  {A.}~\bibnamefont {{Dey}}}, \bibinfo {author} {\bibfnamefont
  {G.}~\bibnamefont {{Dhungana}}}, \bibinfo {author} {\bibfnamefont
  {P.}~\bibnamefont {{Doel}}}, \bibinfo {author} {\bibfnamefont {D.~J.}\
  \bibnamefont {{Eisenstein}}}, \bibinfo {author} {\bibfnamefont
  {N.}~\bibnamefont {{Fahim}}}, \bibinfo {author} {\bibfnamefont
  {J.}~\bibnamefont {{Garcia-Bellido}}}, \bibinfo {author} {\bibfnamefont
  {E.}~\bibnamefont {{Gazta{\~n}aga}}}, \bibinfo {author} {\bibfnamefont
  {S.~G.}\ \bibnamefont {{A Gontcho}}}, \bibinfo {author} {\bibfnamefont
  {G.}~\bibnamefont {{Gutierrez}}}, \bibinfo {author} {\bibfnamefont
  {P.}~\bibnamefont {{H{\"o}rler}}}, \bibinfo {author} {\bibfnamefont
  {R.}~\bibnamefont {{Kehoe}}}, \bibinfo {author} {\bibfnamefont
  {T.}~\bibnamefont {{Kisner}}}, \bibinfo {author} {\bibfnamefont
  {A.}~\bibnamefont {{Kremin}}}, \bibinfo {author} {\bibfnamefont
  {L.}~\bibnamefont {{Kronig}}}, \bibinfo {author} {\bibfnamefont
  {M.}~\bibnamefont {{Landriau}}}, \bibinfo {author} {\bibfnamefont
  {L.}~\bibnamefont {{Le Guillou}}}, \bibinfo {author} {\bibfnamefont
  {P.}~\bibnamefont {{Martini}}}, \bibinfo {author} {\bibfnamefont
  {J.}~\bibnamefont {{Moustakas}}}, \bibinfo {author} {\bibfnamefont
  {N.}~\bibnamefont {{Palanque-Delabrouille}}}, \bibinfo {author}
  {\bibfnamefont {X.}~\bibnamefont {{Peng}}}, \bibinfo {author} {\bibfnamefont
  {W.}~\bibnamefont {{Percival}}}, \bibinfo {author} {\bibfnamefont
  {F.}~\bibnamefont {{Prada}}}, \bibinfo {author} {\bibfnamefont
  {C.}~\bibnamefont {{Allende Prieto}}}, \bibinfo {author} {\bibfnamefont
  {G.~G.}\ \bibnamefont {{de Rivera}}}, \bibinfo {author} {\bibfnamefont
  {E.}~\bibnamefont {{Sanchez}}}, \bibinfo {author} {\bibfnamefont
  {J.}~\bibnamefont {{Sanchez}}}, \bibinfo {author} {\bibfnamefont
  {R.}~\bibnamefont {{Sharples}}}, \bibinfo {author} {\bibfnamefont
  {M.}~\bibnamefont {{Soares-Santos}}}, \bibinfo {author} {\bibfnamefont
  {E.}~\bibnamefont {{Schlafly}}}, \bibinfo {author} {\bibfnamefont {B.~A.}\
  \bibnamefont {{Weaver}}}, \bibinfo {author} {\bibfnamefont {Z.}~\bibnamefont
  {{Zhou}}}, \bibinfo {author} {\bibfnamefont {Y.}~\bibnamefont {{Zhu}}},
  \bibinfo {author} {\bibfnamefont {H.}~\bibnamefont {{Zou}}},\ and\ \bibinfo
  {author} {\bibnamefont {{DESI Collaboration}}},\ }\bibfield  {title}
  {\bibinfo {title} {{The Robotic Multiobject Focal Plane System of the Dark
  Energy Spectroscopic Instrument (DESI)}},\ }\href
  {https://doi.org/10.3847/1538-3881/ac9ab1} {\bibfield  {journal} {\bibinfo
  {journal} {\aj}\ }\textbf {\bibinfo {volume} {165}},\ \bibinfo {eid} {9}
  (\bibinfo {year} {2023})},\ \Eprint {https://arxiv.org/abs/2205.09014}
  {arXiv:2205.09014 [astro-ph.IM]} \BibitemShut {NoStop}%
\bibitem [{\citenamefont {{Miller}}\ \emph {et~al.}(2023)\citenamefont
  {{Miller}}, \citenamefont {{Doel}}, \citenamefont {{Gutierrez}},
  \citenamefont {{Besuner}}, \citenamefont {{Brooks}}, \citenamefont {{Gallo}},
  \citenamefont {{Heetderks}}, \citenamefont {{Jelinsky}}, \citenamefont
  {{Kent}}, \citenamefont {{Lampton}}, \citenamefont {{Levi}}, \citenamefont
  {{Liang}}, \citenamefont {{Meisner}}, \citenamefont {{Sholl}}, \citenamefont
  {{Silber}}, \citenamefont {{Sprayberry}}, \citenamefont {{Aguilar}},
  \citenamefont {{de la Macorra}}, \citenamefont {{Eisenstein}}, \citenamefont
  {{Fanning}}, \citenamefont {{Font-Ribera}}, \citenamefont {{Gaztanaga}},
  \citenamefont {{Gontcho}}, \citenamefont {{Honscheid}}, \citenamefont
  {{Jimenez}}, \citenamefont {{Joyce}}, \citenamefont {{Kehoe}}, \citenamefont
  {{Kisner}}, \citenamefont {{Kremin}}, \citenamefont {{Landriau}},
  \citenamefont {{Le Guillou}}, \citenamefont {{Magneville}}, \citenamefont
  {{Martini}}, \citenamefont {{Miquel}}, \citenamefont {{Moustakas}},
  \citenamefont {{Nie}}, \citenamefont {{Percival}}, \citenamefont {{Poppett}},
  \citenamefont {{Prada}}, \citenamefont {{Rossi}}, \citenamefont {{Schlegel}},
  \citenamefont {{Schubnell}}, \citenamefont {{Seo}}, \citenamefont
  {{Sharples}}, \citenamefont {{Tarle}}, \citenamefont {{Vargas-Magana}},\ and\
  \citenamefont {{Zhou}}}]{2023arXiv230606310M}%
  \BibitemOpen
  \bibfield  {author} {\bibinfo {author} {\bibfnamefont {T.~N.}\ \bibnamefont
  {{Miller}}}, \bibinfo {author} {\bibfnamefont {P.}~\bibnamefont {{Doel}}},
  \bibinfo {author} {\bibfnamefont {G.}~\bibnamefont {{Gutierrez}}}, \bibinfo
  {author} {\bibfnamefont {R.}~\bibnamefont {{Besuner}}}, \bibinfo {author}
  {\bibfnamefont {D.}~\bibnamefont {{Brooks}}}, \bibinfo {author}
  {\bibfnamefont {G.}~\bibnamefont {{Gallo}}}, \bibinfo {author} {\bibfnamefont
  {H.}~\bibnamefont {{Heetderks}}}, \bibinfo {author} {\bibfnamefont
  {P.}~\bibnamefont {{Jelinsky}}}, \bibinfo {author} {\bibfnamefont {S.~M.}\
  \bibnamefont {{Kent}}}, \bibinfo {author} {\bibfnamefont {M.}~\bibnamefont
  {{Lampton}}}, \bibinfo {author} {\bibfnamefont {M.}~\bibnamefont {{Levi}}},
  \bibinfo {author} {\bibfnamefont {M.}~\bibnamefont {{Liang}}}, \bibinfo
  {author} {\bibfnamefont {A.}~\bibnamefont {{Meisner}}}, \bibinfo {author}
  {\bibfnamefont {M.~J.}\ \bibnamefont {{Sholl}}}, \bibinfo {author}
  {\bibfnamefont {J.~H.}\ \bibnamefont {{Silber}}}, \bibinfo {author}
  {\bibfnamefont {D.}~\bibnamefont {{Sprayberry}}}, \bibinfo {author}
  {\bibfnamefont {J.~N.}\ \bibnamefont {{Aguilar}}}, \bibinfo {author}
  {\bibfnamefont {A.}~\bibnamefont {{de la Macorra}}}, \bibinfo {author}
  {\bibfnamefont {D.}~\bibnamefont {{Eisenstein}}}, \bibinfo {author}
  {\bibfnamefont {K.}~\bibnamefont {{Fanning}}}, \bibinfo {author}
  {\bibfnamefont {A.}~\bibnamefont {{Font-Ribera}}}, \bibinfo {author}
  {\bibfnamefont {E.}~\bibnamefont {{Gaztanaga}}}, \bibinfo {author}
  {\bibfnamefont {S.~G.~A.}\ \bibnamefont {{Gontcho}}}, \bibinfo {author}
  {\bibfnamefont {K.}~\bibnamefont {{Honscheid}}}, \bibinfo {author}
  {\bibfnamefont {J.}~\bibnamefont {{Jimenez}}}, \bibinfo {author}
  {\bibfnamefont {D.}~\bibnamefont {{Joyce}}}, \bibinfo {author} {\bibfnamefont
  {R.}~\bibnamefont {{Kehoe}}}, \bibinfo {author} {\bibfnamefont
  {T.}~\bibnamefont {{Kisner}}}, \bibinfo {author} {\bibfnamefont
  {A.}~\bibnamefont {{Kremin}}}, \bibinfo {author} {\bibfnamefont
  {M.}~\bibnamefont {{Landriau}}}, \bibinfo {author} {\bibfnamefont
  {L.}~\bibnamefont {{Le Guillou}}}, \bibinfo {author} {\bibfnamefont
  {C.}~\bibnamefont {{Magneville}}}, \bibinfo {author} {\bibfnamefont
  {P.}~\bibnamefont {{Martini}}}, \bibinfo {author} {\bibfnamefont
  {R.}~\bibnamefont {{Miquel}}}, \bibinfo {author} {\bibfnamefont
  {J.}~\bibnamefont {{Moustakas}}}, \bibinfo {author} {\bibfnamefont
  {J.}~\bibnamefont {{Nie}}}, \bibinfo {author} {\bibfnamefont
  {W.}~\bibnamefont {{Percival}}}, \bibinfo {author} {\bibfnamefont
  {C.}~\bibnamefont {{Poppett}}}, \bibinfo {author} {\bibfnamefont
  {F.}~\bibnamefont {{Prada}}}, \bibinfo {author} {\bibfnamefont
  {G.}~\bibnamefont {{Rossi}}}, \bibinfo {author} {\bibfnamefont
  {D.}~\bibnamefont {{Schlegel}}}, \bibinfo {author} {\bibfnamefont
  {M.}~\bibnamefont {{Schubnell}}}, \bibinfo {author} {\bibfnamefont {H.-J.}\
  \bibnamefont {{Seo}}}, \bibinfo {author} {\bibfnamefont {R.}~\bibnamefont
  {{Sharples}}}, \bibinfo {author} {\bibfnamefont {G.}~\bibnamefont {{Tarle}}},
  \bibinfo {author} {\bibfnamefont {M.}~\bibnamefont {{Vargas-Magana}}},\ and\
  \bibinfo {author} {\bibfnamefont {Z.}~\bibnamefont {{Zhou}}},\ }\bibfield
  {title} {\bibinfo {title} {{The Optical Corrector for the Dark Energy
  Spectroscopic Instrument}},\ }\href
  {https://doi.org/10.48550/arXiv.2306.06310} {\bibfield  {journal} {\bibinfo
  {journal} {arXiv e-prints}\ ,\ \bibinfo {eid} {arXiv:2306.06310}} (\bibinfo
  {year} {2023})},\ \Eprint {https://arxiv.org/abs/2306.06310}
  {arXiv:2306.06310 [astro-ph.IM]} \BibitemShut {NoStop}%
\bibitem [{\citenamefont {{Levi}}\ \emph {et~al.}(2013)\citenamefont {{Levi}},
  \citenamefont {{Bebek}}, \citenamefont {{Beers}}, \citenamefont {{Blum}},
  \citenamefont {{Cahn}}, \citenamefont {{Eisenstein}}, \citenamefont
  {{Flaugher}}, \citenamefont {{Honscheid}}, \citenamefont {{Kron}},
  \citenamefont {{Lahav}}, \citenamefont {{McDonald}}, \citenamefont {{Roe}},
  \citenamefont {{Schlegel}},\ and\ \citenamefont {{representing the DESI
  collaboration}}}]{2013arXiv1308.0847L}%
  \BibitemOpen
  \bibfield  {author} {\bibinfo {author} {\bibfnamefont {M.}~\bibnamefont
  {{Levi}}}, \bibinfo {author} {\bibfnamefont {C.}~\bibnamefont {{Bebek}}},
  \bibinfo {author} {\bibfnamefont {T.}~\bibnamefont {{Beers}}}, \bibinfo
  {author} {\bibfnamefont {R.}~\bibnamefont {{Blum}}}, \bibinfo {author}
  {\bibfnamefont {R.}~\bibnamefont {{Cahn}}}, \bibinfo {author} {\bibfnamefont
  {D.}~\bibnamefont {{Eisenstein}}}, \bibinfo {author} {\bibfnamefont
  {B.}~\bibnamefont {{Flaugher}}}, \bibinfo {author} {\bibfnamefont
  {K.}~\bibnamefont {{Honscheid}}}, \bibinfo {author} {\bibfnamefont
  {R.}~\bibnamefont {{Kron}}}, \bibinfo {author} {\bibfnamefont
  {O.}~\bibnamefont {{Lahav}}}, \bibinfo {author} {\bibfnamefont
  {P.}~\bibnamefont {{McDonald}}}, \bibinfo {author} {\bibfnamefont
  {N.}~\bibnamefont {{Roe}}}, \bibinfo {author} {\bibfnamefont
  {D.}~\bibnamefont {{Schlegel}}},\ and\ \bibinfo {author} {\bibnamefont
  {{representing the DESI collaboration}}},\ }\bibfield  {title} {\bibinfo
  {title} {{The DESI Experiment, a whitepaper for Snowmass 2013}},\ }\href@noop
  {} {\bibfield  {journal} {\bibinfo  {journal} {arXiv e-prints}\ ,\ \bibinfo
  {eid} {arXiv:1308.0847}} (\bibinfo {year} {2013})},\ \Eprint
  {https://arxiv.org/abs/1308.0847} {arXiv:1308.0847 [astro-ph.CO]}
  \BibitemShut {NoStop}%
\bibitem [{\citenamefont {{DESI Collaboration}}\ \emph
  {et~al.}(2016{\natexlab{b}})\citenamefont {{DESI Collaboration}},
  \citenamefont {{Aghamousa}}, \citenamefont {{Aguilar}}, \citenamefont
  {{Ahlen}}, \citenamefont {{Alam}}, \citenamefont {{Allen}}, \citenamefont
  {{Allende Prieto}}, \citenamefont {{Annis}}, \citenamefont {{Bailey}},
  \citenamefont {{Balland}}, \citenamefont {{Ballester}}, \citenamefont
  {{Baltay}}, \citenamefont {{Beaufore}}, \citenamefont {{Bebek}},
  \citenamefont {{Beers}}, \citenamefont {{Bell}}, \citenamefont {{Bernal}},
  \citenamefont {{Besuner}}, \citenamefont {{Beutler}}, \citenamefont
  {{Blake}}, \citenamefont {{Bleuler}}, \citenamefont {{Blomqvist}},
  \citenamefont {{Blum}}, \citenamefont {{Bolton}}, \citenamefont {{Briceno}},
  \citenamefont {{Brooks}}, \citenamefont {{Brownstein}}, \citenamefont
  {{Buckley-Geer}}, \citenamefont {{Burden}}, \citenamefont {{Burtin}},
  \citenamefont {{Busca}}, \citenamefont {{Cahn}}, \citenamefont {{Cai}},
  \citenamefont {{Cardiel-Sas}}, \citenamefont {{Carlberg}}, \citenamefont
  {{Carton}}, \citenamefont {{Casas}}, \citenamefont {{Castander}},
  \citenamefont {{Cervantes-Cota}}, \citenamefont {{Claybaugh}}, \citenamefont
  {{Close}}, \citenamefont {{Coker}}, \citenamefont {{Cole}}, \citenamefont
  {{Comparat}}, \citenamefont {{Cooper}}, \citenamefont {{Cousinou}},
  \citenamefont {{Crocce}}, \citenamefont {{Cuby}}, \citenamefont
  {{Cunningham}}, \citenamefont {{Davis}}, \citenamefont {{Dawson}},
  \citenamefont {{de la Macorra}}, \citenamefont {{De Vicente}}, \citenamefont
  {{Delubac}}, \citenamefont {{Derwent}}, \citenamefont {{Dey}}, \citenamefont
  {{Dhungana}}, \citenamefont {{Ding}}, \citenamefont {{Doel}}, \citenamefont
  {{Duan}}, \citenamefont {{Ealet}}, \citenamefont {{Edelstein}}, \citenamefont
  {{Eftekharzadeh}}, \citenamefont {{Eisenstein}}, \citenamefont {{Elliott}},
  \citenamefont {{Escoffier}}, \citenamefont {{Evatt}}, \citenamefont
  {{Fagrelius}}, \citenamefont {{Fan}}, \citenamefont {{Fanning}},
  \citenamefont {{Farahi}}, \citenamefont {{Farihi}}, \citenamefont {{Favole}},
  \citenamefont {{Feng}}, \citenamefont {{Fernandez}}, \citenamefont
  {{Findlay}}, \citenamefont {{Finkbeiner}}, \citenamefont {{Fitzpatrick}},
  \citenamefont {{Flaugher}}, \citenamefont {{Flender}}, \citenamefont
  {{Font-Ribera}}, \citenamefont {{Forero-Romero}}, \citenamefont {{Fosalba}},
  \citenamefont {{Frenk}}, \citenamefont {{Fumagalli}}, \citenamefont
  {{Gaensicke}}, \citenamefont {{Gallo}}, \citenamefont {{Garcia-Bellido}},
  \citenamefont {{Gaztanaga}}, \citenamefont {{Pietro Gentile Fusillo}},
  \citenamefont {{Gerard}}, \citenamefont {{Gershkovich}}, \citenamefont
  {{Giannantonio}}, \citenamefont {{Gillet}}, \citenamefont
  {{Gonzalez-de-Rivera}}, \citenamefont {{Gonzalez-Perez}}, \citenamefont
  {{Gott}}, \citenamefont {{Graur}}, \citenamefont {{Gutierrez}}, \citenamefont
  {{Guy}}, \citenamefont {{Habib}}, \citenamefont {{Heetderks}}, \citenamefont
  {{Heetderks}}, \citenamefont {{Heitmann}}, \citenamefont {{Hellwing}},
  \citenamefont {{Herrera}}, \citenamefont {{Ho}}, \citenamefont {{Holland}},
  \citenamefont {{Honscheid}}, \citenamefont {{Huff}}, \citenamefont
  {{Hutchinson}}, \citenamefont {{Huterer}}, \citenamefont {{Hwang}},
  \citenamefont {{Illa Laguna}}, \citenamefont {{Ishikawa}}, \citenamefont
  {{Jacobs}}, \citenamefont {{Jeffrey}}, \citenamefont {{Jelinsky}},
  \citenamefont {{Jennings}}, \citenamefont {{Jiang}}, \citenamefont
  {{Jimenez}}, \citenamefont {{Johnson}}, \citenamefont {{Joyce}},
  \citenamefont {{Jullo}}, \citenamefont {{Juneau}}, \citenamefont {{Kama}},
  \citenamefont {{Karcher}}, \citenamefont {{Karkar}}, \citenamefont {{Kehoe}},
  \citenamefont {{Kennamer}}, \citenamefont {{Kent}}, \citenamefont
  {{Kilbinger}}, \citenamefont {{Kim}}, \citenamefont {{Kirkby}}, \citenamefont
  {{Kisner}}, \citenamefont {{Kitanidis}}, \citenamefont {{Kneib}},
  \citenamefont {{Koposov}}, \citenamefont {{Kovacs}}, \citenamefont
  {{Koyama}}, \citenamefont {{Kremin}}, \citenamefont {{Kron}}, \citenamefont
  {{Kronig}}, \citenamefont {{Kueter-Young}}, \citenamefont {{Lacey}},
  \citenamefont {{Lafever}}, \citenamefont {{Lahav}}, \citenamefont
  {{Lambert}}, \citenamefont {{Lampton}}, \citenamefont {{Landriau}},
  \citenamefont {{Lang}}, \citenamefont {{Lauer}}, \citenamefont {{Le Goff}},
  \citenamefont {{Le Guillou}}, \citenamefont {{Le Van Suu}}, \citenamefont
  {{Lee}}, \citenamefont {{Lee}}, \citenamefont {{Leitner}}, \citenamefont
  {{Lesser}}, \citenamefont {{Levi}}, \citenamefont {{L'Huillier}},
  \citenamefont {{Li}}, \citenamefont {{Liang}}, \citenamefont {{Lin}},
  \citenamefont {{Linder}}, \citenamefont {{Loebman}}, \citenamefont
  {{Luki{\'c}}}, \citenamefont {{Ma}}, \citenamefont {{MacCrann}},
  \citenamefont {{Magneville}}, \citenamefont {{Makarem}}, \citenamefont
  {{Manera}}, \citenamefont {{Manser}}, \citenamefont {{Marshall}},
  \citenamefont {{Martini}}, \citenamefont {{Massey}}, \citenamefont
  {{Matheson}}, \citenamefont {{McCauley}}, \citenamefont {{McDonald}},
  \citenamefont {{McGreer}}, \citenamefont {{Meisner}}, \citenamefont
  {{Metcalfe}}, \citenamefont {{Miller}}, \citenamefont {{Miquel}},
  \citenamefont {{Moustakas}}, \citenamefont {{Myers}}, \citenamefont {{Naik}},
  \citenamefont {{Newman}}, \citenamefont {{Nichol}}, \citenamefont {{Nicola}},
  \citenamefont {{Nicolati da Costa}}, \citenamefont {{Nie}}, \citenamefont
  {{Niz}}, \citenamefont {{Norberg}}, \citenamefont {{Nord}}, \citenamefont
  {{Norman}}, \citenamefont {{Nugent}}, \citenamefont {{O'Brien}},
  \citenamefont {{Oh}}, \citenamefont {{Olsen}}, \citenamefont {{Padilla}},
  \citenamefont {{Padmanabhan}}, \citenamefont {{Padmanabhan}}, \citenamefont
  {{Palanque-Delabrouille}}, \citenamefont {{Palmese}}, \citenamefont
  {{Pappalardo}}, \citenamefont {{P{\^a}ris}}, \citenamefont {{Park}},
  \citenamefont {{Patej}}, \citenamefont {{Peacock}}, \citenamefont {{Peiris}},
  \citenamefont {{Peng}}, \citenamefont {{Percival}}, \citenamefont
  {{Perruchot}}, \citenamefont {{Pieri}}, \citenamefont {{Pogge}},
  \citenamefont {{Pollack}}, \citenamefont {{Poppett}}, \citenamefont
  {{Prada}}, \citenamefont {{Prakash}}, \citenamefont {{Probst}}, \citenamefont
  {{Rabinowitz}}, \citenamefont {{Raichoor}}, \citenamefont {{Ree}},
  \citenamefont {{Refregier}}, \citenamefont {{Regal}}, \citenamefont {{Reid}},
  \citenamefont {{Reil}}, \citenamefont {{Rezaie}}, \citenamefont {{Rockosi}},
  \citenamefont {{Roe}}, \citenamefont {{Ronayette}}, \citenamefont
  {{Roodman}}, \citenamefont {{Ross}}, \citenamefont {{Ross}}, \citenamefont
  {{Rossi}}, \citenamefont {{Rozo}}, \citenamefont {{Ruhlmann-Kleider}},
  \citenamefont {{Rykoff}}, \citenamefont {{Sabiu}}, \citenamefont
  {{Samushia}}, \citenamefont {{Sanchez}}, \citenamefont {{Sanchez}},
  \citenamefont {{Schlegel}}, \citenamefont {{Schneider}}, \citenamefont
  {{Schubnell}}, \citenamefont {{Secroun}}, \citenamefont {{Seljak}},
  \citenamefont {{Seo}}, \citenamefont {{Serrano}}, \citenamefont
  {{Shafieloo}}, \citenamefont {{Shan}}, \citenamefont {{Sharples}},
  \citenamefont {{Sholl}}, \citenamefont {{Shourt}}, \citenamefont {{Silber}},
  \citenamefont {{Silva}}, \citenamefont {{Sirk}}, \citenamefont {{Slosar}},
  \citenamefont {{Smith}}, \citenamefont {{Smoot}}, \citenamefont {{Som}},
  \citenamefont {{Song}}, \citenamefont {{Sprayberry}}, \citenamefont
  {{Staten}}, \citenamefont {{Stefanik}}, \citenamefont {{Tarle}},
  \citenamefont {{Sien Tie}}, \citenamefont {{Tinker}}, \citenamefont
  {{Tojeiro}}, \citenamefont {{Valdes}}, \citenamefont {{Valenzuela}},
  \citenamefont {{Valluri}}, \citenamefont {{Vargas-Magana}}, \citenamefont
  {{Verde}}, \citenamefont {{Walker}}, \citenamefont {{Wang}}, \citenamefont
  {{Wang}}, \citenamefont {{Weaver}}, \citenamefont {{Weaverdyck}},
  \citenamefont {{Wechsler}}, \citenamefont {{Weinberg}}, \citenamefont
  {{White}}, \citenamefont {{Yang}}, \citenamefont {{Yeche}}, \citenamefont
  {{Zhang}}, \citenamefont {{Zhao}}, \citenamefont {{Zheng}}, \citenamefont
  {{Zhou}}, \citenamefont {{Zhou}}, \citenamefont {{Zhu}}, \citenamefont
  {{Zou}},\ and\ \citenamefont {{Zu}}}]{2016arXiv161100036D}%
  \BibitemOpen
  \bibfield  {author} {\bibinfo {author} {\bibnamefont {{DESI Collaboration}}},
  \bibinfo {author} {\bibfnamefont {A.}~\bibnamefont {{Aghamousa}}}, \bibinfo
  {author} {\bibfnamefont {J.}~\bibnamefont {{Aguilar}}}, \bibinfo {author}
  {\bibfnamefont {S.}~\bibnamefont {{Ahlen}}}, \bibinfo {author} {\bibfnamefont
  {S.}~\bibnamefont {{Alam}}}, \bibinfo {author} {\bibfnamefont {L.~E.}\
  \bibnamefont {{Allen}}}, \bibinfo {author} {\bibfnamefont {C.}~\bibnamefont
  {{Allende Prieto}}}, \bibinfo {author} {\bibfnamefont {J.}~\bibnamefont
  {{Annis}}}, \bibinfo {author} {\bibfnamefont {S.}~\bibnamefont {{Bailey}}},
  \bibinfo {author} {\bibfnamefont {C.}~\bibnamefont {{Balland}}}, \bibinfo
  {author} {\bibfnamefont {O.}~\bibnamefont {{Ballester}}}, \bibinfo {author}
  {\bibfnamefont {C.}~\bibnamefont {{Baltay}}}, \bibinfo {author}
  {\bibfnamefont {L.}~\bibnamefont {{Beaufore}}}, \bibinfo {author}
  {\bibfnamefont {C.}~\bibnamefont {{Bebek}}}, \bibinfo {author} {\bibfnamefont
  {T.~C.}\ \bibnamefont {{Beers}}}, \bibinfo {author} {\bibfnamefont {E.~F.}\
  \bibnamefont {{Bell}}}, \bibinfo {author} {\bibfnamefont {J.~L.}\
  \bibnamefont {{Bernal}}}, \bibinfo {author} {\bibfnamefont {R.}~\bibnamefont
  {{Besuner}}}, \bibinfo {author} {\bibfnamefont {F.}~\bibnamefont
  {{Beutler}}}, \bibinfo {author} {\bibfnamefont {C.}~\bibnamefont {{Blake}}},
  \bibinfo {author} {\bibfnamefont {H.}~\bibnamefont {{Bleuler}}}, \bibinfo
  {author} {\bibfnamefont {M.}~\bibnamefont {{Blomqvist}}}, \bibinfo {author}
  {\bibfnamefont {R.}~\bibnamefont {{Blum}}}, \bibinfo {author} {\bibfnamefont
  {A.~S.}\ \bibnamefont {{Bolton}}}, \bibinfo {author} {\bibfnamefont
  {C.}~\bibnamefont {{Briceno}}}, \bibinfo {author} {\bibfnamefont
  {D.}~\bibnamefont {{Brooks}}}, \bibinfo {author} {\bibfnamefont {J.~R.}\
  \bibnamefont {{Brownstein}}}, \bibinfo {author} {\bibfnamefont
  {E.}~\bibnamefont {{Buckley-Geer}}}, \bibinfo {author} {\bibfnamefont
  {A.}~\bibnamefont {{Burden}}}, \bibinfo {author} {\bibfnamefont
  {E.}~\bibnamefont {{Burtin}}}, \bibinfo {author} {\bibfnamefont {N.~G.}\
  \bibnamefont {{Busca}}}, \bibinfo {author} {\bibfnamefont {R.~N.}\
  \bibnamefont {{Cahn}}}, \bibinfo {author} {\bibfnamefont {Y.-C.}\
  \bibnamefont {{Cai}}}, \bibinfo {author} {\bibfnamefont {L.}~\bibnamefont
  {{Cardiel-Sas}}}, \bibinfo {author} {\bibfnamefont {R.~G.}\ \bibnamefont
  {{Carlberg}}}, \bibinfo {author} {\bibfnamefont {P.-H.}\ \bibnamefont
  {{Carton}}}, \bibinfo {author} {\bibfnamefont {R.}~\bibnamefont {{Casas}}},
  \bibinfo {author} {\bibfnamefont {F.~J.}\ \bibnamefont {{Castander}}},
  \bibinfo {author} {\bibfnamefont {J.~L.}\ \bibnamefont {{Cervantes-Cota}}},
  \bibinfo {author} {\bibfnamefont {T.~M.}\ \bibnamefont {{Claybaugh}}},
  \bibinfo {author} {\bibfnamefont {M.}~\bibnamefont {{Close}}}, \bibinfo
  {author} {\bibfnamefont {C.~T.}\ \bibnamefont {{Coker}}}, \bibinfo {author}
  {\bibfnamefont {S.}~\bibnamefont {{Cole}}}, \bibinfo {author} {\bibfnamefont
  {J.}~\bibnamefont {{Comparat}}}, \bibinfo {author} {\bibfnamefont {A.~P.}\
  \bibnamefont {{Cooper}}}, \bibinfo {author} {\bibfnamefont {M.~C.}\
  \bibnamefont {{Cousinou}}}, \bibinfo {author} {\bibfnamefont
  {M.}~\bibnamefont {{Crocce}}}, \bibinfo {author} {\bibfnamefont {J.-G.}\
  \bibnamefont {{Cuby}}}, \bibinfo {author} {\bibfnamefont {D.~P.}\
  \bibnamefont {{Cunningham}}}, \bibinfo {author} {\bibfnamefont {T.~M.}\
  \bibnamefont {{Davis}}}, \bibinfo {author} {\bibfnamefont {K.~S.}\
  \bibnamefont {{Dawson}}}, \bibinfo {author} {\bibfnamefont {A.}~\bibnamefont
  {{de la Macorra}}}, \bibinfo {author} {\bibfnamefont {J.}~\bibnamefont {{De
  Vicente}}}, \bibinfo {author} {\bibfnamefont {T.}~\bibnamefont {{Delubac}}},
  \bibinfo {author} {\bibfnamefont {M.}~\bibnamefont {{Derwent}}}, \bibinfo
  {author} {\bibfnamefont {A.}~\bibnamefont {{Dey}}}, \bibinfo {author}
  {\bibfnamefont {G.}~\bibnamefont {{Dhungana}}}, \bibinfo {author}
  {\bibfnamefont {Z.}~\bibnamefont {{Ding}}}, \bibinfo {author} {\bibfnamefont
  {P.}~\bibnamefont {{Doel}}}, \bibinfo {author} {\bibfnamefont {Y.~T.}\
  \bibnamefont {{Duan}}}, \bibinfo {author} {\bibfnamefont {A.}~\bibnamefont
  {{Ealet}}}, \bibinfo {author} {\bibfnamefont {J.}~\bibnamefont
  {{Edelstein}}}, \bibinfo {author} {\bibfnamefont {S.}~\bibnamefont
  {{Eftekharzadeh}}}, \bibinfo {author} {\bibfnamefont {D.~J.}\ \bibnamefont
  {{Eisenstein}}}, \bibinfo {author} {\bibfnamefont {A.}~\bibnamefont
  {{Elliott}}}, \bibinfo {author} {\bibfnamefont {S.}~\bibnamefont
  {{Escoffier}}}, \bibinfo {author} {\bibfnamefont {M.}~\bibnamefont
  {{Evatt}}}, \bibinfo {author} {\bibfnamefont {P.}~\bibnamefont
  {{Fagrelius}}}, \bibinfo {author} {\bibfnamefont {X.}~\bibnamefont {{Fan}}},
  \bibinfo {author} {\bibfnamefont {K.}~\bibnamefont {{Fanning}}}, \bibinfo
  {author} {\bibfnamefont {A.}~\bibnamefont {{Farahi}}}, \bibinfo {author}
  {\bibfnamefont {J.}~\bibnamefont {{Farihi}}}, \bibinfo {author}
  {\bibfnamefont {G.}~\bibnamefont {{Favole}}}, \bibinfo {author}
  {\bibfnamefont {Y.}~\bibnamefont {{Feng}}}, \bibinfo {author} {\bibfnamefont
  {E.}~\bibnamefont {{Fernandez}}}, \bibinfo {author} {\bibfnamefont {J.~R.}\
  \bibnamefont {{Findlay}}}, \bibinfo {author} {\bibfnamefont {D.~P.}\
  \bibnamefont {{Finkbeiner}}}, \bibinfo {author} {\bibfnamefont {M.~J.}\
  \bibnamefont {{Fitzpatrick}}}, \bibinfo {author} {\bibfnamefont
  {B.}~\bibnamefont {{Flaugher}}}, \bibinfo {author} {\bibfnamefont
  {S.}~\bibnamefont {{Flender}}}, \bibinfo {author} {\bibfnamefont
  {A.}~\bibnamefont {{Font-Ribera}}}, \bibinfo {author} {\bibfnamefont {J.~E.}\
  \bibnamefont {{Forero-Romero}}}, \bibinfo {author} {\bibfnamefont
  {P.}~\bibnamefont {{Fosalba}}}, \bibinfo {author} {\bibfnamefont {C.~S.}\
  \bibnamefont {{Frenk}}}, \bibinfo {author} {\bibfnamefont {M.}~\bibnamefont
  {{Fumagalli}}}, \bibinfo {author} {\bibfnamefont {B.~T.}\ \bibnamefont
  {{Gaensicke}}}, \bibinfo {author} {\bibfnamefont {G.}~\bibnamefont
  {{Gallo}}}, \bibinfo {author} {\bibfnamefont {J.}~\bibnamefont
  {{Garcia-Bellido}}}, \bibinfo {author} {\bibfnamefont {E.}~\bibnamefont
  {{Gaztanaga}}}, \bibinfo {author} {\bibfnamefont {N.}~\bibnamefont {{Pietro
  Gentile Fusillo}}}, \bibinfo {author} {\bibfnamefont {T.}~\bibnamefont
  {{Gerard}}}, \bibinfo {author} {\bibfnamefont {I.}~\bibnamefont
  {{Gershkovich}}}, \bibinfo {author} {\bibfnamefont {T.}~\bibnamefont
  {{Giannantonio}}}, \bibinfo {author} {\bibfnamefont {D.}~\bibnamefont
  {{Gillet}}}, \bibinfo {author} {\bibfnamefont {G.}~\bibnamefont
  {{Gonzalez-de-Rivera}}}, \bibinfo {author} {\bibfnamefont {V.}~\bibnamefont
  {{Gonzalez-Perez}}}, \bibinfo {author} {\bibfnamefont {S.}~\bibnamefont
  {{Gott}}}, \bibinfo {author} {\bibfnamefont {O.}~\bibnamefont {{Graur}}},
  \bibinfo {author} {\bibfnamefont {G.}~\bibnamefont {{Gutierrez}}}, \bibinfo
  {author} {\bibfnamefont {J.}~\bibnamefont {{Guy}}}, \bibinfo {author}
  {\bibfnamefont {S.}~\bibnamefont {{Habib}}}, \bibinfo {author} {\bibfnamefont
  {H.}~\bibnamefont {{Heetderks}}}, \bibinfo {author} {\bibfnamefont
  {I.}~\bibnamefont {{Heetderks}}}, \bibinfo {author} {\bibfnamefont
  {K.}~\bibnamefont {{Heitmann}}}, \bibinfo {author} {\bibfnamefont {W.~A.}\
  \bibnamefont {{Hellwing}}}, \bibinfo {author} {\bibfnamefont {D.~A.}\
  \bibnamefont {{Herrera}}}, \bibinfo {author} {\bibfnamefont {S.}~\bibnamefont
  {{Ho}}}, \bibinfo {author} {\bibfnamefont {S.}~\bibnamefont {{Holland}}},
  \bibinfo {author} {\bibfnamefont {K.}~\bibnamefont {{Honscheid}}}, \bibinfo
  {author} {\bibfnamefont {E.}~\bibnamefont {{Huff}}}, \bibinfo {author}
  {\bibfnamefont {T.~A.}\ \bibnamefont {{Hutchinson}}}, \bibinfo {author}
  {\bibfnamefont {D.}~\bibnamefont {{Huterer}}}, \bibinfo {author}
  {\bibfnamefont {H.~S.}\ \bibnamefont {{Hwang}}}, \bibinfo {author}
  {\bibfnamefont {J.~M.}\ \bibnamefont {{Illa Laguna}}}, \bibinfo {author}
  {\bibfnamefont {Y.}~\bibnamefont {{Ishikawa}}}, \bibinfo {author}
  {\bibfnamefont {D.}~\bibnamefont {{Jacobs}}}, \bibinfo {author}
  {\bibfnamefont {N.}~\bibnamefont {{Jeffrey}}}, \bibinfo {author}
  {\bibfnamefont {P.}~\bibnamefont {{Jelinsky}}}, \bibinfo {author}
  {\bibfnamefont {E.}~\bibnamefont {{Jennings}}}, \bibinfo {author}
  {\bibfnamefont {L.}~\bibnamefont {{Jiang}}}, \bibinfo {author} {\bibfnamefont
  {J.}~\bibnamefont {{Jimenez}}}, \bibinfo {author} {\bibfnamefont
  {J.}~\bibnamefont {{Johnson}}}, \bibinfo {author} {\bibfnamefont
  {R.}~\bibnamefont {{Joyce}}}, \bibinfo {author} {\bibfnamefont
  {E.}~\bibnamefont {{Jullo}}}, \bibinfo {author} {\bibfnamefont
  {S.}~\bibnamefont {{Juneau}}}, \bibinfo {author} {\bibfnamefont
  {S.}~\bibnamefont {{Kama}}}, \bibinfo {author} {\bibfnamefont
  {A.}~\bibnamefont {{Karcher}}}, \bibinfo {author} {\bibfnamefont
  {S.}~\bibnamefont {{Karkar}}}, \bibinfo {author} {\bibfnamefont
  {R.}~\bibnamefont {{Kehoe}}}, \bibinfo {author} {\bibfnamefont
  {N.}~\bibnamefont {{Kennamer}}}, \bibinfo {author} {\bibfnamefont
  {S.}~\bibnamefont {{Kent}}}, \bibinfo {author} {\bibfnamefont
  {M.}~\bibnamefont {{Kilbinger}}}, \bibinfo {author} {\bibfnamefont {A.~G.}\
  \bibnamefont {{Kim}}}, \bibinfo {author} {\bibfnamefont {D.}~\bibnamefont
  {{Kirkby}}}, \bibinfo {author} {\bibfnamefont {T.}~\bibnamefont {{Kisner}}},
  \bibinfo {author} {\bibfnamefont {E.}~\bibnamefont {{Kitanidis}}}, \bibinfo
  {author} {\bibfnamefont {J.-P.}\ \bibnamefont {{Kneib}}}, \bibinfo {author}
  {\bibfnamefont {S.}~\bibnamefont {{Koposov}}}, \bibinfo {author}
  {\bibfnamefont {E.}~\bibnamefont {{Kovacs}}}, \bibinfo {author}
  {\bibfnamefont {K.}~\bibnamefont {{Koyama}}}, \bibinfo {author}
  {\bibfnamefont {A.}~\bibnamefont {{Kremin}}}, \bibinfo {author}
  {\bibfnamefont {R.}~\bibnamefont {{Kron}}}, \bibinfo {author} {\bibfnamefont
  {L.}~\bibnamefont {{Kronig}}}, \bibinfo {author} {\bibfnamefont
  {A.}~\bibnamefont {{Kueter-Young}}}, \bibinfo {author} {\bibfnamefont
  {C.~G.}\ \bibnamefont {{Lacey}}}, \bibinfo {author} {\bibfnamefont
  {R.}~\bibnamefont {{Lafever}}}, \bibinfo {author} {\bibfnamefont
  {O.}~\bibnamefont {{Lahav}}}, \bibinfo {author} {\bibfnamefont
  {A.}~\bibnamefont {{Lambert}}}, \bibinfo {author} {\bibfnamefont
  {M.}~\bibnamefont {{Lampton}}}, \bibinfo {author} {\bibfnamefont
  {M.}~\bibnamefont {{Landriau}}}, \bibinfo {author} {\bibfnamefont
  {D.}~\bibnamefont {{Lang}}}, \bibinfo {author} {\bibfnamefont {T.~R.}\
  \bibnamefont {{Lauer}}}, \bibinfo {author} {\bibfnamefont {J.-M.}\
  \bibnamefont {{Le Goff}}}, \bibinfo {author} {\bibfnamefont {L.}~\bibnamefont
  {{Le Guillou}}}, \bibinfo {author} {\bibfnamefont {A.}~\bibnamefont {{Le Van
  Suu}}}, \bibinfo {author} {\bibfnamefont {J.~H.}\ \bibnamefont {{Lee}}},
  \bibinfo {author} {\bibfnamefont {S.-J.}\ \bibnamefont {{Lee}}}, \bibinfo
  {author} {\bibfnamefont {D.}~\bibnamefont {{Leitner}}}, \bibinfo {author}
  {\bibfnamefont {M.}~\bibnamefont {{Lesser}}}, \bibinfo {author}
  {\bibfnamefont {M.~E.}\ \bibnamefont {{Levi}}}, \bibinfo {author}
  {\bibfnamefont {B.}~\bibnamefont {{L'Huillier}}}, \bibinfo {author}
  {\bibfnamefont {B.}~\bibnamefont {{Li}}}, \bibinfo {author} {\bibfnamefont
  {M.}~\bibnamefont {{Liang}}}, \bibinfo {author} {\bibfnamefont
  {H.}~\bibnamefont {{Lin}}}, \bibinfo {author} {\bibfnamefont
  {E.}~\bibnamefont {{Linder}}}, \bibinfo {author} {\bibfnamefont {S.~R.}\
  \bibnamefont {{Loebman}}}, \bibinfo {author} {\bibfnamefont {Z.}~\bibnamefont
  {{Luki{\'c}}}}, \bibinfo {author} {\bibfnamefont {J.}~\bibnamefont {{Ma}}},
  \bibinfo {author} {\bibfnamefont {N.}~\bibnamefont {{MacCrann}}}, \bibinfo
  {author} {\bibfnamefont {C.}~\bibnamefont {{Magneville}}}, \bibinfo {author}
  {\bibfnamefont {L.}~\bibnamefont {{Makarem}}}, \bibinfo {author}
  {\bibfnamefont {M.}~\bibnamefont {{Manera}}}, \bibinfo {author}
  {\bibfnamefont {C.~J.}\ \bibnamefont {{Manser}}}, \bibinfo {author}
  {\bibfnamefont {R.}~\bibnamefont {{Marshall}}}, \bibinfo {author}
  {\bibfnamefont {P.}~\bibnamefont {{Martini}}}, \bibinfo {author}
  {\bibfnamefont {R.}~\bibnamefont {{Massey}}}, \bibinfo {author}
  {\bibfnamefont {T.}~\bibnamefont {{Matheson}}}, \bibinfo {author}
  {\bibfnamefont {J.}~\bibnamefont {{McCauley}}}, \bibinfo {author}
  {\bibfnamefont {P.}~\bibnamefont {{McDonald}}}, \bibinfo {author}
  {\bibfnamefont {I.~D.}\ \bibnamefont {{McGreer}}}, \bibinfo {author}
  {\bibfnamefont {A.}~\bibnamefont {{Meisner}}}, \bibinfo {author}
  {\bibfnamefont {N.}~\bibnamefont {{Metcalfe}}}, \bibinfo {author}
  {\bibfnamefont {T.~N.}\ \bibnamefont {{Miller}}}, \bibinfo {author}
  {\bibfnamefont {R.}~\bibnamefont {{Miquel}}}, \bibinfo {author}
  {\bibfnamefont {J.}~\bibnamefont {{Moustakas}}}, \bibinfo {author}
  {\bibfnamefont {A.}~\bibnamefont {{Myers}}}, \bibinfo {author} {\bibfnamefont
  {M.}~\bibnamefont {{Naik}}}, \bibinfo {author} {\bibfnamefont {J.~A.}\
  \bibnamefont {{Newman}}}, \bibinfo {author} {\bibfnamefont {R.~C.}\
  \bibnamefont {{Nichol}}}, \bibinfo {author} {\bibfnamefont {A.}~\bibnamefont
  {{Nicola}}}, \bibinfo {author} {\bibfnamefont {L.}~\bibnamefont {{Nicolati da
  Costa}}}, \bibinfo {author} {\bibfnamefont {J.}~\bibnamefont {{Nie}}},
  \bibinfo {author} {\bibfnamefont {G.}~\bibnamefont {{Niz}}}, \bibinfo
  {author} {\bibfnamefont {P.}~\bibnamefont {{Norberg}}}, \bibinfo {author}
  {\bibfnamefont {B.}~\bibnamefont {{Nord}}}, \bibinfo {author} {\bibfnamefont
  {D.}~\bibnamefont {{Norman}}}, \bibinfo {author} {\bibfnamefont
  {P.}~\bibnamefont {{Nugent}}}, \bibinfo {author} {\bibfnamefont
  {T.}~\bibnamefont {{O'Brien}}}, \bibinfo {author} {\bibfnamefont
  {M.}~\bibnamefont {{Oh}}}, \bibinfo {author} {\bibfnamefont {K.~A.~G.}\
  \bibnamefont {{Olsen}}}, \bibinfo {author} {\bibfnamefont {C.}~\bibnamefont
  {{Padilla}}}, \bibinfo {author} {\bibfnamefont {H.}~\bibnamefont
  {{Padmanabhan}}}, \bibinfo {author} {\bibfnamefont {N.}~\bibnamefont
  {{Padmanabhan}}}, \bibinfo {author} {\bibfnamefont {N.}~\bibnamefont
  {{Palanque-Delabrouille}}}, \bibinfo {author} {\bibfnamefont
  {A.}~\bibnamefont {{Palmese}}}, \bibinfo {author} {\bibfnamefont
  {D.}~\bibnamefont {{Pappalardo}}}, \bibinfo {author} {\bibfnamefont
  {I.}~\bibnamefont {{P{\^a}ris}}}, \bibinfo {author} {\bibfnamefont
  {C.}~\bibnamefont {{Park}}}, \bibinfo {author} {\bibfnamefont
  {A.}~\bibnamefont {{Patej}}}, \bibinfo {author} {\bibfnamefont {J.~A.}\
  \bibnamefont {{Peacock}}}, \bibinfo {author} {\bibfnamefont {H.~V.}\
  \bibnamefont {{Peiris}}}, \bibinfo {author} {\bibfnamefont {X.}~\bibnamefont
  {{Peng}}}, \bibinfo {author} {\bibfnamefont {W.~J.}\ \bibnamefont
  {{Percival}}}, \bibinfo {author} {\bibfnamefont {S.}~\bibnamefont
  {{Perruchot}}}, \bibinfo {author} {\bibfnamefont {M.~M.}\ \bibnamefont
  {{Pieri}}}, \bibinfo {author} {\bibfnamefont {R.}~\bibnamefont {{Pogge}}},
  \bibinfo {author} {\bibfnamefont {J.~E.}\ \bibnamefont {{Pollack}}}, \bibinfo
  {author} {\bibfnamefont {C.}~\bibnamefont {{Poppett}}}, \bibinfo {author}
  {\bibfnamefont {F.}~\bibnamefont {{Prada}}}, \bibinfo {author} {\bibfnamefont
  {A.}~\bibnamefont {{Prakash}}}, \bibinfo {author} {\bibfnamefont {R.~G.}\
  \bibnamefont {{Probst}}}, \bibinfo {author} {\bibfnamefont {D.}~\bibnamefont
  {{Rabinowitz}}}, \bibinfo {author} {\bibfnamefont {A.}~\bibnamefont
  {{Raichoor}}}, \bibinfo {author} {\bibfnamefont {C.~H.}\ \bibnamefont
  {{Ree}}}, \bibinfo {author} {\bibfnamefont {A.}~\bibnamefont {{Refregier}}},
  \bibinfo {author} {\bibfnamefont {X.}~\bibnamefont {{Regal}}}, \bibinfo
  {author} {\bibfnamefont {B.}~\bibnamefont {{Reid}}}, \bibinfo {author}
  {\bibfnamefont {K.}~\bibnamefont {{Reil}}}, \bibinfo {author} {\bibfnamefont
  {M.}~\bibnamefont {{Rezaie}}}, \bibinfo {author} {\bibfnamefont {C.~M.}\
  \bibnamefont {{Rockosi}}}, \bibinfo {author} {\bibfnamefont {N.}~\bibnamefont
  {{Roe}}}, \bibinfo {author} {\bibfnamefont {S.}~\bibnamefont {{Ronayette}}},
  \bibinfo {author} {\bibfnamefont {A.}~\bibnamefont {{Roodman}}}, \bibinfo
  {author} {\bibfnamefont {A.~J.}\ \bibnamefont {{Ross}}}, \bibinfo {author}
  {\bibfnamefont {N.~P.}\ \bibnamefont {{Ross}}}, \bibinfo {author}
  {\bibfnamefont {G.}~\bibnamefont {{Rossi}}}, \bibinfo {author} {\bibfnamefont
  {E.}~\bibnamefont {{Rozo}}}, \bibinfo {author} {\bibfnamefont
  {V.}~\bibnamefont {{Ruhlmann-Kleider}}}, \bibinfo {author} {\bibfnamefont
  {E.~S.}\ \bibnamefont {{Rykoff}}}, \bibinfo {author} {\bibfnamefont
  {C.}~\bibnamefont {{Sabiu}}}, \bibinfo {author} {\bibfnamefont
  {L.}~\bibnamefont {{Samushia}}}, \bibinfo {author} {\bibfnamefont
  {E.}~\bibnamefont {{Sanchez}}}, \bibinfo {author} {\bibfnamefont
  {J.}~\bibnamefont {{Sanchez}}}, \bibinfo {author} {\bibfnamefont {D.~J.}\
  \bibnamefont {{Schlegel}}}, \bibinfo {author} {\bibfnamefont
  {M.}~\bibnamefont {{Schneider}}}, \bibinfo {author} {\bibfnamefont
  {M.}~\bibnamefont {{Schubnell}}}, \bibinfo {author} {\bibfnamefont
  {A.}~\bibnamefont {{Secroun}}}, \bibinfo {author} {\bibfnamefont
  {U.}~\bibnamefont {{Seljak}}}, \bibinfo {author} {\bibfnamefont {H.-J.}\
  \bibnamefont {{Seo}}}, \bibinfo {author} {\bibfnamefont {S.}~\bibnamefont
  {{Serrano}}}, \bibinfo {author} {\bibfnamefont {A.}~\bibnamefont
  {{Shafieloo}}}, \bibinfo {author} {\bibfnamefont {H.}~\bibnamefont {{Shan}}},
  \bibinfo {author} {\bibfnamefont {R.}~\bibnamefont {{Sharples}}}, \bibinfo
  {author} {\bibfnamefont {M.~J.}\ \bibnamefont {{Sholl}}}, \bibinfo {author}
  {\bibfnamefont {W.~V.}\ \bibnamefont {{Shourt}}}, \bibinfo {author}
  {\bibfnamefont {J.~H.}\ \bibnamefont {{Silber}}}, \bibinfo {author}
  {\bibfnamefont {D.~R.}\ \bibnamefont {{Silva}}}, \bibinfo {author}
  {\bibfnamefont {M.~M.}\ \bibnamefont {{Sirk}}}, \bibinfo {author}
  {\bibfnamefont {A.}~\bibnamefont {{Slosar}}}, \bibinfo {author}
  {\bibfnamefont {A.}~\bibnamefont {{Smith}}}, \bibinfo {author} {\bibfnamefont
  {G.~F.}\ \bibnamefont {{Smoot}}}, \bibinfo {author} {\bibfnamefont
  {D.}~\bibnamefont {{Som}}}, \bibinfo {author} {\bibfnamefont {Y.-S.}\
  \bibnamefont {{Song}}}, \bibinfo {author} {\bibfnamefont {D.}~\bibnamefont
  {{Sprayberry}}}, \bibinfo {author} {\bibfnamefont {R.}~\bibnamefont
  {{Staten}}}, \bibinfo {author} {\bibfnamefont {A.}~\bibnamefont
  {{Stefanik}}}, \bibinfo {author} {\bibfnamefont {G.}~\bibnamefont {{Tarle}}},
  \bibinfo {author} {\bibfnamefont {S.}~\bibnamefont {{Sien Tie}}}, \bibinfo
  {author} {\bibfnamefont {J.~L.}\ \bibnamefont {{Tinker}}}, \bibinfo {author}
  {\bibfnamefont {R.}~\bibnamefont {{Tojeiro}}}, \bibinfo {author}
  {\bibfnamefont {F.}~\bibnamefont {{Valdes}}}, \bibinfo {author}
  {\bibfnamefont {O.}~\bibnamefont {{Valenzuela}}}, \bibinfo {author}
  {\bibfnamefont {M.}~\bibnamefont {{Valluri}}}, \bibinfo {author}
  {\bibfnamefont {M.}~\bibnamefont {{Vargas-Magana}}}, \bibinfo {author}
  {\bibfnamefont {L.}~\bibnamefont {{Verde}}}, \bibinfo {author} {\bibfnamefont
  {A.~R.}\ \bibnamefont {{Walker}}}, \bibinfo {author} {\bibfnamefont
  {J.}~\bibnamefont {{Wang}}}, \bibinfo {author} {\bibfnamefont
  {Y.}~\bibnamefont {{Wang}}}, \bibinfo {author} {\bibfnamefont {B.~A.}\
  \bibnamefont {{Weaver}}}, \bibinfo {author} {\bibfnamefont {C.}~\bibnamefont
  {{Weaverdyck}}}, \bibinfo {author} {\bibfnamefont {R.~H.}\ \bibnamefont
  {{Wechsler}}}, \bibinfo {author} {\bibfnamefont {D.~H.}\ \bibnamefont
  {{Weinberg}}}, \bibinfo {author} {\bibfnamefont {M.}~\bibnamefont {{White}}},
  \bibinfo {author} {\bibfnamefont {Q.}~\bibnamefont {{Yang}}}, \bibinfo
  {author} {\bibfnamefont {C.}~\bibnamefont {{Yeche}}}, \bibinfo {author}
  {\bibfnamefont {T.}~\bibnamefont {{Zhang}}}, \bibinfo {author} {\bibfnamefont
  {G.-B.}\ \bibnamefont {{Zhao}}}, \bibinfo {author} {\bibfnamefont
  {Y.}~\bibnamefont {{Zheng}}}, \bibinfo {author} {\bibfnamefont
  {X.}~\bibnamefont {{Zhou}}}, \bibinfo {author} {\bibfnamefont
  {Z.}~\bibnamefont {{Zhou}}}, \bibinfo {author} {\bibfnamefont
  {Y.}~\bibnamefont {{Zhu}}}, \bibinfo {author} {\bibfnamefont
  {H.}~\bibnamefont {{Zou}}},\ and\ \bibinfo {author} {\bibfnamefont
  {Y.}~\bibnamefont {{Zu}}},\ }\bibfield  {title} {\bibinfo {title} {{The DESI
  Experiment Part I: Science,Targeting, and Survey Design}},\ }\href
  {https://doi.org/10.48550/arXiv.1611.00036} {\bibfield  {journal} {\bibinfo
  {journal} {arXiv e-prints}\ ,\ \bibinfo {eid} {arXiv:1611.00036}} (\bibinfo
  {year} {2016}{\natexlab{b}})},\ \Eprint {https://arxiv.org/abs/1611.00036}
  {arXiv:1611.00036 [astro-ph.IM]} \BibitemShut {NoStop}%
\bibitem [{\citenamefont {{Hahn}}\ \emph {et~al.}(2023)\citenamefont {{Hahn}},
  \citenamefont {{Wilson}}, \citenamefont {{Ruiz-Macias}}, \citenamefont
  {{Cole}}, \citenamefont {{Weinberg}}, \citenamefont {{Moustakas}},
  \citenamefont {{Kremin}}, \citenamefont {{Tinker}}, \citenamefont {{Smith}},
  \citenamefont {{Wechsler}}, \citenamefont {{Ahlen}}, \citenamefont {{Alam}},
  \citenamefont {{Bailey}}, \citenamefont {{Brooks}}, \citenamefont {{Cooper}},
  \citenamefont {{Davis}}, \citenamefont {{Dawson}}, \citenamefont {{Dey}},
  \citenamefont {{Dey}}, \citenamefont {{Eftekharzadeh}}, \citenamefont
  {{Eisenstein}}, \citenamefont {{Fanning}}, \citenamefont {{Forero-Romero}},
  \citenamefont {{Frenk}}, \citenamefont {{Gazta{\~n}aga}}, \citenamefont {{A
  Gontcho}}, \citenamefont {{Guy}}, \citenamefont {{Honscheid}}, \citenamefont
  {{Ishak}}, \citenamefont {{Juneau}}, \citenamefont {{Kehoe}}, \citenamefont
  {{Kisner}}, \citenamefont {{Lan}}, \citenamefont {{Landriau}}, \citenamefont
  {{Le Guillou}}, \citenamefont {{Levi}}, \citenamefont {{Magneville}},
  \citenamefont {{Martini}}, \citenamefont {{Meisner}}, \citenamefont
  {{Myers}}, \citenamefont {{Nie}}, \citenamefont {{Norberg}}, \citenamefont
  {{Palanque-Delabrouille}}, \citenamefont {{Percival}}, \citenamefont
  {{Poppett}}, \citenamefont {{Prada}}, \citenamefont {{Raichoor}},
  \citenamefont {{Ross}}, \citenamefont {{Gaines}}, \citenamefont {{Saulder}},
  \citenamefont {{Schlafly}}, \citenamefont {{Schlegel}}, \citenamefont
  {{Sierra-Porta}}, \citenamefont {{Tarle}}, \citenamefont {{Weaver}},
  \citenamefont {{Y{\`e}che}}, \citenamefont {{Zarrouk}}, \citenamefont
  {{Zhou}}, \citenamefont {{Zhou}},\ and\ \citenamefont
  {{Zou}}}]{2023AJ....165..253H}%
  \BibitemOpen
  \bibfield  {author} {\bibinfo {author} {\bibfnamefont {C.}~\bibnamefont
  {{Hahn}}}, \bibinfo {author} {\bibfnamefont {M.~J.}\ \bibnamefont
  {{Wilson}}}, \bibinfo {author} {\bibfnamefont {O.}~\bibnamefont
  {{Ruiz-Macias}}}, \bibinfo {author} {\bibfnamefont {S.}~\bibnamefont
  {{Cole}}}, \bibinfo {author} {\bibfnamefont {D.~H.}\ \bibnamefont
  {{Weinberg}}}, \bibinfo {author} {\bibfnamefont {J.}~\bibnamefont
  {{Moustakas}}}, \bibinfo {author} {\bibfnamefont {A.}~\bibnamefont
  {{Kremin}}}, \bibinfo {author} {\bibfnamefont {J.~L.}\ \bibnamefont
  {{Tinker}}}, \bibinfo {author} {\bibfnamefont {A.}~\bibnamefont {{Smith}}},
  \bibinfo {author} {\bibfnamefont {R.~H.}\ \bibnamefont {{Wechsler}}},
  \bibinfo {author} {\bibfnamefont {S.}~\bibnamefont {{Ahlen}}}, \bibinfo
  {author} {\bibfnamefont {S.}~\bibnamefont {{Alam}}}, \bibinfo {author}
  {\bibfnamefont {S.}~\bibnamefont {{Bailey}}}, \bibinfo {author}
  {\bibfnamefont {D.}~\bibnamefont {{Brooks}}}, \bibinfo {author}
  {\bibfnamefont {A.~P.}\ \bibnamefont {{Cooper}}}, \bibinfo {author}
  {\bibfnamefont {T.~M.}\ \bibnamefont {{Davis}}}, \bibinfo {author}
  {\bibfnamefont {K.}~\bibnamefont {{Dawson}}}, \bibinfo {author}
  {\bibfnamefont {A.}~\bibnamefont {{Dey}}}, \bibinfo {author} {\bibfnamefont
  {B.}~\bibnamefont {{Dey}}}, \bibinfo {author} {\bibfnamefont
  {S.}~\bibnamefont {{Eftekharzadeh}}}, \bibinfo {author} {\bibfnamefont
  {D.~J.}\ \bibnamefont {{Eisenstein}}}, \bibinfo {author} {\bibfnamefont
  {K.}~\bibnamefont {{Fanning}}}, \bibinfo {author} {\bibfnamefont {J.~E.}\
  \bibnamefont {{Forero-Romero}}}, \bibinfo {author} {\bibfnamefont {C.~S.}\
  \bibnamefont {{Frenk}}}, \bibinfo {author} {\bibfnamefont {E.}~\bibnamefont
  {{Gazta{\~n}aga}}}, \bibinfo {author} {\bibfnamefont {S.~G.}\ \bibnamefont
  {{A Gontcho}}}, \bibinfo {author} {\bibfnamefont {J.}~\bibnamefont {{Guy}}},
  \bibinfo {author} {\bibfnamefont {K.}~\bibnamefont {{Honscheid}}}, \bibinfo
  {author} {\bibfnamefont {M.}~\bibnamefont {{Ishak}}}, \bibinfo {author}
  {\bibfnamefont {S.}~\bibnamefont {{Juneau}}}, \bibinfo {author}
  {\bibfnamefont {R.}~\bibnamefont {{Kehoe}}}, \bibinfo {author} {\bibfnamefont
  {T.}~\bibnamefont {{Kisner}}}, \bibinfo {author} {\bibfnamefont {T.-W.}\
  \bibnamefont {{Lan}}}, \bibinfo {author} {\bibfnamefont {M.}~\bibnamefont
  {{Landriau}}}, \bibinfo {author} {\bibfnamefont {L.}~\bibnamefont {{Le
  Guillou}}}, \bibinfo {author} {\bibfnamefont {M.~E.}\ \bibnamefont {{Levi}}},
  \bibinfo {author} {\bibfnamefont {C.}~\bibnamefont {{Magneville}}}, \bibinfo
  {author} {\bibfnamefont {P.}~\bibnamefont {{Martini}}}, \bibinfo {author}
  {\bibfnamefont {A.}~\bibnamefont {{Meisner}}}, \bibinfo {author}
  {\bibfnamefont {A.~D.}\ \bibnamefont {{Myers}}}, \bibinfo {author}
  {\bibfnamefont {J.}~\bibnamefont {{Nie}}}, \bibinfo {author} {\bibfnamefont
  {P.}~\bibnamefont {{Norberg}}}, \bibinfo {author} {\bibfnamefont
  {N.}~\bibnamefont {{Palanque-Delabrouille}}}, \bibinfo {author}
  {\bibfnamefont {W.~J.}\ \bibnamefont {{Percival}}}, \bibinfo {author}
  {\bibfnamefont {C.}~\bibnamefont {{Poppett}}}, \bibinfo {author}
  {\bibfnamefont {F.}~\bibnamefont {{Prada}}}, \bibinfo {author} {\bibfnamefont
  {A.}~\bibnamefont {{Raichoor}}}, \bibinfo {author} {\bibfnamefont {A.~J.}\
  \bibnamefont {{Ross}}}, \bibinfo {author} {\bibfnamefont {S.}~\bibnamefont
  {{Gaines}}}, \bibinfo {author} {\bibfnamefont {C.}~\bibnamefont {{Saulder}}},
  \bibinfo {author} {\bibfnamefont {E.}~\bibnamefont {{Schlafly}}}, \bibinfo
  {author} {\bibfnamefont {D.}~\bibnamefont {{Schlegel}}}, \bibinfo {author}
  {\bibfnamefont {D.}~\bibnamefont {{Sierra-Porta}}}, \bibinfo {author}
  {\bibfnamefont {G.}~\bibnamefont {{Tarle}}}, \bibinfo {author} {\bibfnamefont
  {B.~A.}\ \bibnamefont {{Weaver}}}, \bibinfo {author} {\bibfnamefont
  {C.}~\bibnamefont {{Y{\`e}che}}}, \bibinfo {author} {\bibfnamefont
  {P.}~\bibnamefont {{Zarrouk}}}, \bibinfo {author} {\bibfnamefont
  {R.}~\bibnamefont {{Zhou}}}, \bibinfo {author} {\bibfnamefont
  {Z.}~\bibnamefont {{Zhou}}},\ and\ \bibinfo {author} {\bibfnamefont
  {H.}~\bibnamefont {{Zou}}},\ }\bibfield  {title} {\bibinfo {title} {{The DESI
  Bright Galaxy Survey: Final Target Selection, Design, and Validation}},\
  }\href {https://doi.org/10.3847/1538-3881/accff8} {\bibfield  {journal}
  {\bibinfo  {journal} {\aj}\ }\textbf {\bibinfo {volume} {165}},\ \bibinfo
  {eid} {253} (\bibinfo {year} {2023})},\ \Eprint
  {https://arxiv.org/abs/2208.08512} {arXiv:2208.08512 [astro-ph.CO]}
  \BibitemShut {NoStop}%
\bibitem [{\citenamefont {{Zhou}}\ \emph
  {et~al.}(2023{\natexlab{a}})\citenamefont {{Zhou}}, \citenamefont {{Dey}},
  \citenamefont {{Newman}}, \citenamefont {{Eisenstein}}, \citenamefont
  {{Dawson}}, \citenamefont {{Bailey}}, \citenamefont {{Berti}}, \citenamefont
  {{Guy}}, \citenamefont {{Lan}}, \citenamefont {{Zou}}, \citenamefont
  {{Aguilar}}, \citenamefont {{Ahlen}}, \citenamefont {{Alam}}, \citenamefont
  {{Brooks}}, \citenamefont {{de la Macorra}}, \citenamefont {{Dey}},
  \citenamefont {{Dhungana}}, \citenamefont {{Fanning}}, \citenamefont
  {{Font-Ribera}}, \citenamefont {{Gontcho}}, \citenamefont {{Honscheid}},
  \citenamefont {{Ishak}}, \citenamefont {{Kisner}}, \citenamefont
  {{Kov{\'a}cs}}, \citenamefont {{Kremin}}, \citenamefont {{Landriau}},
  \citenamefont {{Levi}}, \citenamefont {{Magneville}}, \citenamefont
  {{Manera}}, \citenamefont {{Martini}}, \citenamefont {{Meisner}},
  \citenamefont {{Miquel}}, \citenamefont {{Moustakas}}, \citenamefont
  {{Myers}}, \citenamefont {{Nie}}, \citenamefont {{Palanque-Delabrouille}},
  \citenamefont {{Percival}}, \citenamefont {{Poppett}}, \citenamefont
  {{Prada}}, \citenamefont {{Raichoor}}, \citenamefont {{Ross}}, \citenamefont
  {{Schlafly}}, \citenamefont {{Schlegel}}, \citenamefont {{Schubnell}},
  \citenamefont {{Tarl{\'e}}}, \citenamefont {{Weaver}}, \citenamefont
  {{Wechsler}}, \citenamefont {{Y{\'e}che}},\ and\ \citenamefont
  {{Zhou}}}]{2023AJ....165...58Z}%
  \BibitemOpen
  \bibfield  {author} {\bibinfo {author} {\bibfnamefont {R.}~\bibnamefont
  {{Zhou}}}, \bibinfo {author} {\bibfnamefont {B.}~\bibnamefont {{Dey}}},
  \bibinfo {author} {\bibfnamefont {J.~A.}\ \bibnamefont {{Newman}}}, \bibinfo
  {author} {\bibfnamefont {D.~J.}\ \bibnamefont {{Eisenstein}}}, \bibinfo
  {author} {\bibfnamefont {K.}~\bibnamefont {{Dawson}}}, \bibinfo {author}
  {\bibfnamefont {S.}~\bibnamefont {{Bailey}}}, \bibinfo {author}
  {\bibfnamefont {A.}~\bibnamefont {{Berti}}}, \bibinfo {author} {\bibfnamefont
  {J.}~\bibnamefont {{Guy}}}, \bibinfo {author} {\bibfnamefont {T.-W.}\
  \bibnamefont {{Lan}}}, \bibinfo {author} {\bibfnamefont {H.}~\bibnamefont
  {{Zou}}}, \bibinfo {author} {\bibfnamefont {J.}~\bibnamefont {{Aguilar}}},
  \bibinfo {author} {\bibfnamefont {S.}~\bibnamefont {{Ahlen}}}, \bibinfo
  {author} {\bibfnamefont {S.}~\bibnamefont {{Alam}}}, \bibinfo {author}
  {\bibfnamefont {D.}~\bibnamefont {{Brooks}}}, \bibinfo {author}
  {\bibfnamefont {A.}~\bibnamefont {{de la Macorra}}}, \bibinfo {author}
  {\bibfnamefont {A.}~\bibnamefont {{Dey}}}, \bibinfo {author} {\bibfnamefont
  {G.}~\bibnamefont {{Dhungana}}}, \bibinfo {author} {\bibfnamefont
  {K.}~\bibnamefont {{Fanning}}}, \bibinfo {author} {\bibfnamefont
  {A.}~\bibnamefont {{Font-Ribera}}}, \bibinfo {author} {\bibfnamefont
  {S.~G.~A.}\ \bibnamefont {{Gontcho}}}, \bibinfo {author} {\bibfnamefont
  {K.}~\bibnamefont {{Honscheid}}}, \bibinfo {author} {\bibfnamefont
  {M.}~\bibnamefont {{Ishak}}}, \bibinfo {author} {\bibfnamefont
  {T.}~\bibnamefont {{Kisner}}}, \bibinfo {author} {\bibfnamefont
  {A.}~\bibnamefont {{Kov{\'a}cs}}}, \bibinfo {author} {\bibfnamefont
  {A.}~\bibnamefont {{Kremin}}}, \bibinfo {author} {\bibfnamefont
  {M.}~\bibnamefont {{Landriau}}}, \bibinfo {author} {\bibfnamefont {M.~E.}\
  \bibnamefont {{Levi}}}, \bibinfo {author} {\bibfnamefont {C.}~\bibnamefont
  {{Magneville}}}, \bibinfo {author} {\bibfnamefont {M.}~\bibnamefont
  {{Manera}}}, \bibinfo {author} {\bibfnamefont {P.}~\bibnamefont {{Martini}}},
  \bibinfo {author} {\bibfnamefont {A.~M.}\ \bibnamefont {{Meisner}}}, \bibinfo
  {author} {\bibfnamefont {R.}~\bibnamefont {{Miquel}}}, \bibinfo {author}
  {\bibfnamefont {J.}~\bibnamefont {{Moustakas}}}, \bibinfo {author}
  {\bibfnamefont {A.~D.}\ \bibnamefont {{Myers}}}, \bibinfo {author}
  {\bibfnamefont {J.}~\bibnamefont {{Nie}}}, \bibinfo {author} {\bibfnamefont
  {N.}~\bibnamefont {{Palanque-Delabrouille}}}, \bibinfo {author}
  {\bibfnamefont {W.~J.}\ \bibnamefont {{Percival}}}, \bibinfo {author}
  {\bibfnamefont {C.}~\bibnamefont {{Poppett}}}, \bibinfo {author}
  {\bibfnamefont {F.}~\bibnamefont {{Prada}}}, \bibinfo {author} {\bibfnamefont
  {A.}~\bibnamefont {{Raichoor}}}, \bibinfo {author} {\bibfnamefont {A.~J.}\
  \bibnamefont {{Ross}}}, \bibinfo {author} {\bibfnamefont {E.}~\bibnamefont
  {{Schlafly}}}, \bibinfo {author} {\bibfnamefont {D.}~\bibnamefont
  {{Schlegel}}}, \bibinfo {author} {\bibfnamefont {M.}~\bibnamefont
  {{Schubnell}}}, \bibinfo {author} {\bibfnamefont {G.}~\bibnamefont
  {{Tarl{\'e}}}}, \bibinfo {author} {\bibfnamefont {B.~A.}\ \bibnamefont
  {{Weaver}}}, \bibinfo {author} {\bibfnamefont {R.~H.}\ \bibnamefont
  {{Wechsler}}}, \bibinfo {author} {\bibfnamefont {C.}~\bibnamefont
  {{Y{\'e}che}}},\ and\ \bibinfo {author} {\bibfnamefont {Z.}~\bibnamefont
  {{Zhou}}},\ }\bibfield  {title} {\bibinfo {title} {{Target Selection and
  Validation of DESI Luminous Red Galaxies}},\ }\href
  {https://doi.org/10.3847/1538-3881/aca5fb} {\bibfield  {journal} {\bibinfo
  {journal} {\aj}\ }\textbf {\bibinfo {volume} {165}},\ \bibinfo {eid} {58}
  (\bibinfo {year} {2023}{\natexlab{a}})},\ \Eprint
  {https://arxiv.org/abs/2208.08515} {arXiv:2208.08515 [astro-ph.CO]}
  \BibitemShut {NoStop}%
\bibitem [{\citenamefont {{Zhou}}\ \emph
  {et~al.}(2023{\natexlab{b}})\citenamefont {{Zhou}}, \citenamefont
  {{Ferraro}}, \citenamefont {{White}}, \citenamefont {{DeRose}}, \citenamefont
  {{Sailer}}, \citenamefont {{Aguilar}}, \citenamefont {{Ahlen}}, \citenamefont
  {{Bailey}}, \citenamefont {{Brooks}}, \citenamefont {{Claybaugh}},
  \citenamefont {{Dawson}}, \citenamefont {{de la Macorra}}, \citenamefont
  {{Dey}}, \citenamefont {{Doel}}, \citenamefont {{Font-Ribera}}, \citenamefont
  {{Forero-Romero}}, \citenamefont {{Gontcho A Gontcho}}, \citenamefont
  {{Guy}}, \citenamefont {{Kremin}}, \citenamefont {{Lambert}}, \citenamefont
  {{Le Guillou}}, \citenamefont {{Levi}}, \citenamefont {{Magneville}},
  \citenamefont {{Manera}}, \citenamefont {{Meisner}}, \citenamefont
  {{Miquel}}, \citenamefont {{Moustakas}}, \citenamefont {{Myers}},
  \citenamefont {{Newman}}, \citenamefont {{Nie}}, \citenamefont {{Percival}},
  \citenamefont {{Rezaie}}, \citenamefont {{Rossi}}, \citenamefont {{Sanchez}},
  \citenamefont {{Schlegel}}, \citenamefont {{Schubnell}}, \citenamefont
  {{Seo}}, \citenamefont {{Tarl{\'e}}},\ and\ \citenamefont
  {{Zhou}}}]{Zhou:2023gji}%
  \BibitemOpen
  \bibfield  {author} {\bibinfo {author} {\bibfnamefont {R.}~\bibnamefont
  {{Zhou}}}, \bibinfo {author} {\bibfnamefont {S.}~\bibnamefont {{Ferraro}}},
  \bibinfo {author} {\bibfnamefont {M.}~\bibnamefont {{White}}}, \bibinfo
  {author} {\bibfnamefont {J.}~\bibnamefont {{DeRose}}}, \bibinfo {author}
  {\bibfnamefont {N.}~\bibnamefont {{Sailer}}}, \bibinfo {author}
  {\bibfnamefont {J.}~\bibnamefont {{Aguilar}}}, \bibinfo {author}
  {\bibfnamefont {S.}~\bibnamefont {{Ahlen}}}, \bibinfo {author} {\bibfnamefont
  {S.}~\bibnamefont {{Bailey}}}, \bibinfo {author} {\bibfnamefont
  {D.}~\bibnamefont {{Brooks}}}, \bibinfo {author} {\bibfnamefont
  {T.}~\bibnamefont {{Claybaugh}}}, \bibinfo {author} {\bibfnamefont
  {K.}~\bibnamefont {{Dawson}}}, \bibinfo {author} {\bibfnamefont
  {A.}~\bibnamefont {{de la Macorra}}}, \bibinfo {author} {\bibfnamefont
  {B.}~\bibnamefont {{Dey}}}, \bibinfo {author} {\bibfnamefont
  {P.}~\bibnamefont {{Doel}}}, \bibinfo {author} {\bibfnamefont
  {A.}~\bibnamefont {{Font-Ribera}}}, \bibinfo {author} {\bibfnamefont {J.~E.}\
  \bibnamefont {{Forero-Romero}}}, \bibinfo {author} {\bibfnamefont
  {S.}~\bibnamefont {{Gontcho A Gontcho}}}, \bibinfo {author} {\bibfnamefont
  {J.}~\bibnamefont {{Guy}}}, \bibinfo {author} {\bibfnamefont
  {A.}~\bibnamefont {{Kremin}}}, \bibinfo {author} {\bibfnamefont
  {A.}~\bibnamefont {{Lambert}}}, \bibinfo {author} {\bibfnamefont
  {L.}~\bibnamefont {{Le Guillou}}}, \bibinfo {author} {\bibfnamefont
  {M.}~\bibnamefont {{Levi}}}, \bibinfo {author} {\bibfnamefont
  {C.}~\bibnamefont {{Magneville}}}, \bibinfo {author} {\bibfnamefont
  {M.}~\bibnamefont {{Manera}}}, \bibinfo {author} {\bibfnamefont
  {A.}~\bibnamefont {{Meisner}}}, \bibinfo {author} {\bibfnamefont
  {R.}~\bibnamefont {{Miquel}}}, \bibinfo {author} {\bibfnamefont
  {J.}~\bibnamefont {{Moustakas}}}, \bibinfo {author} {\bibfnamefont {A.~D.}\
  \bibnamefont {{Myers}}}, \bibinfo {author} {\bibfnamefont {J.~A.}\
  \bibnamefont {{Newman}}}, \bibinfo {author} {\bibfnamefont {J.}~\bibnamefont
  {{Nie}}}, \bibinfo {author} {\bibfnamefont {W.}~\bibnamefont {{Percival}}},
  \bibinfo {author} {\bibfnamefont {M.}~\bibnamefont {{Rezaie}}}, \bibinfo
  {author} {\bibfnamefont {G.}~\bibnamefont {{Rossi}}}, \bibinfo {author}
  {\bibfnamefont {E.}~\bibnamefont {{Sanchez}}}, \bibinfo {author}
  {\bibfnamefont {D.}~\bibnamefont {{Schlegel}}}, \bibinfo {author}
  {\bibfnamefont {M.}~\bibnamefont {{Schubnell}}}, \bibinfo {author}
  {\bibfnamefont {H.-J.}\ \bibnamefont {{Seo}}}, \bibinfo {author}
  {\bibfnamefont {G.}~\bibnamefont {{Tarl{\'e}}}},\ and\ \bibinfo {author}
  {\bibfnamefont {Z.}~\bibnamefont {{Zhou}}},\ }\bibfield  {title} {\bibinfo
  {title} {{DESI luminous red galaxy samples for cross-correlations}},\ }\href
  {https://doi.org/10.1088/1475-7516/2023/11/097} {\bibfield  {journal}
  {\bibinfo  {journal} {\jcap}\ }\textbf {\bibinfo {volume} {2023}},\ \bibinfo
  {eid} {097} (\bibinfo {year} {2023}{\natexlab{b}})},\ \Eprint
  {https://arxiv.org/abs/2309.06443} {arXiv:2309.06443 [astro-ph.CO]}
  \BibitemShut {NoStop}%
\bibitem [{\citenamefont {{Naess}}\ \emph {et~al.}(2020)\citenamefont
  {{Naess}}, \citenamefont {{Aiola}}, \citenamefont {{Austermann}},
  \citenamefont {{Battaglia}}, \citenamefont {{Beall}}, \citenamefont
  {{Becker}}, \citenamefont {{Bond}}, \citenamefont {{Calabrese}},
  \citenamefont {{Choi}}, \citenamefont {{Cothard}}, \citenamefont {{Crowley}},
  \citenamefont {{Darwish}}, \citenamefont {{Datta}}, \citenamefont
  {{Denison}}, \citenamefont {{Devlin}}, \citenamefont {{Duell}}, \citenamefont
  {{Duff}}, \citenamefont {{Duivenvoorden}}, \citenamefont {{Dunkley}},
  \citenamefont {{D{\"u}nner}}, \citenamefont {{Fox}}, \citenamefont
  {{Gallardo}}, \citenamefont {{Halpern}}, \citenamefont {{Han}}, \citenamefont
  {{Hasselfield}}, \citenamefont {{Hill}}, \citenamefont {{Hilton}},
  \citenamefont {{Hilton}}, \citenamefont {{Hincks}}, \citenamefont
  {{Hlo{\v{z}}ek}}, \citenamefont {{Ho}}, \citenamefont {{Hubmayr}},
  \citenamefont {{Huffenberger}}, \citenamefont {{Hughes}}, \citenamefont
  {{Kosowsky}}, \citenamefont {{Louis}}, \citenamefont {{Madhavacheril}},
  \citenamefont {{McMahon}}, \citenamefont {{Moodley}}, \citenamefont {{Nati}},
  \citenamefont {{Nibarger}}, \citenamefont {{Niemack}}, \citenamefont
  {{Page}}, \citenamefont {{Partridge}}, \citenamefont {{Salatino}},
  \citenamefont {{Schaan}}, \citenamefont {{Schillaci}}, \citenamefont
  {{Schmitt}}, \citenamefont {{Sherwin}}, \citenamefont {{Sehgal}},
  \citenamefont {{Sif{\'o}n}}, \citenamefont {{Spergel}}, \citenamefont
  {{Staggs}}, \citenamefont {{Stevens}}, \citenamefont {{Storer}},
  \citenamefont {{Ullom}}, \citenamefont {{Vale}}, \citenamefont {{Van
  Engelen}}, \citenamefont {{Van Lanen}}, \citenamefont {{Vavagiakis}},
  \citenamefont {{Wollack}},\ and\ \citenamefont {{Xu}}}]{2020JCAP...12..046N}%
  \BibitemOpen
  \bibfield  {author} {\bibinfo {author} {\bibfnamefont {S.}~\bibnamefont
  {{Naess}}}, \bibinfo {author} {\bibfnamefont {S.}~\bibnamefont {{Aiola}}},
  \bibinfo {author} {\bibfnamefont {J.~E.}\ \bibnamefont {{Austermann}}},
  \bibinfo {author} {\bibfnamefont {N.}~\bibnamefont {{Battaglia}}}, \bibinfo
  {author} {\bibfnamefont {J.~A.}\ \bibnamefont {{Beall}}}, \bibinfo {author}
  {\bibfnamefont {D.~T.}\ \bibnamefont {{Becker}}}, \bibinfo {author}
  {\bibfnamefont {R.~J.}\ \bibnamefont {{Bond}}}, \bibinfo {author}
  {\bibfnamefont {E.}~\bibnamefont {{Calabrese}}}, \bibinfo {author}
  {\bibfnamefont {S.~K.}\ \bibnamefont {{Choi}}}, \bibinfo {author}
  {\bibfnamefont {N.~F.}\ \bibnamefont {{Cothard}}}, \bibinfo {author}
  {\bibfnamefont {K.~T.}\ \bibnamefont {{Crowley}}}, \bibinfo {author}
  {\bibfnamefont {O.}~\bibnamefont {{Darwish}}}, \bibinfo {author}
  {\bibfnamefont {R.}~\bibnamefont {{Datta}}}, \bibinfo {author} {\bibfnamefont
  {E.~V.}\ \bibnamefont {{Denison}}}, \bibinfo {author} {\bibfnamefont
  {M.}~\bibnamefont {{Devlin}}}, \bibinfo {author} {\bibfnamefont {C.~J.}\
  \bibnamefont {{Duell}}}, \bibinfo {author} {\bibfnamefont {S.~M.}\
  \bibnamefont {{Duff}}}, \bibinfo {author} {\bibfnamefont {A.~J.}\
  \bibnamefont {{Duivenvoorden}}}, \bibinfo {author} {\bibfnamefont
  {J.}~\bibnamefont {{Dunkley}}}, \bibinfo {author} {\bibfnamefont
  {R.}~\bibnamefont {{D{\"u}nner}}}, \bibinfo {author} {\bibfnamefont {A.~E.}\
  \bibnamefont {{Fox}}}, \bibinfo {author} {\bibfnamefont {P.~A.}\ \bibnamefont
  {{Gallardo}}}, \bibinfo {author} {\bibfnamefont {M.}~\bibnamefont
  {{Halpern}}}, \bibinfo {author} {\bibfnamefont {D.}~\bibnamefont {{Han}}},
  \bibinfo {author} {\bibfnamefont {M.}~\bibnamefont {{Hasselfield}}}, \bibinfo
  {author} {\bibfnamefont {J.~C.}\ \bibnamefont {{Hill}}}, \bibinfo {author}
  {\bibfnamefont {G.~C.}\ \bibnamefont {{Hilton}}}, \bibinfo {author}
  {\bibfnamefont {M.}~\bibnamefont {{Hilton}}}, \bibinfo {author}
  {\bibfnamefont {A.~D.}\ \bibnamefont {{Hincks}}}, \bibinfo {author}
  {\bibfnamefont {R.}~\bibnamefont {{Hlo{\v{z}}ek}}}, \bibinfo {author}
  {\bibfnamefont {S.-P.~P.}\ \bibnamefont {{Ho}}}, \bibinfo {author}
  {\bibfnamefont {J.}~\bibnamefont {{Hubmayr}}}, \bibinfo {author}
  {\bibfnamefont {K.}~\bibnamefont {{Huffenberger}}}, \bibinfo {author}
  {\bibfnamefont {J.~P.}\ \bibnamefont {{Hughes}}}, \bibinfo {author}
  {\bibfnamefont {A.~B.}\ \bibnamefont {{Kosowsky}}}, \bibinfo {author}
  {\bibfnamefont {T.}~\bibnamefont {{Louis}}}, \bibinfo {author} {\bibfnamefont
  {M.~S.}\ \bibnamefont {{Madhavacheril}}}, \bibinfo {author} {\bibfnamefont
  {J.}~\bibnamefont {{McMahon}}}, \bibinfo {author} {\bibfnamefont
  {K.}~\bibnamefont {{Moodley}}}, \bibinfo {author} {\bibfnamefont
  {F.}~\bibnamefont {{Nati}}}, \bibinfo {author} {\bibfnamefont {J.~P.}\
  \bibnamefont {{Nibarger}}}, \bibinfo {author} {\bibfnamefont {M.~D.}\
  \bibnamefont {{Niemack}}}, \bibinfo {author} {\bibfnamefont {L.}~\bibnamefont
  {{Page}}}, \bibinfo {author} {\bibfnamefont {B.}~\bibnamefont {{Partridge}}},
  \bibinfo {author} {\bibfnamefont {M.}~\bibnamefont {{Salatino}}}, \bibinfo
  {author} {\bibfnamefont {E.}~\bibnamefont {{Schaan}}}, \bibinfo {author}
  {\bibfnamefont {A.}~\bibnamefont {{Schillaci}}}, \bibinfo {author}
  {\bibfnamefont {B.}~\bibnamefont {{Schmitt}}}, \bibinfo {author}
  {\bibfnamefont {B.~D.}\ \bibnamefont {{Sherwin}}}, \bibinfo {author}
  {\bibfnamefont {N.}~\bibnamefont {{Sehgal}}}, \bibinfo {author}
  {\bibfnamefont {C.}~\bibnamefont {{Sif{\'o}n}}}, \bibinfo {author}
  {\bibfnamefont {D.}~\bibnamefont {{Spergel}}}, \bibinfo {author}
  {\bibfnamefont {S.}~\bibnamefont {{Staggs}}}, \bibinfo {author}
  {\bibfnamefont {J.}~\bibnamefont {{Stevens}}}, \bibinfo {author}
  {\bibfnamefont {E.}~\bibnamefont {{Storer}}}, \bibinfo {author}
  {\bibfnamefont {J.~N.}\ \bibnamefont {{Ullom}}}, \bibinfo {author}
  {\bibfnamefont {L.~R.}\ \bibnamefont {{Vale}}}, \bibinfo {author}
  {\bibfnamefont {A.}~\bibnamefont {{Van Engelen}}}, \bibinfo {author}
  {\bibfnamefont {J.}~\bibnamefont {{Van Lanen}}}, \bibinfo {author}
  {\bibfnamefont {E.~M.}\ \bibnamefont {{Vavagiakis}}}, \bibinfo {author}
  {\bibfnamefont {E.~J.}\ \bibnamefont {{Wollack}}},\ and\ \bibinfo {author}
  {\bibfnamefont {Z.}~\bibnamefont {{Xu}}},\ }\bibfield  {title} {\bibinfo
  {title} {{The Atacama Cosmology Telescope: arcminute-resolution maps of 18
  000 square degrees of the microwave sky from ACT 2008-2018 data combined with
  Planck}},\ }\href {https://doi.org/10.1088/1475-7516/2020/12/046} {\bibfield
  {journal} {\bibinfo  {journal} {\jcap}\ }\textbf {\bibinfo {volume} {2020}},\
  \bibinfo {eid} {046} (\bibinfo {year} {2020})},\ \Eprint
  {https://arxiv.org/abs/2007.07290} {arXiv:2007.07290 [astro-ph.IM]}
  \BibitemShut {NoStop}%
\bibitem [{\citenamefont {{Qu}}\ \emph {et~al.}(2024)\citenamefont {{Qu}},
  \citenamefont {{Sherwin}}, \citenamefont {{Madhavacheril}}, \citenamefont
  {{Han}}, \citenamefont {{Crowley}}, \citenamefont {{Abril-Cabezas}},
  \citenamefont {{Ade}}, \citenamefont {{Aiola}}, \citenamefont {{Alford}},
  \citenamefont {{Amiri}}, \citenamefont {{Amodeo}}, \citenamefont {{An}},
  \citenamefont {{Atkins}}, \citenamefont {{Austermann}}, \citenamefont
  {{Battaglia}}, \citenamefont {{Battistelli}}, \citenamefont {{Beall}},
  \citenamefont {{Bean}}, \citenamefont {{Beringue}}, \citenamefont
  {{Bhandarkar}}, \citenamefont {{Biermann}}, \citenamefont {{Bolliet}},
  \citenamefont {{Bond}}, \citenamefont {{Cai}}, \citenamefont {{Calabrese}},
  \citenamefont {{Calafut}}, \citenamefont {{Capalbo}}, \citenamefont
  {{Carrero}}, \citenamefont {{Carron}}, \citenamefont {{Challinor}},
  \citenamefont {{Chesmore}}, \citenamefont {{Cho}}, \citenamefont {{Choi}},
  \citenamefont {{Clark}}, \citenamefont {{C{\'o}rdova Rosado}}, \citenamefont
  {{Cothard}}, \citenamefont {{Coughlin}}, \citenamefont {{Coulton}},
  \citenamefont {{Dalal}}, \citenamefont {{Darwish}}, \citenamefont {{Devlin}},
  \citenamefont {{Dicker}}, \citenamefont {{Doze}}, \citenamefont {{Duell}},
  \citenamefont {{Duff}}, \citenamefont {{Duivenvoorden}}, \citenamefont
  {{Dunkley}}, \citenamefont {{D{\"u}nner}}, \citenamefont {{Fanfani}},
  \citenamefont {{Fankhanel}}, \citenamefont {{Farren}}, \citenamefont
  {{Ferraro}}, \citenamefont {{Freundt}}, \citenamefont {{Fuzia}},
  \citenamefont {{Gallardo}}, \citenamefont {{Garrido}}, \citenamefont
  {{Gluscevic}}, \citenamefont {{Golec}}, \citenamefont {{Guan}}, \citenamefont
  {{Halpern}}, \citenamefont {{Harrison}}, \citenamefont {{Hasselfield}},
  \citenamefont {{Healy}}, \citenamefont {{Henderson}}, \citenamefont
  {{Hensley}}, \citenamefont {{Herv{\'\i}as-Caimapo}}, \citenamefont {{Hill}},
  \citenamefont {{Hilton}}, \citenamefont {{Hilton}}, \citenamefont {{Hincks}},
  \citenamefont {{Hlo{\v{z}}ek}}, \citenamefont {{Ho}}, \citenamefont
  {{Huber}}, \citenamefont {{Hubmayr}}, \citenamefont {{Huffenberger}},
  \citenamefont {{Hughes}}, \citenamefont {{Irwin}}, \citenamefont {{Isopi}},
  \citenamefont {{Jense}}, \citenamefont {{Keller}}, \citenamefont {{Kim}},
  \citenamefont {{Knowles}}, \citenamefont {{Koopman}}, \citenamefont
  {{Kosowsky}}, \citenamefont {{Kramer}}, \citenamefont {{Kusiak}},
  \citenamefont {{La Posta}}, \citenamefont {{Lague}}, \citenamefont {{Lakey}},
  \citenamefont {{Lee}}, \citenamefont {{Li}}, \citenamefont {{Li}},
  \citenamefont {{Limon}}, \citenamefont {{Lokken}}, \citenamefont {{Louis}},
  \citenamefont {{Lungu}}, \citenamefont {{MacCrann}}, \citenamefont
  {{MacInnis}}, \citenamefont {{Maldonado}}, \citenamefont {{Maldonado}},
  \citenamefont {{Mallaby-Kay}}, \citenamefont {{Marques}}, \citenamefont
  {{McMahon}}, \citenamefont {{Mehta}}, \citenamefont {{Menanteau}},
  \citenamefont {{Moodley}}, \citenamefont {{Morris}}, \citenamefont
  {{Mroczkowski}}, \citenamefont {{Naess}}, \citenamefont {{Namikawa}},
  \citenamefont {{Nati}}, \citenamefont {{Newburgh}}, \citenamefont {{Nicola}},
  \citenamefont {{Niemack}}, \citenamefont {{Nolta}}, \citenamefont
  {{Orlowski-Scherer}}, \citenamefont {{Page}}, \citenamefont {{Pandey}},
  \citenamefont {{Partridge}}, \citenamefont {{Prince}}, \citenamefont
  {{Puddu}}, \citenamefont {{Radiconi}}, \citenamefont {{Robertson}},
  \citenamefont {{Rojas}}, \citenamefont {{Sakuma}}, \citenamefont
  {{Salatino}}, \citenamefont {{Schaan}}, \citenamefont {{Schmitt}},
  \citenamefont {{Sehgal}}, \citenamefont {{Shaikh}}, \citenamefont {{Sierra}},
  \citenamefont {{Sievers}}, \citenamefont {{Sif{\'o}n}}, \citenamefont
  {{Simon}}, \citenamefont {{Sonka}}, \citenamefont {{Spergel}}, \citenamefont
  {{Staggs}}, \citenamefont {{Storer}}, \citenamefont {{Switzer}},
  \citenamefont {{Tampier}}, \citenamefont {{Thornton}}, \citenamefont
  {{Trac}}, \citenamefont {{Treu}}, \citenamefont {{Tucker}}, \citenamefont
  {{Ullom}}, \citenamefont {{Vale}}, \citenamefont {{Van Engelen}},
  \citenamefont {{Van Lanen}}, \citenamefont {{van Marrewijk}}, \citenamefont
  {{Vargas}}, \citenamefont {{Vavagiakis}}, \citenamefont {{Wagoner}},
  \citenamefont {{Wang}}, \citenamefont {{Wenzl}}, \citenamefont {{Wollack}},
  \citenamefont {{Xu}}, \citenamefont {{Zago}},\ and\ \citenamefont
  {{Zheng}}}]{2024ApJ...962..112Q}%
  \BibitemOpen
  \bibfield  {author} {\bibinfo {author} {\bibfnamefont {F.~J.}\ \bibnamefont
  {{Qu}}}, \bibinfo {author} {\bibfnamefont {B.~D.}\ \bibnamefont {{Sherwin}}},
  \bibinfo {author} {\bibfnamefont {M.~S.}\ \bibnamefont {{Madhavacheril}}},
  \bibinfo {author} {\bibfnamefont {D.}~\bibnamefont {{Han}}}, \bibinfo
  {author} {\bibfnamefont {K.~T.}\ \bibnamefont {{Crowley}}}, \bibinfo {author}
  {\bibfnamefont {I.}~\bibnamefont {{Abril-Cabezas}}}, \bibinfo {author}
  {\bibfnamefont {P.~A.~R.}\ \bibnamefont {{Ade}}}, \bibinfo {author}
  {\bibfnamefont {S.}~\bibnamefont {{Aiola}}}, \bibinfo {author} {\bibfnamefont
  {T.}~\bibnamefont {{Alford}}}, \bibinfo {author} {\bibfnamefont
  {M.}~\bibnamefont {{Amiri}}}, \bibinfo {author} {\bibfnamefont
  {S.}~\bibnamefont {{Amodeo}}}, \bibinfo {author} {\bibfnamefont
  {R.}~\bibnamefont {{An}}}, \bibinfo {author} {\bibfnamefont {Z.}~\bibnamefont
  {{Atkins}}}, \bibinfo {author} {\bibfnamefont {J.~E.}\ \bibnamefont
  {{Austermann}}}, \bibinfo {author} {\bibfnamefont {N.}~\bibnamefont
  {{Battaglia}}}, \bibinfo {author} {\bibfnamefont {E.~S.}\ \bibnamefont
  {{Battistelli}}}, \bibinfo {author} {\bibfnamefont {J.~A.}\ \bibnamefont
  {{Beall}}}, \bibinfo {author} {\bibfnamefont {R.}~\bibnamefont {{Bean}}},
  \bibinfo {author} {\bibfnamefont {B.}~\bibnamefont {{Beringue}}}, \bibinfo
  {author} {\bibfnamefont {T.}~\bibnamefont {{Bhandarkar}}}, \bibinfo {author}
  {\bibfnamefont {E.}~\bibnamefont {{Biermann}}}, \bibinfo {author}
  {\bibfnamefont {B.}~\bibnamefont {{Bolliet}}}, \bibinfo {author}
  {\bibfnamefont {J.~R.}\ \bibnamefont {{Bond}}}, \bibinfo {author}
  {\bibfnamefont {H.}~\bibnamefont {{Cai}}}, \bibinfo {author} {\bibfnamefont
  {E.}~\bibnamefont {{Calabrese}}}, \bibinfo {author} {\bibfnamefont
  {V.}~\bibnamefont {{Calafut}}}, \bibinfo {author} {\bibfnamefont
  {V.}~\bibnamefont {{Capalbo}}}, \bibinfo {author} {\bibfnamefont
  {F.}~\bibnamefont {{Carrero}}}, \bibinfo {author} {\bibfnamefont
  {J.}~\bibnamefont {{Carron}}}, \bibinfo {author} {\bibfnamefont
  {A.}~\bibnamefont {{Challinor}}}, \bibinfo {author} {\bibfnamefont {G.~E.}\
  \bibnamefont {{Chesmore}}}, \bibinfo {author} {\bibfnamefont {H.-m.}\
  \bibnamefont {{Cho}}}, \bibinfo {author} {\bibfnamefont {S.~K.}\ \bibnamefont
  {{Choi}}}, \bibinfo {author} {\bibfnamefont {S.~E.}\ \bibnamefont {{Clark}}},
  \bibinfo {author} {\bibfnamefont {R.}~\bibnamefont {{C{\'o}rdova Rosado}}},
  \bibinfo {author} {\bibfnamefont {N.~F.}\ \bibnamefont {{Cothard}}}, \bibinfo
  {author} {\bibfnamefont {K.}~\bibnamefont {{Coughlin}}}, \bibinfo {author}
  {\bibfnamefont {W.}~\bibnamefont {{Coulton}}}, \bibinfo {author}
  {\bibfnamefont {R.}~\bibnamefont {{Dalal}}}, \bibinfo {author} {\bibfnamefont
  {O.}~\bibnamefont {{Darwish}}}, \bibinfo {author} {\bibfnamefont {M.~J.}\
  \bibnamefont {{Devlin}}}, \bibinfo {author} {\bibfnamefont {S.}~\bibnamefont
  {{Dicker}}}, \bibinfo {author} {\bibfnamefont {P.}~\bibnamefont {{Doze}}},
  \bibinfo {author} {\bibfnamefont {C.~J.}\ \bibnamefont {{Duell}}}, \bibinfo
  {author} {\bibfnamefont {S.~M.}\ \bibnamefont {{Duff}}}, \bibinfo {author}
  {\bibfnamefont {A.~J.}\ \bibnamefont {{Duivenvoorden}}}, \bibinfo {author}
  {\bibfnamefont {J.}~\bibnamefont {{Dunkley}}}, \bibinfo {author}
  {\bibfnamefont {R.}~\bibnamefont {{D{\"u}nner}}}, \bibinfo {author}
  {\bibfnamefont {V.}~\bibnamefont {{Fanfani}}}, \bibinfo {author}
  {\bibfnamefont {M.}~\bibnamefont {{Fankhanel}}}, \bibinfo {author}
  {\bibfnamefont {G.}~\bibnamefont {{Farren}}}, \bibinfo {author}
  {\bibfnamefont {S.}~\bibnamefont {{Ferraro}}}, \bibinfo {author}
  {\bibfnamefont {R.}~\bibnamefont {{Freundt}}}, \bibinfo {author}
  {\bibfnamefont {B.}~\bibnamefont {{Fuzia}}}, \bibinfo {author} {\bibfnamefont
  {P.~A.}\ \bibnamefont {{Gallardo}}}, \bibinfo {author} {\bibfnamefont
  {X.}~\bibnamefont {{Garrido}}}, \bibinfo {author} {\bibfnamefont
  {V.}~\bibnamefont {{Gluscevic}}}, \bibinfo {author} {\bibfnamefont {J.~E.}\
  \bibnamefont {{Golec}}}, \bibinfo {author} {\bibfnamefont {Y.}~\bibnamefont
  {{Guan}}}, \bibinfo {author} {\bibfnamefont {M.}~\bibnamefont {{Halpern}}},
  \bibinfo {author} {\bibfnamefont {I.}~\bibnamefont {{Harrison}}}, \bibinfo
  {author} {\bibfnamefont {M.}~\bibnamefont {{Hasselfield}}}, \bibinfo {author}
  {\bibfnamefont {E.}~\bibnamefont {{Healy}}}, \bibinfo {author} {\bibfnamefont
  {S.}~\bibnamefont {{Henderson}}}, \bibinfo {author} {\bibfnamefont
  {B.}~\bibnamefont {{Hensley}}}, \bibinfo {author} {\bibfnamefont
  {C.}~\bibnamefont {{Herv{\'\i}as-Caimapo}}}, \bibinfo {author} {\bibfnamefont
  {J.~C.}\ \bibnamefont {{Hill}}}, \bibinfo {author} {\bibfnamefont {G.~C.}\
  \bibnamefont {{Hilton}}}, \bibinfo {author} {\bibfnamefont {M.}~\bibnamefont
  {{Hilton}}}, \bibinfo {author} {\bibfnamefont {A.~D.}\ \bibnamefont
  {{Hincks}}}, \bibinfo {author} {\bibfnamefont {R.}~\bibnamefont
  {{Hlo{\v{z}}ek}}}, \bibinfo {author} {\bibfnamefont {S.-P.~P.}\ \bibnamefont
  {{Ho}}}, \bibinfo {author} {\bibfnamefont {Z.~B.}\ \bibnamefont {{Huber}}},
  \bibinfo {author} {\bibfnamefont {J.}~\bibnamefont {{Hubmayr}}}, \bibinfo
  {author} {\bibfnamefont {K.~M.}\ \bibnamefont {{Huffenberger}}}, \bibinfo
  {author} {\bibfnamefont {J.~P.}\ \bibnamefont {{Hughes}}}, \bibinfo {author}
  {\bibfnamefont {K.}~\bibnamefont {{Irwin}}}, \bibinfo {author} {\bibfnamefont
  {G.}~\bibnamefont {{Isopi}}}, \bibinfo {author} {\bibfnamefont {H.~T.}\
  \bibnamefont {{Jense}}}, \bibinfo {author} {\bibfnamefont {B.}~\bibnamefont
  {{Keller}}}, \bibinfo {author} {\bibfnamefont {J.}~\bibnamefont {{Kim}}},
  \bibinfo {author} {\bibfnamefont {K.}~\bibnamefont {{Knowles}}}, \bibinfo
  {author} {\bibfnamefont {B.~J.}\ \bibnamefont {{Koopman}}}, \bibinfo {author}
  {\bibfnamefont {A.}~\bibnamefont {{Kosowsky}}}, \bibinfo {author}
  {\bibfnamefont {D.}~\bibnamefont {{Kramer}}}, \bibinfo {author}
  {\bibfnamefont {A.}~\bibnamefont {{Kusiak}}}, \bibinfo {author}
  {\bibfnamefont {A.}~\bibnamefont {{La Posta}}}, \bibinfo {author}
  {\bibfnamefont {A.}~\bibnamefont {{Lague}}}, \bibinfo {author} {\bibfnamefont
  {V.}~\bibnamefont {{Lakey}}}, \bibinfo {author} {\bibfnamefont
  {E.}~\bibnamefont {{Lee}}}, \bibinfo {author} {\bibfnamefont
  {Z.}~\bibnamefont {{Li}}}, \bibinfo {author} {\bibfnamefont {Y.}~\bibnamefont
  {{Li}}}, \bibinfo {author} {\bibfnamefont {M.}~\bibnamefont {{Limon}}},
  \bibinfo {author} {\bibfnamefont {M.}~\bibnamefont {{Lokken}}}, \bibinfo
  {author} {\bibfnamefont {T.}~\bibnamefont {{Louis}}}, \bibinfo {author}
  {\bibfnamefont {M.}~\bibnamefont {{Lungu}}}, \bibinfo {author} {\bibfnamefont
  {N.}~\bibnamefont {{MacCrann}}}, \bibinfo {author} {\bibfnamefont
  {A.}~\bibnamefont {{MacInnis}}}, \bibinfo {author} {\bibfnamefont
  {D.}~\bibnamefont {{Maldonado}}}, \bibinfo {author} {\bibfnamefont
  {F.}~\bibnamefont {{Maldonado}}}, \bibinfo {author} {\bibfnamefont
  {M.}~\bibnamefont {{Mallaby-Kay}}}, \bibinfo {author} {\bibfnamefont {G.~A.}\
  \bibnamefont {{Marques}}}, \bibinfo {author} {\bibfnamefont {J.}~\bibnamefont
  {{McMahon}}}, \bibinfo {author} {\bibfnamefont {Y.}~\bibnamefont {{Mehta}}},
  \bibinfo {author} {\bibfnamefont {F.}~\bibnamefont {{Menanteau}}}, \bibinfo
  {author} {\bibfnamefont {K.}~\bibnamefont {{Moodley}}}, \bibinfo {author}
  {\bibfnamefont {T.~W.}\ \bibnamefont {{Morris}}}, \bibinfo {author}
  {\bibfnamefont {T.}~\bibnamefont {{Mroczkowski}}}, \bibinfo {author}
  {\bibfnamefont {S.}~\bibnamefont {{Naess}}}, \bibinfo {author} {\bibfnamefont
  {T.}~\bibnamefont {{Namikawa}}}, \bibinfo {author} {\bibfnamefont
  {F.}~\bibnamefont {{Nati}}}, \bibinfo {author} {\bibfnamefont
  {L.}~\bibnamefont {{Newburgh}}}, \bibinfo {author} {\bibfnamefont
  {A.}~\bibnamefont {{Nicola}}}, \bibinfo {author} {\bibfnamefont {M.~D.}\
  \bibnamefont {{Niemack}}}, \bibinfo {author} {\bibfnamefont {M.~R.}\
  \bibnamefont {{Nolta}}}, \bibinfo {author} {\bibfnamefont {J.}~\bibnamefont
  {{Orlowski-Scherer}}}, \bibinfo {author} {\bibfnamefont {L.~A.}\ \bibnamefont
  {{Page}}}, \bibinfo {author} {\bibfnamefont {S.}~\bibnamefont {{Pandey}}},
  \bibinfo {author} {\bibfnamefont {B.}~\bibnamefont {{Partridge}}}, \bibinfo
  {author} {\bibfnamefont {H.}~\bibnamefont {{Prince}}}, \bibinfo {author}
  {\bibfnamefont {R.}~\bibnamefont {{Puddu}}}, \bibinfo {author} {\bibfnamefont
  {F.}~\bibnamefont {{Radiconi}}}, \bibinfo {author} {\bibfnamefont
  {N.}~\bibnamefont {{Robertson}}}, \bibinfo {author} {\bibfnamefont
  {F.}~\bibnamefont {{Rojas}}}, \bibinfo {author} {\bibfnamefont
  {T.}~\bibnamefont {{Sakuma}}}, \bibinfo {author} {\bibfnamefont
  {M.}~\bibnamefont {{Salatino}}}, \bibinfo {author} {\bibfnamefont
  {E.}~\bibnamefont {{Schaan}}}, \bibinfo {author} {\bibfnamefont {B.~L.}\
  \bibnamefont {{Schmitt}}}, \bibinfo {author} {\bibfnamefont {N.}~\bibnamefont
  {{Sehgal}}}, \bibinfo {author} {\bibfnamefont {S.}~\bibnamefont {{Shaikh}}},
  \bibinfo {author} {\bibfnamefont {C.}~\bibnamefont {{Sierra}}}, \bibinfo
  {author} {\bibfnamefont {J.}~\bibnamefont {{Sievers}}}, \bibinfo {author}
  {\bibfnamefont {C.}~\bibnamefont {{Sif{\'o}n}}}, \bibinfo {author}
  {\bibfnamefont {S.}~\bibnamefont {{Simon}}}, \bibinfo {author} {\bibfnamefont
  {R.}~\bibnamefont {{Sonka}}}, \bibinfo {author} {\bibfnamefont {D.~N.}\
  \bibnamefont {{Spergel}}}, \bibinfo {author} {\bibfnamefont {S.~T.}\
  \bibnamefont {{Staggs}}}, \bibinfo {author} {\bibfnamefont {E.}~\bibnamefont
  {{Storer}}}, \bibinfo {author} {\bibfnamefont {E.~R.}\ \bibnamefont
  {{Switzer}}}, \bibinfo {author} {\bibfnamefont {N.}~\bibnamefont
  {{Tampier}}}, \bibinfo {author} {\bibfnamefont {R.}~\bibnamefont
  {{Thornton}}}, \bibinfo {author} {\bibfnamefont {H.}~\bibnamefont {{Trac}}},
  \bibinfo {author} {\bibfnamefont {J.}~\bibnamefont {{Treu}}}, \bibinfo
  {author} {\bibfnamefont {C.}~\bibnamefont {{Tucker}}}, \bibinfo {author}
  {\bibfnamefont {J.}~\bibnamefont {{Ullom}}}, \bibinfo {author} {\bibfnamefont
  {L.~R.}\ \bibnamefont {{Vale}}}, \bibinfo {author} {\bibfnamefont
  {A.}~\bibnamefont {{Van Engelen}}}, \bibinfo {author} {\bibfnamefont
  {J.}~\bibnamefont {{Van Lanen}}}, \bibinfo {author} {\bibfnamefont
  {J.}~\bibnamefont {{van Marrewijk}}}, \bibinfo {author} {\bibfnamefont
  {C.}~\bibnamefont {{Vargas}}}, \bibinfo {author} {\bibfnamefont {E.~M.}\
  \bibnamefont {{Vavagiakis}}}, \bibinfo {author} {\bibfnamefont
  {K.}~\bibnamefont {{Wagoner}}}, \bibinfo {author} {\bibfnamefont
  {Y.}~\bibnamefont {{Wang}}}, \bibinfo {author} {\bibfnamefont
  {L.}~\bibnamefont {{Wenzl}}}, \bibinfo {author} {\bibfnamefont {E.~J.}\
  \bibnamefont {{Wollack}}}, \bibinfo {author} {\bibfnamefont {Z.}~\bibnamefont
  {{Xu}}}, \bibinfo {author} {\bibfnamefont {F.}~\bibnamefont {{Zago}}},\ and\
  \bibinfo {author} {\bibfnamefont {K.}~\bibnamefont {{Zheng}}},\ }\bibfield
  {title} {\bibinfo {title} {{The Atacama Cosmology Telescope: A Measurement of
  the DR6 CMB Lensing Power Spectrum and Its Implications for Structure
  Growth}},\ }\href {https://doi.org/10.3847/1538-4357/acfe06} {\bibfield
  {journal} {\bibinfo  {journal} {\apj}\ }\textbf {\bibinfo {volume} {962}},\
  \bibinfo {eid} {112} (\bibinfo {year} {2024})},\ \Eprint
  {https://arxiv.org/abs/2304.05202} {arXiv:2304.05202 [astro-ph.CO]}
  \BibitemShut {NoStop}%
\bibitem [{\citenamefont {{Hadzhiyska}}\ \emph {et~al.}(2023)\citenamefont
  {{Hadzhiyska}}, \citenamefont {{Ferraro}}, \citenamefont {{Pakmor}},
  \citenamefont {{Bose}}, \citenamefont {{Delgado}}, \citenamefont
  {{Hern{\'a}ndez-Aguayo}}, \citenamefont {{Kannan}}, \citenamefont
  {{Springel}}, \citenamefont {{White}},\ and\ \citenamefont
  {{Hernquist}}}]{2023MNRAS.526..369H}%
  \BibitemOpen
  \bibfield  {author} {\bibinfo {author} {\bibfnamefont {B.}~\bibnamefont
  {{Hadzhiyska}}}, \bibinfo {author} {\bibfnamefont {S.}~\bibnamefont
  {{Ferraro}}}, \bibinfo {author} {\bibfnamefont {R.}~\bibnamefont {{Pakmor}}},
  \bibinfo {author} {\bibfnamefont {S.}~\bibnamefont {{Bose}}}, \bibinfo
  {author} {\bibfnamefont {A.~M.}\ \bibnamefont {{Delgado}}}, \bibinfo {author}
  {\bibfnamefont {C.}~\bibnamefont {{Hern{\'a}ndez-Aguayo}}}, \bibinfo {author}
  {\bibfnamefont {R.}~\bibnamefont {{Kannan}}}, \bibinfo {author}
  {\bibfnamefont {V.}~\bibnamefont {{Springel}}}, \bibinfo {author}
  {\bibfnamefont {S.~D.~M.}\ \bibnamefont {{White}}},\ and\ \bibinfo {author}
  {\bibfnamefont {L.}~\bibnamefont {{Hernquist}}},\ }\bibfield  {title}
  {\bibinfo {title} {{Interpreting Sunyaev-Zel'dovich observations with
  MillenniumTNG: mass and environment scaling relations}},\ }\href
  {https://doi.org/10.1093/mnras/stad2751} {\bibfield  {journal} {\bibinfo
  {journal} {\mnras}\ }\textbf {\bibinfo {volume} {526}},\ \bibinfo {pages}
  {369} (\bibinfo {year} {2023})},\ \Eprint {https://arxiv.org/abs/2305.00992}
  {arXiv:2305.00992 [astro-ph.CO]} \BibitemShut {NoStop}%
\bibitem [{\citenamefont {{White}}(2015)}]{2015MNRAS.450.3822W}%
  \BibitemOpen
  \bibfield  {author} {\bibinfo {author} {\bibfnamefont {M.}~\bibnamefont
  {{White}}},\ }\bibfield  {title} {\bibinfo {title} {{Reconstruction within
  the Zeldovich approximation}},\ }\href {https://doi.org/10.1093/mnras/stv842}
  {\bibfield  {journal} {\bibinfo  {journal} {\mnras}\ }\textbf {\bibinfo
  {volume} {450}},\ \bibinfo {pages} {3822} (\bibinfo {year} {2015})},\ \Eprint
  {https://arxiv.org/abs/1504.03677} {arXiv:1504.03677 [astro-ph.CO]}
  \BibitemShut {NoStop}%
\bibitem [{\citenamefont {{Hadzhiyska}}\ \emph
  {et~al.}(2024{\natexlab{b}})\citenamefont {{Hadzhiyska}}, \citenamefont
  {{Ferraro}}, \citenamefont {{Ried Guachalla}},\ and\ \citenamefont
  {{Schaan}}}]{2024PhRvD.109j3534H}%
  \BibitemOpen
  \bibfield  {author} {\bibinfo {author} {\bibfnamefont {B.}~\bibnamefont
  {{Hadzhiyska}}}, \bibinfo {author} {\bibfnamefont {S.}~\bibnamefont
  {{Ferraro}}}, \bibinfo {author} {\bibfnamefont {B.}~\bibnamefont {{Ried
  Guachalla}}},\ and\ \bibinfo {author} {\bibfnamefont {E.}~\bibnamefont
  {{Schaan}}},\ }\bibfield  {title} {\bibinfo {title} {{Velocity reconstruction
  in the era of DESI and Rubin/LSST. II. Realistic samples on the light
  cone}},\ }\href {https://doi.org/10.1103/PhysRevD.109.103534} {\bibfield
  {journal} {\bibinfo  {journal} {\prd}\ }\textbf {\bibinfo {volume} {109}},\
  \bibinfo {eid} {103534} (\bibinfo {year} {2024}{\natexlab{b}})},\ \Eprint
  {https://arxiv.org/abs/2312.12434} {arXiv:2312.12434 [astro-ph.CO]}
  \BibitemShut {NoStop}%
\bibitem [{\citenamefont {{Coulton}}\ \emph {et~al.}(2024)\citenamefont
  {{Coulton}}, \citenamefont {{Schutt}}, \citenamefont {{Maniyar}},
  \citenamefont {{Schaan}}, \citenamefont {{An}}, \citenamefont {{Atkins}},
  \citenamefont {{Battaglia}}, \citenamefont {{Bond}}, \citenamefont
  {{Calabrese}}, \citenamefont {{Choi}}, \citenamefont {{Devlin}},
  \citenamefont {{Duivenvoorden}}, \citenamefont {{Dunkley}}, \citenamefont
  {{Ferraro}}, \citenamefont {{Gluscevic}}, \citenamefont {{Hill}},
  \citenamefont {{Hilton}}, \citenamefont {{Hincks}}, \citenamefont
  {{Kosowsky}}, \citenamefont {{Kramer}}, \citenamefont {{Kusiak}},
  \citenamefont {{La Posta}}, \citenamefont {{Louis}}, \citenamefont
  {{Madhavacheril}}, \citenamefont {{Marques}}, \citenamefont {{McCarthy}},
  \citenamefont {{McMahon}}, \citenamefont {{Moodley}}, \citenamefont
  {{Naess}}, \citenamefont {{Page}}, \citenamefont {{Partridge}}, \citenamefont
  {{Qu}}, \citenamefont {{Sehgal}}, \citenamefont {{Sherwin}}, \citenamefont
  {{Sif{\'o}n}}, \citenamefont {{Spergel}}, \citenamefont {{Staggs}},
  \citenamefont {{Van Engelen}}, \citenamefont {{Vargas}},\ and\ \citenamefont
  {{Wollack}}}]{2024arXiv240113033C}%
  \BibitemOpen
  \bibfield  {author} {\bibinfo {author} {\bibfnamefont {W.~R.}\ \bibnamefont
  {{Coulton}}}, \bibinfo {author} {\bibfnamefont {T.}~\bibnamefont {{Schutt}}},
  \bibinfo {author} {\bibfnamefont {A.~S.}\ \bibnamefont {{Maniyar}}}, \bibinfo
  {author} {\bibfnamefont {E.}~\bibnamefont {{Schaan}}}, \bibinfo {author}
  {\bibfnamefont {R.}~\bibnamefont {{An}}}, \bibinfo {author} {\bibfnamefont
  {Z.}~\bibnamefont {{Atkins}}}, \bibinfo {author} {\bibfnamefont
  {N.}~\bibnamefont {{Battaglia}}}, \bibinfo {author} {\bibfnamefont {J.~R.}\
  \bibnamefont {{Bond}}}, \bibinfo {author} {\bibfnamefont {E.}~\bibnamefont
  {{Calabrese}}}, \bibinfo {author} {\bibfnamefont {S.~K.}\ \bibnamefont
  {{Choi}}}, \bibinfo {author} {\bibfnamefont {M.~J.}\ \bibnamefont
  {{Devlin}}}, \bibinfo {author} {\bibfnamefont {A.~J.}\ \bibnamefont
  {{Duivenvoorden}}}, \bibinfo {author} {\bibfnamefont {J.}~\bibnamefont
  {{Dunkley}}}, \bibinfo {author} {\bibfnamefont {S.}~\bibnamefont
  {{Ferraro}}}, \bibinfo {author} {\bibfnamefont {V.}~\bibnamefont
  {{Gluscevic}}}, \bibinfo {author} {\bibfnamefont {J.~C.}\ \bibnamefont
  {{Hill}}}, \bibinfo {author} {\bibfnamefont {M.}~\bibnamefont {{Hilton}}},
  \bibinfo {author} {\bibfnamefont {A.~D.}\ \bibnamefont {{Hincks}}}, \bibinfo
  {author} {\bibfnamefont {A.}~\bibnamefont {{Kosowsky}}}, \bibinfo {author}
  {\bibfnamefont {D.}~\bibnamefont {{Kramer}}}, \bibinfo {author}
  {\bibfnamefont {A.}~\bibnamefont {{Kusiak}}}, \bibinfo {author}
  {\bibfnamefont {A.}~\bibnamefont {{La Posta}}}, \bibinfo {author}
  {\bibfnamefont {T.}~\bibnamefont {{Louis}}}, \bibinfo {author} {\bibfnamefont
  {M.~S.}\ \bibnamefont {{Madhavacheril}}}, \bibinfo {author} {\bibfnamefont
  {G.~A.}\ \bibnamefont {{Marques}}}, \bibinfo {author} {\bibfnamefont
  {F.}~\bibnamefont {{McCarthy}}}, \bibinfo {author} {\bibfnamefont
  {J.}~\bibnamefont {{McMahon}}}, \bibinfo {author} {\bibfnamefont
  {K.}~\bibnamefont {{Moodley}}}, \bibinfo {author} {\bibfnamefont
  {S.}~\bibnamefont {{Naess}}}, \bibinfo {author} {\bibfnamefont {L.~A.}\
  \bibnamefont {{Page}}}, \bibinfo {author} {\bibfnamefont {B.}~\bibnamefont
  {{Partridge}}}, \bibinfo {author} {\bibfnamefont {F.~J.}\ \bibnamefont
  {{Qu}}}, \bibinfo {author} {\bibfnamefont {N.}~\bibnamefont {{Sehgal}}},
  \bibinfo {author} {\bibfnamefont {B.~D.}\ \bibnamefont {{Sherwin}}}, \bibinfo
  {author} {\bibfnamefont {C.}~\bibnamefont {{Sif{\'o}n}}}, \bibinfo {author}
  {\bibfnamefont {D.~N.}\ \bibnamefont {{Spergel}}}, \bibinfo {author}
  {\bibfnamefont {S.~T.}\ \bibnamefont {{Staggs}}}, \bibinfo {author}
  {\bibfnamefont {A.}~\bibnamefont {{Van Engelen}}}, \bibinfo {author}
  {\bibfnamefont {C.}~\bibnamefont {{Vargas}}},\ and\ \bibinfo {author}
  {\bibfnamefont {E.~J.}\ \bibnamefont {{Wollack}}},\ }\bibfield  {title}
  {\bibinfo {title} {{The Atacama Cosmology Telescope: Detection of Patchy
  Screening of the Cosmic Microwave Background}},\ }\href
  {https://doi.org/10.48550/arXiv.2401.13033} {\bibfield  {journal} {\bibinfo
  {journal} {arXiv e-prints}\ ,\ \bibinfo {eid} {arXiv:2401.13033}} (\bibinfo
  {year} {2024})},\ \Eprint {https://arxiv.org/abs/2401.13033}
  {arXiv:2401.13033 [astro-ph.CO]} \BibitemShut {NoStop}%
\bibitem [{\citenamefont {{Nelson}}\ \emph {et~al.}(2019)\citenamefont
  {{Nelson}}, \citenamefont {{Springel}}, \citenamefont {{Pillepich}},
  \citenamefont {{Rodriguez-Gomez}}, \citenamefont {{Torrey}}, \citenamefont
  {{Genel}}, \citenamefont {{Vogelsberger}}, \citenamefont {{Pakmor}},
  \citenamefont {{Marinacci}}, \citenamefont {{Weinberger}}, \citenamefont
  {{Kelley}}, \citenamefont {{Lovell}}, \citenamefont {{Diemer}},\ and\
  \citenamefont {{Hernquist}}}]{2019ComAC...6....2N}%
  \BibitemOpen
  \bibfield  {author} {\bibinfo {author} {\bibfnamefont {D.}~\bibnamefont
  {{Nelson}}}, \bibinfo {author} {\bibfnamefont {V.}~\bibnamefont
  {{Springel}}}, \bibinfo {author} {\bibfnamefont {A.}~\bibnamefont
  {{Pillepich}}}, \bibinfo {author} {\bibfnamefont {V.}~\bibnamefont
  {{Rodriguez-Gomez}}}, \bibinfo {author} {\bibfnamefont {P.}~\bibnamefont
  {{Torrey}}}, \bibinfo {author} {\bibfnamefont {S.}~\bibnamefont {{Genel}}},
  \bibinfo {author} {\bibfnamefont {M.}~\bibnamefont {{Vogelsberger}}},
  \bibinfo {author} {\bibfnamefont {R.}~\bibnamefont {{Pakmor}}}, \bibinfo
  {author} {\bibfnamefont {F.}~\bibnamefont {{Marinacci}}}, \bibinfo {author}
  {\bibfnamefont {R.}~\bibnamefont {{Weinberger}}}, \bibinfo {author}
  {\bibfnamefont {L.}~\bibnamefont {{Kelley}}}, \bibinfo {author}
  {\bibfnamefont {M.}~\bibnamefont {{Lovell}}}, \bibinfo {author}
  {\bibfnamefont {B.}~\bibnamefont {{Diemer}}},\ and\ \bibinfo {author}
  {\bibfnamefont {L.}~\bibnamefont {{Hernquist}}},\ }\bibfield  {title}
  {\bibinfo {title} {{The IllustrisTNG simulations: public data release}},\
  }\href {https://doi.org/10.1186/s40668-019-0028-x} {\bibfield  {journal}
  {\bibinfo  {journal} {Computational Astrophysics and Cosmology}\ }\textbf
  {\bibinfo {volume} {6}},\ \bibinfo {eid} {2} (\bibinfo {year} {2019})},\
  \Eprint {https://arxiv.org/abs/1812.05609} {arXiv:1812.05609 [astro-ph.GA]}
  \BibitemShut {NoStop}%
\bibitem [{\citenamefont {{Nelson}}\ \emph {et~al.}(2015)\citenamefont
  {{Nelson}}, \citenamefont {{Pillepich}}, \citenamefont {{Genel}},
  \citenamefont {{Vogelsberger}}, \citenamefont {{Springel}}, \citenamefont
  {{Torrey}}, \citenamefont {{Rodriguez-Gomez}}, \citenamefont {{Sijacki}},
  \citenamefont {{Snyder}}, \citenamefont {{Griffen}}, \citenamefont
  {{Marinacci}}, \citenamefont {{Blecha}}, \citenamefont {{Sales}},
  \citenamefont {{Xu}},\ and\ \citenamefont
  {{Hernquist}}}]{2015A&C....13...12N}%
  \BibitemOpen
  \bibfield  {author} {\bibinfo {author} {\bibfnamefont {D.}~\bibnamefont
  {{Nelson}}}, \bibinfo {author} {\bibfnamefont {A.}~\bibnamefont
  {{Pillepich}}}, \bibinfo {author} {\bibfnamefont {S.}~\bibnamefont
  {{Genel}}}, \bibinfo {author} {\bibfnamefont {M.}~\bibnamefont
  {{Vogelsberger}}}, \bibinfo {author} {\bibfnamefont {V.}~\bibnamefont
  {{Springel}}}, \bibinfo {author} {\bibfnamefont {P.}~\bibnamefont
  {{Torrey}}}, \bibinfo {author} {\bibfnamefont {V.}~\bibnamefont
  {{Rodriguez-Gomez}}}, \bibinfo {author} {\bibfnamefont {D.}~\bibnamefont
  {{Sijacki}}}, \bibinfo {author} {\bibfnamefont {G.~F.}\ \bibnamefont
  {{Snyder}}}, \bibinfo {author} {\bibfnamefont {B.}~\bibnamefont {{Griffen}}},
  \bibinfo {author} {\bibfnamefont {F.}~\bibnamefont {{Marinacci}}}, \bibinfo
  {author} {\bibfnamefont {L.}~\bibnamefont {{Blecha}}}, \bibinfo {author}
  {\bibfnamefont {L.}~\bibnamefont {{Sales}}}, \bibinfo {author} {\bibfnamefont
  {D.}~\bibnamefont {{Xu}}},\ and\ \bibinfo {author} {\bibfnamefont
  {L.}~\bibnamefont {{Hernquist}}},\ }\bibfield  {title} {\bibinfo {title}
  {{The illustris simulation: Public data release}},\ }\href
  {https://doi.org/10.1016/j.ascom.2015.09.003} {\bibfield  {journal} {\bibinfo
   {journal} {Astronomy and Computing}\ }\textbf {\bibinfo {volume} {13}},\
  \bibinfo {pages} {12} (\bibinfo {year} {2015})},\ \Eprint
  {https://arxiv.org/abs/1504.00362} {arXiv:1504.00362 [astro-ph.CO]}
  \BibitemShut {NoStop}%
\bibitem [{\citenamefont {{Alonso}}\ \emph {et~al.}(2016)\citenamefont
  {{Alonso}}, \citenamefont {{Hadzhiyska}},\ and\ \citenamefont
  {{Strauss}}}]{2016MNRAS.460..256A}%
  \BibitemOpen
  \bibfield  {author} {\bibinfo {author} {\bibfnamefont {D.}~\bibnamefont
  {{Alonso}}}, \bibinfo {author} {\bibfnamefont {B.}~\bibnamefont
  {{Hadzhiyska}}},\ and\ \bibinfo {author} {\bibfnamefont {M.~A.}\ \bibnamefont
  {{Strauss}}},\ }\bibfield  {title} {\bibinfo {title} {{Recovering the tidal
  field in the projected galaxy distribution}},\ }\href
  {https://doi.org/10.1093/mnras/stw919} {\bibfield  {journal} {\bibinfo
  {journal} {\mnras}\ }\textbf {\bibinfo {volume} {460}},\ \bibinfo {pages}
  {256} (\bibinfo {year} {2016})},\ \Eprint {https://arxiv.org/abs/1512.03402}
  {arXiv:1512.03402 [astro-ph.CO]} \BibitemShut {NoStop}%
\bibitem [{\citenamefont {{G{\'o}rski}}\ \emph {et~al.}(2005)\citenamefont
  {{G{\'o}rski}}, \citenamefont {{Hivon}}, \citenamefont {{Banday}},
  \citenamefont {{Wandelt}}, \citenamefont {{Hansen}}, \citenamefont
  {{Reinecke}},\ and\ \citenamefont {{Bartelmann}}}]{2005ApJ...622..759G}%
  \BibitemOpen
  \bibfield  {author} {\bibinfo {author} {\bibfnamefont {K.~M.}\ \bibnamefont
  {{G{\'o}rski}}}, \bibinfo {author} {\bibfnamefont {E.}~\bibnamefont
  {{Hivon}}}, \bibinfo {author} {\bibfnamefont {A.~J.}\ \bibnamefont
  {{Banday}}}, \bibinfo {author} {\bibfnamefont {B.~D.}\ \bibnamefont
  {{Wandelt}}}, \bibinfo {author} {\bibfnamefont {F.~K.}\ \bibnamefont
  {{Hansen}}}, \bibinfo {author} {\bibfnamefont {M.}~\bibnamefont
  {{Reinecke}}},\ and\ \bibinfo {author} {\bibfnamefont {M.}~\bibnamefont
  {{Bartelmann}}},\ }\bibfield  {title} {\bibinfo {title} {{HEALPix: A
  Framework for High-Resolution Discretization and Fast Analysis of Data
  Distributed on the Sphere}},\ }\href {https://doi.org/10.1086/427976}
  {\bibfield  {journal} {\bibinfo  {journal} {\apj}\ }\textbf {\bibinfo
  {volume} {622}},\ \bibinfo {pages} {759} (\bibinfo {year} {2005})},\ \Eprint
  {https://arxiv.org/abs/astro-ph/0409513} {arXiv:astro-ph/0409513 [astro-ph]}
  \BibitemShut {NoStop}%
\bibitem [{\citenamefont {{Ried Guachalla}}\ \emph {et~al.}(2024)\citenamefont
  {{Ried Guachalla}}, \citenamefont {{Schaan}}, \citenamefont {{Hadzhiyska}},\
  and\ \citenamefont {{Ferraro}}}]{2024PhRvD.109j3533R}%
  \BibitemOpen
  \bibfield  {author} {\bibinfo {author} {\bibfnamefont {B.}~\bibnamefont
  {{Ried Guachalla}}}, \bibinfo {author} {\bibfnamefont {E.}~\bibnamefont
  {{Schaan}}}, \bibinfo {author} {\bibfnamefont {B.}~\bibnamefont
  {{Hadzhiyska}}},\ and\ \bibinfo {author} {\bibfnamefont {S.}~\bibnamefont
  {{Ferraro}}},\ }\bibfield  {title} {\bibinfo {title} {{Velocity
  reconstruction in the era of DESI and Rubin/LSST. I. Exploring spectroscopic,
  photometric, and hybrid samples}},\ }\href
  {https://doi.org/10.1103/PhysRevD.109.103533} {\bibfield  {journal} {\bibinfo
   {journal} {\prd}\ }\textbf {\bibinfo {volume} {109}},\ \bibinfo {eid}
  {103533} (\bibinfo {year} {2024})},\ \Eprint
  {https://arxiv.org/abs/2312.12435} {arXiv:2312.12435 [astro-ph.CO]}
  \BibitemShut {NoStop}%
\bibitem [{\citenamefont {{Wechsler}}\ and\ \citenamefont
  {{Tinker}}(2018)}]{2018ARA&A..56..435W}%
  \BibitemOpen
  \bibfield  {author} {\bibinfo {author} {\bibfnamefont {R.~H.}\ \bibnamefont
  {{Wechsler}}}\ and\ \bibinfo {author} {\bibfnamefont {J.~L.}\ \bibnamefont
  {{Tinker}}},\ }\bibfield  {title} {\bibinfo {title} {{The Connection Between
  Galaxies and Their Dark Matter Halos}},\ }\href
  {https://doi.org/10.1146/annurev-astro-081817-051756} {\bibfield  {journal}
  {\bibinfo  {journal} {\araa}\ }\textbf {\bibinfo {volume} {56}},\ \bibinfo
  {pages} {435} (\bibinfo {year} {2018})},\ \Eprint
  {https://arxiv.org/abs/1804.03097} {arXiv:1804.03097 [astro-ph.GA]}
  \BibitemShut {NoStop}%
\bibitem [{\citenamefont {{Lee}}\ and\ \citenamefont
  {{Erdogdu}}(2007)}]{2007ApJ...671.1248L}%
  \BibitemOpen
  \bibfield  {author} {\bibinfo {author} {\bibfnamefont {J.}~\bibnamefont
  {{Lee}}}\ and\ \bibinfo {author} {\bibfnamefont {P.}~\bibnamefont
  {{Erdogdu}}},\ }\bibfield  {title} {\bibinfo {title} {{The Alignments of the
  Galaxy Spins with the Real-Space Tidal Field Reconstructed from the 2MASS
  Redshift Survey}},\ }\href {https://doi.org/10.1086/523351} {\bibfield
  {journal} {\bibinfo  {journal} {\apj}\ }\textbf {\bibinfo {volume} {671}},\
  \bibinfo {pages} {1248} (\bibinfo {year} {2007})},\ \Eprint
  {https://arxiv.org/abs/0706.1412} {arXiv:0706.1412 [astro-ph]} \BibitemShut
  {NoStop}%
\bibitem [{\citenamefont {{Tempel}}\ \emph {et~al.}(2013)\citenamefont
  {{Tempel}}, \citenamefont {{Stoica}},\ and\ \citenamefont
  {{Saar}}}]{2013MNRAS.428.1827T}%
  \BibitemOpen
  \bibfield  {author} {\bibinfo {author} {\bibfnamefont {E.}~\bibnamefont
  {{Tempel}}}, \bibinfo {author} {\bibfnamefont {R.~S.}\ \bibnamefont
  {{Stoica}}},\ and\ \bibinfo {author} {\bibfnamefont {E.}~\bibnamefont
  {{Saar}}},\ }\bibfield  {title} {\bibinfo {title} {{Evidence for spin
  alignment of spiral and elliptical/S0 galaxies in filaments}},\ }\href
  {https://doi.org/10.1093/mnras/sts162} {\bibfield  {journal} {\bibinfo
  {journal} {\mnras}\ }\textbf {\bibinfo {volume} {428}},\ \bibinfo {pages}
  {1827} (\bibinfo {year} {2013})},\ \Eprint {https://arxiv.org/abs/1207.0068}
  {arXiv:1207.0068 [astro-ph.CO]} \BibitemShut {NoStop}%
\bibitem [{\citenamefont {{Codis}}\ \emph {et~al.}(2015)\citenamefont
  {{Codis}}, \citenamefont {{Gavazzi}}, \citenamefont {{Dubois}}, \citenamefont
  {{Pichon}}, \citenamefont {{Benabed}}, \citenamefont {{Desjacques}},
  \citenamefont {{Pogosyan}}, \citenamefont {{Devriendt}},\ and\ \citenamefont
  {{Slyz}}}]{2015MNRAS.448.3391C}%
  \BibitemOpen
  \bibfield  {author} {\bibinfo {author} {\bibfnamefont {S.}~\bibnamefont
  {{Codis}}}, \bibinfo {author} {\bibfnamefont {R.}~\bibnamefont {{Gavazzi}}},
  \bibinfo {author} {\bibfnamefont {Y.}~\bibnamefont {{Dubois}}}, \bibinfo
  {author} {\bibfnamefont {C.}~\bibnamefont {{Pichon}}}, \bibinfo {author}
  {\bibfnamefont {K.}~\bibnamefont {{Benabed}}}, \bibinfo {author}
  {\bibfnamefont {V.}~\bibnamefont {{Desjacques}}}, \bibinfo {author}
  {\bibfnamefont {D.}~\bibnamefont {{Pogosyan}}}, \bibinfo {author}
  {\bibfnamefont {J.}~\bibnamefont {{Devriendt}}},\ and\ \bibinfo {author}
  {\bibfnamefont {A.}~\bibnamefont {{Slyz}}},\ }\bibfield  {title} {\bibinfo
  {title} {{Intrinsic alignment of simulated galaxies in the cosmic web:
  implications for weak lensing surveys}},\ }\href
  {https://doi.org/10.1093/mnras/stv231} {\bibfield  {journal} {\bibinfo
  {journal} {\mnras}\ }\textbf {\bibinfo {volume} {448}},\ \bibinfo {pages}
  {3391} (\bibinfo {year} {2015})},\ \Eprint {https://arxiv.org/abs/1406.4668}
  {arXiv:1406.4668 [astro-ph.CO]} \BibitemShut {NoStop}%
\bibitem [{\citenamefont {{Wang}}\ \emph {et~al.}(2018)\citenamefont {{Wang}},
  \citenamefont {{Guo}}, \citenamefont {{Kang}},\ and\ \citenamefont
  {{Libeskind}}}]{2018ApJ...866..138W}%
  \BibitemOpen
  \bibfield  {author} {\bibinfo {author} {\bibfnamefont {P.}~\bibnamefont
  {{Wang}}}, \bibinfo {author} {\bibfnamefont {Q.}~\bibnamefont {{Guo}}},
  \bibinfo {author} {\bibfnamefont {X.}~\bibnamefont {{Kang}}},\ and\ \bibinfo
  {author} {\bibfnamefont {N.~I.}\ \bibnamefont {{Libeskind}}},\ }\bibfield
  {title} {\bibinfo {title} {{The Spin Alignment of Galaxies with the
  Large-scale Tidal Field in Hydrodynamic Simulations}},\ }\href
  {https://doi.org/10.3847/1538-4357/aae20f} {\bibfield  {journal} {\bibinfo
  {journal} {\apj}\ }\textbf {\bibinfo {volume} {866}},\ \bibinfo {eid} {138}
  (\bibinfo {year} {2018})},\ \Eprint {https://arxiv.org/abs/1810.04581}
  {arXiv:1810.04581 [astro-ph.GA]} \BibitemShut {NoStop}%
\bibitem [{\citenamefont {{Abdalla}}\ \emph {et~al.}(2022)\citenamefont
  {{Abdalla}}, \citenamefont {{Abell{\'a}n}}, \citenamefont {{Aboubrahim}},
  \citenamefont {{Agnello}}, \citenamefont {{Akarsu}}, \citenamefont
  {{Akrami}}, \citenamefont {{Alestas}}, \citenamefont {{Aloni}}, \citenamefont
  {{Amendola}}, \citenamefont {{Anchordoqui}}, \citenamefont {{Anderson}},
  \citenamefont {{Arendse}}, \citenamefont {{Asgari}}, \citenamefont
  {{Ballardini}}, \citenamefont {{Barger}}, \citenamefont {{Basilakos}},
  \citenamefont {{Batista}}, \citenamefont {{Battistelli}}, \citenamefont
  {{Battye}}, \citenamefont {{Benetti}}, \citenamefont {{Benisty}},
  \citenamefont {{Berlin}}, \citenamefont {{de Bernardis}}, \citenamefont
  {{Berti}}, \citenamefont {{Bidenko}}, \citenamefont {{Birrer}}, \citenamefont
  {{Blakeslee}}, \citenamefont {{Boddy}}, \citenamefont {{Bom}}, \citenamefont
  {{Bonilla}}, \citenamefont {{Borghi}}, \citenamefont {{Bouchet}},
  \citenamefont {{Braglia}}, \citenamefont {{Buchert}}, \citenamefont
  {{Buckley-Geer}}, \citenamefont {{Calabrese}}, \citenamefont {{Caldwell}},
  \citenamefont {{Camarena}}, \citenamefont {{Capozziello}}, \citenamefont
  {{Casertano}}, \citenamefont {{Chen}}, \citenamefont {{Chluba}},
  \citenamefont {{Chen}}, \citenamefont {{Chen}}, \citenamefont {{Chudaykin}},
  \citenamefont {{Cicoli}}, \citenamefont {{Copi}}, \citenamefont {{Courbin}},
  \citenamefont {{Cyr-Racine}}, \citenamefont {{Czerny}}, \citenamefont
  {{Dainotti}}, \citenamefont {{D'Amico}}, \citenamefont {{Davis}},
  \citenamefont {{de Cruz P{\'e}rez}}, \citenamefont {{de Haro}}, \citenamefont
  {{Delabrouille}}, \citenamefont {{Denton}}, \citenamefont {{Dhawan}},
  \citenamefont {{Dienes}}, \citenamefont {{Di Valentino}}, \citenamefont
  {{Du}}, \citenamefont {{Eckert}}, \citenamefont {{Escamilla-Rivera}},
  \citenamefont {{Fert{\'e}}}, \citenamefont {{Finelli}}, \citenamefont
  {{Fosalba}}, \citenamefont {{Freedman}}, \citenamefont {{Frusciante}},
  \citenamefont {{Gazta{\~n}aga}}, \citenamefont {{Giar{\`e}}}, \citenamefont
  {{Giusarma}}, \citenamefont {{G{\'o}mez-Valent}}, \citenamefont {{Handley}},
  \citenamefont {{Harrison}}, \citenamefont {{Hart}}, \citenamefont {{Hazra}},
  \citenamefont {{Heavens}}, \citenamefont {{Heinesen}}, \citenamefont
  {{Hildebrandt}}, \citenamefont {{Hill}}, \citenamefont {{Hogg}},
  \citenamefont {{Holz}}, \citenamefont {{Hooper}}, \citenamefont
  {{Hosseininejad}}, \citenamefont {{Huterer}}, \citenamefont {{Ishak}},
  \citenamefont {{Ivanov}}, \citenamefont {{Jaffe}}, \citenamefont {{Jang}},
  \citenamefont {{Jedamzik}}, \citenamefont {{Jimenez}}, \citenamefont
  {{Joseph}}, \citenamefont {{Joudaki}}, \citenamefont {{Kamionkowski}},
  \citenamefont {{Karwal}}, \citenamefont {{Kazantzidis}}, \citenamefont
  {{Keeley}}, \citenamefont {{Klasen}}, \citenamefont {{Komatsu}},
  \citenamefont {{Koopmans}}, \citenamefont {{Kumar}}, \citenamefont
  {{Lamagna}}, \citenamefont {{Lazkoz}}, \citenamefont {{Lee}}, \citenamefont
  {{Lesgourgues}}, \citenamefont {{Levi Said}}, \citenamefont {{Lewis}},
  \citenamefont {{L'Huillier}}, \citenamefont {{Lucca}}, \citenamefont
  {{Maartens}}, \citenamefont {{Macri}}, \citenamefont {{Marfatia}},
  \citenamefont {{Marra}}, \citenamefont {{Martins}}, \citenamefont {{Masi}},
  \citenamefont {{Matarrese}}, \citenamefont {{Mazumdar}}, \citenamefont
  {{Melchiorri}}, \citenamefont {{Mena}}, \citenamefont {{Mersini-Houghton}},
  \citenamefont {{Mertens}}, \citenamefont {{Milakovi{\'c}}}, \citenamefont
  {{Minami}}, \citenamefont {{Miranda}}, \citenamefont {{Moreno-Pulido}},
  \citenamefont {{Moresco}}, \citenamefont {{Mota}}, \citenamefont {{Mottola}},
  \citenamefont {{Mozzon}}, \citenamefont {{Muir}}, \citenamefont
  {{Mukherjee}}, \citenamefont {{Mukherjee}}, \citenamefont {{Naselsky}},
  \citenamefont {{Nath}}, \citenamefont {{Nesseris}}, \citenamefont
  {{Niedermann}}, \citenamefont {{Notari}}, \citenamefont {{Nunes}},
  \citenamefont {{{\'O} Colg{\'a}in}}, \citenamefont {{Owens}}, \citenamefont
  {{{\"O}z{\"u}lker}}, \citenamefont {{Pace}}, \citenamefont {{Paliathanasis}},
  \citenamefont {{Palmese}}, \citenamefont {{Pan}}, \citenamefont {{Paoletti}},
  \citenamefont {{Perez Bergliaffa}}, \citenamefont {{Perivolaropoulos}},
  \citenamefont {{Pesce}}, \citenamefont {{Pettorino}}, \citenamefont
  {{Philcox}}, \citenamefont {{Pogosian}}, \citenamefont {{Poulin}},
  \citenamefont {{Poulot}}, \citenamefont {{Raveri}}, \citenamefont {{Reid}},
  \citenamefont {{Renzi}}, \citenamefont {{Riess}}, \citenamefont {{Sabla}},
  \citenamefont {{Salucci}}, \citenamefont {{Salzano}}, \citenamefont
  {{Saridakis}}, \citenamefont {{Sathyaprakash}}, \citenamefont {{Schmaltz}},
  \citenamefont {{Sch{\"o}neberg}}, \citenamefont {{Scolnic}}, \citenamefont
  {{Sen}}, \citenamefont {{Sehgal}}, \citenamefont {{Shafieloo}}, \citenamefont
  {{Sheikh-Jabbari}}, \citenamefont {{Silk}}, \citenamefont {{Silvestri}},
  \citenamefont {{Skara}}, \citenamefont {{Sloth}}, \citenamefont
  {{Soares-Santos}}, \citenamefont {{Sol{\`a} Peracaula}}, \citenamefont
  {{Songsheng}}, \citenamefont {{Soriano}}, \citenamefont {{Staicova}},
  \citenamefont {{Starkman}}, \citenamefont {{Szapudi}}, \citenamefont
  {{Teixeira}}, \citenamefont {{Thomas}}, \citenamefont {{Treu}}, \citenamefont
  {{Trott}}, \citenamefont {{van de Bruck}}, \citenamefont {{Vazquez}},
  \citenamefont {{Verde}}, \citenamefont {{Visinelli}}, \citenamefont {{Wang}},
  \citenamefont {{Wang}}, \citenamefont {{Wang}}, \citenamefont {{Watkins}},
  \citenamefont {{Watson}}, \citenamefont {{Webb}}, \citenamefont {{Weiner}},
  \citenamefont {{Weltman}}, \citenamefont {{Witte}}, \citenamefont {{Wojtak}},
  \citenamefont {{Yadav}}, \citenamefont {{Yang}}, \citenamefont {{Zhao}},\
  and\ \citenamefont {{Zumalac{\'a}rregui}}}]{2022JHEAp..34...49A}%
  \BibitemOpen
  \bibfield  {author} {\bibinfo {author} {\bibfnamefont {E.}~\bibnamefont
  {{Abdalla}}}, \bibinfo {author} {\bibfnamefont {G.~F.}\ \bibnamefont
  {{Abell{\'a}n}}}, \bibinfo {author} {\bibfnamefont {A.}~\bibnamefont
  {{Aboubrahim}}}, \bibinfo {author} {\bibfnamefont {A.}~\bibnamefont
  {{Agnello}}}, \bibinfo {author} {\bibfnamefont {{\"O}.}~\bibnamefont
  {{Akarsu}}}, \bibinfo {author} {\bibfnamefont {Y.}~\bibnamefont {{Akrami}}},
  \bibinfo {author} {\bibfnamefont {G.}~\bibnamefont {{Alestas}}}, \bibinfo
  {author} {\bibfnamefont {D.}~\bibnamefont {{Aloni}}}, \bibinfo {author}
  {\bibfnamefont {L.}~\bibnamefont {{Amendola}}}, \bibinfo {author}
  {\bibfnamefont {L.~A.}\ \bibnamefont {{Anchordoqui}}}, \bibinfo {author}
  {\bibfnamefont {R.~I.}\ \bibnamefont {{Anderson}}}, \bibinfo {author}
  {\bibfnamefont {N.}~\bibnamefont {{Arendse}}}, \bibinfo {author}
  {\bibfnamefont {M.}~\bibnamefont {{Asgari}}}, \bibinfo {author}
  {\bibfnamefont {M.}~\bibnamefont {{Ballardini}}}, \bibinfo {author}
  {\bibfnamefont {V.}~\bibnamefont {{Barger}}}, \bibinfo {author}
  {\bibfnamefont {S.}~\bibnamefont {{Basilakos}}}, \bibinfo {author}
  {\bibfnamefont {R.~C.}\ \bibnamefont {{Batista}}}, \bibinfo {author}
  {\bibfnamefont {E.~S.}\ \bibnamefont {{Battistelli}}}, \bibinfo {author}
  {\bibfnamefont {R.}~\bibnamefont {{Battye}}}, \bibinfo {author}
  {\bibfnamefont {M.}~\bibnamefont {{Benetti}}}, \bibinfo {author}
  {\bibfnamefont {D.}~\bibnamefont {{Benisty}}}, \bibinfo {author}
  {\bibfnamefont {A.}~\bibnamefont {{Berlin}}}, \bibinfo {author}
  {\bibfnamefont {P.}~\bibnamefont {{de Bernardis}}}, \bibinfo {author}
  {\bibfnamefont {E.}~\bibnamefont {{Berti}}}, \bibinfo {author} {\bibfnamefont
  {B.}~\bibnamefont {{Bidenko}}}, \bibinfo {author} {\bibfnamefont
  {S.}~\bibnamefont {{Birrer}}}, \bibinfo {author} {\bibfnamefont {J.~P.}\
  \bibnamefont {{Blakeslee}}}, \bibinfo {author} {\bibfnamefont {K.~K.}\
  \bibnamefont {{Boddy}}}, \bibinfo {author} {\bibfnamefont {C.~R.}\
  \bibnamefont {{Bom}}}, \bibinfo {author} {\bibfnamefont {A.}~\bibnamefont
  {{Bonilla}}}, \bibinfo {author} {\bibfnamefont {N.}~\bibnamefont {{Borghi}}},
  \bibinfo {author} {\bibfnamefont {F.~R.}\ \bibnamefont {{Bouchet}}}, \bibinfo
  {author} {\bibfnamefont {M.}~\bibnamefont {{Braglia}}}, \bibinfo {author}
  {\bibfnamefont {T.}~\bibnamefont {{Buchert}}}, \bibinfo {author}
  {\bibfnamefont {E.}~\bibnamefont {{Buckley-Geer}}}, \bibinfo {author}
  {\bibfnamefont {E.}~\bibnamefont {{Calabrese}}}, \bibinfo {author}
  {\bibfnamefont {R.~R.}\ \bibnamefont {{Caldwell}}}, \bibinfo {author}
  {\bibfnamefont {D.}~\bibnamefont {{Camarena}}}, \bibinfo {author}
  {\bibfnamefont {S.}~\bibnamefont {{Capozziello}}}, \bibinfo {author}
  {\bibfnamefont {S.}~\bibnamefont {{Casertano}}}, \bibinfo {author}
  {\bibfnamefont {G.~C.~F.}\ \bibnamefont {{Chen}}}, \bibinfo {author}
  {\bibfnamefont {J.}~\bibnamefont {{Chluba}}}, \bibinfo {author}
  {\bibfnamefont {A.}~\bibnamefont {{Chen}}}, \bibinfo {author} {\bibfnamefont
  {H.-Y.}\ \bibnamefont {{Chen}}}, \bibinfo {author} {\bibfnamefont
  {A.}~\bibnamefont {{Chudaykin}}}, \bibinfo {author} {\bibfnamefont
  {M.}~\bibnamefont {{Cicoli}}}, \bibinfo {author} {\bibfnamefont {C.~J.}\
  \bibnamefont {{Copi}}}, \bibinfo {author} {\bibfnamefont {F.}~\bibnamefont
  {{Courbin}}}, \bibinfo {author} {\bibfnamefont {F.-Y.}\ \bibnamefont
  {{Cyr-Racine}}}, \bibinfo {author} {\bibfnamefont {B.}~\bibnamefont
  {{Czerny}}}, \bibinfo {author} {\bibfnamefont {M.}~\bibnamefont
  {{Dainotti}}}, \bibinfo {author} {\bibfnamefont {G.}~\bibnamefont
  {{D'Amico}}}, \bibinfo {author} {\bibfnamefont {A.-C.}\ \bibnamefont
  {{Davis}}}, \bibinfo {author} {\bibfnamefont {J.}~\bibnamefont {{de Cruz
  P{\'e}rez}}}, \bibinfo {author} {\bibfnamefont {J.}~\bibnamefont {{de
  Haro}}}, \bibinfo {author} {\bibfnamefont {J.}~\bibnamefont
  {{Delabrouille}}}, \bibinfo {author} {\bibfnamefont {P.~B.}\ \bibnamefont
  {{Denton}}}, \bibinfo {author} {\bibfnamefont {S.}~\bibnamefont {{Dhawan}}},
  \bibinfo {author} {\bibfnamefont {K.~R.}\ \bibnamefont {{Dienes}}}, \bibinfo
  {author} {\bibfnamefont {E.}~\bibnamefont {{Di Valentino}}}, \bibinfo
  {author} {\bibfnamefont {P.}~\bibnamefont {{Du}}}, \bibinfo {author}
  {\bibfnamefont {D.}~\bibnamefont {{Eckert}}}, \bibinfo {author}
  {\bibfnamefont {C.}~\bibnamefont {{Escamilla-Rivera}}}, \bibinfo {author}
  {\bibfnamefont {A.}~\bibnamefont {{Fert{\'e}}}}, \bibinfo {author}
  {\bibfnamefont {F.}~\bibnamefont {{Finelli}}}, \bibinfo {author}
  {\bibfnamefont {P.}~\bibnamefont {{Fosalba}}}, \bibinfo {author}
  {\bibfnamefont {W.~L.}\ \bibnamefont {{Freedman}}}, \bibinfo {author}
  {\bibfnamefont {N.}~\bibnamefont {{Frusciante}}}, \bibinfo {author}
  {\bibfnamefont {E.}~\bibnamefont {{Gazta{\~n}aga}}}, \bibinfo {author}
  {\bibfnamefont {W.}~\bibnamefont {{Giar{\`e}}}}, \bibinfo {author}
  {\bibfnamefont {E.}~\bibnamefont {{Giusarma}}}, \bibinfo {author}
  {\bibfnamefont {A.}~\bibnamefont {{G{\'o}mez-Valent}}}, \bibinfo {author}
  {\bibfnamefont {W.}~\bibnamefont {{Handley}}}, \bibinfo {author}
  {\bibfnamefont {I.}~\bibnamefont {{Harrison}}}, \bibinfo {author}
  {\bibfnamefont {L.}~\bibnamefont {{Hart}}}, \bibinfo {author} {\bibfnamefont
  {D.~K.}\ \bibnamefont {{Hazra}}}, \bibinfo {author} {\bibfnamefont
  {A.}~\bibnamefont {{Heavens}}}, \bibinfo {author} {\bibfnamefont
  {A.}~\bibnamefont {{Heinesen}}}, \bibinfo {author} {\bibfnamefont
  {H.}~\bibnamefont {{Hildebrandt}}}, \bibinfo {author} {\bibfnamefont {J.~C.}\
  \bibnamefont {{Hill}}}, \bibinfo {author} {\bibfnamefont {N.~B.}\
  \bibnamefont {{Hogg}}}, \bibinfo {author} {\bibfnamefont {D.~E.}\
  \bibnamefont {{Holz}}}, \bibinfo {author} {\bibfnamefont {D.~C.}\
  \bibnamefont {{Hooper}}}, \bibinfo {author} {\bibfnamefont {N.}~\bibnamefont
  {{Hosseininejad}}}, \bibinfo {author} {\bibfnamefont {D.}~\bibnamefont
  {{Huterer}}}, \bibinfo {author} {\bibfnamefont {M.}~\bibnamefont {{Ishak}}},
  \bibinfo {author} {\bibfnamefont {M.~M.}\ \bibnamefont {{Ivanov}}}, \bibinfo
  {author} {\bibfnamefont {A.~H.}\ \bibnamefont {{Jaffe}}}, \bibinfo {author}
  {\bibfnamefont {I.~S.}\ \bibnamefont {{Jang}}}, \bibinfo {author}
  {\bibfnamefont {K.}~\bibnamefont {{Jedamzik}}}, \bibinfo {author}
  {\bibfnamefont {R.}~\bibnamefont {{Jimenez}}}, \bibinfo {author}
  {\bibfnamefont {M.}~\bibnamefont {{Joseph}}}, \bibinfo {author}
  {\bibfnamefont {S.}~\bibnamefont {{Joudaki}}}, \bibinfo {author}
  {\bibfnamefont {M.}~\bibnamefont {{Kamionkowski}}}, \bibinfo {author}
  {\bibfnamefont {T.}~\bibnamefont {{Karwal}}}, \bibinfo {author}
  {\bibfnamefont {L.}~\bibnamefont {{Kazantzidis}}}, \bibinfo {author}
  {\bibfnamefont {R.~E.}\ \bibnamefont {{Keeley}}}, \bibinfo {author}
  {\bibfnamefont {M.}~\bibnamefont {{Klasen}}}, \bibinfo {author}
  {\bibfnamefont {E.}~\bibnamefont {{Komatsu}}}, \bibinfo {author}
  {\bibfnamefont {L.~V.~E.}\ \bibnamefont {{Koopmans}}}, \bibinfo {author}
  {\bibfnamefont {S.}~\bibnamefont {{Kumar}}}, \bibinfo {author} {\bibfnamefont
  {L.}~\bibnamefont {{Lamagna}}}, \bibinfo {author} {\bibfnamefont
  {R.}~\bibnamefont {{Lazkoz}}}, \bibinfo {author} {\bibfnamefont {C.-C.}\
  \bibnamefont {{Lee}}}, \bibinfo {author} {\bibfnamefont {J.}~\bibnamefont
  {{Lesgourgues}}}, \bibinfo {author} {\bibfnamefont {J.}~\bibnamefont {{Levi
  Said}}}, \bibinfo {author} {\bibfnamefont {T.~R.}\ \bibnamefont {{Lewis}}},
  \bibinfo {author} {\bibfnamefont {B.}~\bibnamefont {{L'Huillier}}}, \bibinfo
  {author} {\bibfnamefont {M.}~\bibnamefont {{Lucca}}}, \bibinfo {author}
  {\bibfnamefont {R.}~\bibnamefont {{Maartens}}}, \bibinfo {author}
  {\bibfnamefont {L.~M.}\ \bibnamefont {{Macri}}}, \bibinfo {author}
  {\bibfnamefont {D.}~\bibnamefont {{Marfatia}}}, \bibinfo {author}
  {\bibfnamefont {V.}~\bibnamefont {{Marra}}}, \bibinfo {author} {\bibfnamefont
  {C.~J.~A.~P.}\ \bibnamefont {{Martins}}}, \bibinfo {author} {\bibfnamefont
  {S.}~\bibnamefont {{Masi}}}, \bibinfo {author} {\bibfnamefont
  {S.}~\bibnamefont {{Matarrese}}}, \bibinfo {author} {\bibfnamefont
  {A.}~\bibnamefont {{Mazumdar}}}, \bibinfo {author} {\bibfnamefont
  {A.}~\bibnamefont {{Melchiorri}}}, \bibinfo {author} {\bibfnamefont
  {O.}~\bibnamefont {{Mena}}}, \bibinfo {author} {\bibfnamefont
  {L.}~\bibnamefont {{Mersini-Houghton}}}, \bibinfo {author} {\bibfnamefont
  {J.}~\bibnamefont {{Mertens}}}, \bibinfo {author} {\bibfnamefont
  {D.}~\bibnamefont {{Milakovi{\'c}}}}, \bibinfo {author} {\bibfnamefont
  {Y.}~\bibnamefont {{Minami}}}, \bibinfo {author} {\bibfnamefont
  {V.}~\bibnamefont {{Miranda}}}, \bibinfo {author} {\bibfnamefont
  {C.}~\bibnamefont {{Moreno-Pulido}}}, \bibinfo {author} {\bibfnamefont
  {M.}~\bibnamefont {{Moresco}}}, \bibinfo {author} {\bibfnamefont {D.~F.}\
  \bibnamefont {{Mota}}}, \bibinfo {author} {\bibfnamefont {E.}~\bibnamefont
  {{Mottola}}}, \bibinfo {author} {\bibfnamefont {S.}~\bibnamefont {{Mozzon}}},
  \bibinfo {author} {\bibfnamefont {J.}~\bibnamefont {{Muir}}}, \bibinfo
  {author} {\bibfnamefont {A.}~\bibnamefont {{Mukherjee}}}, \bibinfo {author}
  {\bibfnamefont {S.}~\bibnamefont {{Mukherjee}}}, \bibinfo {author}
  {\bibfnamefont {P.}~\bibnamefont {{Naselsky}}}, \bibinfo {author}
  {\bibfnamefont {P.}~\bibnamefont {{Nath}}}, \bibinfo {author} {\bibfnamefont
  {S.}~\bibnamefont {{Nesseris}}}, \bibinfo {author} {\bibfnamefont
  {F.}~\bibnamefont {{Niedermann}}}, \bibinfo {author} {\bibfnamefont
  {A.}~\bibnamefont {{Notari}}}, \bibinfo {author} {\bibfnamefont {R.~C.}\
  \bibnamefont {{Nunes}}}, \bibinfo {author} {\bibfnamefont {E.}~\bibnamefont
  {{{\'O} Colg{\'a}in}}}, \bibinfo {author} {\bibfnamefont {K.~A.}\
  \bibnamefont {{Owens}}}, \bibinfo {author} {\bibfnamefont {E.}~\bibnamefont
  {{{\"O}z{\"u}lker}}}, \bibinfo {author} {\bibfnamefont {F.}~\bibnamefont
  {{Pace}}}, \bibinfo {author} {\bibfnamefont {A.}~\bibnamefont
  {{Paliathanasis}}}, \bibinfo {author} {\bibfnamefont {A.}~\bibnamefont
  {{Palmese}}}, \bibinfo {author} {\bibfnamefont {S.}~\bibnamefont {{Pan}}},
  \bibinfo {author} {\bibfnamefont {D.}~\bibnamefont {{Paoletti}}}, \bibinfo
  {author} {\bibfnamefont {S.~E.}\ \bibnamefont {{Perez Bergliaffa}}}, \bibinfo
  {author} {\bibfnamefont {L.}~\bibnamefont {{Perivolaropoulos}}}, \bibinfo
  {author} {\bibfnamefont {D.~W.}\ \bibnamefont {{Pesce}}}, \bibinfo {author}
  {\bibfnamefont {V.}~\bibnamefont {{Pettorino}}}, \bibinfo {author}
  {\bibfnamefont {O.~H.~E.}\ \bibnamefont {{Philcox}}}, \bibinfo {author}
  {\bibfnamefont {L.}~\bibnamefont {{Pogosian}}}, \bibinfo {author}
  {\bibfnamefont {V.}~\bibnamefont {{Poulin}}}, \bibinfo {author}
  {\bibfnamefont {G.}~\bibnamefont {{Poulot}}}, \bibinfo {author}
  {\bibfnamefont {M.}~\bibnamefont {{Raveri}}}, \bibinfo {author}
  {\bibfnamefont {M.~J.}\ \bibnamefont {{Reid}}}, \bibinfo {author}
  {\bibfnamefont {F.}~\bibnamefont {{Renzi}}}, \bibinfo {author} {\bibfnamefont
  {A.~G.}\ \bibnamefont {{Riess}}}, \bibinfo {author} {\bibfnamefont {V.~I.}\
  \bibnamefont {{Sabla}}}, \bibinfo {author} {\bibfnamefont {P.}~\bibnamefont
  {{Salucci}}}, \bibinfo {author} {\bibfnamefont {V.}~\bibnamefont
  {{Salzano}}}, \bibinfo {author} {\bibfnamefont {E.~N.}\ \bibnamefont
  {{Saridakis}}}, \bibinfo {author} {\bibfnamefont {B.~S.}\ \bibnamefont
  {{Sathyaprakash}}}, \bibinfo {author} {\bibfnamefont {M.}~\bibnamefont
  {{Schmaltz}}}, \bibinfo {author} {\bibfnamefont {N.}~\bibnamefont
  {{Sch{\"o}neberg}}}, \bibinfo {author} {\bibfnamefont {D.}~\bibnamefont
  {{Scolnic}}}, \bibinfo {author} {\bibfnamefont {A.~A.}\ \bibnamefont
  {{Sen}}}, \bibinfo {author} {\bibfnamefont {N.}~\bibnamefont {{Sehgal}}},
  \bibinfo {author} {\bibfnamefont {A.}~\bibnamefont {{Shafieloo}}}, \bibinfo
  {author} {\bibfnamefont {M.~M.}\ \bibnamefont {{Sheikh-Jabbari}}}, \bibinfo
  {author} {\bibfnamefont {J.}~\bibnamefont {{Silk}}}, \bibinfo {author}
  {\bibfnamefont {A.}~\bibnamefont {{Silvestri}}}, \bibinfo {author}
  {\bibfnamefont {F.}~\bibnamefont {{Skara}}}, \bibinfo {author} {\bibfnamefont
  {M.~S.}\ \bibnamefont {{Sloth}}}, \bibinfo {author} {\bibfnamefont
  {M.}~\bibnamefont {{Soares-Santos}}}, \bibinfo {author} {\bibfnamefont
  {J.}~\bibnamefont {{Sol{\`a} Peracaula}}}, \bibinfo {author} {\bibfnamefont
  {Y.-Y.}\ \bibnamefont {{Songsheng}}}, \bibinfo {author} {\bibfnamefont
  {J.~F.}\ \bibnamefont {{Soriano}}}, \bibinfo {author} {\bibfnamefont
  {D.}~\bibnamefont {{Staicova}}}, \bibinfo {author} {\bibfnamefont {G.~D.}\
  \bibnamefont {{Starkman}}}, \bibinfo {author} {\bibfnamefont
  {I.}~\bibnamefont {{Szapudi}}}, \bibinfo {author} {\bibfnamefont {E.~M.}\
  \bibnamefont {{Teixeira}}}, \bibinfo {author} {\bibfnamefont
  {B.}~\bibnamefont {{Thomas}}}, \bibinfo {author} {\bibfnamefont
  {T.}~\bibnamefont {{Treu}}}, \bibinfo {author} {\bibfnamefont
  {E.}~\bibnamefont {{Trott}}}, \bibinfo {author} {\bibfnamefont
  {C.}~\bibnamefont {{van de Bruck}}}, \bibinfo {author} {\bibfnamefont
  {J.~A.}\ \bibnamefont {{Vazquez}}}, \bibinfo {author} {\bibfnamefont
  {L.}~\bibnamefont {{Verde}}}, \bibinfo {author} {\bibfnamefont
  {L.}~\bibnamefont {{Visinelli}}}, \bibinfo {author} {\bibfnamefont
  {D.}~\bibnamefont {{Wang}}}, \bibinfo {author} {\bibfnamefont {J.-M.}\
  \bibnamefont {{Wang}}}, \bibinfo {author} {\bibfnamefont {S.-J.}\
  \bibnamefont {{Wang}}}, \bibinfo {author} {\bibfnamefont {R.}~\bibnamefont
  {{Watkins}}}, \bibinfo {author} {\bibfnamefont {S.}~\bibnamefont {{Watson}}},
  \bibinfo {author} {\bibfnamefont {J.~K.}\ \bibnamefont {{Webb}}}, \bibinfo
  {author} {\bibfnamefont {N.}~\bibnamefont {{Weiner}}}, \bibinfo {author}
  {\bibfnamefont {A.}~\bibnamefont {{Weltman}}}, \bibinfo {author}
  {\bibfnamefont {S.~J.}\ \bibnamefont {{Witte}}}, \bibinfo {author}
  {\bibfnamefont {R.}~\bibnamefont {{Wojtak}}}, \bibinfo {author}
  {\bibfnamefont {A.~K.}\ \bibnamefont {{Yadav}}}, \bibinfo {author}
  {\bibfnamefont {W.}~\bibnamefont {{Yang}}}, \bibinfo {author} {\bibfnamefont
  {G.-B.}\ \bibnamefont {{Zhao}}},\ and\ \bibinfo {author} {\bibfnamefont
  {M.}~\bibnamefont {{Zumalac{\'a}rregui}}},\ }\bibfield  {title} {\bibinfo
  {title} {{Cosmology intertwined: A review of the particle physics,
  astrophysics, and cosmology associated with the cosmological tensions and
  anomalies}},\ }\href {https://doi.org/10.1016/j.jheap.2022.04.002} {\bibfield
   {journal} {\bibinfo  {journal} {Journal of High Energy Astrophysics}\
  }\textbf {\bibinfo {volume} {34}},\ \bibinfo {pages} {49} (\bibinfo {year}
  {2022})},\ \Eprint {https://arxiv.org/abs/2203.06142} {arXiv:2203.06142
  [astro-ph.CO]} \BibitemShut {NoStop}%
\end{thebibliography}%

\end{document}